\documentclass[superscriptaddress,
reprint,longbibliography,twocolumns,
preprintnumbers,nofootinbib,
amsmath,amssymb,amsfonts,prd]{revtex4-1}

\pdfoutput=1

\usepackage[utf8]{inputenc} 
\usepackage[T1]{fontenc} 
\usepackage{CJKutf8}
\usepackage{microtype}
\usepackage{mathrsfs}
\usepackage{amsmath,amsfonts,amssymb}
\usepackage{graphicx}

\usepackage[dvipsnames]{xcolor}

\usepackage{empheq}
\usepackage[most]{tcolorbox}

\newtcbox{\mymath}[1][]{%
    nobeforeafter, width=\linewidth,math upper, tcbox raise base,
    enhanced, colframe=blue!30!black,
    colback=blue!30,outer arc=0pt,
arc=0.3pt,
left=0.3pt,
right=0.3pt,
top=0.3pt,
bottom=0.3pt, boxrule=1pt,
    #1}

\usepackage{dcolumn}
\usepackage{bm}

\usepackage{xcolor,colortbl}

\definecolor{lightgreen}{rgb}{0.56, 0.93, 0.56}
\definecolor{wheat}{rgb}{0.96, 0.87, 0.7}

\definecolor{classicrose}{rgb}{0.98, 0.8, 0.91}

\definecolor{LightCyan}{rgb}{0.88,1,1}

\newcolumntype{a}{>{\columncolor{wheat}}c}
\newcolumntype{b}{>{\columncolor{LightCyan}}c}
\newcolumntype{d}{>{\columncolor{classicrose}}c}
\definecolor{lightsalmonpink}{rgb}{1.0, 0.6, 0.6}
\newcolumntype{e}{>{\columncolor{lightsalmonpink}}c}

\usepackage{slashed}
\usepackage[colorlinks = true,
            linkcolor = blue,
            urlcolor  = blue,
            citecolor =red,
            anchorcolor = blue]{hyperref}
\usepackage{verbatim}
\usepackage{color,ulem}
\usepackage[english]{babel}

\newtcbox{\myboxnew}{on line,
  colframe=orange,colback=orange!10!white,
  boxrule=0.5pt,arc=4pt,boxsep=0pt,left=6pt,right=6pt,top=6pt,bottom=6pt}

\definecolor{lightkhaki}{rgb}{0.94, 0.9, 0.55}

\usepackage[scr]{rsfso}

\usepackage{tikz}

\usepackage{pifont}

\usepackage{epigraph}
\newcommand{\cP}{\mathcal{P}}
\newcommand{\nn}{\nonumber}

\def\be{\begin{equation}}
\def\ee{\end{equation}}
\def\ba{\begin{eqnarray}}
\def\ea{\end{eqnarray}}

\usepackage{kotex}

\def\cP{{\cal P}}

\usepackage[strict]{changepage}

\def\IR{\relax{\rm I\kern-.18em R}}

\usepackage{CJKutf8}

\def\IR{\relax{\rm I\kern-.18em R}}
\def\IL{\relax{\rm I\kern-.18em L}}

\def\inv{^{\raise.15ex\hbox{${\scriptscriptstyle -}$}\kern-.05em 1}}

\usepackage{mathrsfs}

\def\bea{\begin{eqnarray}}
\def\eea{\end{eqnarray}}

\def\nn{\nonumber}

\definecolor{markcolor2}{rgb}{1,0,0}

\definecolor{markcolor3}{rgb}{0,1,0}

\usepackage{blindtext}

\begin{document}

\title{\Huge Holographic Axion Model:\\
\Large A simple gravitational tool for quantum matter}

\author{Matteo Baggioli}%
 \email{b.matteo@sjtu.edu.cn}
\affiliation{Wilczek Quantum Center, School of Physics and Astronomy, Shanghai Jiao Tong University, Shanghai 200240, China}
\affiliation{Shanghai Research Center for Quantum Sciences, Shanghai 201315, China.}
\affiliation{Instituto de Fisica Teorica UAM/CSIC,
Universidad Autonoma de Madrid, Madrid 28049, Spain.}
\author{Keun-Young Kim}%
 \email{fortoe@gist.ac.kr}
\affiliation{ School of Physics and Chemistry, Gwangju Institute of Science and Technology, \\
Gwangju 61005, Korea.}
\author{Li Li}%
 \email{liliphy@itp.ac.cn}
\affiliation{CAS Key Laboratory of Theoretical Physics, Institute of Theoretical Physics,
Chinese Academy of Sciences, Beijing 100190, China}

\affiliation{School of Physical Sciences, University of Chinese Academy of Sciences,\\
Beijing 100049, China}

\affiliation{School of Fundamental Physics and Mathematical Sciences, Hangzhou Institute for Advanced Study, \\
University of Chinese Academy of Sciences, Hangzhou 310024, China}
\author{Wei-Jia Li}%
 \email{weijiali@dlut.edu.cn }
\affiliation{Institute of Theoretical Physics, School of Physics, Dalian University of Technology,
Dalian 116024, China.
}

\begin{abstract}
This is a complete and exhaustive review on the so-called holographic axion model -- a bottom-up holographic system characterized by the presence of a set of shift symmetric scalar bulk fields whose profiles are taken to be linear in the spatial coordinates. This simple model implements the breaking of translational invariance of the dual field theory by retaining the homogeneity of the background geometry and therefore allowing for controllable and fast computations. The usages of this model are very vast and they are a proof of the spectacular versatility of the framework. In this review, we touch upon all the up-to-date aspects of this model from its connection with massive gravity and effective field theories, to its role in modeling momentum dissipation and elastic properties ending with all the phenomenological features and its hydrodynamic description. In summary, this is a complete guide to one of the most used models in Applied Holography and a must-read for any researcher entering this field.\\


\end{abstract}

\maketitle

\tableofcontents

\section{Introduction}
\label{sec:intro}
The Holographic Correspondence (or equivalently \textbf{Holography}, AdS-CFT or Gauge-Gravity duality) is nowadays a respected and widely used tool for applications, ranging from QCD and condensed matter to hydrodynamics and quantum information ~\cite{ammon2015gauge,Hartnoll:2016apf,Hartnoll:2009sz,zaanen2015holographic,CasalderreySolana:2011us,Baggioli:2019rrs,Heller:2016gbp,Florkowski:2017olj}. For this scope, it is often used in its \textbf{bottom-up} version, indeed agnostic of its historical stringy origins ~\cite{Maldacena:1997re} and detached from any issues related with quantum gravity~\cite{Aharony:1999ti}. On the contrary, it is treated as an efficient and powerful playground to learn about physical situations in which other more \textit{Kosher} methods are of no help. In particular, it appears to be extremely advantageous (if not even the only available tool) for systems at strong coupling (where perturbative methods fail), situations dominated by a many-body collective dynamics and no well-defined elementary excitations (where the single-particle approximation fails) and dissipative systems (where a suitable finite temperature field-theory formulation is far from obvious).

With this applied (and if one wants less fundamental) task in mind, it is clear that the most important challenge is to make this playground as close \textbf{as possible to the reality}, or in other words to the realistic physical situation to which we want to apply it. A representative epitomic case is the comparison between QCD, a $SU(3)$ non-abelian gauge theory, and $\mathcal{N}=4$ supersymmetric Yang-Mills theory in the large $N$ limit. The scope is to move as close as possible to reality without losing the solvability and the analytic control on the (possibly toy) model.

In condensed matter, the reality is obviously not Poincar\'e invariant. Inevitably, both \textbf{translations and rotations are broken} (at least spontaneously, \textbf{SSB}). This is the key behind the ``rigidity'' of matter, the theory of elasticity, the propagation of sound in materials and the thermodynamics of solids. Not only that, but in most of the situations, such as electronic transport, translations are broken explicitly (\textbf{EXB}), giving rise to the finite conductivity measured in all common metals. Finally, there are also several situations in which translations are broken both explicitly and spontaneously, in what is called the ``\textit{pseudo-spontaneous}'' limit. This is indeed the case for pinned charge density waves~\cite{RevModPhys.60.1129}, where impurities pin the phason Goldstone mode producing a peculiar finite frequency peak in the optical conductivity.

As a consequence, in order to have a realistic description, it is imperative to introduce and understand in detail the breaking of spatial translations in the dual boundary theory. The early days of Applied Holography focused in particular on the questions around Quark Gluon Plasma (QGP) and its strongly coupled hydrodynamic description~\cite{Policastro:2002se}. An exemplary result is the famous \textbf{Kovtun-Son-Starinets (KSS) bound} on the viscosity to entropy ratio \cite{Kovtun:2004de}. In that context, the role of translations is minimal, if not even negligible. Nevertheless, in the last decade, due to the increasing interest around strongly coupled phases of matter with no quasiparticles and no standard solid state theory description (e.g. Non-Fermi liquids, strange metals, High-Tc superconductors), the need for \textbf{holographic setups with no translational invariance} has become unavoidable~\cite{Hartnoll:2016apf}.

Historically, this program has started with the ``\textit{brute force}'' attempt of embedding into the standard holographic models bulk fields with spatially dependent boundary conditions, mimicking an explicit lattice source. After introducing a gravitational background lattice by adding a periodic source for a neutral scalar, the model of~\cite{Horowitz:2012ky} was able to dissipate the momentum of the dual field theory. At the same time, concomitant works~\cite{Nakamura:2009tf,Donos:2011bh} describing a possible mechanism for the spontaneous breaking of translations in presence of finite charge density have appeared. Despite the validity and novelty of those works, no much progress has been done using those models until the recent days, mainly because of the technical difficulties associated with them.

On the contrary, a totally new fresh wave on the topic has been initiated by the so-called \textbf{homogeneous models}, holographic setups in which translations are broken but the background geometry remains homogeneous~\cite{Vegh:2013sk,Andrade:2013gsa,Donos:2013eha,Donos:2012js}. Among them, a particular subset emerges, because of the possibility of having a closed-form analytical background. This subset is represented by the \textbf{holographic axion model}~\cite{Andrade:2013gsa,Baggioli:2014roa}, which is equivalent to (or better which incorporates) the original massive gravity proposals~\cite{Alberte:2015isw}.

This model has dominated the scene of Applied Holography without translational invariance and it is the \textbf{topic of this review}. It is nowadays a well-known and widely used model which represents mandatory knowledge for any researchers in the field. Because of this reason, and the immense progress made in the last decade around this model, we have found it timely to collect all this material in a single and self-contained review where all the fundamental points will be described. This review attempts to be as exhaustive as possible, covering all the directions in which  this model has been utilised and all the main features to understand it in plain. It is intended both for early researchers starting to work with the model, but also for more advanced \textbf{``holographers''} who will find through the text several open questions and unfinished tasks to think over.

\subsection{Scope of this review}
This review was born as a collective effort to organize and collect in a single self-consistent manuscript all the information about the \textbf{holographic axion model}, from its origins to the most recent developments. This work is intended for a very diverse audience, ranging from young students up to the most experienced researchers in the field.

There is certainly a gap between a series of cutting-edge research papers (in this case started around 2012) and the full understanding of the questions behind them, which only time can close. This review, in a sense, wants to close such a gap (after approximately 10 years of studies). We would like also to take advantage of this opportunity to clarify some points which are very often confused in the literature and taught in the wrong way to early researchers. In particular, we want to emphasize that:
\begin{itemize}
    \item The holographic axion model is not just an \textit{ad-hoc} tool to break translations, but its structure can be consistently mapped to and derived from the standard effective field theory formulations;
    \item Holographic massive gravity (intended as the original dRGT construction~\cite{Vegh:2013sk}) and the holographic axion model are not different beasts, as often conveyed in the literature, but they are exactly the same theory written in a different gauge\,\footnote{To be more precise, dRGT is just a particular choice of the potential in the holographic axion model \cite{BaggioliSolid}.};
    \item The presence of bulk axion fields with profile $\phi^I=x^I$ does not necessarily imply the breaking of momentum conservation but it can lead to a much richer structure of theories.
\end{itemize}
Finally, we have devoted a final part of the review to stimulate the more experienced researchers in the field with some open questions which, to the best of our knowledge, are yet not resolved.\\

The organization of this review is as follows. In Section~\ref{sec:intro}, we introduce the topics of this review and we provide the motivations behind it. We describe the simplest holographic axion model which captures the key features
of the explicit breaking of translations and its physical consequences in Section~\ref{sec:linear}. Section~\ref{sec:spontaneous} generalizes the original model to the case that breaks translations spontaneously and discusses the associated physics. We compare the holographic results to the hydrodynamic description in Section~\ref{hydrodynamics}. Some universal bounds extracted from holographic axion models are discussed in Section~\ref{sec:bounds}. Section~\ref{pinned} makes a step forward and combines the explicit and spontaneous breaking of translations in the pseudo-spontaneous regime. We proceed to give a list of phenomena and topics for which the holographic axion models have been applied in Sections~\ref{sec:Pheomen} and~\ref{sec:add}. We conclude this review with a number of open questions related to the holographic axion models and a short conclusion in Section~\ref{sec:outlook}. The symbols and notations used in this review are summarized in appendix.

\subsection{The Drude model}
A first important scenario where the role of \textbf{translations} appears fundamental is in the determination of the transport properties of metals, \emph{e.g.} the electric conductivity. Let us imagine a simple model for electric conduction and represent our conducting electrons as simple spherical balls non-interacting within each-other and following a classical Newtonian dynamics. Whenever an external and frequency independent (DC) electric field $\Vec{E}$ is switched on, the electrons will be accelerated by a force $\Vec{F}=q\,\Vec{E}$, with $q$ being the electron charge. Assuming the momentum of the electrons being conserved, the electrons will flow unaffected forever and the corresponding electric conductivity $\sigma= J/E$ will result to be infinite. We would be able to have a finite electric current at late time even when the electric field is removed ( exactly like in a superconductor, but for a different reason). This is the same situation that we would encounter if we kick a marble on a table and we would neglect any friction effect between the two; the marble will simply roll forever.

\begin{figure}[ht]
\centering
\includegraphics[width=0.7\linewidth]{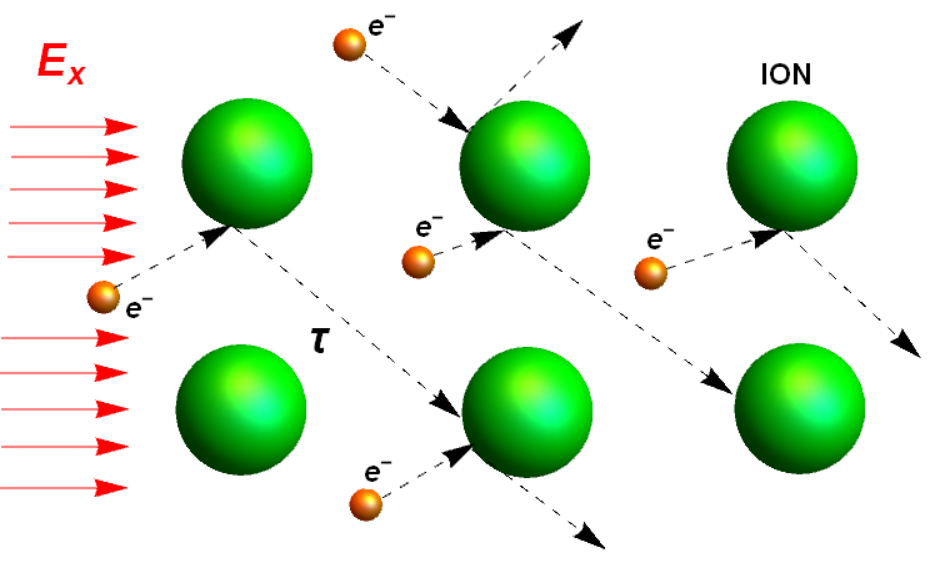}
\caption{A schematic illustration for the Drude model. In orange the electrons, while in green the immobile ions. The red arrows identify the direction of a constant applied electric field. The average time between collisions is given by $\tau$.}
\label{fig:drude1}
\end{figure}

This is obviously not a truthful representation of the reality since all metals have a finite DC conductivity -- \emph{i.e.} a finite conductivity at zero frequency $\omega=0$, in response to a static electric field. In order to recover this well-known experimental fact, the non-conservation or dissipation of the electron momentum has to be considered. This can be done by following the simple \textbf{Drude model} introduced in 1900 (only three years after the discovery of the electron by the British physicist J.~J.~Thomson) by Paul Drude~\cite{w1976solid,abrikosov1988fundamentals,kittel2004introduction}. Drude borrowed the basic elements of his theory from the \textit{kinetic theory of gases} and he simply imagined a metal as a dilute gas of free electrons. Nevertheless, he made a step forward and considered the presence in a metal of also heavy and immobile ions around which the electrons are moving driven by the external electric field (see Fig.~\ref{fig:drude1}).

The electrons, during their motion, collide against the heavier ions losing their momenta and deflecting their trajectories. From an effective field theory (\textbf{EFT}) point of view, the dynamics of the electrons, or more specifically of their average momentum, is determined by the simple equation:
\begin{equation}
    \frac{d}{dt}\,\langle p_x(t)\rangle\,=\,q\,E_x\,-\,\frac{1}{\tau}\,\langle p_x(t) \rangle\,,\label{or1}
\end{equation}
where for simplicity we have considered an isotropic system and aligned the external electric field along the spatial $x$ direction. The first term in the r.h.s. is the standard driving force induced by the external electric field. The second, and more important,  is an effective term which induces a relaxation of the average momentum at a constant rate $\Gamma \equiv 1/\tau$. The timescale $\tau$ is an effective parameter which corresponds to the average time between consecutive collisions and it determines ``how fast'' momentum gets lost. From a more theoretical perspective, this second term encodes the effects due to the explicit breaking of translations.

\begin{figure}[ht]
\centering
\includegraphics[width=0.7\linewidth]{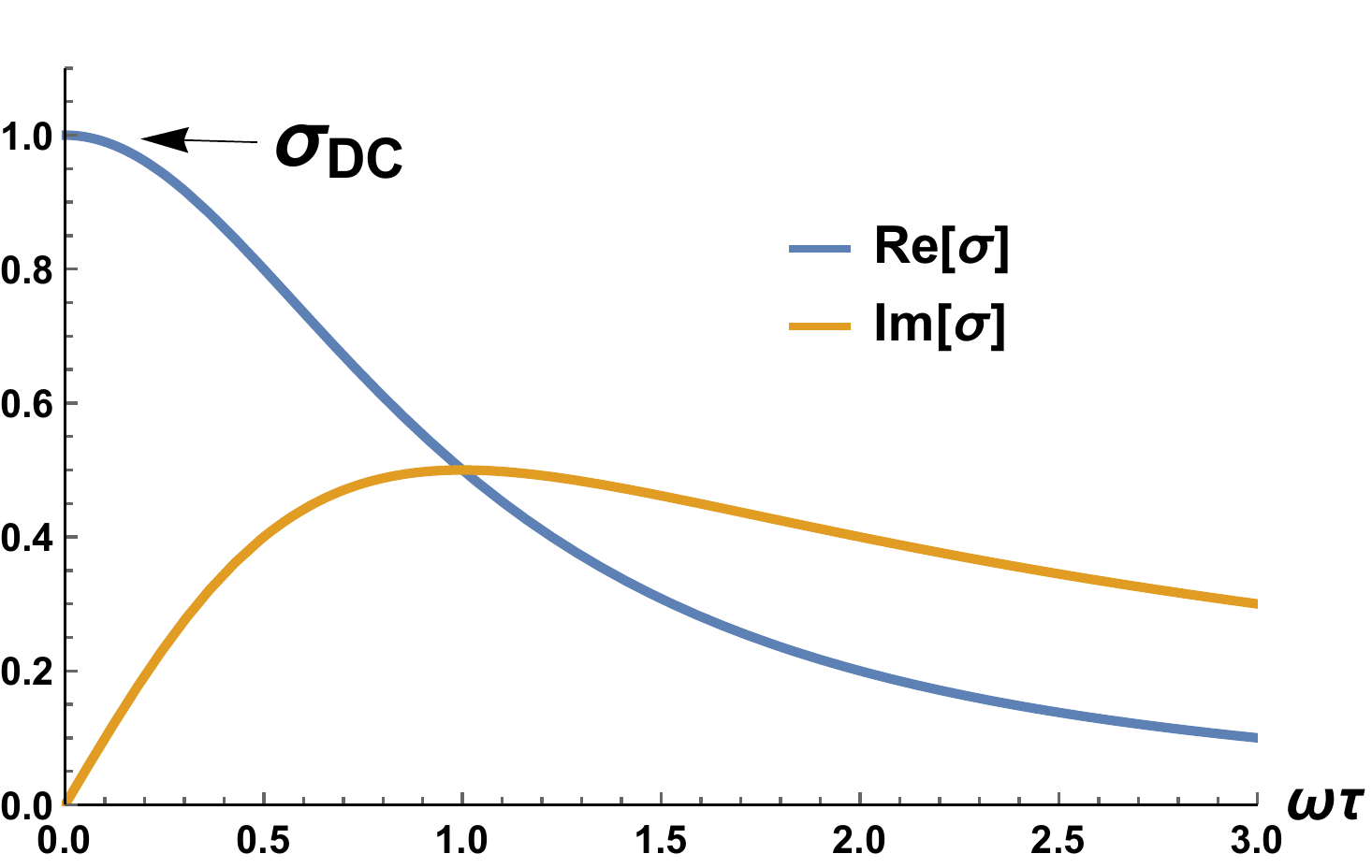}
    
\vspace{0.4cm}
    
\includegraphics[width=0.8\linewidth]{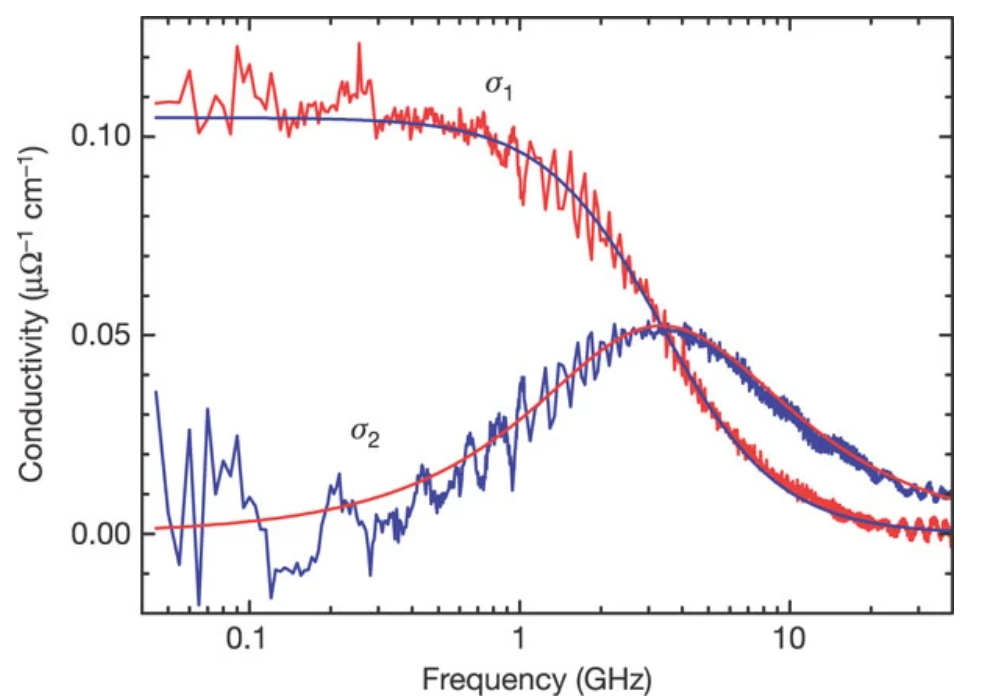}
\caption{\textbf{Top: }The optical conductivity in the Drude model. For simplicity we have fixed $\sigma_{DC}=1$. \textbf{Bottom: } The excellent agreement between the Drude model and the experimental data in UPD$_2$Al$_3$ at $T=2.75$ K taken from~\cite{Scheffler2005}. Here $\sigma_1= \mathrm{Re}\left[\sigma\right]$ and $\sigma_2=\mathrm{Im}\left[\sigma\right]$.}
\label{fig:drude2}
\end{figure}

Using classical identities, we can write down the average momentum of the electrons and the relative electric current generated in terms of their average velocity:
\begin{equation}
    \langle p_x(t) \rangle\,=\,m\,\langle v_x(t) \rangle\,,\quad \quad \langle J_x(t)\rangle\,=\,n\,q\,\langle v_x(t)\rangle\,,\label{def1}
\end{equation}
where $m$ and $n$ are respectively the electron mass and number density.
Using these relations in the dynamical equation~\eqref{or1}, and a standard Fourier decomposition, we immediately get
\begin{equation}
    -\,i\,\omega\,\langle v_x\rangle\,=\,\frac{q}{m}\,E_x\,-\,\frac{1}{\tau}\,\langle v_x \rangle\,.
\end{equation}
Finally, utilizing the definition for the electric conductivity, we obtain the expression for the low-frequency conductivity in the Drude model, which reads
\begin{equation}
    \sigma_{xx}\,=\,\frac{J_x}{E_x}\,=\,\frac{\sigma_{DC}}{1\,-i\,\omega\,\tau}\,,\quad \quad \sigma_{DC}\,=\,\frac{n\,q^2\,\tau}{m}\,. \label{drudecond}
\end{equation}
This is Drude's main result. Several observations are in order. (I) The \textbf{DC conductivity}, $\sigma_{DC}\equiv\sigma_{xx}(\omega=0)$, is \textbf{finite} because of momentum dissipation ($=$ finite $\tau$). In the limit in which momentum is conserved ($\tau \rightarrow \infty$), we recover the previously mentioned infinite result. (II) The faster momentum is dissipated, the lower the DC conductivity; the material conducts less and less. (III) The Drude model implies the presence of a relaxation mode, usually labelled \textbf{Drude pole}, $\omega=-i/\tau$, which incorporates the effect of momentum dissipation. This results in the so-called \textbf{Drude peak}, a peak of the real part of the conductivity located at $\omega=0$, whose width is determined by the relaxation rate $\Gamma\equiv \tau^{-1}$ (see Fig.~\ref{fig:drude2}). In the limit of $\tau= \infty$, the Drude peak reduces to a delta function at $\omega=0$.

The physics of the Drude model and especially the role of momentum conservation can be approached from a more formal point of view. The starting point is the realization that, within \textbf{linear response theory}~\cite{fetter2003quantum}, the electric conductivity can be written in terms of the retarded current-current two-points function as:
\begin{equation}
    \sigma_{ij}(\omega)\,=\,\frac{1}{i\,\omega}\,\langle J_i J_j\rangle\,(\omega,k=0)\label{def2}\,.
\end{equation}
This relation can be derived by considering an external source for the current operator $J^\mu$:
\begin{equation}
    S\,\,\rightarrow\,\,S\,+\,\int d^dx\,A_\mu\,J^\mu\,,
\end{equation}
from which the two-point function of the current is derived as:
\begin{equation}
    \langle J_\mu J_\nu \rangle \,\equiv\,\frac{\delta^2\,S}{\delta A^\mu \delta A^\nu}\,=\,\frac{\langle J_\mu \rangle}{A^\nu}\,,
\end{equation}
where in the last step we have assumed the linear response approximation. Taking into account that $E_i\,\equiv\,i\,\omega\,A_i$ and $\langle J_i\rangle =\sigma_{ij}\,E^j$, then Eq.~\eqref{def2} follows. Despite its simplicity, the Drude model is in perfect agreement with experiments in ultra-clean metals (see Fig.~\ref{fig:drude2}).

After writing the conductivity as a Green's function, we can then apply the \textbf{memory matrix} methods~\cite{Hartnoll:2012rj,Lucas:2015pxa} (in particular see~\cite{andylectures}). The main statement is that whenever the momentum operator overlaps with the current operator (at finite charge density), and the momentum is a conserved operator, then the conductivity contains a pole at zero frequency and its DC component is therefore infinite. Mathematically, this implies that
\begin{equation}
   \langle \Vec{J}\,\Vec{J}\,\rangle\,\supset\,\frac{\chi_{\Vec{J}\Vec{p}}^2}{\chi_{\Vec{p}\Vec{p}}}\,\frac{1}{\Gamma}\,, \label{re2}
\end{equation}
where $\chi_{\Vec{J}\Vec{p}}$ is the off-diagonal susceptibility establishing the mixing between the two operators (and in this case simply coinciding with the charge density). Moreover, $\chi_{\Vec{p}\Vec{p}}$ is the momentum susceptibility determining the relation between momentum $\vec{p}$ and velocity $\vec{v}$. The latter coincides with $\mathcal{E}+\mathfrak{p}$ (energy $+$ pressure) in relativistic systems~\cite{Kovtun:2012rj} and it is simply the mass density $ \varrho$ in non-relativistic ones~\cite{d1989theory}. Finally, $\Gamma$ is the momentum relaxation rate, defined as
\begin{equation}
    \Gamma\,=\,\lim_{\omega \rightarrow 0}\,\frac{M_{\Vec{p}\Vec{p}}}{\chi_{\Vec{p}\Vec{p}}}\,,
\end{equation}
with $M_{AB}$ being the memory matrix (see \cite{andylectures} for more details). This rate being non-zero stems directly from the fact that
\begin{equation}
    \left[\,H,\,\vec{p}\,\right]\,\neq\,0\,,
\end{equation}
namely there is an operator in the theory which explicitly breaks translational invariance. Notice that the r.h.s. of \eqref{re2} reproduces exactly what is known as \textbf{Drude Weight} which is highly discussed in the context of many-body physics (see, for example, the Mazur-Susuki bound~\cite{PhysRevB.55.11029} and its holographic counterpart~\cite{Garcia-Garcia:2015ooc}).

\subsection{Effective field theories for solids and fluids}
Another situation in which translational invariance plays a fundamental role is in the definition of \textbf{solids} and in the study of \textbf{elasticity}~\cite{d1989theory,chaikin2000principles,PhysRevA.6.2401}. A solid is a system with long-range order. From a more fundamental perspective, it is a configuration in which spatial translations are spontaneously broken (\textbf{SSB}). This is tantamount to say that a solid selects a preferred length-scale. The corresponding Goldstone bosons are the (acoustic) \textbf{phonons}~\cite{Leutwyler:1996er}. Despite the standard condensed matter description of solids is not introduced with this language, but rather via more phenomenological models of springs and atoms, an effective field theory description of solids and elasticity is definitely helpful and welcome~\cite{matteo}.

The standard formulation of spontaneous symmetry breaking (think, for example, about superconductivity) is done in terms of Ginzburg-Landau theory and the well-known double-well potential \cite{Beekman:2019pmi}. Despite attempts of this kind have been pursued for spacetime symmetries and phonons~\cite{Nitta:2017mgk,Gudnason:2018bqb,Musso:2018wbv,Musso:2019kii}, the most successful framework in this case~\cite{Nicolis:2015sra} appears slightly different. The main idea is rather simple. Despite Lorentz invariance and the associated Poincar\'e group are fundamental pillars for the description of our world at high energy (\emph{e.g.} special relativity), all phases of matter at low energy are obviously not respecting these rules. Matter always selects a preferred reference frame, being the velocity of a fluid or the lattice structure of a crystal, and it therefore breaks spontaneously part of the Poincar\'e group. Classifying the possible symmetry breaking patterns of the Poincar\'e group is therefore equivalent to classify the possible different phases of matter at low energy. Once this principle is accepted, all the methods relative to SSB (\emph{e.g.} the Coset construction~\cite{Nicolis:2013lma}) are applicable and useful to perform a full ``zoology'' of matter. Because of spacetime limitations, we will describe in detail only the EFT formulation of solids and fluids, putting aside superfluids, supersolids, framids, etc.
\begin{figure}[ht]
    \centering
    \includegraphics[width=\linewidth]{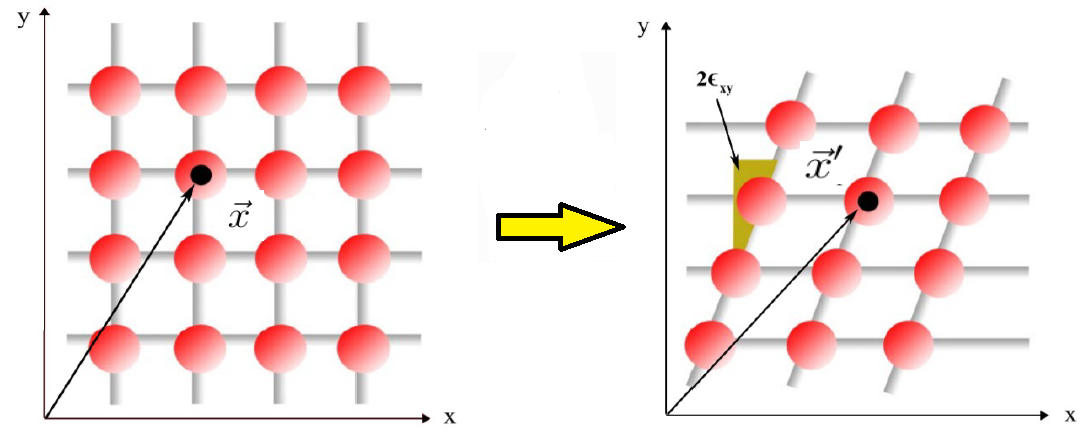}
    \caption{A pure shear deformation and its effects on a square 2D lattice.}
    \label{fig:eft1}
\end{figure}

Before moving to the modern EFT framework, let us briefly review the basics of the theory of elasticity~\cite{d1989theory,chaikin2000principles,PhysRevA.6.2401}. The theory of elasticity describes the dynamics of objects under mechanical deformations and it is based on the so-called (infinitesimal) \textbf{displacements}, the geometrical deviations from equilibrium (see Fig.~\ref{fig:eft1}):
\begin{equation}
    \vec{u}\,\equiv\,\vec{x}\,-\,\vec{x}_{eq}\,.
\end{equation}
The fundamental object describing mechanical deformations is the \textbf{strain tensor}, which is defined as the symmetrized derivative of the displacement:
\begin{equation}
    \varepsilon_{ij}\,=\,\partial_i\,u_j\,+\,\partial_j\,u_i\,,
\end{equation}
from which the final position $x_i$ can be written as $x_i\,=\,{x_{i}}_{eq}+\varepsilon_{ij}\,dx^j$ \footnote{In this review, we will not consider the possibility of having non-affine displacements and incompatible deformations. See \cite{kleinert1989gauge} for more details.}. Once the strain tensor is defined, one needs to use the \textbf{constitutive relation} which at linear level relates the strain tensor to the \textbf{stress tensor} $\sigma_{ij}$:
 \begin{equation}
     \sigma_{ij}\,=\,C_{ijkl}\,\varepsilon^{kl}\,+\,\dots\,,
 \end{equation}
 with $C_{ijkl}$ being the \textbf{elastic tensor}. For an isotropic system in $d$-spatial dimensions, we have
 \begin{equation}
     \sigma_{ij}\,=\,K\,\delta_{ij}\,\varepsilon_{kk}\,+\,2\,G\,\left(\varepsilon_{ij}\,-\,\frac{1}{d}\,\delta_{ij}\,\varepsilon_{kk}\right)\,,\label{const}
 \end{equation}
 where $K,G$ are respectively the \textbf{bulk and shear elastic moduli} and $\varepsilon_{kk}$  the bulk strain, defined as the trace of the strain tensor. Finally, we can write down the equation of \textbf{elasto-dynamics} (which is simply the Newton's equation $\vec{F}= m \vec{a}$):
 \begin{equation}
     \varrho\,\ddot{u}_i\,=\,f_i\,=\,\nabla^j\,\sigma_{ij}\,,\label{dyn}
 \end{equation}
 which constitutes the missing piece to find the full dynamics of the system. Here $\varrho$ stands for the mass density and $f_i$ for the force density. By plugging Eq.~\eqref{const} into Eq.~\eqref{dyn}, and after decomposing the modes into transverse and longitudinal with respect to the momentum $\vec{k}$, one obtains two sets of propagating sound modes:
 \begin{equation}
     \omega\,=\,\pm\,v_{T,L}\,k\,,
 \end{equation}
 which are indeed our transverse (or shear) and longitudinal phonons. One can also derive that the phonons propagation speeds are directly related to the elastic moduli. In particular, in two spatial dimension, one finds
 \begin{equation}
     v_T^2\,=\,\frac{G}{\varrho}\,,\quad \quad v_L^2\,=\,\frac{G\,+\,K}{\varrho}\,.
 \end{equation}
 This is a beautiful result which is obtained only by using symmetries. Nevertheless, to make the role of symmetries, and in particular translations, more evident we need to pass to a more field theory inspired formalism.
 
\begin{figure}[ht]
    \centering
    \includegraphics[width=0.55\linewidth]{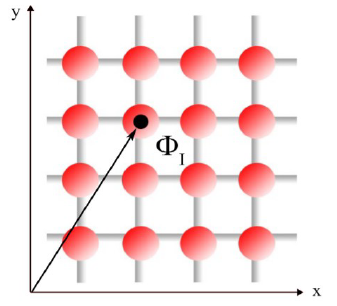}
    \caption{The EFT parametrization in terms of a set of scalar fields $\Phi^I$. The equilibrium configuration is clearly $\Phi_{eq}^I=x^I$.}
    \label{fig:eft2}
\end{figure}

 The main idea consists in introducing a \textbf{set of real scalar fields}
 \begin{equation}
    \Phi^I\,,\quad \quad I\,=\,1\,,\dots\,,d_{spatial}\,,
\end{equation}
one for each of the spatial directions. These scalar fields act as a set of \textbf{co-moving coordinates} and they select a preferred reference frame
\begin{equation}
    \langle \Phi^I \rangle \,\equiv\,\Phi^I_{eq}\,=\,x^I\,,
\end{equation}
so that, at equilibrium, they are identified with the spatial coordinates themselves (see Fig.~\ref{fig:eft2}). The mechanical deformations are then associated to the fluctuations of these scalar fields around equilibrium:
\begin{equation}
    \Phi^I\,=\,\Phi^I_{eq}\,+\,\pi^I\,,
\end{equation}
where, as we will see, the fluctuations $\pi^I$ are exactly the Goldstone modes associated with translational invariance -- the phonons.

\begin{figure}[ht]
\centering
\includegraphics[width=0.9\linewidth]{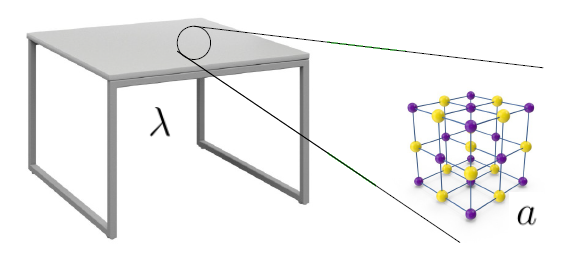}
\caption{A pictorial representation of the homogeneity assumption. Any system, at length-scales $\lambda \gg\,a$ ($a$ being the characteristic microscopic scale), looks homogeneous.}
\label{fig:eft3}
\end{figure}

In order to build an effective field theory for the scalars $\Phi^I$, we need to establish which are the fundamental symmetries of our system. For simplicity, we will consider only isotropic solids, imposing therefore invariance under
\begin{equation}
    \mathbb{R}\,:\,\Phi^I\,\rightarrow\,{R^I}_J\,\Phi^J\,.
\end{equation}
and assuming the equilibrium configuration to be $\Phi^I\,=\,\delta^I_j\,x^j$. More importantly, we will assume that at large scales, scales $\lambda$ much larger than the microscopic characteristic distance $a$, the physics is homogeneous (see \cite{PhysRevB.42.7345}). This assumption appears to be very natural and it is related to the fact that every solid (imagine, for example, the table you are sit at) looks like homogeneous as far as you do not probe it at distances comparable to its crystal structure (see Fig.~\ref{fig:eft3}).

This is obviously connected to the continuous description and to the fact that our EFT breaks down when we reach the microscopic scale $a$ (at which, for example, phonons are not well defined anymore). The microscopic scale $a$, in this case the lattice spacing, represents the UV cutoff of our effective theory. In fluids, the microscopic scale is given in terms of the inter-molecular distance which plays exactly the same cutoff role (see, \emph{e.g.}~\cite{Baggioli:2020loj}).

In order to retain homogeneity at large scales, we need to also impose invariance under the internal global shifts
\begin{equation}
\mathbb{S}\,:\,\Phi^I\,\rightarrow\,\Phi^I\,+\,a^I\,. \label{ss}
\end{equation}
It follows that the equilibrium configuration $\Phi^I_{eq}=x^I$ not only spontaneously breaks the spatial translations
\begin{equation}
\mathbb{T}\,:\, x^I\,\rightarrow\,x^I+b^I\,,
\end{equation}
but it breaks them into the diagonal subgroup combination
\begin{equation}
\mathbb{S}\,\times\,\mathbb{T}\,\rightarrow\,\left(\mathbb{S}\,\times\,\mathbb{T}\right)_{diag}\,\,\,\,\left[a^I\,=\,-b^I\,\right]\,\,.
\end{equation}
This is the symmetry breaking pattern for an isotropic solid.
    
To obey the requirement of invariance under internal shifts~\eqref{ss}, the effective action can include only derivative terms. At leading order in derivatives, the only object which one can build is the following matrix
\begin{equation}
\mathcal{I}^{IJ}\,\equiv\,\partial_\mu\,\Phi^I\,\partial^\mu\,\Phi^J\,,\label{bb}
\end{equation}
where $I,J$ indicate spatial coordinates, while $\mu$ spacetime ones. In two spatial dimensions, the only independent scalar objects built in terms of~\eqref{bb} are
\begin{equation}
X\,=\,\mathrm{Tr}\,\mathcal{I}^{IJ}\,,\quad Z\,=\,\mathrm{det}\,\mathcal{I}^{IJ}\,,
\end{equation}
or equivalently the trace of $\mathcal{I}^{IJ}$ and the trace squared. In higher dimensions, more terms are allowed; in fact all the higher traces of $\mathcal{I}^{IJ}$. All in all, the most generic action, respecting the required symmetries in two spatial dimensions, takes the form of
\begin{equation}
\mathrm{S}\,=\,\int d^3x\,\sqrt{-g}\,\,V(X,Z)\,, \label{EFT1}
\end{equation}
with $g_{\mu\nu}$ a fictitious metric which will always be set to the Minkowski one and $g$ its determinant.
\eqref{EFT1} is the most generic $T=0$ effective action for two-dimensional isotropic solids (and fluids).

To convince ourselves that this is indeed the case, we need to proceed as before and obtain the effective action for the fluctuations $\pi^I$. Such action will govern the full dynamics of the Goldstone modes and it will tell us everything about the elasticity property of the solids and the propagation of sound in them. We will follow closely the notations of~\cite{Alberte:2018doe} (and~\cite{Baggioli:2019elg}).
    
By varying the action~\eqref{EFT1} with respect to the curved spacetime metric $g_{\mu\nu}$ and evaluating it on the Minkowski background, $g_{\mu\nu}=\eta_{\mu\nu}$, we obtain the corresponding stress-energy tensor:
\begin{align}
T_{\mu\nu}\,=&\,-\,\frac{2}{\sqrt{-g}}\,\frac{\delta S}{\delta g^{\mu\nu}}\,\Big|_{g=\eta}=\,-\,\eta_{\mu\nu}\,V\,+\,2\,\partial_\mu \Phi^I \partial_\nu \Phi_I\,V_X\,\nonumber \\&\,+\,2\,\left(\partial_\mu \Phi^I \partial_\nu \Phi_I\,X\,-\,\partial_\mu \Phi^I \partial_\nu \Phi^J \,\mathcal{I}_{IJ}\right)\,V_Z\,.
\end{align}
where $V_X\equiv\partial V /\partial X$ and $V_Z\equiv\partial V /\partial Z$. For any time independent scalar field configurations, the stress-energy tensor components are
\begin{align}\label{rho}
&T^t_t\,\equiv\,\mathcal{E}\,=\,V\,,\\\label{pressure}
& T^x_x\,\equiv\,-\,\mathfrak{p}\,=\,V\,-\,X\,V_X\,-\,2\,Z\,V_Z\,,\\\label{txy}
&T^x_{y}\,=\,2\,\partial_x \Phi^I \partial_y \Phi^I\,V_X\,,
\end{align}
where $\mathcal{E}$ is the energy density and $\mathfrak{p}$ the mechanical pressure. Notice that in the equilibrium configuration $\Phi^I_{eq}=x^I$ we have $T^x_y=0$, as expected from isotropy.

In terms of the scalar fields, the strain tensor is simply:
\begin{equation}
\varepsilon_{ij}\,=\,\partial_i\,\Phi_j\,+\,\partial_j\,\Phi_i\,.
\end{equation}
Using the constitutive relation for an isotropic solid~\eqref{const}, where now $\sigma_{ij}$ has to be identified with the high-energy physics notation $T_{ij}$, we can immediately extract the elastic moduli in terms of the unknown potential $V(X,Z)$:
\begin{align}
    &G= 2\,V_X\,,\\
    &{K}\,=\,2ZV_Z+4Z^2 V_{ZZ}+4XZ V_{XZ}+X^2 V_{XX}\,.
\end{align}
where $V_{ZZ}\equiv\partial^2 V /\partial Z^2$, etc.
To conclude, we can expand the original action~\eqref{EFT1} in terms of the fluctuations $\pi^I$, and after separating them into longitudinal and transverse components (see~\cite{Alberte:2018doe} for details), we obtain again two propagating sound modes
\begin{equation}
    \omega\,=\,\pm\,v_{T,L}\,k\,,
\end{equation}
with
\begin{equation}
v_T\,=\,\sqrt{\frac{{G}}{\mathcal{E}+\mathfrak{p}}}\,,\qquad v_L\,=\,\sqrt{\frac{K+G}{\mathcal{E}+\mathfrak{p}}}\,,
\label{speed0}
\end{equation}
as expected for a relativistic solid system.

The field theory allows for a much simpler description of the non-linear extension of elasticity theory~\cite{Alberte:2018doe,Baggioli:2020qdg}, which will be described in the next sections. Moreover, it provides a fundamental step forward in distinguishing solids and fluids from the point of view of symmetries.
\begin{figure}[ht]
    \centering
    \includegraphics[width=0.9\linewidth]{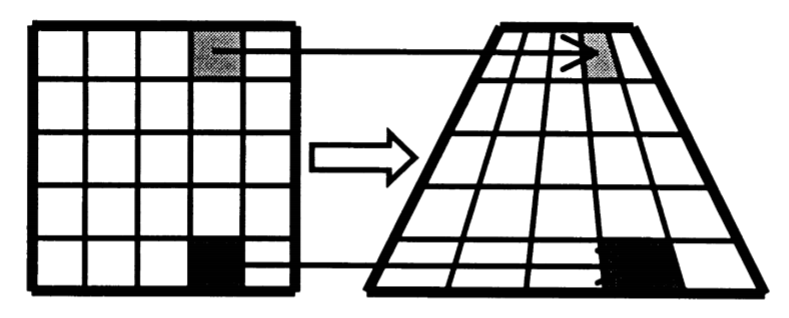}
    \caption{The action of a volume preserving diffeomorphism \eqref{tt}. The total volume remains unchanged.}
    \label{fig:eft4}
\end{figure}\\
As already anticipated, a naive (see~\cite{Baggioli:2019jcm} to learn why it is naive) distinction between solids and fluids relies on the presence of propagating shear waves (transverse phonons). From the field theory we just constructed, it is clear that for $V_X=0$ the transverse phonons speed is zero, and therefore the action is representing a fluid rather than a solid. Interestingly, the condition $V_X=0$ is protected by a specific symmetry which is known as \textit{volume-preserving diffeomorphisms}(VPD):
\begin{equation}
   \Phi^a\,\rightarrow\,\xi^b(\Phi)\,\,\,,\,\,\det\,\frac{\partial \xi^b}{\partial \Phi^a}\,=\,1\,. \label{tt}
\end{equation}
The action of such a symmetry is a coordinates transformation for the mapping $\Phi^I$ which does not change the volume of the system. In other words, invariance under~\eqref{tt} is the mathematical formulation of the fact that fluids do take the shape of the container while solids do not.

In conclusion, the effective action
\begin{equation}
    S\,=\,\int d^3x \,V(Z)\,,
\end{equation}
is the correct description for fluids. Not surprisingly, it bears important relationships with the holographic description of fluids \cite{deBoer:2015ija}.\\

The story becomes highly more complicated when the theory is promoted to the full non-linear dynamics and fluctuations are taken into account \cite{Glorioso:2018wxw}.

\subsection{Gauge-Gravity duality briefing}
The \textbf{AdS-CFT correspondence}, known also as Holography or Gauge-Gravity duality, was originally discovered in 1998 by J.Maldacena~\cite{Maldacena:1997re} (see also~\cite{Witten:1998qj}) and it stands by now as one of the most powerful tools in theoretical physics, providing a deep and fundamental connection between quantum field theory (QFT) and gravity. We refer to the literature~\cite{McGreevy:2016myw,McGreevy:2009xe,Nastase:2007kj,Ramallo:2013bua,Ammon:2015:GDF:2834415,Aharony:1999ti,CasalderreySolana:2011us,Zaffaroni_2000,Polchinski:2010hw,Natsuume:2014sfa,zaanen2015holographic,Hartnoll:2016apf,Hartnoll:2009sz} for a more detailed introduction of the correspondence. 

In one sentence, the slogan of the Gauge-Gravity duality could be phrased as:
\begin{equation}
\text{quantum field theory} \,\,(d\text{-dim})  = \text{gravity}\,\,( d+1\text{-dim})\,,
\label{ddual}
\end{equation}
where the $=$ sign has to be translated as ``dual to''.
\begin{figure}[ht]
\centering
\includegraphics[width=\linewidth]{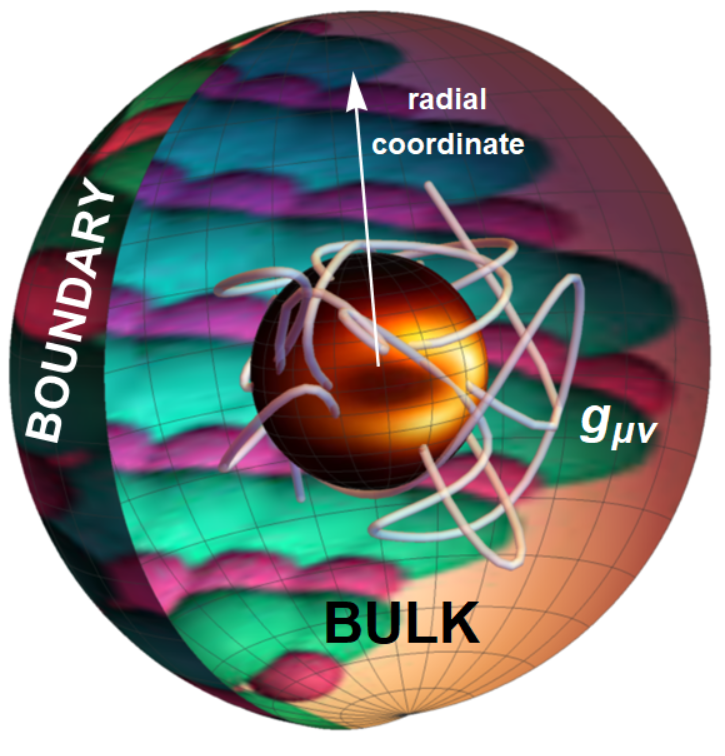}
\caption{An artistic representation of the Gauge-gravity duality. The bulk contains a black hole object dual to a finite temperature thermal state. The bulk spacetime terminates at the so-called boundary where the dual field theory ``lives''. The bulk description contains an extra-dimension, usually denoted as radial coordinate, which describes the energy scale of the dual field theory. The dynamics of the bulk fields, including the metric $g_{\mu\nu}$ happens in a $(d+1)$-dimensional curved spacetime which is asymptotically AdS. In this picture, the boundary field theory lives on the surface of the colored sphere and the bulk region is represented by the 3D region enclosed by such a surface.\color{black}}
\label{fig:adscft}
\end{figure}
In particular, the abstract relation~\eqref{ddual} indicates the existence of a  duality between a gravitational description in $d+1$ dimensions and a QFT one in $d$ dimensions. This idea is artistically represented in Fig.~\ref{fig:adscft} and it can be formally interpreted as:
\begin{align}
&\,\Big\langle\,e^{\int\,\phi_0(x,t)\,\mathcal{O}}\,\Big\rangle_{QFT}\,=\,\mathcal{Z}_{gravity}\left[\phi_0(x,t)\equiv \phi(x,t,u)_{\partial \Sigma}\right]\,,
\label{master}
\end{align}
which is known as the GPKW (Gubser, Polyakov, Klebanov, Witten) \textbf{master rule}~\cite{Witten1,GPKW1}
and its the pillar of the ``dictionary'' defining the $=$ sign in Eq.~\eqref{ddual}. Here $\partial \Sigma$ indicates the boundary of the gravitational spacetime $\Sigma$ at which the QFT source $\phi_0$ is identified using the holographic dictionary.

\begin{figure}[ht]
\centering
\includegraphics[width=0.9\linewidth]{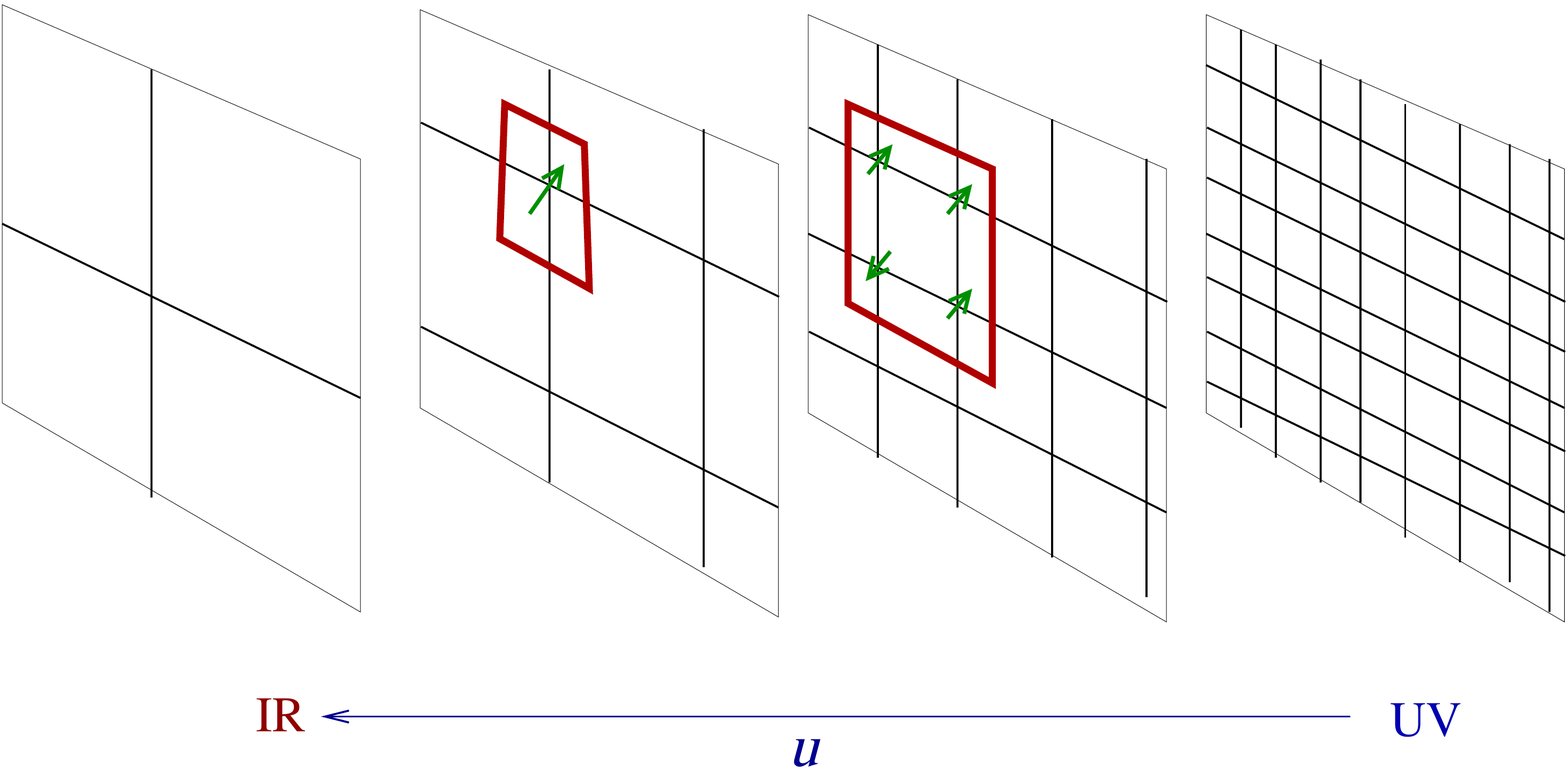}
\includegraphics[width=0.8\linewidth]{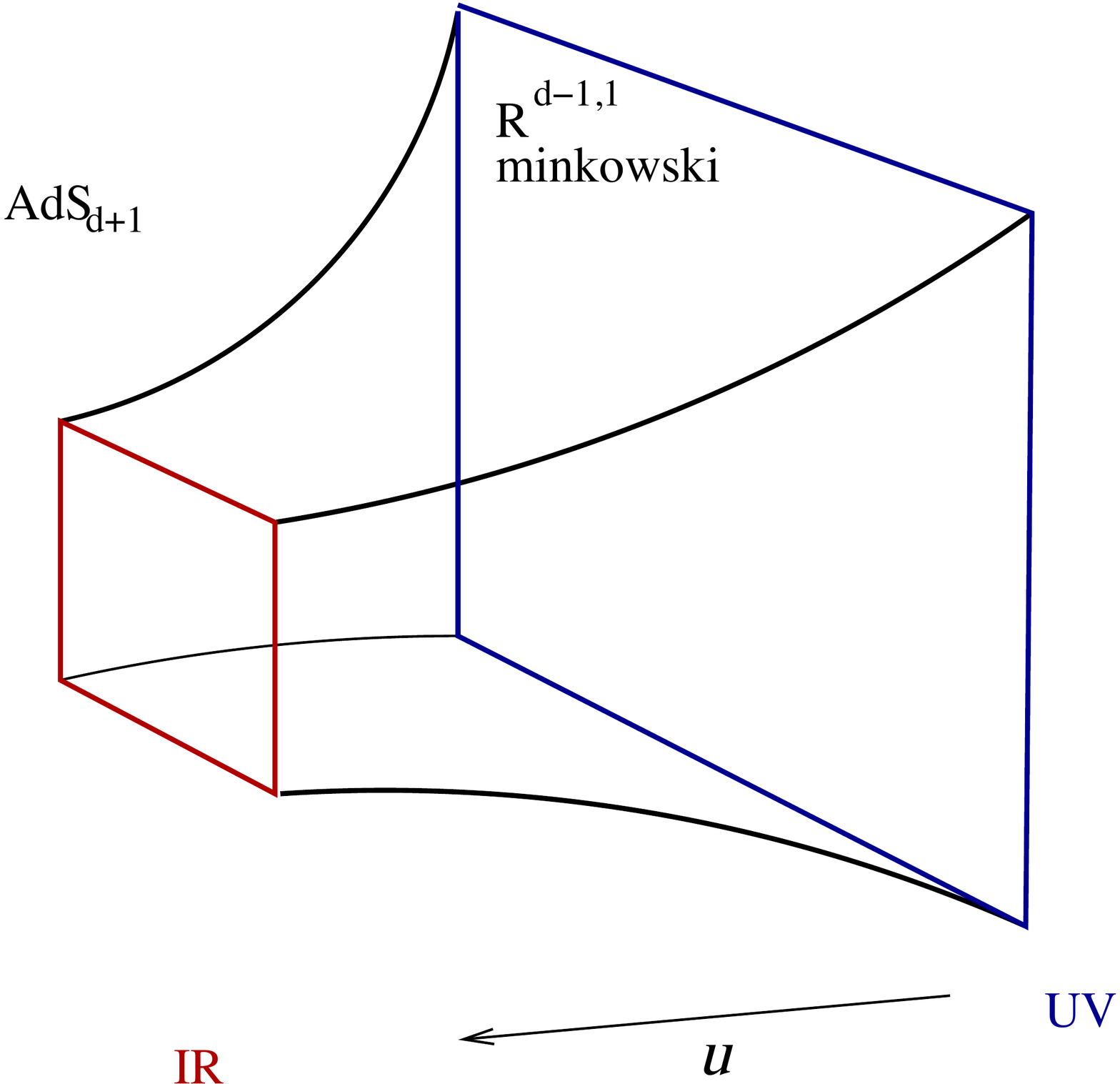}
\caption{Holography provides a geometric representation of RG flow. \textbf{Top: }A series of block spin transformations (coarse-graining process) labeled by the length scale $u$. \textbf{Bottom: }a cartoon of AdS space, where the radial coordinate $u$ plays the role of energy scale of the dual system. Excitations with different energy scale get put in different place in the bulk. Figures updated from~\cite{McGreevy:2009xe}.}
\label{fig:RGflow}
\end{figure}

The core of framework is a $(d+1)$ dimensional bulk where all the bulk fields, including the metric $g_{\mu\nu}$, live and fluctuate. Their dynamics is controlled by a bulk action $\mathcal{S}_{\text{bulk}}[\phi(x,u),g_{\mu\nu}(x,u)\dots]$ defined on a specific bulk geometry. In the limit of large $N$ and infinite coupling for the dual field theory, the gravitational dynamics can be assumed to be classical and stringy corrections can be consistently neglected. This is the limit in which the size of the spacetime geometry $l$ is much larger than the Planck scale $l_p$ and than the string length $l_s$. For all our purposes, we will not deviate from such regime. In our examples, the structure of the background geometry can be written as follows:
\begin{equation}
ds^2_{\text{bulk}_{(d+1)}}\,=\,\frac{L^2}{u^2}\left(\frac{du^2}{g(u)}\,+\,\underbrace{-f(u)\,dt^2+\tilde{g}_{ij}\,dx^idx^j}_{\text{d-dimensional}}\,\right)\,,
\label{bulkgeom}
\end{equation}
where $L$ denotes the AdS radius\footnote{In most case, we set $L\equiv 1$ for simplicity.},  $u$ takes the name of radial-coordinate or holographic coordinate and it plays a very fundamental role in the holographic construction. In particular, this extra-dimension describes the energy scale of the dual system, providing a nice geometric realization of the renormalization flow (RG) of the dual field theory (see Fig.~\ref{fig:RGflow}).\footnote{To be precise, the $u$ coordinate appearing in \eqref{bulkgeom} coincides with the inverse of the energy scale of the dual field theory.} 

The radial coordinate of~\eqref{bulkgeom} spans from $[0,u_h]$ where
\begin{equation}
  u=0\,\,\,:\,\,\,\text{conformal boundary}\,,
\end{equation}
and
\begin{equation}
g(u_h)=f(u_h)=0\,\,\,,\,\,\,u_h:\text{black hole horizon}\,.
\end{equation}
More precisely, $u=0$ is the (conformal) boundary of the asymptotically Anti-de-Sitter (AdS) bulk geometry \eqref{bulkgeom} which is equipped with (a normally flat) metric $(-1,\tilde{g}_{ij})$. The other extreme, $u=u_h$, is the location of the black hole horizon which provides the temperature for the dual field theory, technically given by the surface gravity at its horizon. Another very popular convention in the literature is to use $r \equiv L^2/u$ in which the horizon is set at $r=r_h$ and the conformal boundary at $r=\infty$. The two choices are related by a simple coordinates transformation.

The gravitational bulk action appearing in~\eqref{master} is uniquely defined by choosing boundary conditions (b.c.s) for the various bulk fields. At the horizon $u=u_h$, the appropriate b.c.s. are simply given by the regularity of the solution. At the boundary $u=0$ the b.c.s. uniquely determine the dual field theory and, in particular, the sources with which we deform it. In particular, given a concrete bulk field $\phi(t,x,u)$, its asymptotic expansion in the standard quantization scheme is generally given by
\begin{equation}
\phi(t,x,u)\,=\,\phi_0\,u^{\Delta_L}(1+\dots)+\langle \mathcal{O}\rangle\,u^{\Delta_S}(1+\dots)\,,
\label{expansion}
\end{equation}
where by definition $\Delta_L<\Delta_S$ such that the first term is the ``leading term'' (the one falling-off more slowly towards the boundary) and the second the subleading one. The coefficient of the leading term determines the source $\phi_0$ for the dual operator $\mathcal{O}$ living in the dual field theory. The subleading term determines its vacuum expectation value (vev) $\langle \mathcal{O} \rangle$. The powers $\Delta_{L,S}$ are uniquely determined in terms of the spacetime dimension $d$ and the conformal dimension $\Delta$ of the field theory operator $\mathcal{O}$. Once the sources and the vevs are identified, the gravitational picture can be mapped into a dual field theory:
\begin{equation}
\mathcal{S}\,=\,\mathcal{S}_{CFT}\,+\,\sum_i\,\int d^dx\,\phi_0^i\,\langle \mathcal{O}^i\rangle\,,
\end{equation}
and the correlation functions for the various operators can be obtained using the standard variational prescription.

This is a very brief explanation of how the duality works. For space limitations, we have skipped several important features which the interested Reader can find in the literature mentioned above.  Since the field of applied holography is a vast subject spanning decades of research, we limit this review to recent developments and understandings on strongly coupled quantum matter using holographic axion models. Other active areas of applied holography include condensed matter~\cite{Hartnoll:2009sz,plumbers,Hartnoll:2016apf,Cai:2015cya,Landsteiner:2019kxb}, nuclear physics~\cite{adscftMateos}, quantum information~\cite{Rangamani:2016dms}, non-equilibrium physics~\cite{Hubeny:2010ry,Liu:2018crr} and so on. It is likely to have even wider applicability in the future.

\subsection{Holographic axion model}
When we discuss the \textbf{holographic axion model}, we refer (unless clearly stated otherwise) to an action of the form\footnote{Another popular convention is to take the Einstein-Maxwell part of the action to be:
\begin{equation}
\mathcal{S}=\int d^4x\sqrt{-g}\left[R-2\Lambda-\frac{Y(X,Z)}{4\,e^2} F^2\,+\,\dots\right]\,. 
\label{conven}
\end{equation}
This amounts to a constant re-scaling of the boundary chemical potential $\mu$ and charge density $\rho$. In this review, we will try to keep the notations as uniform as possible. In any case, this constant re-scaling does not affect any of the physical qualitative features of the model and it is in a sense harmless.}
\begin{equation}
\mathcal{S}=\int d^4x\sqrt{-g}\left[{R\over 2}-\Lambda-\frac{Y(X,Z)}{4\,e^2} F^2-m^2 V(X,Z)\right]\,.
\label{action}
\end{equation}
Here $R$ is the Ricci scalar, $\Lambda$ the cosmological constant, $e$ the electric charge. Furthermore, we have defined $\mathcal{I}^{IJ}=g^{\mu\nu}\partial_\mu\phi^I\partial_\nu\phi^J$, with  $X=\frac{1}{2}\,\mathrm{Tr}\,\mathcal{I}^{IJ}$, $Z=\mathrm{det}\,\mathcal{I}^{IJ}$ and $F^2=F_{\mu\nu}F^{\mu\nu}$, where as usual $F=dA$. In the rest of the manuscript, we fix the charge unit to one, $e=1$ and the cosmological constant to $\Lambda=-3$.

The background geometry is defined as
\begin{equation}\label{metric0}
ds^2 = \frac{1}{u^2}\left[-f(u)dt^2+\frac{1}{f(u)}\,du^2+dx^2+dy^2\right]\,,
\end{equation}
where $u$ is the radial bulk coordinates spanning from $u=0$ (the asymptotic AdS boundary) to $u=u_h$ (the black brane horizon radius). The blackening function $f(u)$ displays the following asymptotic behaviours:
\begin{equation}
f(0)\,=\,1\,,\quad f(u)\,=\,-4\,\pi\,T\,\left(u-u_h\right)+\dots\,,
\end{equation}
For simplicity, in most of the review we will focus on two spatial dimensions $x,y$ but the generalizations to three is totally straightforward.

The fields $\phi^I$ are responsible for the breaking of translational symmetry in the $\{x,y\}$ directions of the CFT and their bulk profile is chosen to be:
\begin{equation}\label{linear}
\phi^I=\alpha \,\delta^I_i x^i\,\,,\,\,\,\,\,I=\{x,y\}\,.
\end{equation}
This is the choice which respects the SO(2) rotational symmetry of the dual field theory. This assumption of isotropy could be relaxed and one could consider more complicated anisotropic models of the type:
\begin{equation}\label{linear}
\phi^x=\alpha_x\,x\,,\quad \phi^y=\alpha_y\,y\,.
\end{equation}
For simplicity, we do not consider these situations. See \emph{e.g.}~\cite{Jain:2014vka,Ge:2014aza,Jain:2015txa} for discussions about this case.

Moreover, for monomial potentials, the parameters $\alpha$ and $m$ are redundant but it is anyway good practice to keep both since their origin is rather different. Nevertheless, in few sections where we consider the linear model $V(X)=X$ we will use $m$ and $\alpha$ interchangeably. Finally, the background solution is completed by
\begin{align}
&f(u)=-u^3\int_{u}^{u_h} \left(\frac{\rho^2}{2 Y\left(\bar{X},\bar{Z}\right)}+\frac{m^2\, V\left(\bar{X},\bar{Z}\right)}{\Xi^4}+\frac{\Lambda }{\Xi^4}\right)d\Xi\,,\\
&A_t(u)\,=\,\rho\,\int_{u}^{u_h}\frac{1}{Y(\bar{X},\bar{Z})}\,d\Xi\,,
\label{ansatz}
\end{align}
where $\bar{X}(\Xi)=\alpha^2 \Xi^2$ and $\bar{Z}(\Xi)\,=\,\alpha^4 \Xi^4$.

Furthermore, the temperature of the background geometry reads
\begin{equation}
T\,=\,-\frac{\rho^2\,u_h^3}{8\, \pi \, Y\left(\bar{X}_h,\bar{Z}_h\right)}-\frac{m^2\, V\left(\bar{X}_h,\bar{Z}_h\right)}{4\, \pi  \,u_h}-\frac{\Lambda}{4\, \pi  \,u_h}\,,
\end{equation}
with $\bar{X}_h=\bar{X}(\Xi=u_h)$ and $\bar{Z}_h=\bar{Z}(\Xi=u_h)$. The entropy density is given by
\begin{equation}
    s\,=\,\frac{2\,\pi}{u_h^2}\,.
\end{equation}
In case additional ingredients or couplings are used, they will be explicitly indicated and described. 

\subsection{From inhomogeneous lattices to massive gravity and homogeneous models}
Following the historical path, the holographic axion model has been originally constructed to remedy to the infinite DC conductivity of the Reissner-Nordstrom (RN) solution. Indeed, in its original formulation it was dubbed ``\textit{a simple holographic model for momentum relaxation}''~\cite{Andrade:2013gsa}. Despite the model, as we will see, is much more than that, we find it interesting and instructive to revisit its initial steps as they actually happened.

\begin{figure}[ht]
    \centering
    \includegraphics[width=1.0\linewidth]{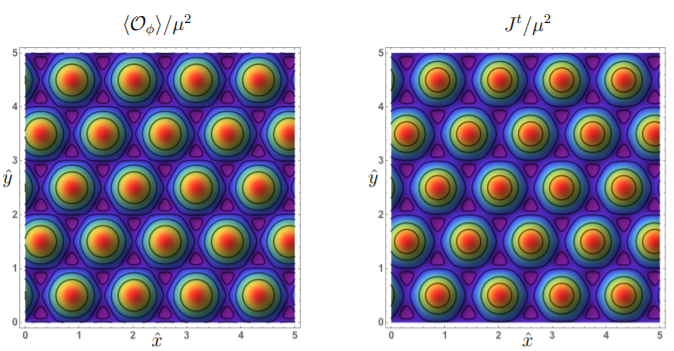}
    \caption{A holographic example of highly inhomogeneous 2D solutions. Figure taken from~\cite{Donos:2015eew}.}
    \label{fig:inho1}
\end{figure}

\begin{figure}[ht]
    \centering
    \includegraphics[width=1.0\linewidth]{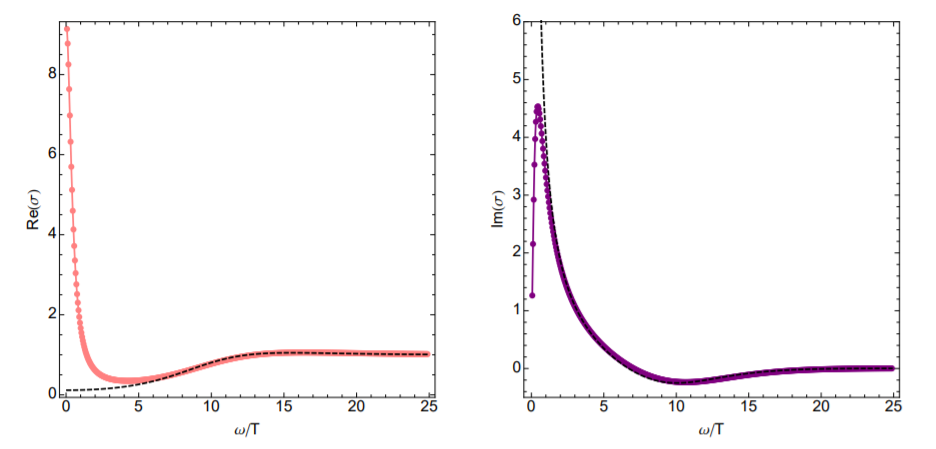}
    \caption{The first holographic computation showing a finite DC electric conductivity in a inhomogeneous periodic lattice. Figure taken from~\cite{Horowitz:2012ky}.}
    \label{fig:inho2}
\end{figure}

An obvious way to relax momentum consists in considering \textbf{inhomogeneous models} where a certain operator (represented by its dual bulk field) displays a spatially dependent expectation value (see Fig.~\ref{fig:inho1} for a specific example), \emph{e.g.}
\begin{equation}
    \langle \mathcal{O}(x) \rangle\,=\,\mathcal{A}\,\cos(kx)\,,
\end{equation}
or a spatially dependent source is introduced
\begin{equation}
    \phi_0(x)\,=\,\mathcal{B}\,\sin(kx)\,.
\end{equation}
In both cases, the resulting geometry will not remain homogeneous and Einstein's equation will result in complicated partial differential equations (PDEs) whose solution might involve very complicated numerical routines~\cite{Dias:2015nua,Krikun:2018ufr,Andrade:2017jmt}.
\\
Despite the validity of these inhomogeneous models, which were, for example, the first to give rise to a finite holographic conductivity (see Fig.~\ref{fig:inho2}), handling them is very complicated and for this reason very few results are available.

A possible way to overcome the difficulties of the inhomogeneous models is to consider simpler models which retain some of their major features (such as the symmetry breaking patterns) but allow for much more reliable and fast computations (which sometimes are even analytical). This is exactly the way the \textbf{homogeneous models with broken translations} became famous and spread around the holographic community. As we will investigate in detail, these models, despite their simplicity, will recover most of the features of the more complicated counterparts and they will reveal extremely useful and rich phenomena.

There is more than that! The homogeneous models, and in particular \textbf{massive gravity} in its general formulation, emerge as the universal low-energy description for \underline{any} holographic models with broken translations. All holographic models with broken translations provide in a way or in another a mass to the graviton (or at least some of its components) and this is nothing else that a universal statement regarding the Ward-identity for translations. By identifying the translations at the boundary with the diffeomorphisms in the bulk, it appears obvious that any model with broken translations must involve a gravitational picture where diffeomorphisms are broken and therefore the graviton being massive.

This statement has been shown explicitly for a concrete lattice construction in~\cite{Blake:2013owa} making a beautiful connection between the more realistic lattice models and the more useful homogeneous relatives. Let us briefly revisit the fundamental steps. Let us take a simple gravitational bulk action in four dimensions:
\begin{equation}
S=\int d^4x\,\sqrt{-g}\,\left[R\,+\frac{6}{L^2}\,-\,\frac{1}{4}F^2\,-\,\frac{1}{2}\partial_\mu \phi \partial^\mu \phi\,-\,\frac{m^2}{2}\phi^2\right]\,,
\end{equation}
where the mass of the scalar is chosen $m^2<0$ in order to have the dual operator $\mathcal{O}$ marginally relevant.\footnote{Notice this is not a problem in curved spacetime as far as the Breitenlohner-Freedman (BF) bound \cite{Breitenlohner:1982bm} is respected.} In general, the associated holographic conductivity would be infinite because of translational invariance. Nevertheless, when spatially dependent boundary conditions are introduced, this is not anymore the case. The authors of~\cite{Blake:2013owa} did that perturbatively by introducing a source for the scalar operator:
\begin{equation}
\phi_0(x)\,=\,\epsilon\,\cos(k_L\,x)\,,
\label{sourc1}
\end{equation}
where $\epsilon \ll 1$ is taken to be infinitesimal. This source mimics the effect of a periodic lattice with wave-vector $k_L$. The boundary source~\eqref{sourc1} corresponds to a bulk profile of the type $\phi(u,x)\,=\,\phi_u(u)\,\phi_0(x)$ with $u$ the holographic radial coordinate. The main idea is then to solve perturbatively the bulk equations of motion up to order $\mathcal{O}(\epsilon^2)$ by using an appropriate expansion for the various bulk fields $g_{\mu\nu},A_\mu$.

The most important result is that the effective action at order $\mathcal{O}(\epsilon^2)$ contains a term 
\begin{equation}
S_{\text{eff}}^{(\epsilon^2)}\,=\,\frac{1}{2}\,\int d^4x\,\sqrt{g}\,M^2(u)\,g^{xx}\,,
\end{equation}
where
\begin{equation}
M^2(u)\,=\,\frac{1}{2}\,\epsilon^2\,k_L^2\,\phi_u^2(u)\,.
\end{equation}
By performing standard perturbation techniques, this new effective term gives a mass to the graviton components $\delta g_{tx},\delta g_{rx}$, as already anticipated. The fact that the vector components of the graviton become massive leads to the expected finite DC conductivity. Most importantly, this simple computation shows directly the universal appearance of an effective graviton mass as a result of a inhomogeneous holographic lattice. In other words, it constitutes strong evidence that \textit{massive gravity is the universal low energy effective holographic description for systems with broken translations.}

\subsection{Other holographic homogeneous models}
In a broad sense, we define a holographic model ``homogeneous'' if the background geometry does not depend on the boundary spacetime coordinates $(t,x^i)$. In the context of translational symmetry breaking, the holographic axion models are not the only homogeneous setups available in the market. In fact, one could define at least three distinct classes of homogenous setups: (I) the axion models discussed in this review, (II) the Q-lattice models~\cite{Donos:2013eha} and (III) the Bianchi VII helical models~\cite{Donos:2012js}. 

These three different classes differ only in terms of the bulk global symmetry used to retain homogeneity. In the axion models, the bulk symmetry is a global shift symmetry which acts on the axion fields as:
\begin{equation}
    \phi^I\,\rightarrow \phi^I+c^I\,,
\end{equation}
with $c$ a constant vector. In order to respect this global symmetry, the axions action contains only derivative terms in the fields. The Q-lattice models are slightly more complicated and they are written in terms of a set of complex fields $\psi^I$ with background profile:
\begin{equation}
    \psi^I(u,x^I)\,=\,\Psi(u)\,e^{i k x^I}\,.
\end{equation}
with $k$ a constant. The corresponding global symmetry is a global U(1) transformation which acts on the complex fields as a phase shift:
\begin{equation}
\psi^I\,\rightarrow \psi^I\,e^{i \varphi}\,, \end{equation}
where $\varphi$ is a constant phase. Again, in order to respect this symmetry, the Q-lattice action is a function only of the absolute value of the scalar fields $|\psi^I|$. Finally, the helical model are more complicated systems whose global symmetry is given by the Bianchi VII group~\cite{Iizuka:2012iv}. This symmetry group is a combination of rotations and translations which geometrically can be represented by a helix.

Despite the different details, mostly regarding the implementation of the bulk global symmetry, all these models display very similar features and their low-energy dynamics is in a sense universal. Nevertheless, it is important to notice that only the axion models allow for a fully analytical background solution. Because of this fact, they are the simplest and most powerful homogeneous models. In this review we will only consider the axion models. All the features present in the most complicated Q-lattice and helical models can be also found in this simpler setup.

\color{black}
\section{A simple model for momentum relaxation}\label{sec:linear}
\subsection{The origins}
The simplest version of the holographic axion model, known as the \textbf{linear axion model}, was introduced in 2013 by Andrade and Withers~\cite{Andrade:2013gsa}. The original intuition came by looking at the following Ward's identity for translations:
\begin{equation}
\nabla_\mu \,\langle T^{\mu j} \,\rangle\,=\,\nabla^j\,\phi^{(0)}\,\langle \mathcal{O} \rangle\,+\,F^{(0)j\mu}\,\langle J_\mu \rangle\,,
\label{ward0}
\end{equation}
where $\mathcal{O},J$ are some unspecified scalar and vector operators and $\phi^{(0)},F^{(0)}$ their external sources. By looking at Eq.~\eqref{ward0}, the authors of~\cite{Andrade:2013gsa} noticed that considering shift-symmetric scalars and turning on sources for them linear in the boundary spatial coordinates
\begin{equation}
\phi^{(0),I}\,\sim\,x^I\,,
\end{equation}
would result in an explicit breaking of the stress tensor conservation. Moreover, given that the bulk stress tensor associated to the scalar fields contains only two derivatives terms, the corresponding geometry would remain homogeneous, i.e. independent on the spatial coordinates $x^i$.

These gravitational theories have already been studied, in a totally different context, in~\cite{Bardoux:2012aw}.
For simplicity, \cite{Andrade:2013gsa} considered the simplest bulk action which preserves the scalars shift-symmetry:
\begin{equation}
    V(X,Z)\,=\,X\,,\quad Y(X,Z)\,=\,\frac{1}{2}\,,\quad m^2=\,\frac{1}{2},\label{linearaxion0}
\end{equation}
from which the name ``linear'' axion model.With this choice, the background solution becomes particularly simple and it reads
\begin{align}
&f(u)=\frac{\mu ^2 u^4}{4 u_h^2}-\frac{u^3}{u_h^3}+\frac{\alpha ^2
   u^3}{2u_h}-\frac{\mu ^2 u^3}{4 u_h}-\frac{\alpha ^2 u^2}{2}+1\\
&A_t(u)\,=\,\mu\,\left(1\,-\,\frac{u}{u_h}\right)\,,\quad \rho=\frac{\mu}{u_h}\,.
\end{align}
Notice that here we have re-scaled the chemical potential $\mu\rightarrow \mu/2$ with respect to Eq.\eqref{ansatz} to match the notations of \cite{Andrade:2013gsa}.\\
Before moving to the phenomenology related to this model, let us spend some words about a few developments appeared after~\cite{Andrade:2013gsa}. In particular, in~\cite{Andrade:2013gsa} it was noticed that the equations for the fluctuations are very similar to those found few months before in massive gravity theories~\cite{Vegh:2013sk}, but not exactly. This point was analyzed further in~\cite{Taylor:2014tka} which considered a square-root deformation of the original model:
\begin{equation}
\mathcal{L}_\phi\,=-\,a_{1/2}\,\sum_I\,\sqrt{X}\,,\qquad X \equiv \frac{1}{2}\partial_\mu \phi^I \partial^\mu \phi^I\,,
\end{equation}
and in~\cite{Baggioli:2014roa} which built an even more generic action:
\begin{equation}
\mathcal{L}_\phi\,=\,-\,m^2\,V(X)\,.
\end{equation}
Nevertheless, the equivalence with the dRGT massive gravity theory was shown only later in~\cite{Alberte:2015isw}. It is important to take in mind that the holographic axion model, written in its more general formulation, is much richer and more general than the dRGT original model of~\cite{Vegh:2013sk}.
\begin{figure}[ht]
\centering
\includegraphics[width=\linewidth]{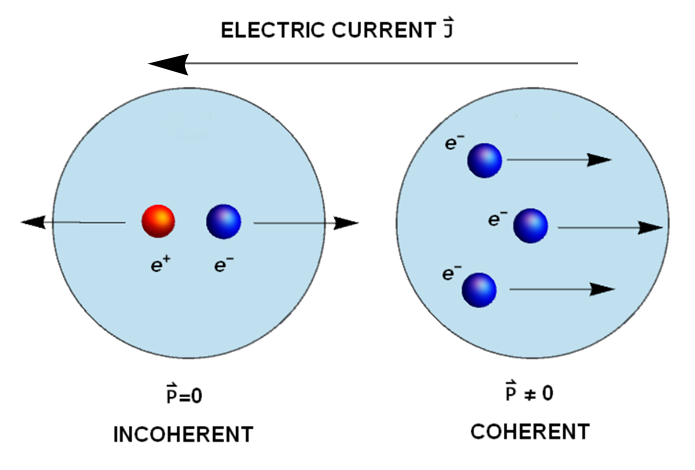}
\caption{The incoherent and coherent contributions to the electric current. The incoherent processes transport charge but they do not transport momentum (\emph{e.g.} particle-antiparticle pair).}
\label{fig:cond}
\end{figure}

\subsection{A holographic Drude model}
The most important physical result of ~\cite{Andrade:2013gsa} is that the DC conductivity of the dual field theory becomes finite. In particular, it takes the simple form:
\begin{equation}
\sigma_{DC}\,\equiv\,\sigma(\omega=0)\,=\,{u_h^{3-d}}\left(1\,+\,(d-2)^2\,\frac{\mu^2}{\alpha^2}\right)\,,
\label{condu0}
\end{equation}
where $\mu$ is the chemical potential of the dual field theory. We will describe in detail how to obtain this result (at least for $d=3$) in Section \ref{DCdatasec}. Expression~\eqref{condu0} displays a very specific structure which is in common of all the holographic models. In particular, the full DC conductivity can be split into two contributions:
\begin{equation}
\sigma_{DC}\,=\,\sigma_{DC}^{(0)}\,+\,\sigma_{DC}^{\text{Drude}}\,.
\end{equation}
The first contribution, which in this simple case is just {$\sigma_{DC}^{(0)}=u_h^{3-d}$}, coincides in the limit of strong momentum relaxation with the  \textit{incoherent conductivity}~\cite{Davison:2015taa}:
\begin{equation}
    \sigma_{DC}^{\text{incoherent}}\,=\,\left(\frac{s\,T}{s\,T\,+\,\mu\rho}\right)^2\,\left(\frac{s}{4\,\pi}\right)^{\frac{d-2}{d}}
\end{equation}
which can be derived by considering the incoherent current \begin{equation}
    J^{\text{incoherent}}\,=\,J\,-\,\frac{\chi_{PJ}}{\chi_{PP}}\,P
\end{equation}
where, here, both the momentum $P$ and the currents $J$ are intended as operators\footnote{We thank Blaise Gouteraux for clarifying this point to us.}.\\
The incoherent conductivity relates to the part of the electric current $J$ which does not overlap with the momentum operator and it is therefore insensitive to any momentum relaxing mechanism (in this case independent of $\alpha$). This contribution is finite even in absence of momentum dissipation and it corresponds to the probe limit result (with no backreaction of the bulk fields on the background metric) in the limits of strong momentum dissipation or zero charge density.

The second contribution corresponds to the part of the electric current which transports also momentum (see Fig.~\ref{fig:cond}) and it is infinite in the absence of momentum dissipation $\alpha \rightarrow 0$). It is the equivalent of the Drude result~\eqref{drudecond} and it vanishes in the limit $\mu=0$, at which electric current and momentum decouple.

\begin{figure}
    \centering
    \includegraphics[width=0.8\linewidth]{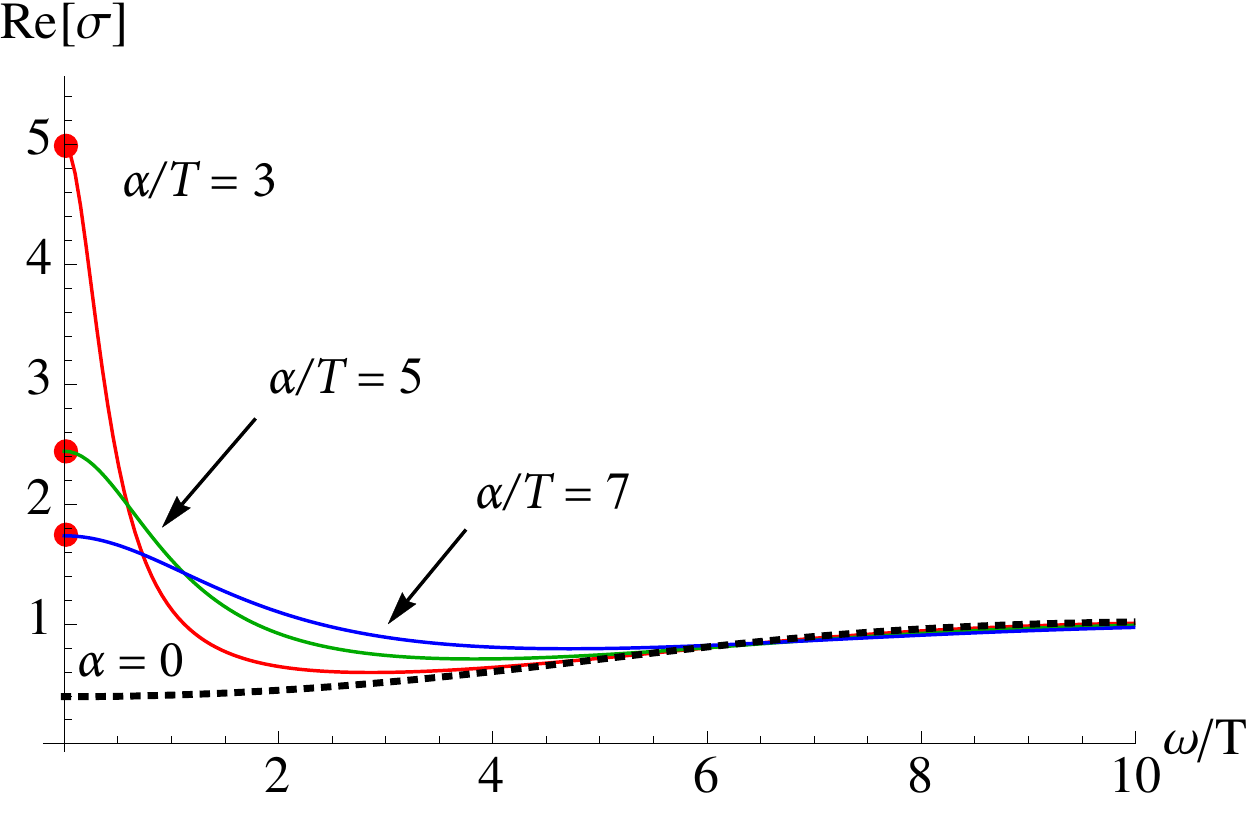}
    
    \vspace{0.2cm}

    \includegraphics[width=0.8\linewidth]{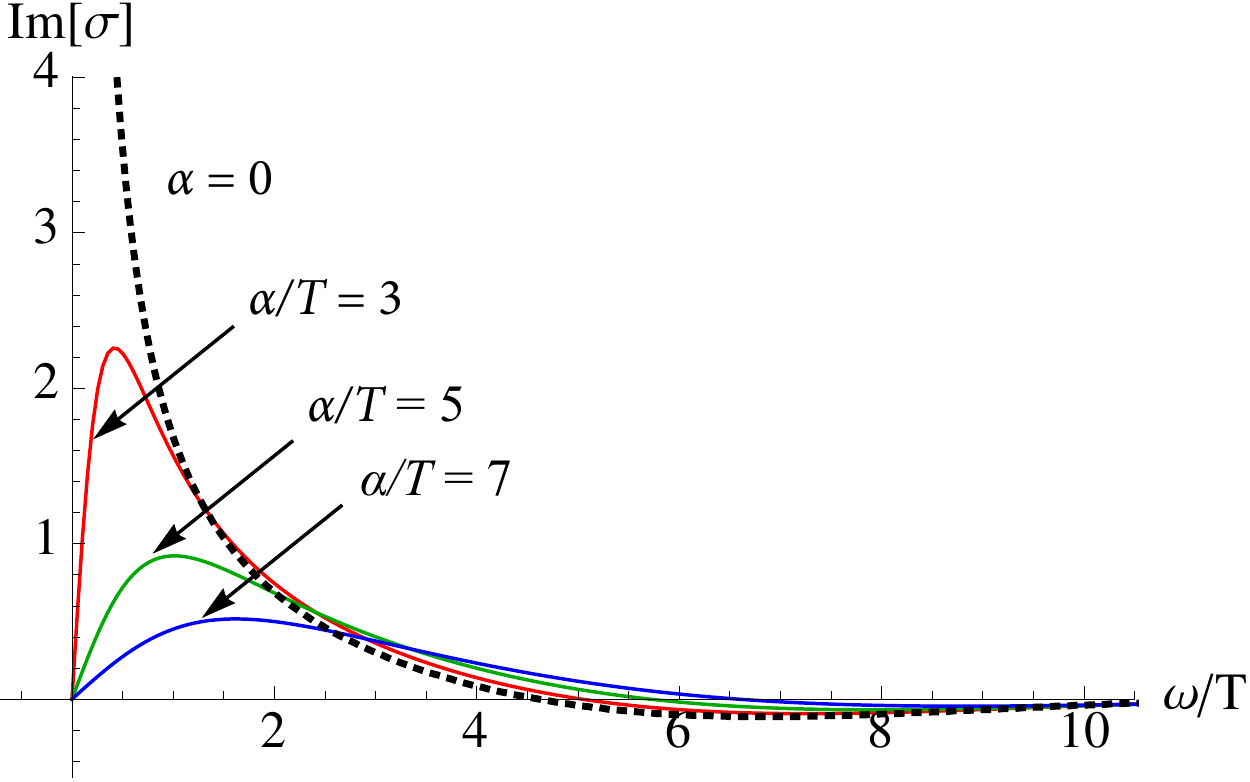}
    \caption{The optical conductivity of the linear axion model for various values of the momentum dissipation rate $\alpha/\mu$. Figure taken from \cite{Kim:2014bza}. }
    \label{fig:sigmalinear}
\end{figure}

One can do more and compute also the AC -- frequency dependent -- electric conductivity. In order to do that, one has to switch on fluctuations for the gauge field, the metric and the scalar fields. A consistent truncation at zero momentum ($k=0$) is given by
\begin{align}
   & \delta A_x\,=\,e^{-i \omega t}\,a_x(u)\,,\,\,\,\delta g_{tx}\,=\,e^{-i \omega t}\,h_{tx}(u)\,,\,\,\,\nonumber\\
   &\delta\phi\,=\,e^{-i \omega t}\,\varphi(u)\,,
\end{align}
and the corresponding equations of motion can be found in the original work~\cite{Andrade:2013gsa}. Following the standard procedure to compute the holographic conductivity (see~\cite{Baggioli:2019rrs} and~\cite{tonglectures}), one can finally obtain numerically $\sigma(\omega)$. The AC conductivity was originally presented in~\cite{Kim:2014bza} and it is here reproduced in Fig.~\ref{fig:sigmalinear}.

\begin{figure}
    \centering
    \includegraphics[width=0.8\linewidth]{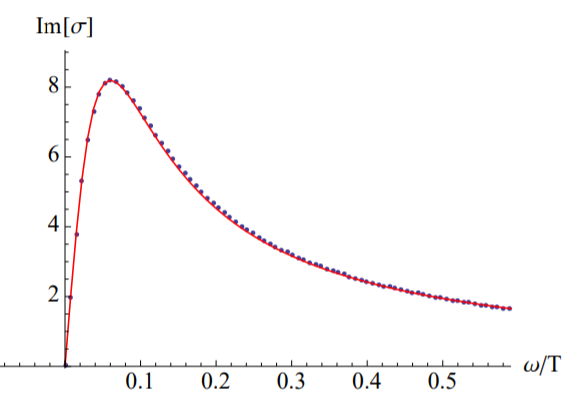}
    
    \vspace{0.2cm}
    
    \includegraphics[width=0.8\linewidth]{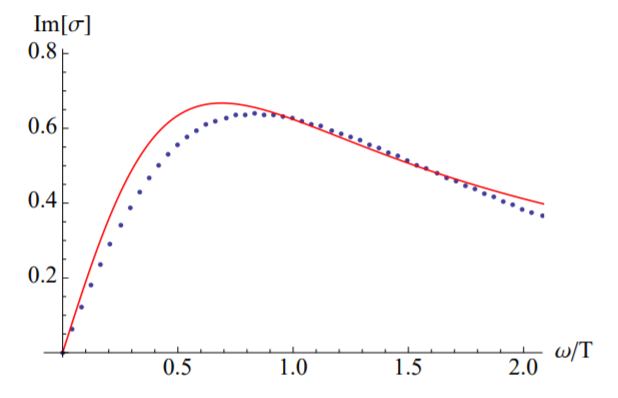}
    \caption{The optical conductivity of the linear axion model compared with the Drude model formulas. Top panel is for $\alpha/\mu=0.25$, bottom panel for $\alpha/\mu=1$. Figure taken from \cite{Kim:2014bza}.}
    \label{fig:comp}
\end{figure}

The first important result is that the DC conductivity is finite and it appears in perfect agreement with the analytic formula \eqref{condu0}. Moreover, at slow momentum relaxation, $\alpha/T \ll 1$, the conductivity shows a nice Drude peak. Indeed, one can fits the numerical data with the Drude formula very well (see Fig.~\ref{fig:comp}). See~\cite{Andrade:2015hpa} for a study in large $D$ (spatial dimensions). This is not anymore true at large momentum dissipation, where the relaxation-time approximation of the Drude model fails because the corresponding relaxation rate becomes too large. In this limit, the holographic model goes beyond the Drude model and momentum is not anymore an almost conserved operator. This brings us directly to the next section.

\subsection{Coherent-incoherent transition}
\label{co-in trans}
\begin{figure}[ht]
\centering
\includegraphics[width=\linewidth]{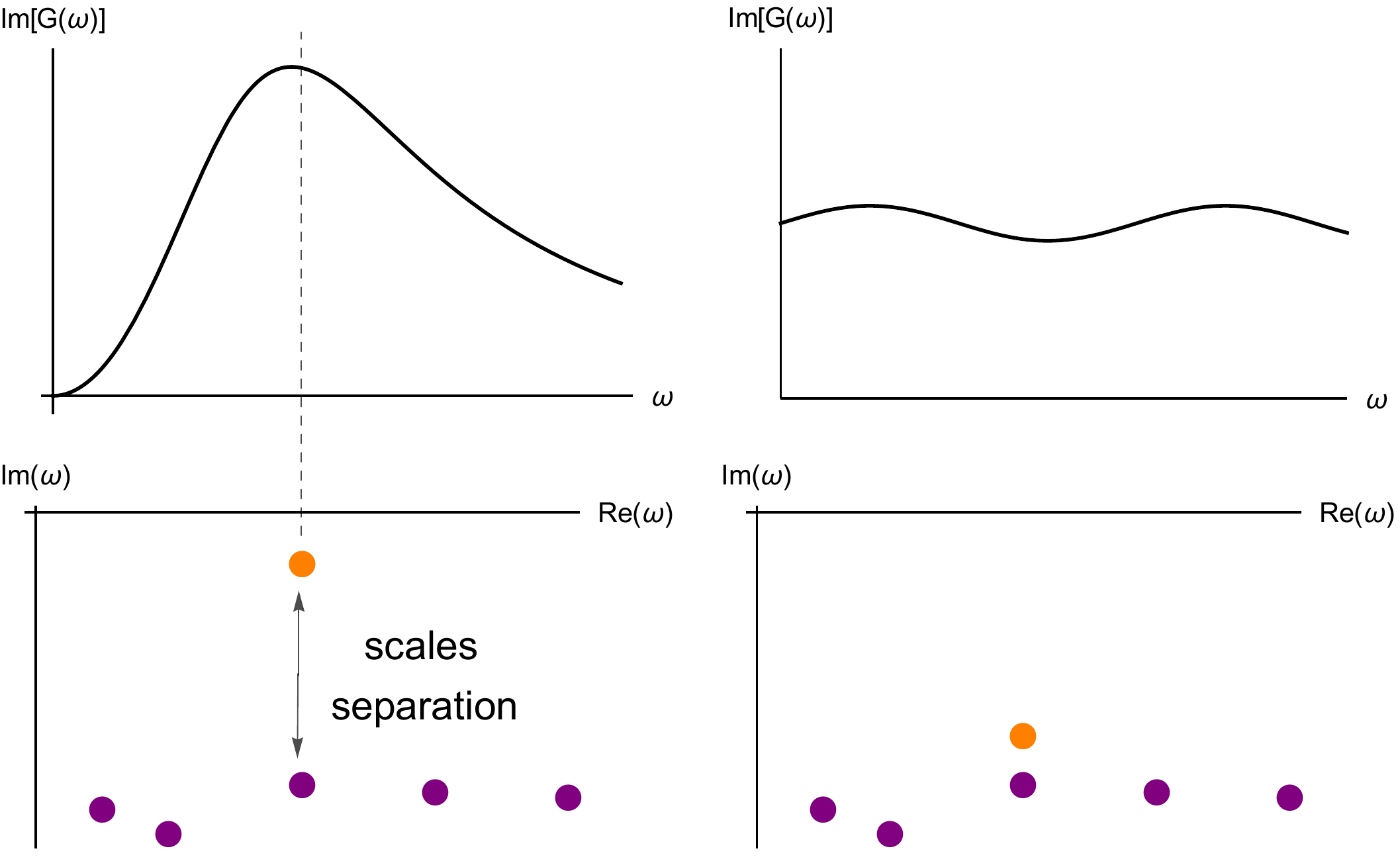}
\caption{\textbf{Left Panel: }the coherent regime, where the lowest mode (in orange) is well separated from the rest of the excitations. The corresponding response function displays a nice Lorentzian peak at the position of the lowest mode. \textbf{Right Panel: }the incoherent regime where the separation of scales is lost. The response function is featureless and cannot be well approximated by using only the first (orange) mode.}
\label{tworeg}
\end{figure}

In the holographic linear axion model, and in general in any model containing a relaxational mode, one can distinguish two regimes (see Fig.~\ref{tworeg}). The first regime is known as the \textbf{coherent regime} and it appears for slow momentum dissipation, $\Gamma/T\ll 1$. In this case, the Drude pole $\omega=-i \Gamma$ is well-separated from the rest of the excitations, in the sense that is parametrically more long-living than any other mode in the system. This regime is obtained at finite values of $\alpha$, \emph{i.e.} $\alpha/T \ll 1$, at which the optical conductivity displays a nice Drude peak. This is also the regime in which the Drude model well describes the frequency dependent conductivity, since the momentum relaxation time is large.

A second regime, known as the \textbf{incoherent regime}, appears at very large values of the momentum dissipation rate, $\Gamma/T \gg 1$, where the Drude pole becomes very short living and its lifetime becomes comparable with the rest of the excitations (see Fig.~\ref{tworeg}). In this regime, the Drude model is not anymore a good description and the optical conductivity becomes featureless and flat.

The coherent-incoherent transition in the linear axion model has been studied in detail in~\cite{Kim:2014bza,Davison:2014lua,Davison:2015bea}. For simplicity, we will consider the model at zero charge density in two spatial dimensions. Let us start by the coherent regime, in which $\Gamma/T \ll 1$. In this case, the Green's functions for the momentum density $p_i\equiv T^{t\,i}$ parallel and transverse two the wave-number $k$ have the following structure~\cite{Davison:2014lua}:
\begin{align}
&\mathcal{G}^R_{p_\parallel p_\parallel}\,=\,\frac{(\mathcal{E}+\mathfrak{p})\left[k^2\,\frac{\partial \mathfrak{p}}{\partial \mathcal{E}}-i \omega \left(\Gamma+k^2\,\frac{\eta}{\mathcal{E}+\mathfrak{p}}\right)\right]}{i \omega\left(-i \omega+\Gamma+k^2\,\frac{\eta}{\mathcal{E}+\mathfrak{p}}\right)-k^2\,\frac{\partial \mathfrak{p} }{\partial \mathcal{E}}}\,,\\
&\mathcal{G}^R_{p_\perp p_\perp}\,=\,\frac{-\left(\mathcal{E}+\mathfrak{p}\right)\left(\Gamma+k^2\,\frac{\eta}{\mathcal{E}+\mathfrak{p}}\right)}{-i \omega+\Gamma+k^2\,\frac{\eta}{\mathcal{E}+\mathfrak{p}}}\,,
\end{align}
where $\mathcal{E}$, $\mathfrak{p}$ and $\eta$ are energy density, pressure and shear viscosity, respectively.
The (longitudinal) thermal conductivity $\kappa(\omega)$ reads\footnote{This conductivity can be read directly from the longitudinal Green's function using the Kubo formula, 
 $\kappa(\omega, k)\,\equiv\,\frac{i}{\omega\,T} \left[\mathcal{G}^R_{p_\parallel\,p_\parallel}(\omega\,,\,k)-\mathcal{G}^R_{p_\parallel\,p_\parallel}(0\,,\,k)\right]$.}
\begin{equation}
    \kappa(\omega, k)\,=\,\frac{i \omega\,s}{i \omega\left(-i \omega+\Gamma+k^2\,\frac{\eta}{\mathcal{E}+\mathfrak{p}}\right)-k^2\,\frac{\partial \mathfrak{p} }{\partial \mathcal{E}}}\,,
\end{equation}
such that its DC component is simply:
\begin{equation}
    \kappa_{DC}\,=\,\frac{s}{\Gamma}\,,
\end{equation}
and it is controlled by the momentum relaxation rate $\Gamma$, as expected.

\begin{figure}[h]
\centering
\includegraphics[width=\linewidth]{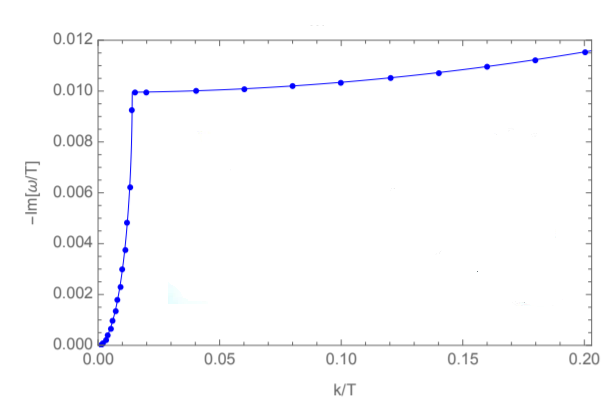}
\caption{The imaginary part of the lowest mode in the longitudinal sector of the linear axion model with $\alpha/T=1/2$ in the coherent regime showing the diffusive-to-propagating crossover. Figure adapted from~\cite{Davison:2014lua}.}
\label{cross}
\end{figure}

Now, by looking at the poles of the parallel Green function, we can find the dispersion relation of the lowest modes in the longitudinal spectrum:
\begin{align}
\omega\,=\,&\pm\,k\,\sqrt{\frac{\partial \mathfrak{p}}{\partial \mathcal{E}}-\frac{1}{4}\left(\Gamma  k^{-1}+\frac{\eta}{\mathcal{E}+\mathfrak{p}}\,k \right)^2}\nonumber\\&-\frac{i}{2}\left(\Gamma+\frac{\eta}{\mathcal{E}+\mathfrak{p}}\,k^2\right)\,,
\label{ww}
\end{align}
which already indicates that the original longitudinal sound mode is destroyed by the presence of momentum dissipation. Moreover, there is an interesting crossover between diffusive-like behaviour at small momentum and propagating like at high one. More precisely, for $k/\Gamma \gg 1$, Eq.~\eqref{ww} gives two propagating sound modes:
\begin{equation}
\omega\,=\,\pm\,k\,\sqrt{\frac{\partial \mathfrak{p}}{\partial \mathcal{E}}}-\frac{i}{2}\left(\Gamma+\frac{\eta}{\mathcal{E}+\mathfrak{p}}k^2\right)\,,
\end{equation}
while at large distances, $k/\Gamma \ll 1$, there are two separated modes, one diffusive and one damped Drude-like:
\begin{align}
    & \omega\,=\,-\,i\,\frac{\partial \mathfrak{p}}{\partial \mathcal{E}}\,\Gamma^{-1}\,k^2\,+\,\dots\,,\\
    &\omega\,=\,-\,i\,\Gamma\,+\,i\,k^2\,\left(\frac{\partial \mathfrak{p}}{\partial \mathcal{E}}\,\Gamma^{-1}-\frac{\eta}{\mathcal{E}+\mathfrak{p}}\right)\,+\,\dots\,,
\end{align}
This means that heat is transported ballistically at short distances but diffusively at long ones. The crossover happens exactly at
\begin{equation}
\Gamma\,k^{-1}\,+\,\frac{\eta}{\mathcal{E}+\mathfrak{p}}\,k\,=\,2\,\sqrt{\frac{\partial \mathfrak{p}}{\partial \mathcal{E}}}\,,
\end{equation}
and it is shown in Fig.~\ref{cross}.

\begin{figure}[h]
\centering
\includegraphics[width=\linewidth]{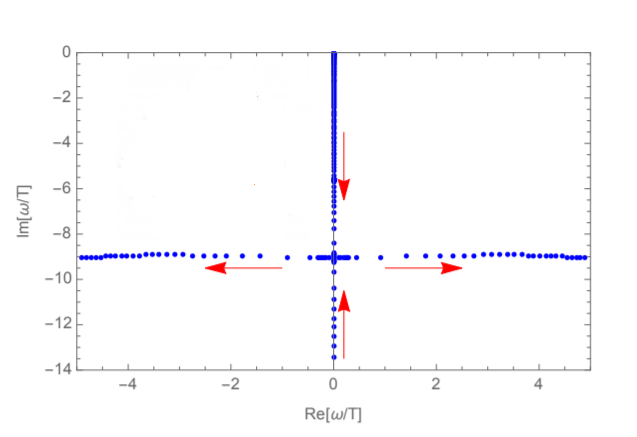}
\caption{The modes collision associated to the coherent-incoherent transition in the linear axion model. Here the wave-number $k$ is taken to be zero and the parameter $\alpha/T$ is increased from 0 to 12 in the direction of the arrows. There is a Drude-like pole near the origin at weak momentum dissipation rate. As $\alpha$ increases, it moves down the imaginary axis and collides with another purely imaginary pole at $\alpha/T \approx 9.5$, producing two off-axis poles. Figure adapted from~\cite{Davison:2014lua}.}
\label{boom}
\end{figure}

\begin{figure}
\centering
\includegraphics[width=\linewidth]{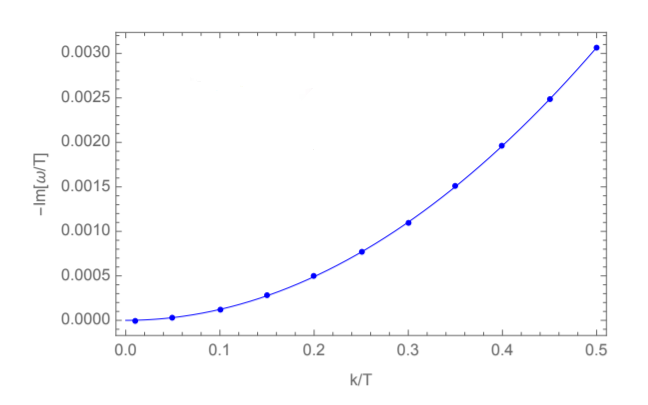}
    
\vspace{0.2cm}
    
\includegraphics[width=\linewidth]{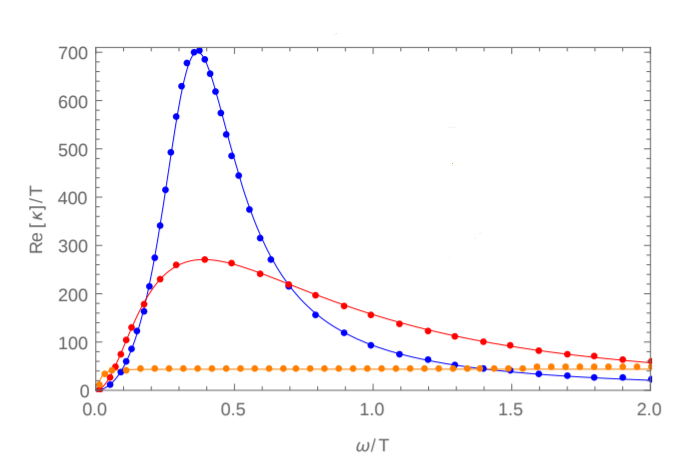}
\caption{\textbf{Top panel: }The numerical confirmation of the energy diffusion mode in the incoherent regime with $\alpha/T=100$. \textbf{Bottom panel: } the coherent-incoherent transition in the frequency dependent thermal conductivity for $k/T=1/2$. Values of $\alpha/T(=2,7/2,20)$ from top to bottom. Figures adapted from~\cite{Davison:2014lua}.}
\label{inco}
\end{figure}

From the coherent regime where $\Gamma/T\ll 1$ with
\begin{equation}
\Gamma\,=\,\frac{\alpha^2}{4\,\pi\,T}\,,
\end{equation}
we can increase further the axions strength $\alpha$. At a certain point, $\Gamma/T\sim\,\mathcal{O}(1)$, the Drude pole collides on the imaginary axes with a secondary pole coming up and it produces to off-axes poles with finite real part which at this point are not anymore well detached from the rest of the excitations. This collision is shown explicitly in Fig.~\ref{boom}.
Once the incoherent regime is reached, the only conserved, and therefore long-living, quantity is the energy density $\mathcal{E}$. Its Green's function takes the form~\cite{Davison:2014lua}:
\begin{equation}
\mathcal{G}^R_{\mathcal{E}\mathcal{E}}\,=\,T^2\,\frac{\partial s}{\partial T}\,\frac{D_\mathcal{E}\,k^2}{i \omega-D_\mathcal{E}k^2}\,,
\end{equation}
where $D_\mathcal{E}$ is the energy diffusion constant. Finally, the DC thermal conductivity is
\begin{equation}
\kappa_{DC}\,=\,T\,\frac{\partial s}{\partial T}\,D_\mathcal{E}\,=\,c_v\,D_\mathcal{E}\,,
\end{equation}
and it obeys the well-known Einstein's relation. As already mentioned before, the frequency dependent thermal conductivity passes from displaying a well-defined coherent peak to a flat incoherent response. These features are shown in Fig.~\ref{inco}.

The same phenomenology has been later found also in holographic axion model with fluid symmetry~\cite{Baggioli:2019abx} confirming its universal character.

\subsection{DC conductivities from horizon data}\label{DCdatasec}
Historically, the first motivation behind the holographic axion model was to render the DC conductivities finite. As such, after the introduction of the model and the first studies big part of the community focused on studying the transport properties of holographic models with broken translations. A fundamental step in this direction is represented by the seminal work by Donos and Gauntlett \cite{Donos:2014cya} which provided a fast and very general way of deriving the DC transport coefficients from horizon data, by generalizing the idea of the \textit{membrane paradigm} \cite{Iqbal:2008by}.
In this section, we show how the methods of \cite{Donos:2014cya} apply to the holographic linear axion model (see also \cite{Baggioli:2019rrs} for explanations about this procedure). 

To illustrate the main idea, let us consider the homogeneous and isotropic background with the metric taking the form of~\eqref{metric0}.
Importantly, a full knowledge of the blackening factor $f(u)$ is not needed and therefore this method can be applied also to background solutions which are not analytical or expressible in close form. For linear axion models, we perturb the black hole background with the following fluctuations:
%
\begin{equation}\label{pertDC}
\begin{split}
&\delta g_{tx}(u,t)\,=\,\frac{1}{u^2}\,\left(-t\,\zeta\,f(u)+\,h_{tx}(u)\,\right),\\
& \delta g_{xu}(u)\,=\,\frac{1}{u^2}\,h_{xu}(u),\qquad \delta \phi^1(u,t)\,=\,\chi(u)\,,\\
&\delta A_x(t,u)\,=\,t\,\left(-E_x\,+\,\zeta\,A_t(u)\right)\,+\,a_x(u)\,,
\end{split}
\end{equation}
%
where $E_x\equiv F_{xt}$ is an external electric field in the $x$ direction and $\zeta$ a thermal gradient. Given this set of external sources, we can now compute the full matrix of thermoelectric conductivities using
\begin{equation}
\left(
    \begin{array}{c}
      \mathcal{J} \\
      \mathcal{Q}
    \end{array}
  \right)\,=\,\left(
    \begin{array}{cc}
      \sigma & \mathscr{A}\,T\\
      \bar{\mathscr{A}}\,T & \bar{\kappa}\,T
    \end{array}
  \right)\left(
    \begin{array}{c}
      E \\
      -\frac{\nabla T}{T}
    \end{array}
  \right)\,,
\end{equation}
where $\mathcal{J}^I$ is the electric current and $\mathcal{Q}^i= T^{ti}\,-\,\mu\,\mathcal{J}^i$ the thermal/energy current. The various coefficients appearing in the expression above are the electric conductivity ($\sigma$), thermal conductivity ($\bar{\kappa}$) and thermoelectric conductivities ($\mathscr{A},\bar{\mathscr{A}}$). These four objects codify the response of the system under an external electric field $E$ and a thermal gradient $\nabla T/T$.

Using the perturbations defined in Eq.~\eqref{pertDC}, we can obtain that the bulk Maxwell equation takes the form of a conservation equation
\begin{equation}
\partial_u\, \mathcal{J}^{\text{bulk}}(u)\,=\,0\,,
\end{equation}
with
\begin{equation}
\mathcal{J}^{\text{bulk}}(u)\,=\,f\,\delta A_x'(u)\,-\,\frac{u^2\,\mu}{u_h}\,h_{tx}(u)\,,
\end{equation}
which, at the boundary $u=0$, gives nothing but the electric current $\mathcal{J}$ of the dual field theory. Given that $\mathcal{J}^{\text{bulk}}(u)$ is radially conserved, one can decide to compute it at any location in the bulk and in particular at the black hole horizon $u=u_h$. In order to do that, we need to find out the constraints to have the perturbations well-behaving -- non-singular -- at the horizon, which are\,\footnote{The simplest way to find them is by using Eddington-Finkelstein coordinates.}:
\begin{equation}
\delta A_x'(u) \,=\,-\,\frac{E_x}{f(u)} \,,\qquad h_{tx}(u)\,=\,\frac{f(u)}{u_h^2}\,h_{xu}(u)\,,
\end{equation}
as $u \rightarrow u_h$.

Moreover, one notices that the $uu$-component of Einstein's equations is a constraint equation, which at the horizon reduces to
\begin{equation}
h_{xu}(u)\,=\,\frac{2\,\mu\,u^2\,E_x\,u_h-u_h^2\,\zeta\,f'(u)}{\alpha^2\,u_h^2\,f(u)}\,.
\end{equation}
All in all, we can evaluate the bulk current at the horizon and obtain that
\begin{equation}
\mathcal{J}=\mathcal{J}^{\text{bulk}}(u_h)\,=\left(1+\frac{\mu^2}{\alpha^2}\right)E_x\,-\,\frac{\mu\,f'(u_h)}{\alpha^2\,u_h}\,\zeta\,.
\end{equation}
From above equation, we can obtain
\begin{align}
& \sigma\,\equiv\,\frac{\partial \mathcal{J}}{\partial E_x}\,=\,1\,+\,\frac{\mu^2}{\alpha^2}\,,\\
& \bar{\mathscr{A}}\,=\,\mathscr{A}\,=\,\frac{1}{T}\,\frac{\partial \mathcal{J}}{\partial \zeta}\,=\,\frac{4\,\pi\,\mu}{\alpha^2\,u_h}\,,
\end{align}
where the two off-diagonal terms are equivalent because of the Onsager's relation. It is interesting to notice that the transport coefficients above obey the Kelvin's formula:
\begin{equation}
\frac{\mathscr{A}}{\sigma}\Big{|}_{T=0}\,=\,\lim_{T \rightarrow 0}\,\frac{\partial s}{\partial \rho}\Big{|}_T
\end{equation}
with $\rho$ the charge density, as observed in~\cite{Davison:2016ngz}.

In order to compute thermal transport, we have to work a bit harder. The key observation is that the bulk equations of motion hiddenly imply the conservation of another combination of bulk fields:
\begin{equation}
\mathcal{Q}(u)\,=\,f^2(u)\,\left(\frac{\delta g_{tx}(u)}{f(u)}\right)'\,-\,A_t(u)\,\mathcal{J}(u)
\end{equation}
which reduces at the boundary $u=0$ to the thermal/energy current of the dual field theory. This fact can be derived ``brute-force'' or in a more elegant way using the properties of the solution as done in~\cite{Donos:2014cya}. By following the same procedure, one finally obtain
\begin{equation}
\bar{\kappa}\,=\,\frac{(4\,\pi)^2\,T}{\alpha^2\,u_h^2}\,.
\end{equation}
After this initial finding, the thermoelectric transport has been computed in many holographic models with and without an external magnetic field. See~\cite{Amoretti:2014mma,Zhou:2015dha,Lucas:2015lna,Banks:2015wha,Ge:2016sel,Wu:2016jjd,Blake:2014yla,Donos:2014yya,Cheng:2014tya,Ge:2014aza,Lucas:2015vna,Amoretti:2015gna,Blake:2015ina,Kim:2015wba,Zhou:2015dha,Donos:2015bxe,Cremonini:2016avj} for a subset of the related developments.

\subsection{Thermoelectric transport}
In analogy to the electric conductivity, one can compute also the frequency dependence of the other thermoelectric transport coefficients using the Kubo formulas for the stress tensor and the electric current.
\begin{figure}[ht]
\centering
\includegraphics[width=0.8\linewidth]{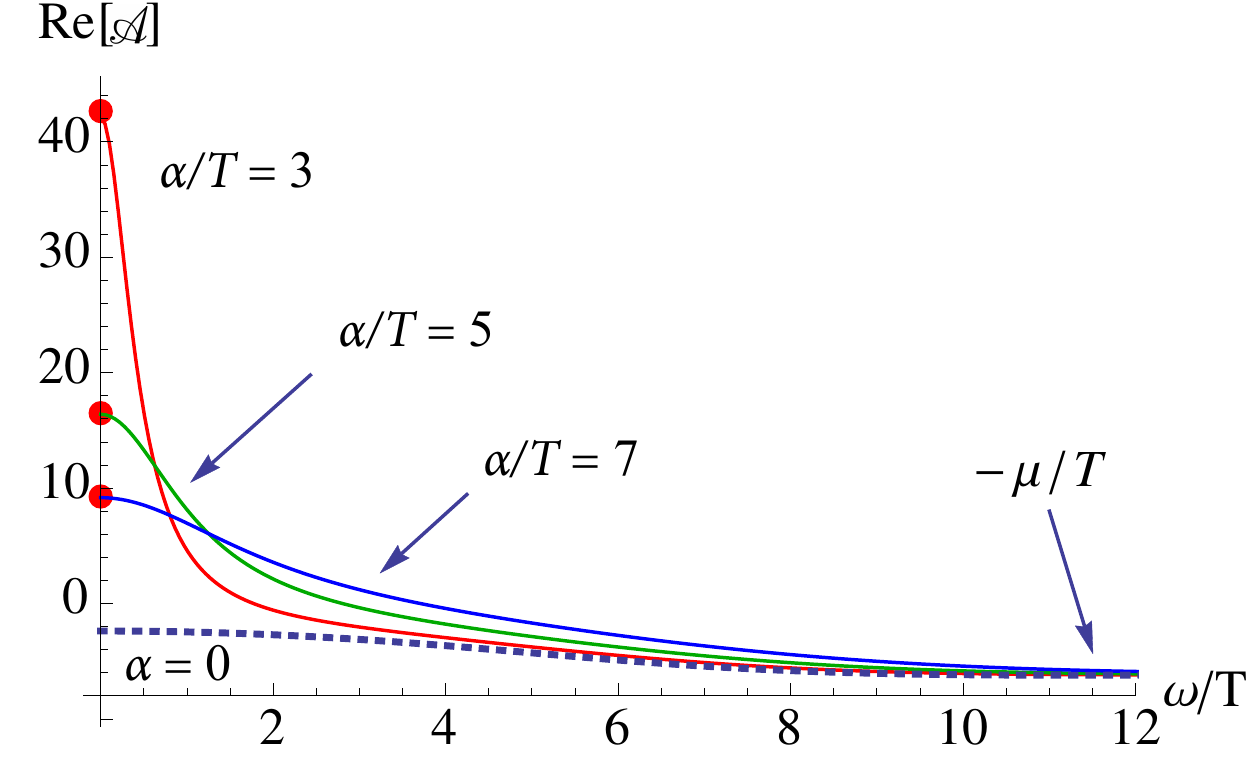}
    
\vspace{0.2cm}
    
\includegraphics[width=0.8\linewidth]{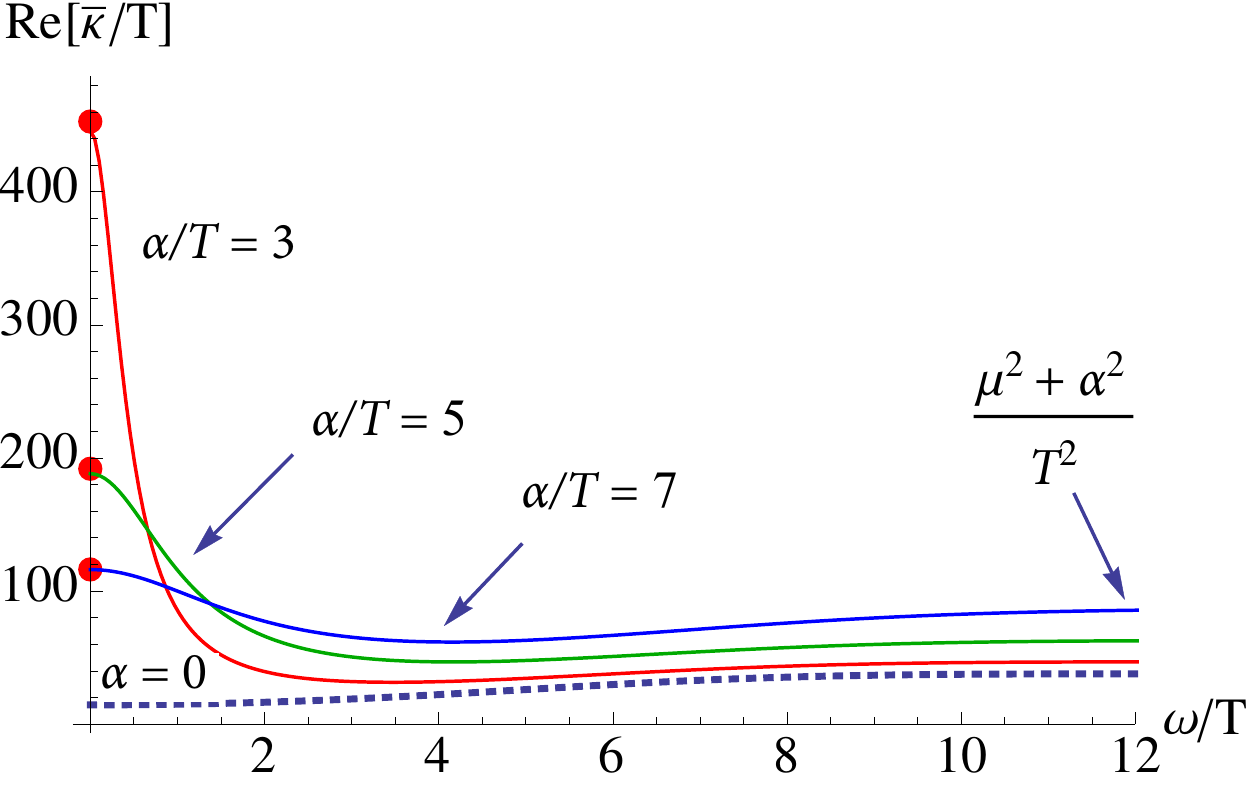}
\caption{The frequency dependent thermoelectric coefficients for the linear axion model~\eqref{linearaxion0} with $d=3$. Figures taken from~\cite{Kim:2014bza}. }
\label{figthermo}
\end{figure}

The results for the linear axion model~\eqref{linearaxion0} with $d=3$ are shown in Fig.~\ref{figthermo} and they display a behaviour very similar to the electric conductivity $\sigma$. First, at small $\alpha$, there is a nice Drude peak which transits to a flat incoherent response for large momentum dissipation. The DC values are in perfect agreement with the analytic results shown in the previous section. Also, the value of the conductivities at large frequency $\omega/T \rightarrow \infty$ can be obtained using the Ward's identities~\cite{Kim:2015sma} and in the present case we have
\begin{equation}
    \mathscr{A} \rightarrow -\frac{\mu}{T}\,,\qquad \frac{\bar{\kappa}}{T}\rightarrow\frac{\mu^2\,+\,\alpha^2}{T^2}\,.
\end{equation}

\section{Breaking translations spontaneously}\label{sec:spontaneous}
\subsection{Axion model 2.0}
As explained in the previous sections, the linear axion model~\cite{Andrade:2013gsa}, despite its simplicity, captures the key features of the EXB of translations and for that reason it has been widely used in the holographic community. Nevertheless, the axion model is much more powerful than that. In this section, we will generalize the model of~\cite{Andrade:2013gsa} in order to consider the spontaneous breaking of translations and study the associated physics.

We start by writing the most general Einstein-Maxwell-axions action 
\footnote{Note that there can be other possible couplings between the axion sector and the gauge one. For example, one could introduce a term of the type $\partial_{\mu}\phi^I\partial^{\nu}\phi^I {F_{\nu}}^\rho {F_{\rho}}^\mu$, which cannot be written in a shorthand with our notation. Introducing such kind of couplings does not change the background solution. Nevertheless, it does have a finite contribution to the linearized equations for the fluctuations. For more details, we refer to~\cite{Gouteraux:2016wxj,Baggioli:2016pia}. One could also couple in a Horndeski fashion the axionic fields to the curvature tensors (see, for example,~\cite{Baggioli:2017ojd}).} %
\begin{equation}
\mathcal{S}=\int d^4x\sqrt{-g}\left[{R\over 2}-\Lambda-\mathcal{W}(X,Z,F^2)\, \right]\,,
\label{effaction}
\end{equation}
where $\mathcal{W}$ is a generic scalar function.
Expanding the action~\eqref{effaction} to leading order in the field strength $F$ and ignoring non-scalar couplings between the various sectors, the generic expression~\eqref{effaction} reduces to~\eqref{action} which is general enough to discuss all the important physical feature of the holographic axion model. Self-consistency of action~\eqref{action} imposes precise constraints on the scalar functions $Y(X,Z)$ and $V(X,Z)$. An analysis of the transverse fluctuations showed that we should require that $V'(\bar{X},\bar{Z})>0$, $Y(\bar{X},\bar{Z})>0$ and $Y'(\bar{X},\bar{Z})<0$ to avoid ghosty instability~\cite{Baggioli:2016oqk}. Note that in more complicated backgrounds, for instance, turning on an external magnetic field, the constraints on $V$ and $Y$ will become tighter but still equivalent to impose the positivity of the electric conductivity~\cite{An:2020tkn}. 

Now, let us explain the physical interpretation of the $V$-term and $YF^2$-term in (\ref{action}) from the point of view of the dual field theory side, respectively.

\begin{itemize}
    \item Setting $Y(X,Z)=0$, the system is neutral. The axions configuration~\eqref{linear} breaks the spatial translations explicitly (as the simple axion model) or spontaneously (which is the focus of this section). In analogy to the EFT description~\eqref{EFT1}, $V(X)$ provides an effective description for solids holographically, while $V(Z)$ is related to fluids and we will come to this later.
    \item The coupling $YF^2$ can be viewed as the holographic dual of some charged disorders or charge lattices, depending on the form of $Y(X,Z)$. In the SSB pattern, it might be viewed as an analogy to charge density waves (CDWs). The simple linear axion model behaves like a metal. But the presence of such a  coupling  can significantly change the charge transport of the system and finally a metal-insulator transition (MIT) may come as the result.
\end{itemize}

We shall compare the differences of the low energy spectrum in solids and fluids in subsection~\ref{solidfluidsec}, and we will systematically investigate the charge transport and MIT in section~\ref{MITsection}.

\subsection{From explicit breaking to spontaneous breaking}
We continue by considering a simpler solid action of the type:
\begin{equation}
\mathcal{S}=\int d^4x\sqrt{-g}\left[{R\over 2}-\Lambda-m^2\,V(X)\, \right],
\label{solid}
\end{equation}
which reduces to the linear axion model~\cite{Andrade:2013gsa} for $V(X)=X$. As always, we will fix the background solution for the axion fields to be $\phi^I=x^I$. It is now important to analyze what this background solution means from the dual field theory point of view. This argument has been originally discussed in~\cite{Alberte:2017oqx}. Considering for simplicity a monomial potential\,\footnote{The argument could be actually generalized to any potential $V(X)$ where $N$ is the leading power in the expansion of $V(X)$ close to the boundary $X\rightarrow 0$.} $V(X)=X^N$, the expansion of the scalar fields close to the boundary $u=0$ takes the general form:
\begin{equation}
\phi(u,t,x)\,=\,\phi_0(t,x)\,\left(1+\dots\right)+\,\phi_1(t,x)\,u^{5-2N}\,\left(1+\dots\right)\,.\label{expa}
\end{equation}
Now, sticking to the standard quantization procedure\,\footnote{See~\cite{Armas:2019sbe,Ammon:2020xyv} for discussions about the alternative quantization possibility and implementation.}, the leading term in such expansion has to be identified with an external source for the operator $\mathcal{O}$ dual to the bulk field $\phi$, while the subleading term with its expectation value $\langle \mathcal{O} \rangle$. Therefore,
 \begin{itemize}
     \item for $N<5/2$ (\emph{e.g.} the linear axion model~\cite{Andrade:2013gsa}) the leading term in the expansion~\eqref{expa} is given by a constant in $u$ term and consequently $\phi^I_0(t,x)=x^I$. This is equivalent to say that we are introducing into our field theory an $x$-dependent source and therefore breaking translations explicitly.
     \item For $N>5/2$ (\emph{e.g.} the models considered in~\cite{Alberte:2017oqx}),  the story is reversed and the constant term $\phi^I=x^I$ is this time an $x-$dependent expectation value of our dual field theory, breaking therefore translational invariance spontaneously with
     \begin{equation}
        \langle \mathcal{O}(t,x) \rangle\,=\,x^I\,.
     \end{equation}
\end{itemize}
\begin{figure}
\centering
\includegraphics[width=\linewidth]{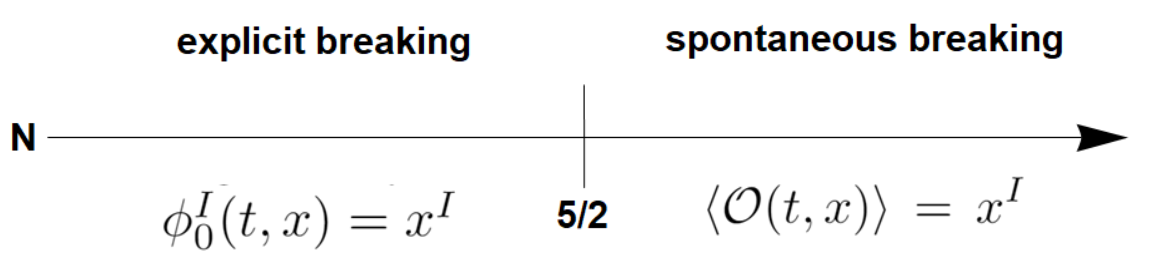}
\caption{The different symmetry breaking patterns depending on the power of the potential $V(X)=X^N$.}
\label{fig:break}
\end{figure}
In summary, the idea is that the bulk solution for the axions always break translations in the dual field theory, but the nature of this breaking is uniquely (up to the quantization scheme chosen) determined by the boundary asymptotic expansion, which can be modified by considering different bulk actions (see Fig.~\ref{fig:break}). In this review, we will focus on the original ideas of~\cite{Alberte:2017oqx} described above. Nevertheless, introducing more bulk fields (\emph{e.g.} dilaton, gauge field, \dots), it is possible to achieve the SSB in different ways. See~\cite{Amoretti:2017frz,Li:2018vrz,Li2021} for more details.\,\footnote{Spontaneously generated inhomogeneous lattices for density wave phases, such as charge density wave and pair density wave, can be found, \emph{e.g.} in~\cite{Ling:2014saa,Cremonini:2016rbd,Cremonini:2017usb,Cai:2017qdz}.}
 
\subsection{Elastic black holes}
A key difference between solids and fluids is that solids are resistant against shear deformations while fluids are not. Then, the excitations moving inside a solid are waves propagating in an elastic medium.

\begin{figure}
\centering
\includegraphics[width=0.85 \linewidth]{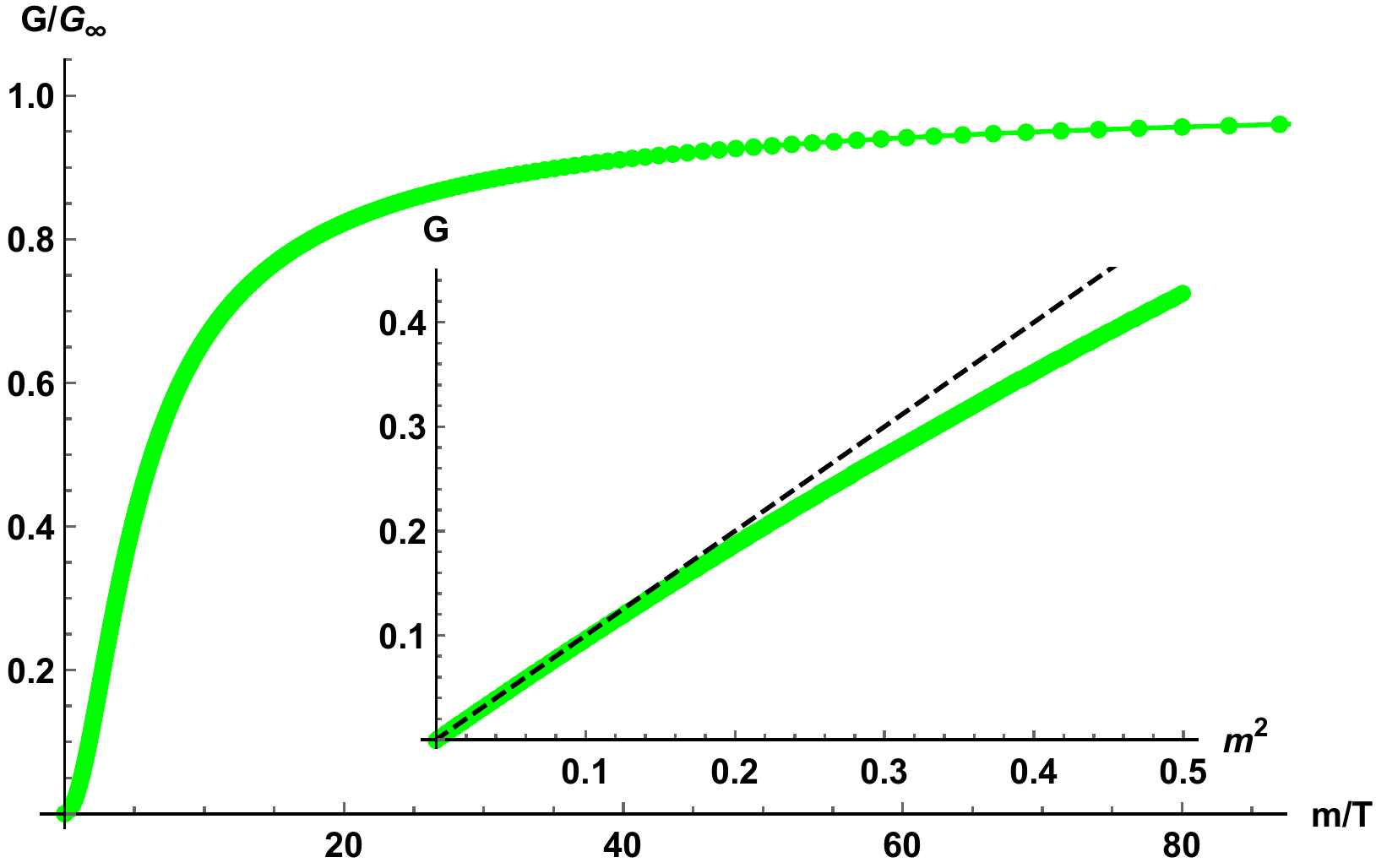}
\caption{The shear modulus $G$ normalized by its zero temperature value $G_\infty\equiv G(m/T\rightarrow \infty)$ as a function of the dimensionless quantity $m/T$. The dashed line comes from the analytic formula~\eqref{modulus} for small values of $m$.}
\label{fig:shearmodulus}
\end{figure}

To see why the background solution given by the model (\ref{solid}) is dual to some elastic medium, we look at the spin-$2$ perturbations $h_{xy}$ which encodes the information about the Green's function of the stress tensor $T_{xy}$ on the boundary. Interestingly, the shear equation is massive:
\begin{equation}
\Box \,h^{x}_{y}\,=\,M^2(u)\,h^{x}_{y}\,,
\end{equation}
where the effective mass of graviton becomes
\begin{equation}
u^2\,M^2(u)=m^2\,V_X(\bar{X})\,,
\end{equation}
and $V_X\equiv d V/dX$. Near the AdS boundary, we have the following expansion,
\begin{equation}
h^{x}_{y}\,=\,h_{(0)}(1+\dots)\,+\,h_{(3)}u^3(1+\dots)\,,
\end{equation}
where $h_{(0)}$ and $h_{(3)}$ are $u$-independent coefficients. 
Imposing the infalling condition at the horizon and fixing the leading coefficient $h_{(0)}$, this differential equation can be solved numerically, or even analytically for small $\omega$ and $m^2$ in Fourier space by using the perturbative methods~\cite{Alberte:2016xja,Baggioli:2019rrs}. According to the holographic dictionary, the Green's function of the stress tensor reads
\begin{equation}
\mathcal{G}^{(R)}_{T_{xy}T_{xy}}=\frac{3}{2}\frac{h_{(3)}}{h_{(0)}}\,,
\end{equation}
up to a contact term. In the low frequency expansion, we obtain that
\begin{equation}
\mathcal{G}^{(R)}_{T_{xy}T_{xy}}(\omega)\Big|_{k=0}=G-i\,\omega\,\eta+\mathcal{O}(\omega^2)\,,
\end{equation}
where $G$ is the shear modulus and $\eta$ the shear viscosity. In the massive gravity case, the non-zero effective mass brings a non-trivial contribution to the real part of the Green's function. As a result, for small $m$, this gives
\begin{equation}\label{eqforG}
G=m^2\,\int^{u_h}_{0}\frac{V_X(u^2)}{u^2}\,du+\mathcal{O}(m^4)\,.
\end{equation}
Choosing $V(X)=X^N$, we get
\begin{equation}\label{modulus}
G=\frac{N}{2N-3}\,u_h^{2N-3}\,m^2 +\mathcal{O}(m^4)\,.
\end{equation}
Then, it is clearly seen that the dynamical stability of the system requires $N>3/2$. And for all SSB cases, $N>5/2$ ensures the existence of a solid state that is dynamically stable. For general values of $m/T$, we plot the numeric data in Fig.~\ref{fig:shearmodulus}. 

\subsection{Holographic phonons}\label{subsec:phonon}
In the next, we turn to study the low energy excitations in the SSB pattern of translations. In holography, the spectrum of various excitations can be read by computing the quasi-normal modes (QNMs) of the black hole~\cite{Berti:2009kk}.

Note that in the translationally invariant case (which is simply the Schwarzschild black hole here) there are two sound modes in the longitudinal channel that are related to the fluctuations of energy density $\delta\,\varepsilon^\parallel$ as well as the momentum $\delta\,\pi^\parallel$. On the contrary, in the transverse channel, there is only one diffusive momentum mode $\delta\,\pi^\perp$ whose diffusion constant relates to the minimal shear viscosity, $\eta/s=1/4\pi$. 

The case of explicit breaking of translations with $N=1$ has already been analyzed in subsection~\ref{co-in trans}. The momentum relaxation destroys the longitudinal sound which becomes diffusive (in the hydrodynamic regime). Moreover the shear diffusive mode is pushed downwards along the imaginary axis to form the Drude pole.

\begin{figure}
\centering
\includegraphics[width=0.62 \linewidth]{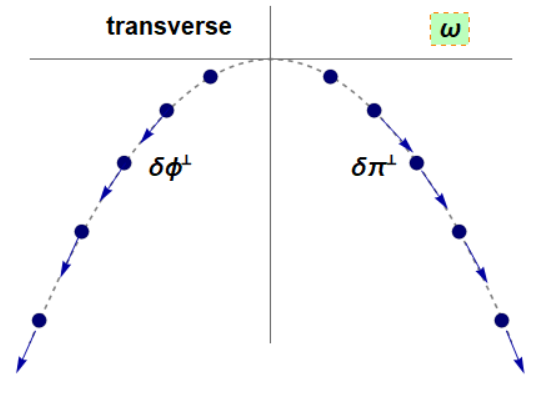}
    
\vspace{0.5cm}
    
\includegraphics[width=0.62 \linewidth]{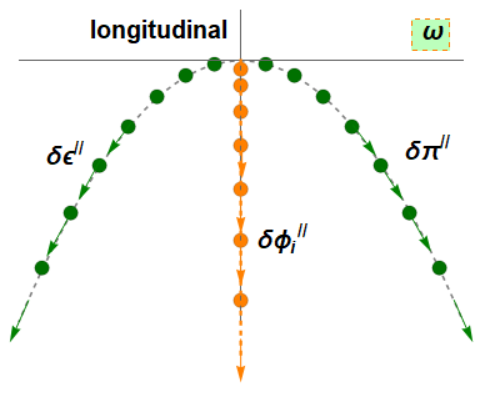}
\caption{The spectrum of hydrodynamic modes can be read from the QNMs of the black hole. \textbf{Top:} Gapless modes propagating at the sound speed $v_T$ in the transverse channel, which are related to the transverse phonons and transverse momentum in the dual field theory. \textbf{Bottom:} Gapless sound modes propagating at the speed $v_L$ in the longitudinal channel, related to the longitudinal phonons and longitudinal momentum. In addition, there is an unexpected diffusive mode(orange dots), which we will call it crystal diffusion mode hereafter.}
\label{fig:SSBmodes}
\end{figure}

When the translations are broken spontaneously, there exist massless Goldstones in the low energy description which are the acoustic phonons. In the holographic axion model with $N>5/2$, the two modes $\delta\,\varepsilon^\parallel$ and $\delta\,\pi^\parallel$ still remains sound like, albeit propagating at an enhanced speed comparing to that in fluids. We call them \textbf{longitudinal phonons}. Furthermore, there is an extra longitudinal diffusive mode -- \textbf{crystal diffusion}. We will explore the physical nature of this mode later, in subsection~\ref{phasonsection}. Of the most interest to us here are the two sound modes emerging in the transverse channel which is related to $\phi^\perp$ and $\pi^\perp$. We call them \textbf{transverse phonons}. All of these can be clearly seen in Fig.~\ref{fig:SSBmodes}. It is well-known that transverse sounds can never survive inside a fluid (at low momentum). The appearance of shear sounds again reflect the fact that the dual boundary system under study is a solid. 

\begin{figure}[htp]
\begin{center}
\includegraphics[width=0.63\linewidth]{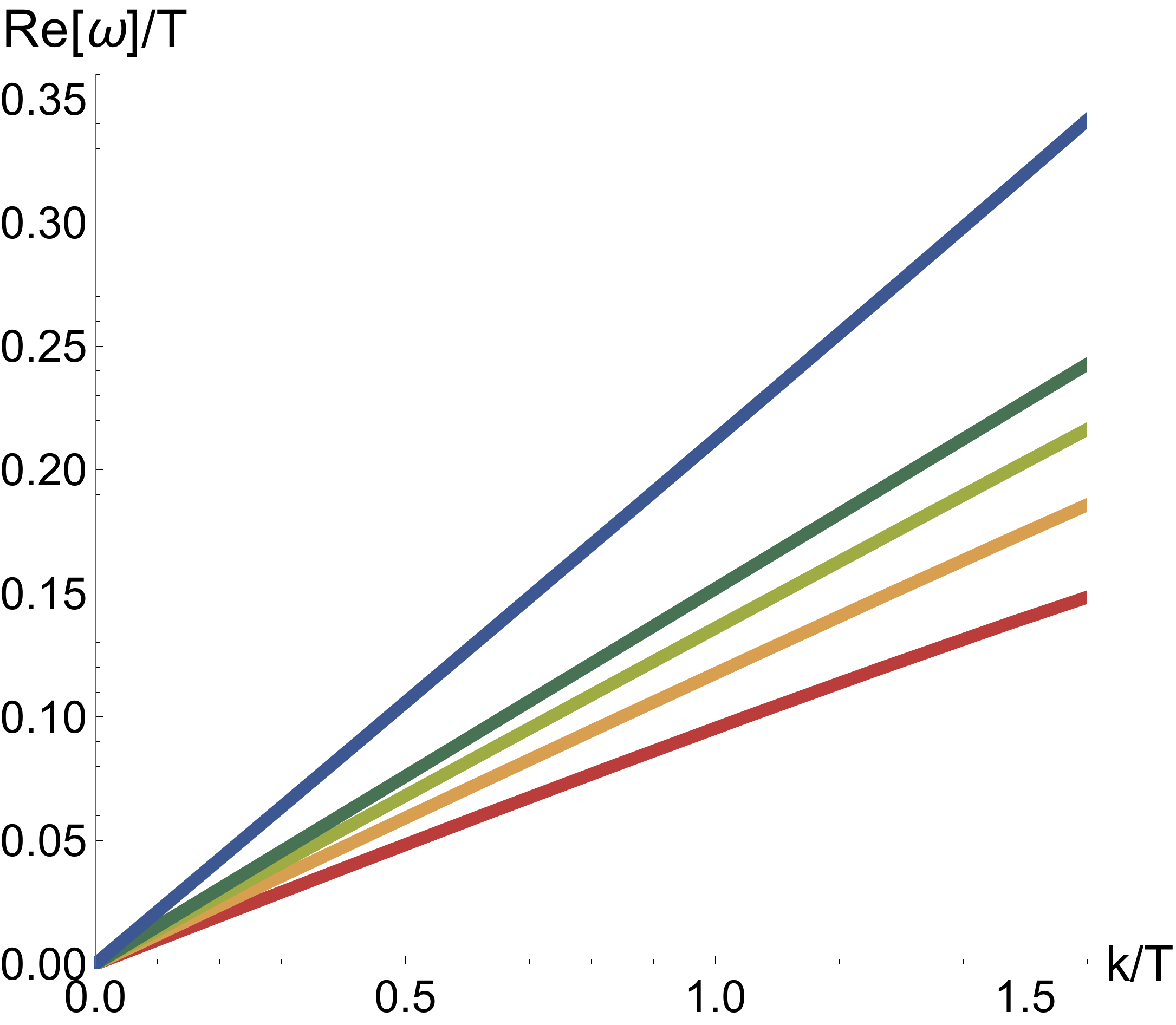}

\vspace{0.4cm}

\includegraphics[width=0.65\linewidth]{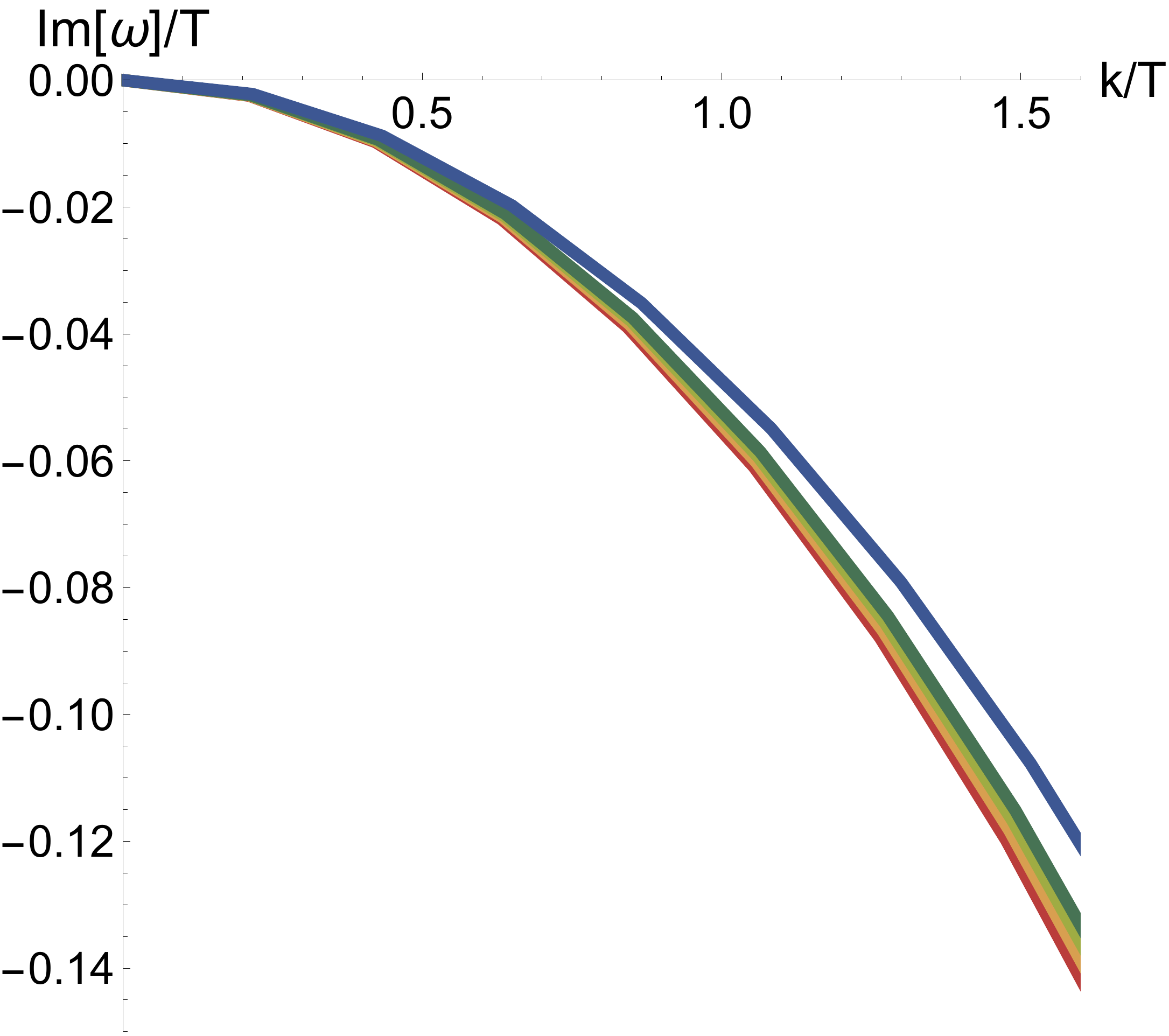}
\caption{The dispersion relation of the transverse phonons in the holographic axion model with $V(X)=X^5$. $m/T$ increases from the red line to the blue one. Figure taken from~\cite{Alberte:2017oqx}.}
 \label{fig:ph1}
\end{center}
\end{figure}

\begin{figure}
\centering
\includegraphics[width=0.7\linewidth]{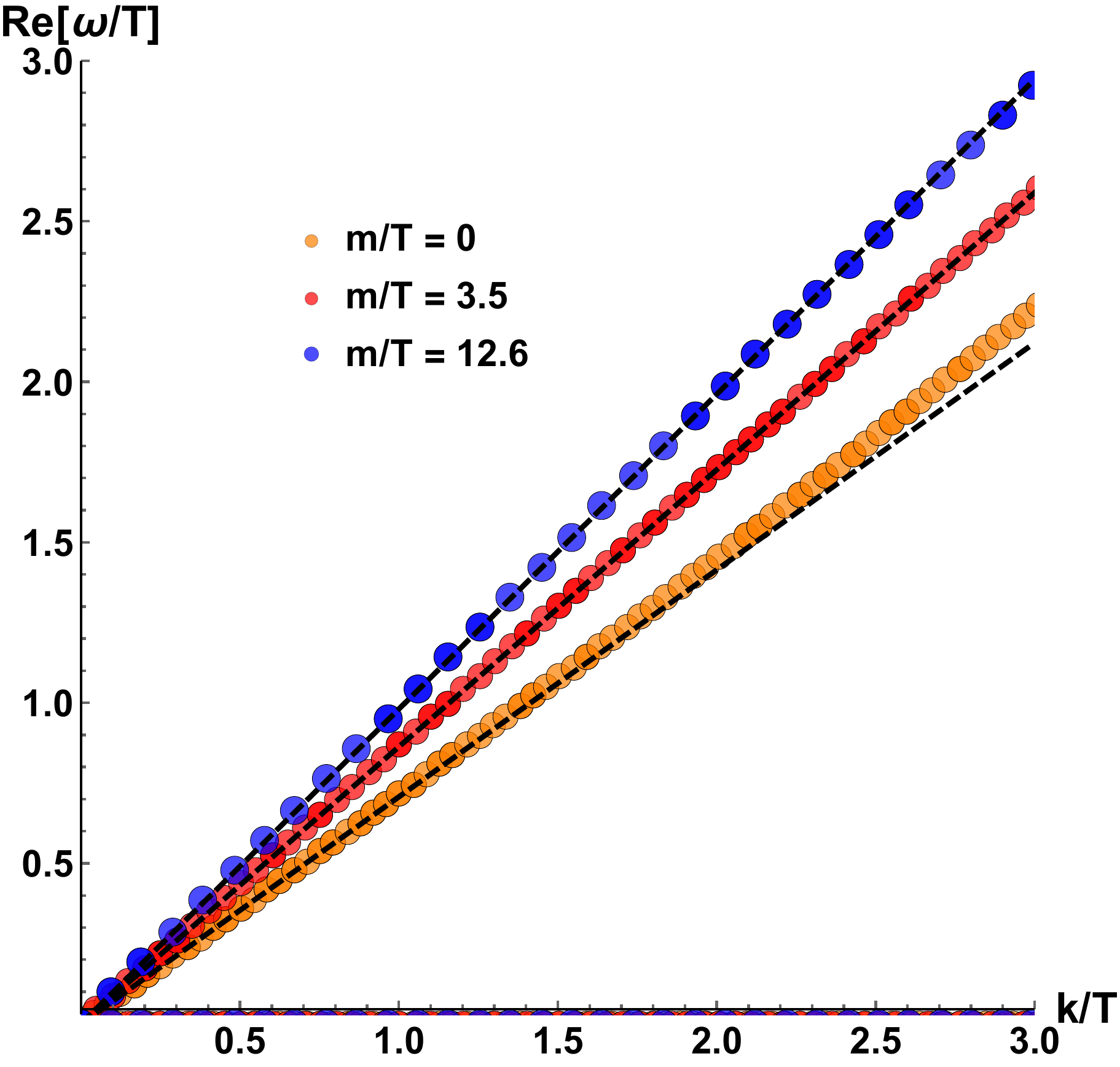}
\caption{The real part of the dispersion relation of the longitudinal phonons in the holographic axion model. Figure taken from~\cite{Ammon:2019apj}.}
\label{fig:ph2}
\end{figure}

For $N=3$, we have plotted the dispersion relations for the transverse and longitudinal phonons in Fig.~\ref{fig:ph1} and Fig.~\ref{fig:ph2}, respectively. Our results show that at leading order in $k$,
\begin{equation}
\omega_{L,T}\,=\,\pm\,v_{L,T}\,k\,-\,\frac{i\,\Gamma_{L,T}}{2}\,k^2+\dots\,.
\label{sounddisp}
\end{equation}
Besides the linear dispersion, there is also an attenuation factor due to the background viscosities. This is a dissipative term due to finite temperature effects and can be formulated in the standard hydrodynamic approach, but dealing with it in the framework of EFT is challenging. The exact forms of $\Gamma_{T,L}$ will be discussed in section~\ref{hydrodynamics}.

\begin{figure}
\centering
\includegraphics[width=0.85\linewidth]{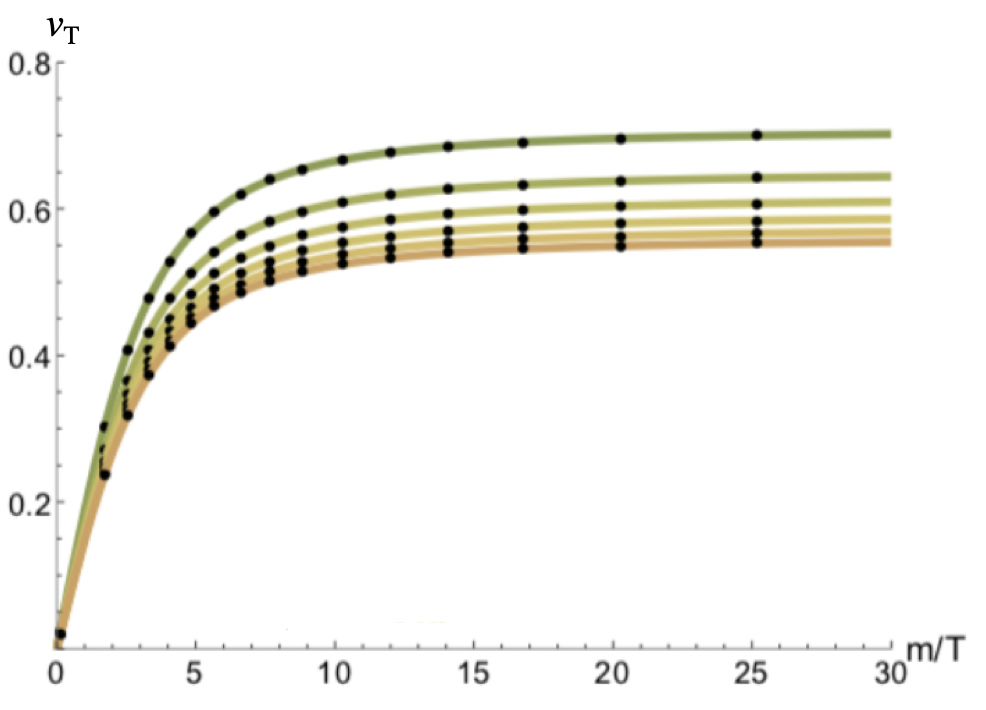}
\caption{$v_T$ extracted from QNMs (black dots) and computed by the formula $v_T=\sqrt{G/\chi_{\pi\pi}}$ from elasticity (solid lines) for $n\in[3,8]$ (green-orange). Figure taken from~\cite{Alberte:2017oqx}.}
\label{fig:Tvelocity}
\end{figure}

One can further check that the numerical data from the holographic model are in perfect agreement with the prediction of the elasticity theory, \emph{i.e.}
\begin{equation}
v_{T}=\sqrt{\frac{G}{\chi_{\pi\pi}}},\,\,\,\,\,\,\,\,\,v_{L}=\sqrt{\frac{K+G}{\chi_{\pi\pi}}}\,.
\end{equation}
Here, the momentum susceptibility $\chi_{\pi\pi}\equiv \frac{\delta T_{ti}}{\delta v^i}=\mathcal{E}+\mathfrak{p}$. One can see a comparison of $v_T$ extracted from the QNMs and the prediction of the elasticity theory in Fig.~\ref{fig:Tvelocity}.

Finally, in a conformal solid, $v_L$ and $v_T$ are not independent of each other. One can verify this by explicitly computing the bulk modulus which is given by $K=\frac{3}{4}\mathcal{E}$~\cite{Ammon:2019apj} or using the EFT method of conformal solids~\cite{Esposito:2017qpj}. As a result, we have that
\begin{equation}
v_{L}^2=\frac{1}{2}+v_{T}^2\,,
\end{equation}
which represents a further validity check for the holographic model. 

\subsection{Zoology of solids and fluids}\label{solidfluidsec}
Let us move to a (reduced) model with the following action
\begin{equation}
\mathcal{S}=\int d^4x\sqrt{-g}\left[{R\over 2}-\Lambda-m^2\,V(Z)\, \right],\,\,\,\,\,V(Z)=Z^n\,,
\label{fluids}
\end{equation}
and compare its hydro-spectrum with that of the $V(X)$ model in previous subsection~\ref{subsec:phonon}. Note that since $V_X=0$ in this case, the spin-2 graviton is massless in contrast to the solid model~\cite{Alberte:2015isw} (see also Eq.~\eqref{eqforG}). Then, the (static) shear modulus $G$ vanishes and the system is not resisting anymore to static shear deformations. This reflects the fact that the $V(Z)$ model is related to a fluid system. One can further check that the action \eqref{fluids}, with generic potential $V(Z)$, remains unchanged under the VPD transformation~\eqref{tt}. Analyzing the UV expansion of the axion fields, it is found that, in this case, to have SSB, we should require that $n>5/4$.

\begin{figure}
\centering
\includegraphics[width=0.75\linewidth]{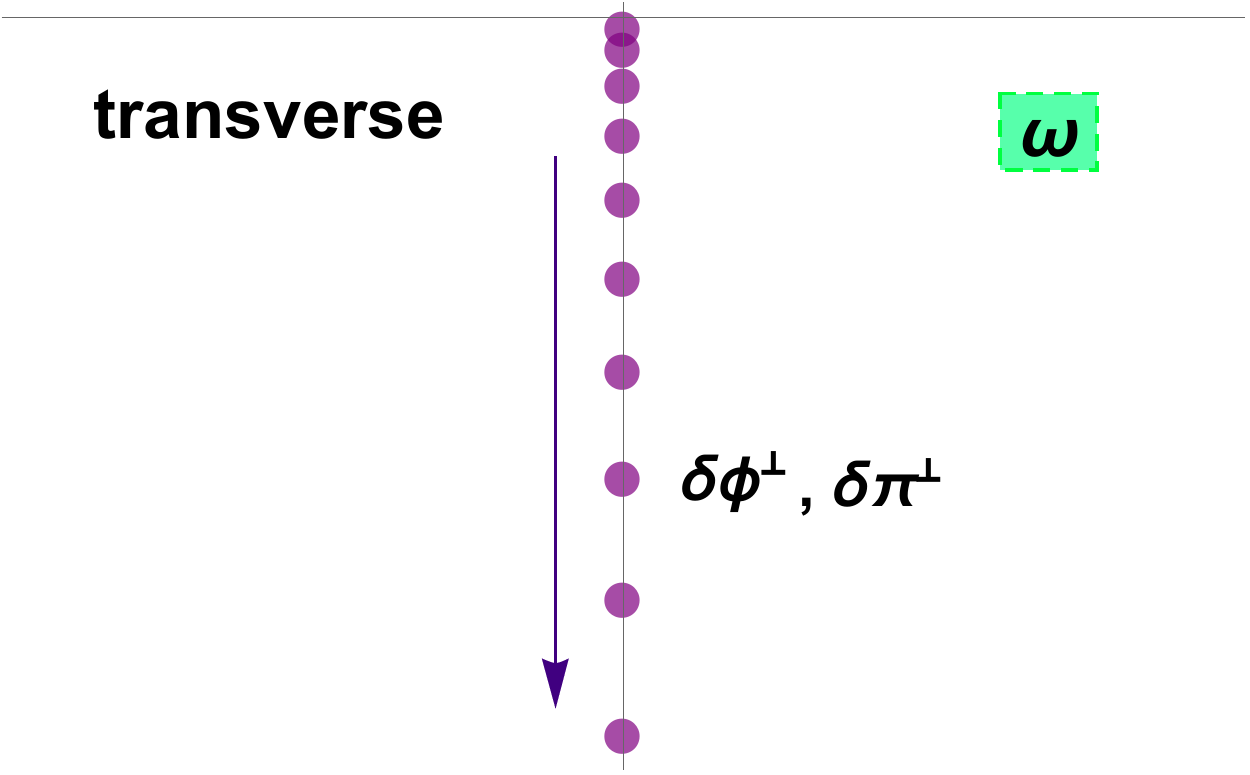}
\caption{The hydrodynamic diffusive modes in the transverse of the holographic fluid model~\eqref{fluids}.}
\label{fig:transdiffu}
\end{figure}

The absence of $G$ means that there are no propagating phonons in the transverse channel. Then, the leading behavior of the dispersion relation~\eqref{sounddisp} becomes diffusive, \emph{i.e,} we have
\begin{equation}
\omega_{T}=-i\,D_T\,k^2+\dots,\,\,\,\,\,\,D_T=\frac{\eta}{\chi_{\pi\pi}}+\dots\,,
\end{equation}
where $\eta=\frac{s}{4\pi}$. See Fig.~\ref{fig:transdiffu} for illustration.

The longitudinal spectrum of fluids share the similar features as those of solids: there are two sounds and one crystal diffusion mode. Since $G=0$, the longitudinal sound speed is now given by
\begin{equation}
v_{L}=\sqrt{\frac{K}{\chi_{\pi\pi}}}\,.
\end{equation}
In the present model, $K=\frac{3}{4}\mathcal{E}$ and $\chi_{\pi\pi}=\frac{3}{2} \mathcal{E}$. It turns out to be
\begin{equation}
v_{L}=\frac{1}{\sqrt{2}}\equiv v_c\,,
\end{equation}
where in the last step we introduce the conformal value of sound speed which is defined by
\begin{equation}
v_c\equiv \frac{1}{\sqrt{d-1}}\,,
\end{equation}
for general $(d-1)$ spatial dimensions. 

For a much more detailed discussion of the holographic fluid models see~\cite{Baggioli:2019abx}.

In conclusion, the holographic homogeneous models with axions provide us a simple effective description for a wide class of solids as well as fluids with no translational invariance, perfectly capturing the viscoelastic property of the system and the expected spectrum of low energy excitations, etc. So far, we have not examined how the system will be influenced in presence of finite charge density or EXB of translations. These problems will be discussed in section~\ref{pinned} and subsection~\ref{MITsection}.

\subsection{The dual view}
So far, we have focused on \textit{bona-fide} axion models in which the common ingredient was the presence of a set of shift invariant scalar fields with background profile $\phi^I\sim x^I$. In terms of this construction, it is almost straightforward to implement the physics of momentum dissipation and explicit breaking of translations but it is much harder and less intuitive to obtain the theory of elasticity and the dynamics of the SSB.

In order to achieve this second task, it might be convenient to use a \textbf{dual picture} in which the scalar fields $\phi^I$ are substituted by a set of \textbf{two-forms} $J^I_{\mu\nu}$. This is what has been put forward in~\cite{PhysRevD.97.106005} and later re-utilized in~\cite{Armas:2019sbe}. The idea is very interesting and it originates from the study of generalized higher form symmetries~\cite{Gaiotto:2014kfa} in analogy to the electromagnetism case~\cite{Grozdanov:2016tdf}. Let us go back to the field theory description of elasticity. We can re-introduce our set of scalar fields $\Phi^I$, labelling the co-moving coordinates and providing a preferred frame for us. The dynamics of these field is simply governed by the conservation of momentum:
\begin{equation}
\partial_\mu \,P^\mu_I\,=\,0\,,\quad P^\mu_I\,=\,C^{\mu\nu}_{IJ}\,\partial_\nu \Phi^J\,, \label{cons1a}
\end{equation}
where $C^{ij}_{IJ}$ is the elastic tensor and $C^{tt}_{IJ}=\varrho\,\delta_{IJ}$ with $\varrho$ being the mass density. Eq.~\eqref{cons1a} is equivalent to the conservation of the stress tensor. Nevertheless, in a solid without defects, there is another hidden symmetry encoded in the conservation of a set of two-form currents:
\begin{equation}
\partial_{\mu_1}\,J_I^{\mu_1\,\dots\,\mu_d}\,=\,0\,,\quad  J_I^{\mu_1\,\dots\,\mu_d}\,=\,\epsilon^{\mu_1\dots \mu_d\,\nu}\,\partial_\nu \Phi^I\,.
\label{cons1}
\end{equation}
This conservation is somehow trivial if the $\Phi^I$ fields are single valued. It is a topological constraint and plays the role of the Bianchi identity.

All of these mean that the theory of elasticity can be formulated in a dual formalism where instead of considering the stress tensor $T^{\mu\nu}$ and the Goldstone modes $\Phi^I$ (together with their Josephson relation), one considers the conserved stress tensor and a conserved set of higher-forms $J_I^{\mu_1\dots \mu_d}$. The conservation of both these objects results into a new description of elasticity which recovers all the previously known results.

It is then immediate to translate this language into holography, by assuming a theory with a set of conserved two-form currents. The appropriate bulk action reads
\begin{equation}
S\,=\,\int d^4x \sqrt{-g}\,\left[R-2\Lambda-\frac{1}{12}\sum_I H_{I,abc}H^{abc}_I\right]\,,
\end{equation}
with $H=dB$ being the field strength of the two-form $B_{\mu\nu}$ dual to the operator $J_{\mu\nu}$ mentioned before. The local bulk gauge symmetry imposes immediately the conservation~\eqref{cons1}. Imposing the appropriate boundary conditions, we can show that such holographic model gives rise to the dual formulation of elasticity theory. A similar possibility, which is basically equivalent to that, is to work with the original scalar fields and impose alternative boundary conditions~\cite{Armas:2019sbe,Ammon:2020xyv}. Unfortunately, this second option results in bad instabilities and it has not been successfully employed.

This dual formulation has been studied only in~\cite{PhysRevD.97.106005} and it definitely deserves more attention in the near future.

\section{On the hydrodynamic description}\label{hydrodynamics}
\subsection{A puzzle}\label{IVA}
\begin{figure}[ht]
\centering
\includegraphics[width=0.7\linewidth]{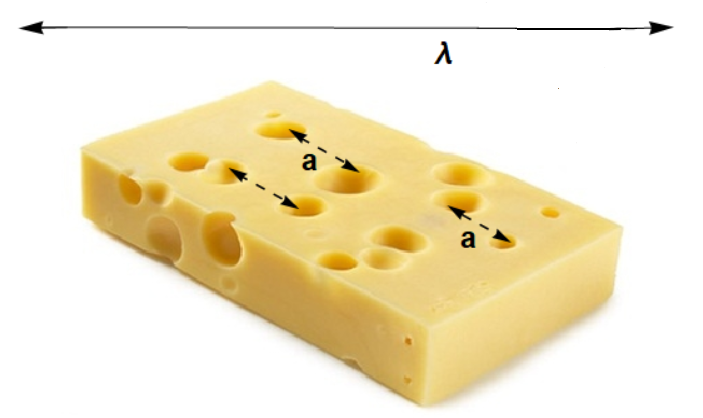}
\caption{The hydrodynamic limit $\lambda \gg a$, with $a$ being the characteristic microscopic scale and $\lambda$ the length-scale at which we are probing out system. This regime is equivalent to the standard small momentum regime $k/T \ll 1$.}
\label{fig:hydro1}
\end{figure}

\textbf{Hydrodynamics} is a universal effective field theory which describes the late-time and large-scale dynamics of any physical systems (see Fig.~\ref{fig:hydro1}). It is expected to be valid at low frequency and momentum and it represents a continuum description which clearly breaks down when the microscopic characteristic scale of the system is reached (\emph{e.g.} the lattice spacing in solids or the inter-molecular distance in liquids~\cite{Baggioli:2020loj}).

Despite the disorientating oxymoron, a \textbf{hydrodynamic theory of solids} (not to be confused with fluid-dynamics in the sense of Navier-Stokes equations) has been derived several decades ago \cite{PhysRevA.6.2401} (see also~\cite{PhysRevLett.41.121}). The interest about a hydrodynamic theory of solids has re-appeared more recently in the context of systems with no quasiparticles, for which the underlying Galilean invariance is obviously gone. In particular, a precise study  of hydrodynamics in presence of explicit and spontaneous breaking of translations has been done in~\cite{Delacretaz:2017zxd} following some previous discussions regarding the role of such theory for the phenomenology of bad metals~\cite{Delacretaz:2016ivq} (see also~\cite{Amoretti:2019buu} for a preliminary attempt to connect it with experimental data).

The main new aspect in building a hydrodynamic theory for solids is the introduction of additional degrees of freedom -- the Goldstone modes (the phonons). The formalism appears to be very similar to that required to construct superfluid hydrodynamics, with the only difference that the Goldstone mode here is not associated to an internal $U(1)$ symmetry but rather to translational invariance.

Neglecting the presence of a finite charge density, the hydrodynamics is governed by the conservation of the stress tensor $\partial_\mu T^{\mu\nu}=0$ (unless any explicit breaking source is introduced) and by the Josephson equation for the Goldstone mode which simply corresponds to
\begin{equation}
    \dot \Phi\,=\,\left[H\,,\Phi\right]\,.
\end{equation}
Following the standard Martin-Kadanoff procedure~\cite{KADANOFF1963419}, one can extract directly the Green functions of the system via Kubo formulas and the corresponding hydrodynamic excitations. Neglecting the details of these computations, which can be found in~\cite{Delacretaz:2017zxd}, the final hydrodynamic spectrum of a solid contains:
\begin{align}
& \textbf{transverse sound}\,\,\,:\,\,\omega\,=\,\pm\,v_T\,k\,-\,\frac{i\,\Gamma_T}{2}\,k^2\,+\,\dots\,, \\
& \textbf{longitudinal sound}\,\,\,:\,\,\omega\,=\,\pm\,v_L\,k\,-\,\frac{i\,\Gamma_L}{2}\,k^2\,+\,\dots\,, \\
& \textbf{crystal diffusion}\,\,\,:\,\,\omega\,=\,-\,i\,D_\phi\,k^2\,+\,\dots\,. 
\end{align}
The first two sets of modes are the standard phononic sound modes which are now attenuated with the characteristic diffusive-like damping due to the viscosity of the system. The third mode is more interesting and maybe unusual. We will discuss in much more detail the physical nature of this mode in Section~\ref{phasonsection}.
\begin{figure}[ht]
\centering
\includegraphics[width=0.8\linewidth]{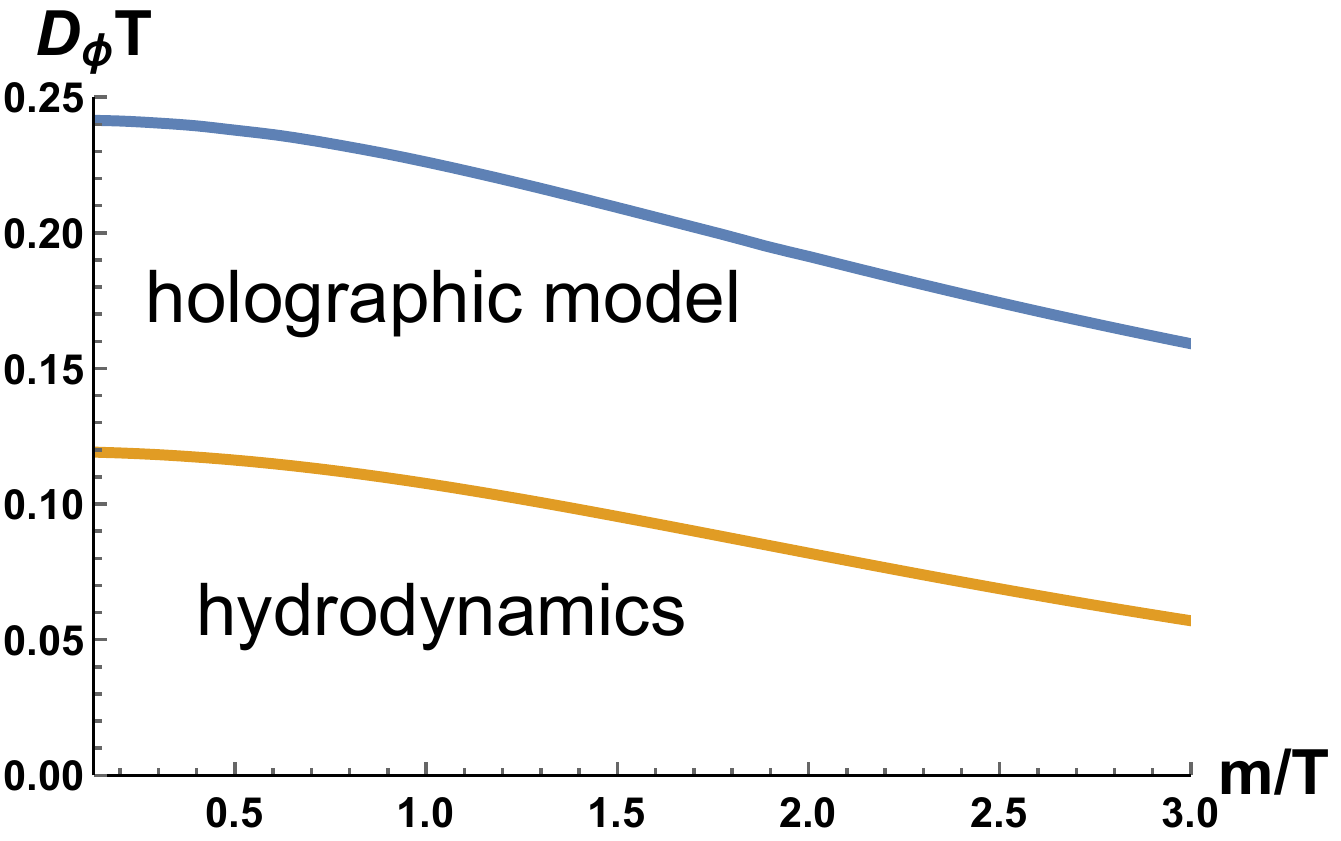}
\caption{The discrepancy between the hydrodynamic theory of~\cite{Delacretaz:2017zxd} and the holographic model of~\cite{Alberte:2017oqx} reported in~\cite{Ammon:2019apj}. The diffusion constant of the longitudinal crystal diffusion mode is denoted as $D_\phi$.}
\label{fig:discr}
\end{figure}

The hydrodynamic theory indicates a concrete expression for the diffusion constant $D_\phi$ which at leading order reads
\begin{equation}
D_\phi\,=\,\left(G+\,K\right)\,\xi\,+\,\dots\,,
\label{diffhydro}
\end{equation}
where $G,K$ are respectively the shear and bulk elastic moduli. The new transport coefficient $\xi$ is a dissipative term which controls the Goldstone diffusion and which appears in the Goldstone's two-points function:
\begin{equation}
    \mathcal{G}_{\Phi^I\Phi^J}\,=\,\left(-\,\frac{1}{\omega^2\,\chi_{\pi\pi}}\,+\,\xi\,\frac{i}{\omega}\right)\,\delta^{IJ}\,+\,\dots\,.
\end{equation}
All the transport coefficients $(G,K,\xi)$ can be computed independently using the corresponding Kubo formulas at strictly zero momentum $k=0$. On the contrary, the dispersion relation of the hydrodynamics modes can be obtained via a more complicated numerical computation of the QNMs of the system at finite momentum (see \cite{Grieninger:2020wsb} for details).

The comparison between the two results was performed for a large class of holographic axion models in~\cite{Ammon:2019apj} and presented a surprising outcome. The numerical data, extracted from the dispersion relation of the crystal diffusion mode, were not well described by the hydrodynamics formula~\eqref{diffhydro}. As evident from Fig.~\ref{fig:discr}, the hydrodynamics prediction is completely off with respect to the numerical holographic data. To be more concrete, the hydrodynamic framework of~\cite{Delacretaz:2017zxd} does not correctly describe the low-energy physics of the holographic axion models~\cite{Alberte:2017oqx}. What is happening and what causes this \textbf{discrepancy}?

\subsection{Strain pressure and its resolution}
In order to understand the discrepancy discussed in the previous subsection, we have to re-consider the hydrodynamic description of~\cite{Delacretaz:2017zxd} in more detail. This was done in~\cite{Armas:2019sbe} (and later~\cite{Armas:2020bmo} for the charged case)  using a slightly different formalism which we will follow in this section.

The fundamental ingredients of the hydrodynamic theory are the fluid velocity $u^\mu$, temperature $T$, and translation Goldstone bosons $\Phi^I$. To proceed, we define the one-form $e^I_\mu = \partial_\mu\Phi^I$, the crystal metric tensor $h^{IJ} = e^I_\mu e^{J\mu}$, $e_{I\mu} = h^{-1}_{IJ}e^J_\mu$, $h_{\mu\nu} = h^{-1}_{IJ}e^I_\mu e^J_\nu$, and the strain tensor $u_{\mu\nu} = \frac12 (h^{-1}_{IJ} - \delta_{IJ}/\alpha^2) e^I_\mu e^J_\nu$, where $\alpha$ is simply a constant. The constitutive relations for an isotropic viscoelastic medium are 
\begin{align}
\label{eq:consti}
T^{\mu\nu}
&= \left(\mathcal{E} + \mathfrak{p}
+ T\mathcal{P}' u^{\lambda}{}_{\!\!\lambda} \right) u^\mu u^\nu
+ \left( \mathfrak{p} + \mathcal{P} u^{\lambda}{}_{\!\!\lambda} \right) \eta^{\mu\nu}
+ \mathcal{P} h^{\mu\nu} \nn\\
- \eta\, &\sigma^{\mu\nu}
- \zeta\, P^{\mu\nu} \partial_\rho u^\rho
- 2G\, u^{\mu\nu}
- (\mathfrak{K}-G)\, u^{\lambda}{}_{\!\!\lambda} h^{\mu\nu}\,,
\end{align}
together with the thermodynamic identities $\mathrm{d}\mathfrak{p} = s\, \mathrm{d}T$, $\mathcal{E} = Ts - \mathfrak{p}$ and $P^{\mu\nu} = \eta^{\mu\nu} + u^\mu u^\nu$. Here, $\mathfrak{K}$ is the part of the total bulk modulus depending on the SSB strength -- its ``solid'' contribution -- not to be confused with the total bulk modulus $K=-V d\langle T^{xx}\rangle/dV$. Additionally, $\sigma^{\mu\nu} = 2P^{\rho(\mu}P^{\nu)\sigma}\partial_{\rho}u_{\sigma} - P^{\mu\nu} \partial_\rho u^\rho$ is the standard fluid shear tensor encoding the dissipative/viscous part of the response, while $\eta$ and $\zeta$ are shear and bulk viscosities. The most important and new parameter entering here is the \textbf{strain pressure} $\mathcal{P}$ with $\mathcal{P}'=\partial_T \mathcal{P}$.

The dynamics of the system is governed by the stress-energy tensor conservation:
\begin{equation}
\partial_\mu T^{\mu\nu}\,=\,0\,,
\end{equation}
and by the Josephson's relation:
\begin{equation}
\label{eq:config}
u^\mu e^I_\mu= \frac{h^{IJ}}{\Sigma}\partial_\mu \left( \mathcal{P} e^{\mu}_J\,{-}\, (K-G) u^{\lambda}{}_{\!\!\lambda} e^\mu_J 
\,{-}\, 2G u^{\mu\nu} e_{J\nu}\right),
\end{equation}
where $\Sigma$ is a dissipative coefficient characteristic of spontaneously broken translations.
\begin{figure}[ht]
\centering
\includegraphics[width=0.8\linewidth]{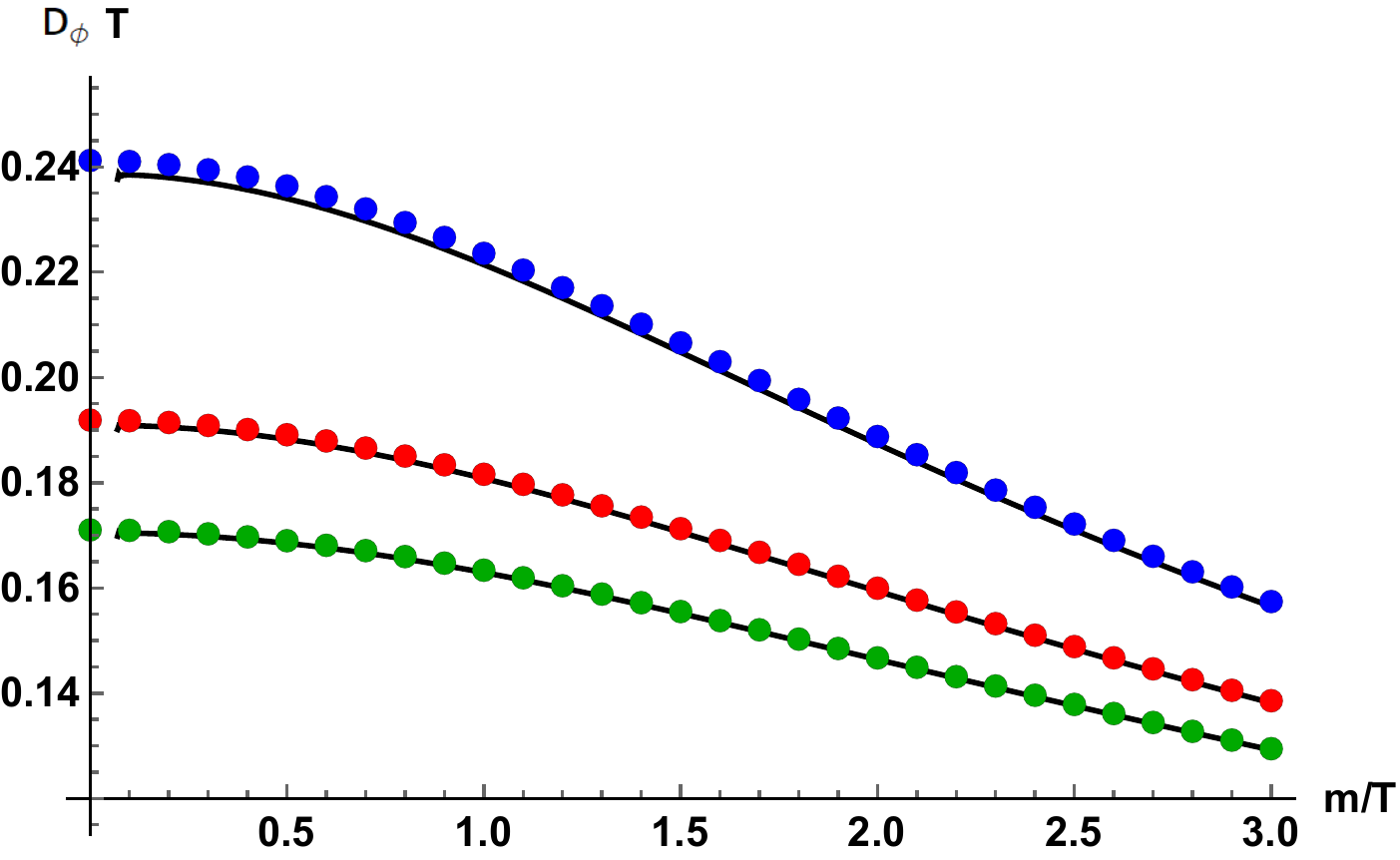}
\caption{The comparison of the crystal diffusion constant predicted by hydrodynamics~\eqref{eq:modes} with the holographic results. The previous discrepancy is now successfully resolved. Figure taken from~\cite{Ammon:2020xyv}.}
\label{match}
\end{figure}

We expand the equations above around an equilibrium state with $u^\mu = \delta^\mu_t$, $T=T_0$, and $\Phi^I = \alpha\, x^I$, and we obtain the following hydrodynamic modes:
\begin{equation}\label{disp}
\omega = \pm v_{\parallel,\perp} k -\frac{i}{2}\Gamma_{\parallel,\perp} k^2+ \ldots\,,\quad 
\omega = -iD_\phi k^2 + \ldots\,,
\end{equation}
including two sets of propagating sound modes and a longitudinal diffusive mode. The various coefficients entering in the dispersion relations are given by
%
\begin{equation}\label{eq:modes}
\begin{split}
 &  v_\perp^2= \frac{G}{\chi_{\pi\pi}}\,,\qquad v_\parallel^2 
    = \frac{(s+\mathcal{P}')^2}{s'\chi_{\pi\pi}}
    + \frac{K+G-\mathcal{P}}{\chi_{\pi\pi}}\,,\\
   &\Gamma_\perp
  = \frac{\eta}{\chi_{\pi\pi}}
  +\frac{G}{\sigma}\frac{s^2 T^2}{\chi_{\pi\pi}^2}\,,\quad  D_\phi
  = \frac{s^2}{\sigma s'}
  \frac{K+G-\cP}{\chi_{\pi\pi}v_\parallel^2}\,,\\
   &\Gamma_\parallel
  = \frac{\eta+\zeta}{\chi_{\pi\pi}}
  +\frac{T^2s^2v_\parallel^2}{\sigma\chi_{\pi\pi}}
  \left(1 - \frac{s+\cP'}{Ts'v_\parallel^2}\right)^2 \,.
\end{split}
\end{equation}
%
The various transport coefficients can be obtained using linear response approach via the following Kubo's formulas:
\begin{equation} \label{Kubo123}
\begin{split}
    &\mathcal{E} = \langle T^{tt} \rangle\,, \quad
    \mathfrak{p} = -\,\Omega\,, \quad
    \mathcal{P} = \langle T^{xx} \rangle + \Omega\,,\\
    &\chi_{\pi\pi}v_\parallel^2 = \lim_{\omega\to0}\lim_{k\to0}
    \mathrm{Re}\,\mathcal{G}^R_{T^{xx}T^{xx}}\,,\\
    &G = \chi_{\pi\pi}\,v_\perp^2 = \lim_{\omega\to0}\lim_{k\to0}
    \mathrm{Re}\,\mathcal{G}^R_{T^{xy}T^{xy}}\,, \\
    &\eta = -\lim_{\omega\to0}\lim_{k\to0}
    \frac{1}{\omega}\mathrm{Im}\,\mathcal{G}^R_{T^{xy}T^{xy}}\,,\\
    & \frac{(\mathcal{E}+\mathfrak{p})^2}{\Sigma\,\chi_{\pi\pi}^2} 
    = \lim_{\omega\to0}\lim_{k\to0}
    \omega\,\mathrm{Im}\,\mathcal{G}^R_{\Phi^{x}\Phi^{x}}\,.
\end{split}
\end{equation}
where $\Omega$ is the free energy.

Additionally, in presence of conformal invariance (which is typical of the holographic models considered in this review), we have the following constraints:
\begin{equation}
\mathcal{E} = 2(\mathfrak{p}+\cP)\,,\quad 
T\cP' = 3\,\cP-2\,K\,, \quad \zeta = 0\,.
\end{equation}
The hydrodynamic relations~\eqref{eq:modes} match perfectly the numerical data for the holographic axion model (see Fig.~\ref{match}). Therefore, we can confidently say that the hydrodynamic theory of~\cite{Armas:2019sbe} is the correct low energy effective description of the holographic axion model of~\cite{Alberte:2017oqx}.

Where did the hydrodynamic theory of~\cite{Delacretaz:2017zxd} fail and why? It failed for two reasons. First and most importantly, the holographic models have a finite strain pressure $\mathcal{P}=\langle T^{xx}\rangle-\mathfrak{p}$, which was not taken into account in~\cite{Delacretaz:2017zxd}. This term was neglected because in all ground states which are thermodynamically stable it must be zero~\cite{Donos:2013cka}. Unfortunately, the generic solution of the holographic axion model is not a preferred solution from the thermodynamic point of view -- it is equipped with a background strain.

\begin{figure}[ht]
\centering
\includegraphics[width=0.8\linewidth]{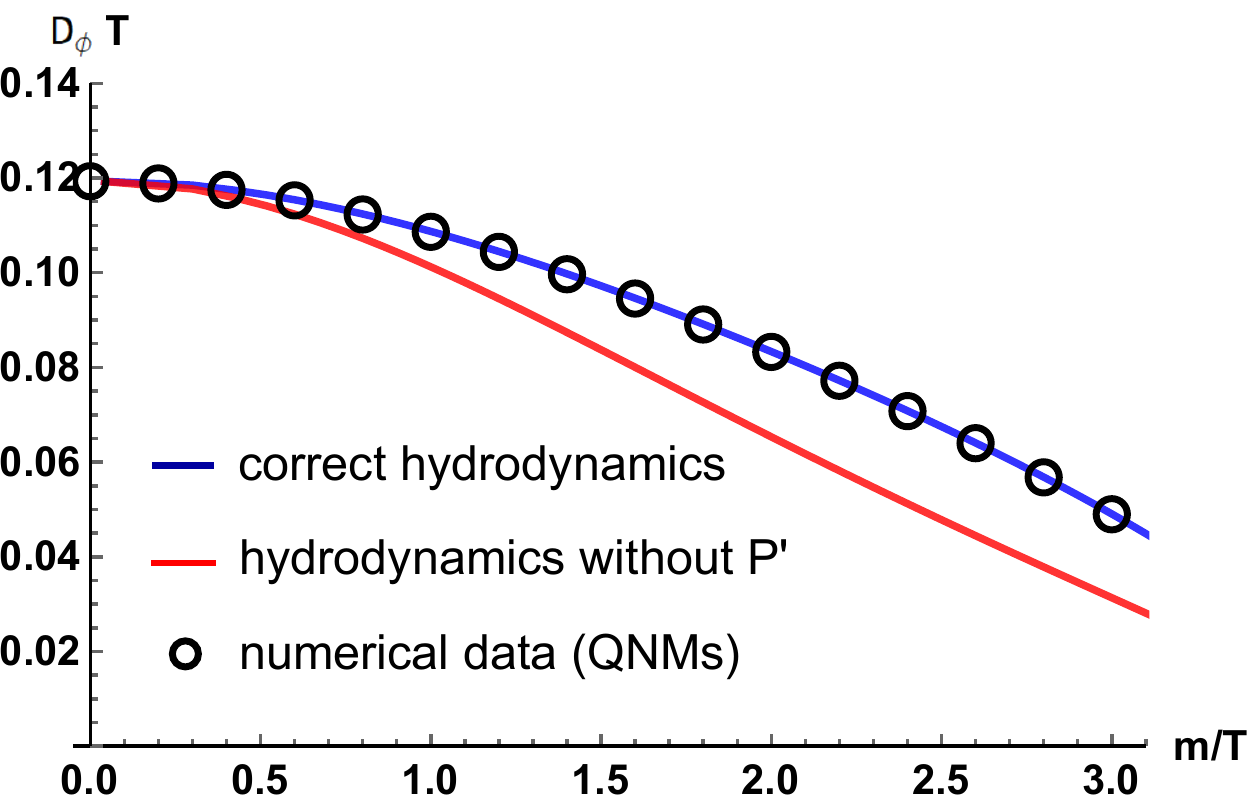}
\caption{The proof that the description of~\cite{Delacretaz:2017zxd} is still inaccurate even for holographic models with zero strain pressure --thermodynamically favourable. Figure taken from~\cite{Ammon:2020xyv}.}
\label{incorrect}
\end{figure}

With some fine tuning, one could nevertheless set the strain pressure $\mathcal{P}=0$ by choosing a specific potential for the axion fields~\cite{Ammon:2020xyv}. Even in that case, the predictions of~\cite{Delacretaz:2017zxd} are incorrect (see Fig.~\ref{incorrect}). The reason, this time, is that the authors of~\cite{Delacretaz:2017zxd} have (consciously) neglected some off-diagonal susceptibilities playing the role of $\mathcal{P}'$, which are fundamental to match the holographic data and cannot be discarded.

Fortunately, when all the correct terms are considered, the predictions from hydrodynamics are in perfect agreement with the holographic results for the axions model. This constitutes a further proof of the solidity and validity of the holographic axion model as the gravity dual of a strongly coupled viscoelastic medium.

\subsection{The hydrodynamics of phonons}
After having discussed at length the hydrodynamic description of the holographic axion model with spontaneously broken translations, it is timely to give a concrete example of the success of such description. In particular, we can focus for simplicity on the dispersion relation of the transverse phonons. As we have already repeated several times, at small momentum the transverse phonons exhibit a linear dispersion relation of the type:
\begin{equation}
\omega_\pm\,=\,\pm\,v_T\,k\,-\,i\,\frac{\Gamma_T}{2}\,k^2\,,\qquad v_T^2\,=\,\frac{G}{\mathcal{E}+\mathfrak{p}}\,.
\end{equation}
This dynamics was successfully confirmed in the seminal work of~\cite{Alberte:2017oqx}. Nevertheless, the hydrodynamic framework is more powerful than that. In particular, following the methods of~\cite{Delacretaz:2017zxd}, one could extend the dispersion relation of the phonons at higher momenta obtaining:
\begin{equation}
\omega_{\pm}=-\frac{i}{2}\,k^2\,\left(\xi\,+\,\frac{\eta}{\chi_{\pi\pi}}\right)\,\pm\,k\,\sqrt{v_T^2\,-\,\frac{k^2}{4}\,\left(\frac{\eta}{\chi_{\pi\pi}}\,-\,\xi \right)^2}\,,
\label{hydroformula}
\end{equation}
in which the meaning of all the parameters has been already explained in the previous section.

\begin{figure}
\centering
\includegraphics[width=0.7\linewidth]{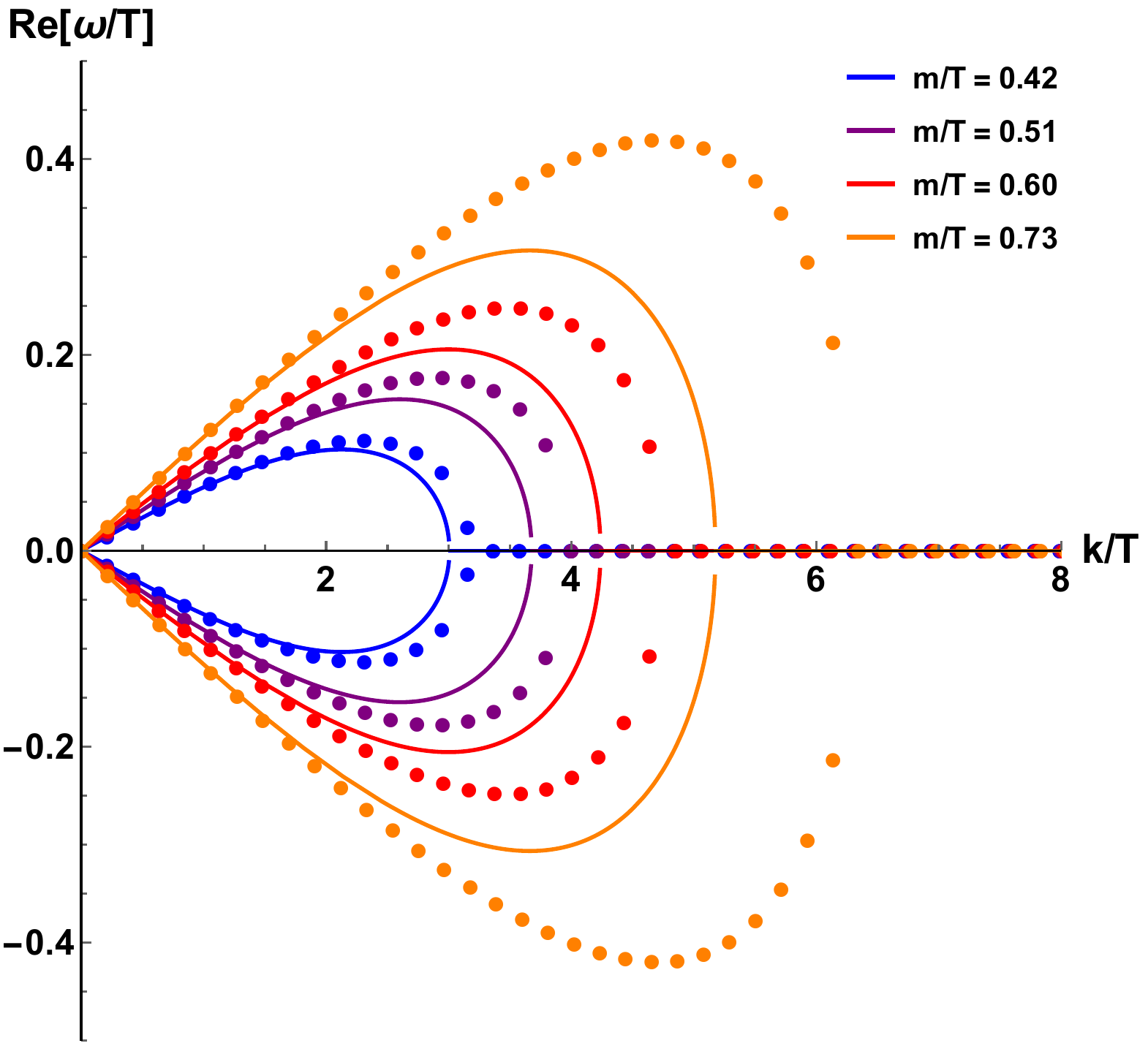}
    
\vspace{0.5cm}
    
\includegraphics[width=0.7 \linewidth]{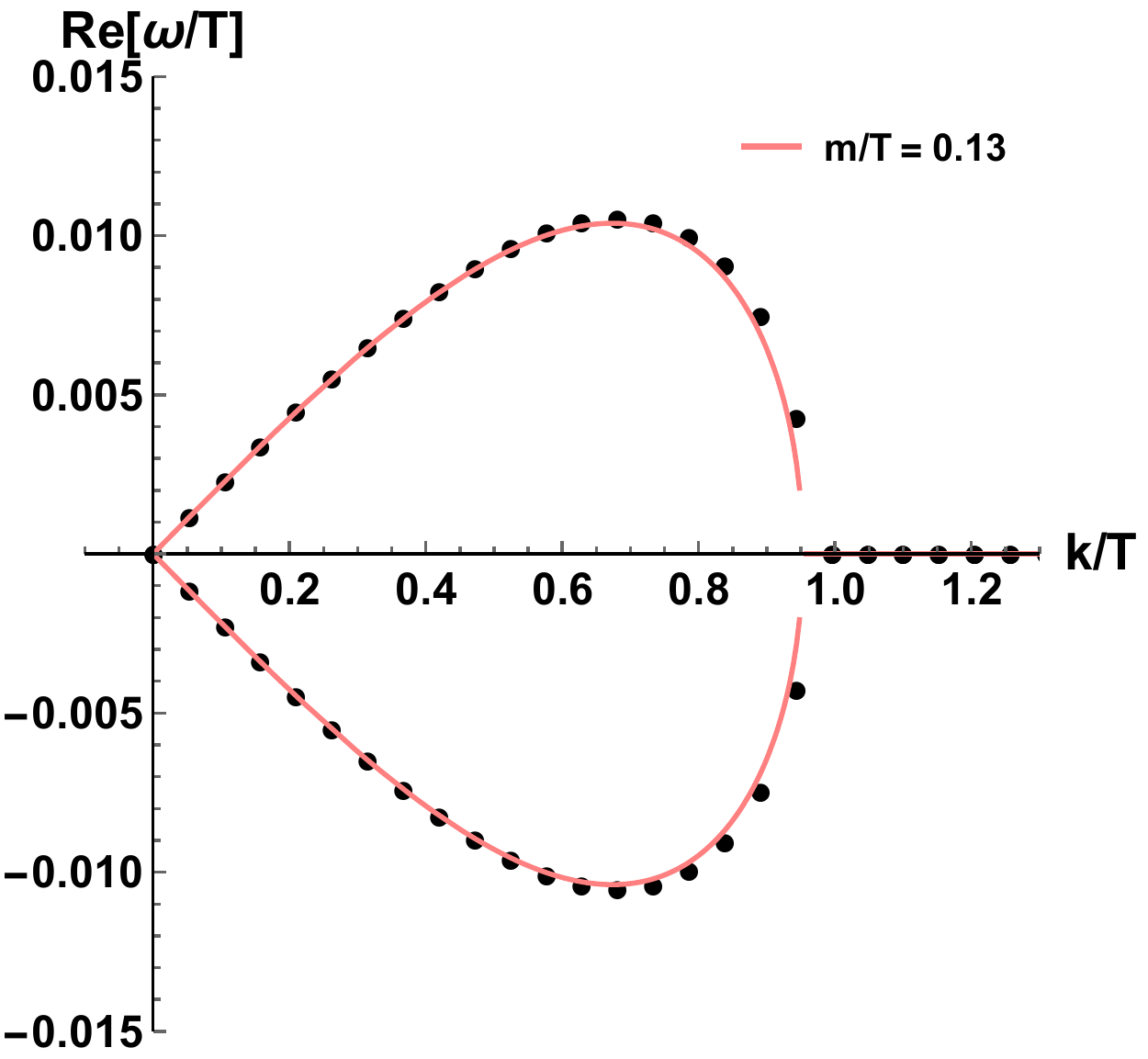}
\caption{The comparison between the hydrodynamic formula \eqref{hydroformula} (solid lines) and the holographic results (dots). Figure taken from~\cite{Ammon:2019wci}.}
\label{comp1}
\end{figure}

The comparison with the holographic results is shown in Fig.~\ref{comp1} and it is very successful~\cite{Ammon:2019wci}. In particular, the agreement becomes better and better at low $m/T$. Interestingly, at least for small $m/T$, there is a critical momentum at which the real part of the phonons dispersion relation goes to zero. This seems to indicate a soft mode instability which is so far not totally understood. It is fair to say that this feature is surely related to the viscoelastic nature of the system and it resembles closely the idea of Ioffe-Regel crossover~\cite{PhysRevB.61.12031,taraskin2002vector,beltukov2013ioffe,Shintani2008}.

\subsection{Zero strain pressure and stability}
As explained in detail in the previous subsection, the presence of a finite strain pressure $\mathcal{P}$ is equivalent to say that the model is not in a thermodynamically favorable phase, which would require $\langle T^{xx}\rangle =\mathfrak{p}$~\cite{Donos:2013cka}. This is tantamount to say that the model does not describe a ground state, but rather an excited one. Despite the equivalence between linear dynamic stability and thermodynamic stability is far from obvious, one would then expect these models to be unstable. Nevertheless, such instability was never found and all the linear hydrodynamics modes are well-behaving~\cite{Alberte:2017oqx,Ammon:2019apj,Baggioli:2019abx,Ammon:2020xyv,Baggioli:2020edn}. This opens the path towards several options: (I) these models are stable (and then one has to understand why), (II) the instability is suppressed by the large $N$ limit and it would re-appear when $1/N$ corrections are considered, (III) the instability appears only at non-linear level (despite it was not seen in~\cite{Baggioli:2019mck}) and (IV) the instability is driven by non-homogeneous modes which are not captured by the standard perturbative analysis.

\begin{figure}[ht]
\centering
\includegraphics[width=0.85\linewidth]{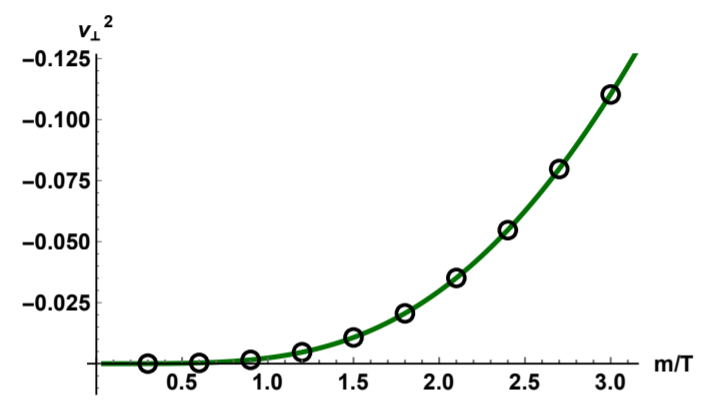}
\caption{The linear instability for the axions model $V(X)=X+X^2/2$ with zero strain pressure $\mathcal{P}=0$. Plot taken from~\cite{Ammon:2020xyv}.}
\label{uns}
\end{figure}

The story is even funnier! One could ``cook-up'' very fine tuned models where the strain pressure is vanishing~\cite{Amoretti:2017frz,Armas:2019sbe,Ammon:2020xyv}. In these cases, the solution is a real thermodynamic ground state. Now everything should be fine and stable. Unfortunately, Nature is not so kind. All these cases suffer a terrible linear instability; the shear modulus is negative and as such the shear sound mode is unstable (see Fig.~\ref{uns}). It is hard to believe that this result is a pure coincidence and not a physical feature of models sustained by axion-like scalar fields $\phi^I=x^I$. The question is open!

\subsection{Phasons dynamics}\label{phasonsection}
\begin{figure}
\centering
\includegraphics[width=0.8\linewidth]{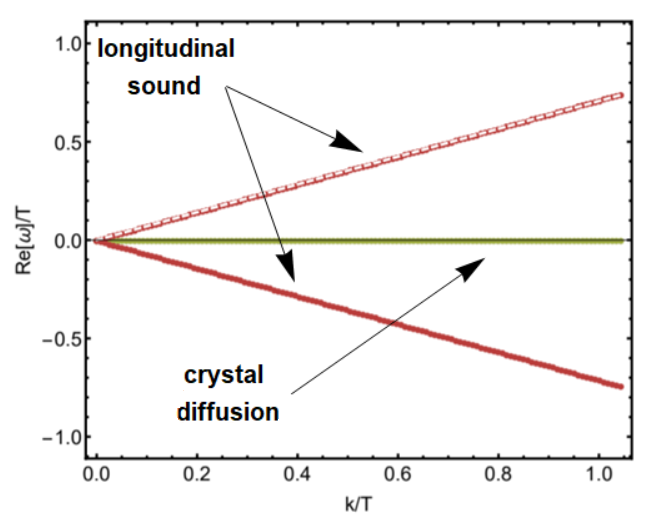}
    
\vspace{0.5cm}
    
\includegraphics[width=0.8\linewidth]{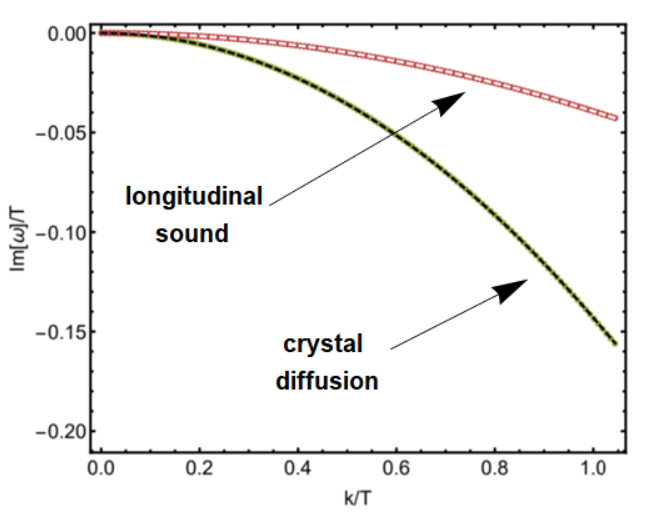}
\caption{The hydrodynamics modes in the longitudinal spectrum of one of the holographic axion model with SSB. The figure is taken from~\cite{Baggioli:2019abx}.}
\label{longfig}
\end{figure}

As emphasized in Section \ref{IVA}, in addition to the expected phonon modes, the holographic axion model contains an \textbf{extra longitudinal diffusive mode} -- \textit{crystal diffusion} (see Fig.~\ref{longfig}). The nature of this mode is very interesting and it has been the subject of a long standing (and still running) debate. 

Despite several hydrodynamic setups~\cite{PhysRevA.6.2401,Delacretaz:2017zxd,Armas:2019sbe} predicted an extra diffusive mode in the longitudinal sector, after the first holographic identification in \cite{Ammon:2019apj}, the interest around this excitation has rapidly increased. A turning point in the story has been put forward almost at the same time by Donos et Al.~\cite{Donos:2019txg} and Amoretti et Al.~\cite{Amoretti:2018tzw}. The \textit{crystal diffusion} mode is (I) a Goldstone mode, (II) coming from the spontaneous breaking of the \textbf{internal global shift symmetry} $\phi \rightarrow \phi+a$. This is surprising in several ways. First, what is a \textbf{diffusive Goldstone boson}? Second, this mode is totally unrelated to spacetime symmetries and their spontaneous breaking. What is this mode? Is it an artifact of the simplified assumption of homogeneity or not? 

Taking into account the results of~\cite{Donos:2019txg,Amoretti:2018tzw}, the first step is to understand what is the physical role of the internal space and shifts along that. A recent idea~\cite{Baggioli:2020nay} tried to connect the dynamics of this mode with the physics of \textbf{quasicrystals} -- systems with long range order but without periodicity. Quasicyrstals have a long and curious history~\cite{steinhardt2019second} and several reviews are available~\cite{janssen1988aperiodic,divincenzo1999quasicrystals,janot1997quasicrystals}. Curiously enough, just because they lack the periodicity of standard crystalline structure, these systems display an additional diffusive longitudinal mode which is known as \textbf{phason}. This mode appears both in the hydrodynamic description~\cite{PhysRevB.32.7444,PhysRevLett.49.468,PhysRevB.32.7412,Currat2002,PhysRevLett.49.1833,AGIASOFITOU2014923} and it is also observed directly in experiments~\cite{PhysRevLett.91.225501,durand1991investigation,QUILICHINI19831011}.

Using the superspace formalism~\cite{janssen2014aperiodic}, one can show that this mode arises indeed from an extra dynamics related to an internal symmetry of the system. In particular, from the formal mathematical point of view, any aperiodic structure in $d$ dimension can be seen as a periodic structure in $(d+n)$ dimensions cut at an irrational angle (see~\cite{Baggioli:2020nay,Baggioli:2020haa} for details). The phason mode is the Goldstone mode related to the rigid internal shift of this cut within the extra-dimensional picture and it is not related in any way to spacetime symmetries. This hypothesis seems to be confirmed by the fact that the full dynamics found in the holographic models can be re-obtained from an effective field theory of quasicrystals built using Keldysh-Scwhinger techniques~\cite{Baggioli:2020haa}. Finally, it was recently proven \cite{Andrade:2020hpu} that the dynamics of the phason is not peculiar of the homogeneous holographic models, but can be found also in more realistic inhomogeneous constructions. It is therefore a ``real'' physical feature and not an artifact of the description.

\subsection{The dynamics of shear waves with momentum dissipation}
\begin{figure}[h]
\centering
\includegraphics[width=0.7\linewidth]{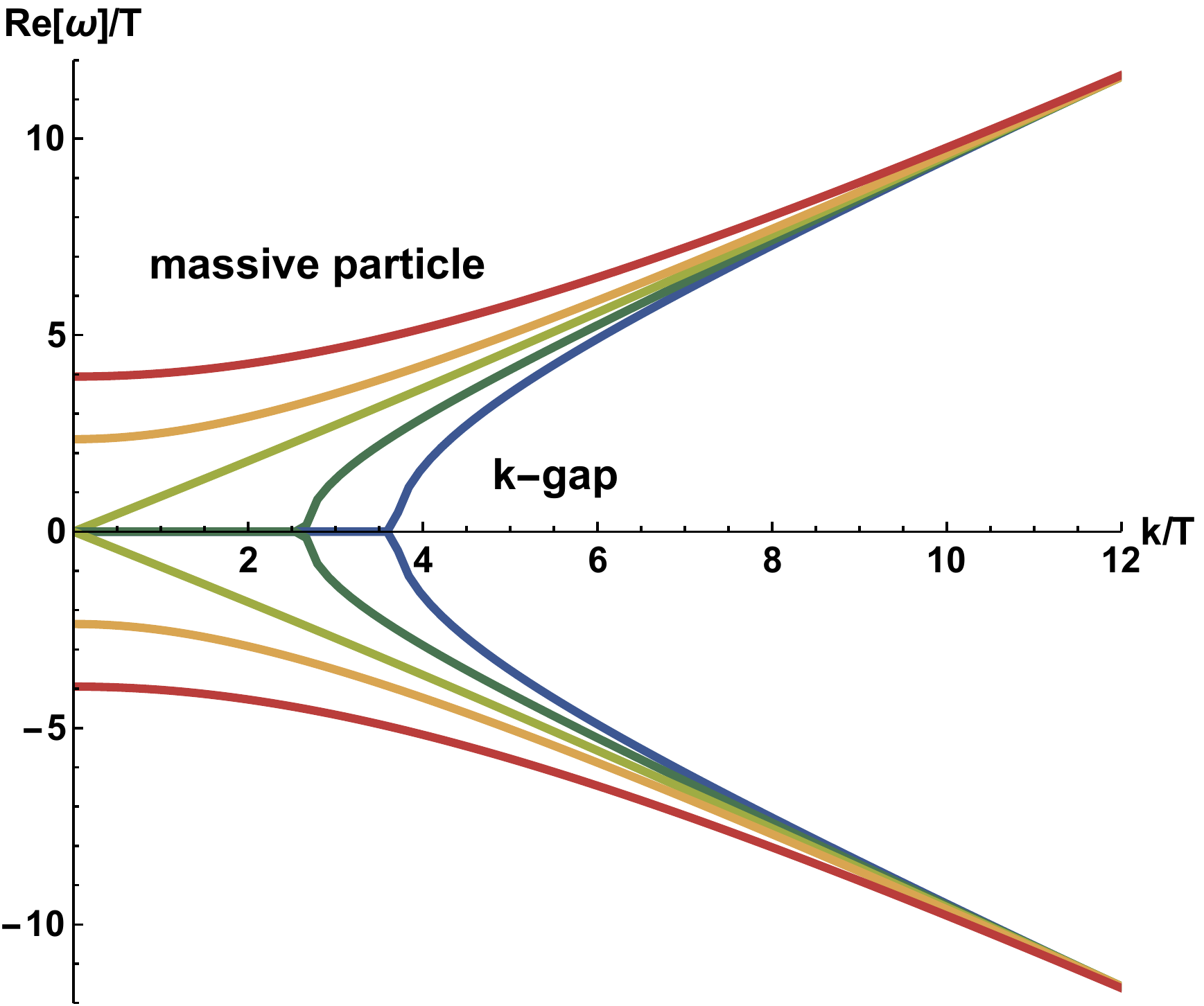}
    
\vspace{0.5cm}
    
\includegraphics[width=0.7\linewidth]{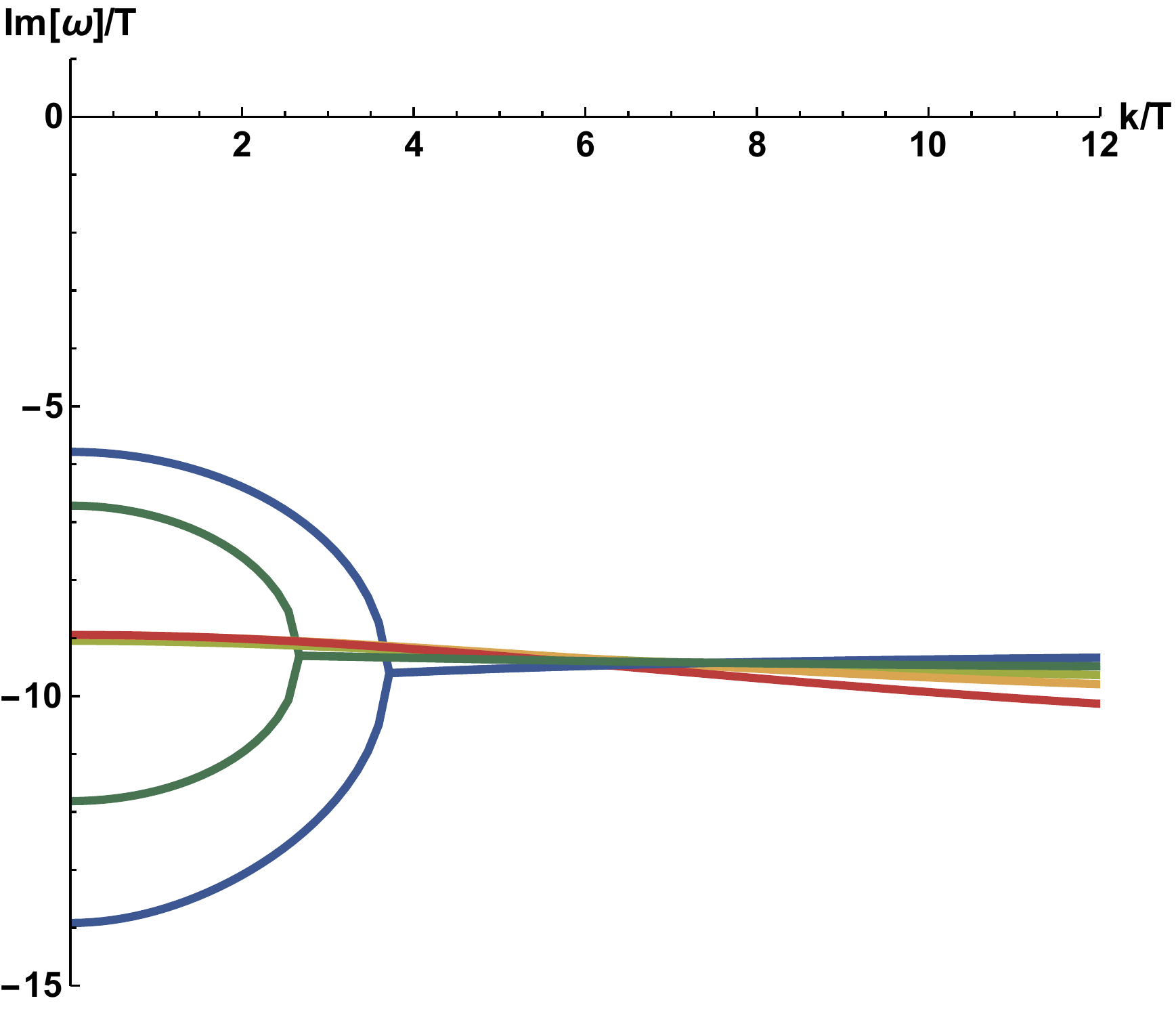}
\caption{The transverse spectrum of excitations for the linear model $V(X)=X$. $m/T$ increases from the blue line to the red one. Figures taken from~\cite{Baggioli:2018vfc}.}
\label{figlin}
\end{figure}

As we have argued in the previous subsections, the hydrodynamic description of the holographic axion model in the regime of spontaneous symmetry breaking is far from trivial and it keeps giving headache to the community. On the contrary, the low energy description of the models in the regime of explicit breaking is far way simple and it has been nicely described in~\cite{Davison:2013jba} (see also~\cite{Davison:2014lua}). The low-frequency dynamics is governed by a so-called \textbf{Drude pole}, $\omega=-i \Gamma$, which stems from the introduction of momentum dissipation. The parameter $\Gamma$ is indeed the momentum dissipation rate, the inverse of the relaxation time $\tau$. This result comes from the simple fact that the stress-energy tensor conservation is now modified at leading order into:
\begin{align}
& \partial_t\,T^{tt}\,=\,0\,,\\
&\partial_i\,T^{it}\,=\,-\,\frac{1}{\tau}\,T^{ti}\,,
\label{ward}
\end{align}
where the first line is just energy conservation while the second one is the relativistic version of the Drude equation \eqref{or1}. From a holographic perspective, the momentum relaxation rate is given by the graviton mass computed at the horizon:
\begin{equation}
\Gamma\,\sim\,m_g^2\,|_{\text{horizon}}\,.
\label{mass}
\end{equation}
Despite the dynamics at zero momentum is not surprising, the story becomes richer and more interesting looking at the dispersion relation of the low-energy transverse modes at finite momentum. The full dance of the modes is shown in fig.~\ref{figlin} and it has been analyzed in detail in~\cite{Baggioli:2018vfc,Baggioli:2018nnp}.

One obvious and evident feature is that the ``would be'' shear diffusion mode, which arises because of (transverse) momentum conservation, acquires now a finite lifetime given by the timescale $\tau$. This implies a modified dispersion relation of the type:
\begin{equation}
\omega\,=\,-\,i\,\Gamma\,-\,i\,D_\pi\,k^2\,+\,\dots\,,
\label{dru}
\end{equation}
where, in the holographic models, $\Gamma$ grows with the graviton mass as in Eq.~\eqref{mass}. Notice that this description is valid only when $\Gamma/T$ is small and therefore the leading order symmetry breaking effect can be described as in Eq.~\eqref{ward}.

Let us first have a look at the real part of the dispersion relation of the lowest modes. At small but finite $\Gamma$, the real part is zero until  a cutoff wave-number which we call \textbf{k-gap}, $k_g$. Above the k-gap, for $k>k_g$, the real part grows in a square root fashion and at very large $k$ it asymptotes a linearly dispersing mode. The emerging speed is given by the speed of light as required by the UV relativistic fixed point of the theory. Importantly, this feature arises because of the collision of the Drude pole \eqref{dru} with a second non-hydrodynamic mode. This collision is actually a manifestation of the coherent-incoherent transition described in~\cite{Davison:2014lua,Kim:2014bza}.

\begin{figure}
\centering
\includegraphics[width=0.6\linewidth]{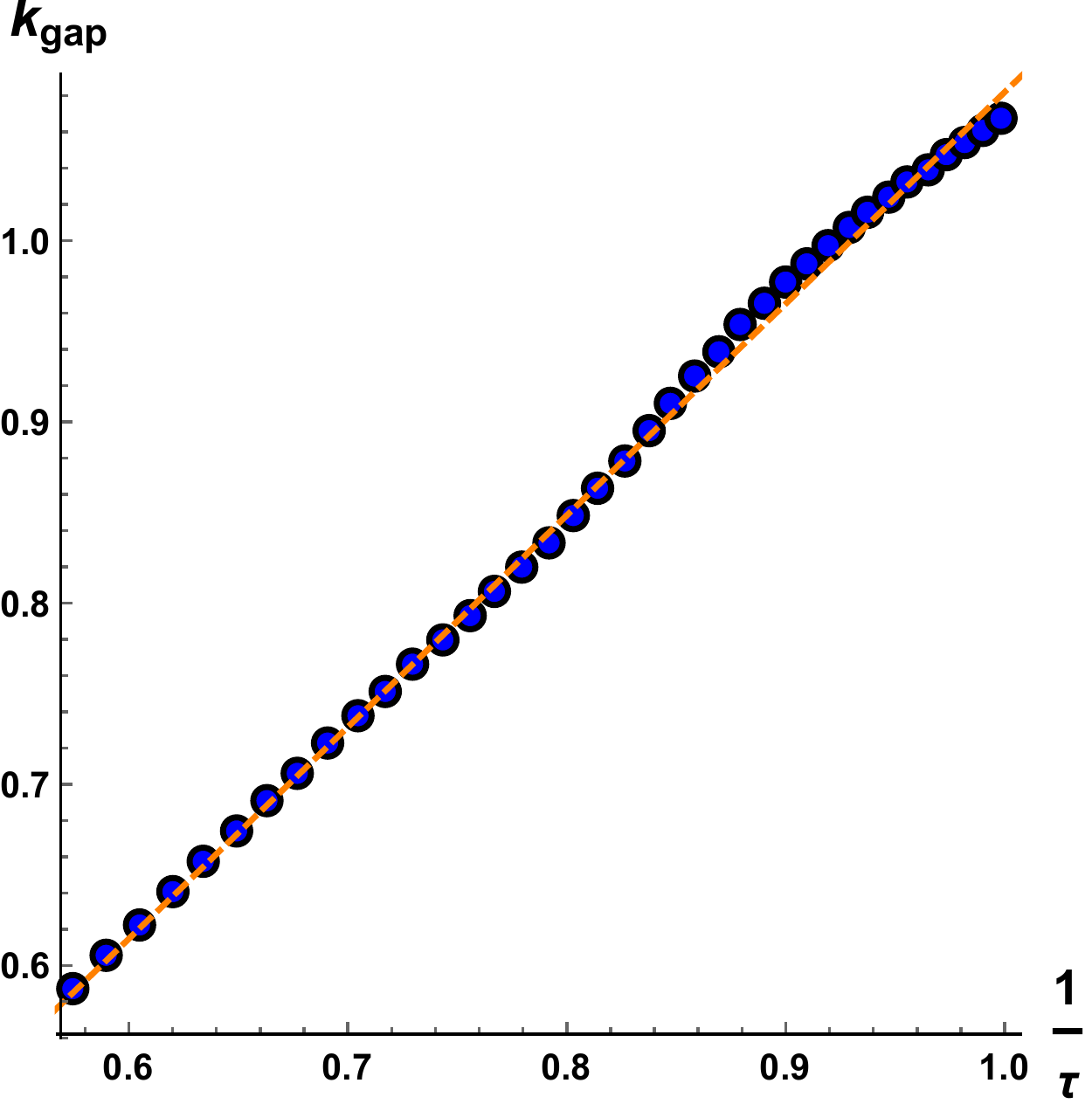}
    
\vspace{0.5cm}
    
\includegraphics[width=0.65\linewidth]{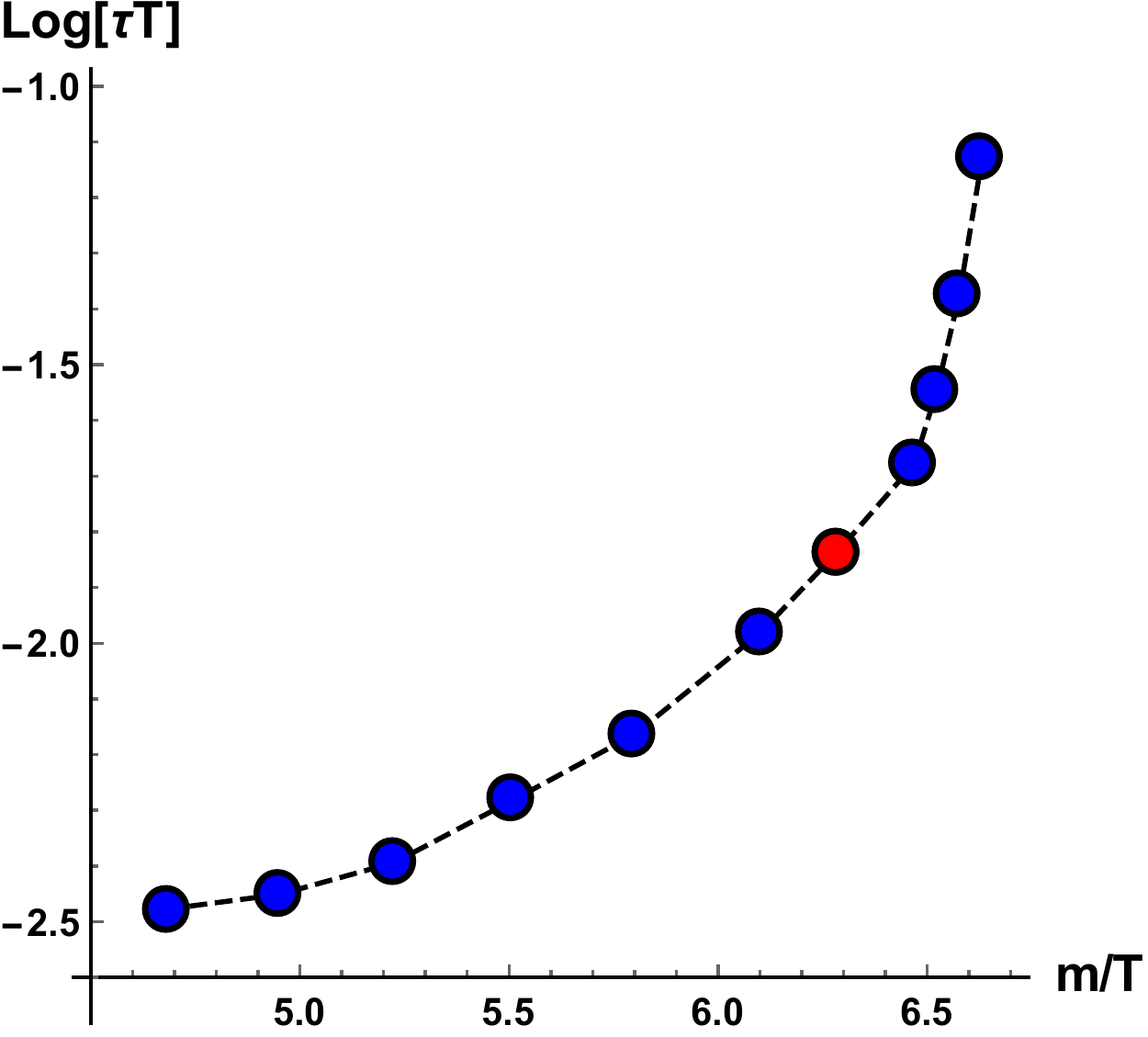}
\caption{\textbf{Top: } The numerical confirmation that $k_g \sim 1/\tau$. \textbf{Bottom: }The behaviour of the relaxation time in function of the inverse of the temperature which is in qualitative agreement with the Arrhenius law \cite{arr} . Figures taken from~\cite{Baggioli:2018vfc}.}
\label{figk}
\end{figure}

When momentum dissipation becomes very strong, the $k-$gap approaches the origin and the dispersion relation becomes of the massive particle type, $\mathrm{Re}[\omega]^2=k^2+M^2$, with an approximately constant lifetime given by $\Gamma^{-1}$. This feature, of having an emerging transverse shear waves at low frequency is very interesting since it resembles what is happening in realistic liquids. The presence of these emerging propagating phonons is tested indirectly in recent experiments~\cite{Noirez_2012,Kume2020,kume2020unexpected} (see also~\cite{PhysRevLett.62.2616}) and it can be explained by the so-called k-gap or \textbf{telegraph equation}~\cite{BAGGIOLI20201,Baggioli:2020whu}. In a specific limit of the linear axions theory the corresponding dispersion relation can be found analytically~\cite{Davison:2014lua}. Interestingly, this k-gap dynamics is shared by several holographic models~\cite{Grozdanov:2018ewh,Hofman:2017vwr,Arias:2014msa,Baggioli:2019sio,Gran:2018vdn,Baggioli:2019aqf} and it can be explained  by using the \textbf{quasi-hydrodynamic} theory of~\cite{Grozdanov:2018fic}. Finally, we can verify that the critical momentum $k_g$ is as expected inversely proportional to the  relaxation time $\tau$ (top panel of Fig.~\ref{figk}) and compare the latter with the famous Arrhenius law for fluids \cite{arr} (bottom panel of Fig.~\ref{figk}). 

\section{Bounds from hydrodynamics and holography}\label{sec:bounds}
\subsection{The violation of the KSS bound}
Continuing with the hydrodynamic description of the axion model, we cannot avoid mentioning one of the most surprising, and perhaps less understood outcome. The main character of this story is the \textbf{shear viscosity}, defined via the Kubo formula:
\begin{equation}
\eta\,\equiv\,-\,\lim_{\omega \rightarrow 0}\,\frac{1}{\omega}\,\mathrm{Im}\langle T_{xy}T_{xy}\rangle\,.
\end{equation}

One of the most remarkable achievements of the holographic duality is the discovery of the so-called \textbf{Kovtun-Son-Starintes(KSS) bound}~\cite{Kovtun:2004de}. The statement of the bound is that the ratio of shear viscosity and entropy density should be bounded below by a universal constant,
\begin {equation}
\frac{\eta}{s}\ge\frac{1}{4\pi}\left(\frac{\hbar}{k_B}\right)\,,
\end {equation}
where the Planck constant $\hbar$ and the Boltzmann constant $k_B$ will be set to unit hereinafter. In the strong coupling limit (which is usually assumed in the holographic computations), the equality holds and the inequality is saturated. For translationally invariant systems, this bound brings us a very limitation on the transport of momentum. The validity regime of the bound has been widely checked not only in holographic theories but also experimentally in Quark-Gluon Plasma, cold atoms, graphene, etc.~\cite{Schafer:2009dj,Cremonini:2011iq,Luzum:2008cw,Nagle:2011uz,Shen:2011eg} It has explicitly been shown that, in strongly coupled many-body systems with translational invariance, the viscosity bound always holds.\,\footnote{It is known that the finite $1/N$ coupling corrections can introduce a mild violation of the KSS bound and push the ratio $\eta/s$ below $1/4\pi$~\cite{Buchel:2004di,Buchel:2008wy,Myers:2008yi,Buchel:2008ae,Ghodsi:2009hg,Buchel:2008sh,Brigante:2007nu,Brigante:2008gz,Kats:2007mq}. As a result of that, there is an alternative bound $\eta/s\ge \#$, where $\#$ is still an $\mathcal{O}(1)$ number. In this paper, we will focus only on holographic models in the large $N$ limit and infinite coupling limit.} 

\begin{figure}[ht]
\centering
\includegraphics[width=0.8\linewidth]{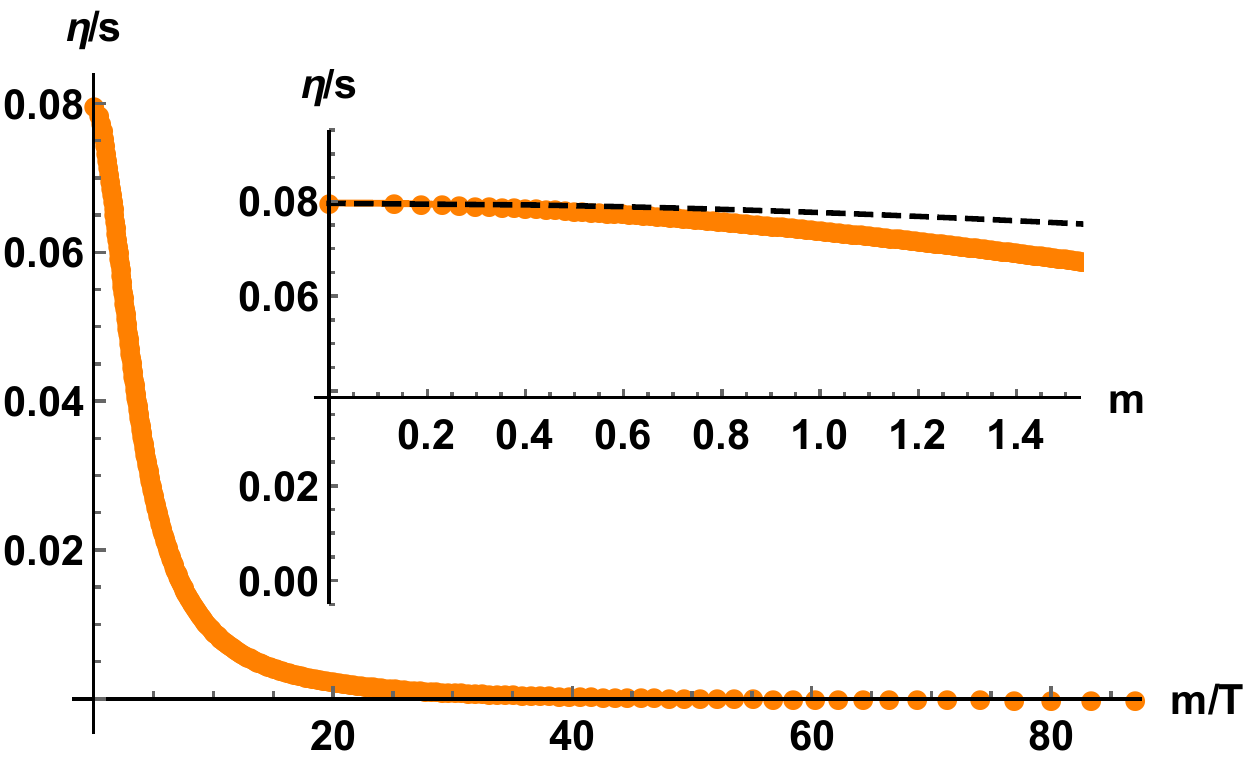}
    
\vspace{0.5cm}
    
\includegraphics[width=0.8\linewidth]{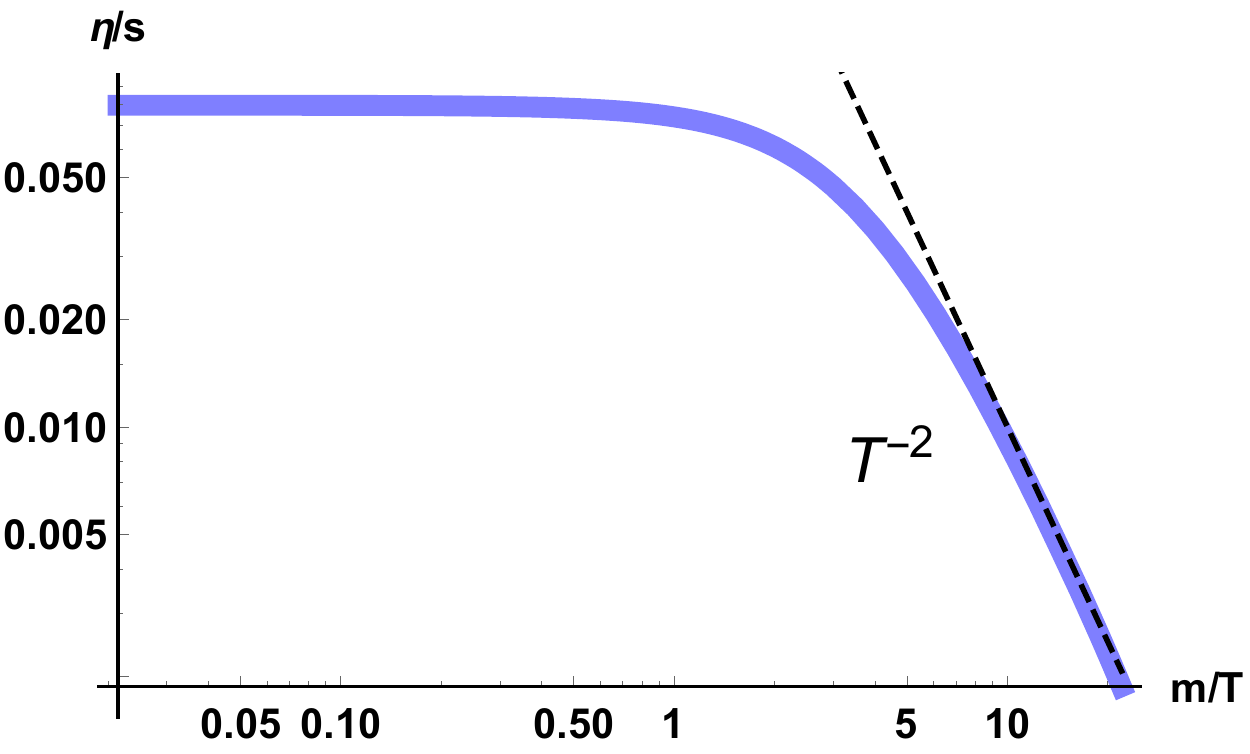}
\caption{The violation of the KSS bound in the linear axion model. Figures taken from~\cite{Baggioli:2019rrs}.}
\label{visco}
\end{figure}

On the contrary, this bound can be parametrically violated in presence of broken translations.\footnote{The KSS bound can also be violated in presence of broken rotations \cite{Rebhan:2011vd,Jain:2015txa,Giataganas:2013lga,Jahnke:2014vwa} (notice how these models use also a single scalar field with linear profile as introduced in \cite{Mateos:2011ix,Mateos:2011tv})\footnote{See also \cite{Giataganas:2012zy,Giataganas:2013hwa,Giataganas:2017koz} for further studies.} or an external magnetic field~\cite{Finazzo:2016mhm} (this is true only in $3$ spatial dimensions but not in $2$ in which the magnetic field is just a scalar field \cite{Buchbinder:2008nf}), which however will not be touched in detail in this paper.} Let us follow a historical timeline. The first results came from~\cite{Hartnoll:2016tri}. There the authors proved that:
\begin{enumerate}
    \item [(I)] The KSS bound $\eta/s=1/4\pi$ is \textbf{violated} in the linear axion model of~\cite{Andrade:2013gsa}. This can be shown both numerically and analytically in a perturbative scheme (see Fig.~\ref{visco}).
    \item [(II)] Any holographic model in which the shear equation is \textbf{massive}:
    \begin{equation}
      \Box \,h^{x}_{y}\,=\,M^2(r)\,h^{x}_{y}\,,
    \end{equation}
    with $M^2>0$, violates automatically the KSS bound.
    \item [(II)] At low temperature , the viscosity to entropy ratio scales like:
    \begin{equation}
        \frac{\eta}{s}\,\sim\,\left(\frac{T}{M}\right)^{\Delta}\,\,\rightarrow\,\,0\,,
    \end{equation}
    where $\Delta$ is directly given by the conformal dimension of the $T_{xy}$ operator at the IR extremal fixed point. In the case of~\cite{Andrade:2013gsa}, $\Delta=2$ (see Fig.~\ref{visco}). See~\cite{Ling:2016ien} for generalizations.
    \item[(IV)] Despite translations are explicitly broken in this model, the ratio $\eta/s$ can be still identified as the coefficient determining the rate of entropy production when the equilibrium state is subjected to a slowly varying homogeneous source, a strain.
\end{enumerate}
Just one day after, \cite{Alberte:2016xja} appeared, showing that:
\begin{itemize}
    \item The violation reported in~\cite{Hartnoll:2016tri} is very general but the breaking of translations does not necessarily imply it. Indeed, one can have holographic models which break translations explicitly but for which $\eta/s=1/4\pi$ -- the so-called fluid theories.
    \item The violation appears independently of whether translations are broken explicitly and spontaneously. This opens a much more difficult question since the excuse that momentum is dissipated (see~\cite{Burikham:2016roo} for discussions on this point) and therefore the viscosity is not well defined falls apart.
\end{itemize}
Regardless of the large activity after that, not much progress in understanding this feature has been done so far. Nevertheless, it is worth mentioning that similar violations have appeared outside the realm of holography~\cite{Ge:2018lzo,Gochan:2018eez}.

The most interesting aspect is the one related to the holographic models claiming to be dual of conformal solids (in which the breaking is spontaneous). In realistic situations, moving from a liquid to a solid phase, the viscosity definitely grows; this is evident if you leave your honey in the fridge. In these holographic models, on the contrary, by making them solids (with larger shear modulus for example) the viscosity decreases. Is that physical? And where does it come from? The suspicion is that this is related to the fact that the viscosity in the holographic models grows with temperature as in gas and not as in liquids.

 In the next section, we will take a different approach and discuss whether or not $\eta/s$ is at all the quantity to bound. The answer will be ``likely not''. For example, it has been shown in~\cite{PhysRevLett.99.021602} that in non-relativistic setups one can provide a simple counterexample of the KSS bound just by increasing the number of species.

\subsection{From viscosity to diffusion}
 As was pointed in the previous subsection, the KSS bound is definitely violated in certain holographic models. Nevertheless, it still remains controversial whether the momentum transport in strongly coupled systems is universal or not. In the following, we will argue that a (more) universal and general bound, valid also with broken spacetime symmetries, can be obtained by considering the \textbf{momentum diffusivity} instead of the $\eta/s$ ratio.

The idea was firstly introduced in~\cite{Hartnoll:2014lpa} and recently elaborated in~\cite{Baggiolili2005}. Consider a relativistic neutral system and its diffusive hydrodynamic mode
in the shear channel, known as \textit{shear diffusion}, with the following dispersion relation~\cite{Kovtun:2012rj}
\begin {equation}
\omega=-i\,D_\pi\,k^2+\dots,\,\,\,D_\pi=\frac{\eta}{\chi_{\pi\pi}}\,(c^2\equiv1),
\end {equation}
where $\chi_{\pi\pi}= s T$ is the momentum susceptibility. The KSS bound can then be reformulated as 
\begin {equation}
D_\pi \ge\frac{c^2}{4\pi T}\sim c^2\tau_{pl}\,,
\end {equation}
where $\tau_{pl}\sim \hbar/k_B T$ is called the Planckian time, supposed to be the minimal relaxation timescale in the Nature~\cite{Zaanen2004,10.21468/SciPostPhys.6.5.061}.
In presence of EXB, the momentum of the system is not conserved. As a result of that, the diffusive mode acquires a finite relaxation rate $\Gamma$ and the dispersion becomes
\begin {equation}
\omega=-i\,\Gamma-i\,D_\pi\,k^2\,.
\end {equation}

\begin{figure}[t]
    \centering
    \includegraphics[width=0.65\linewidth]{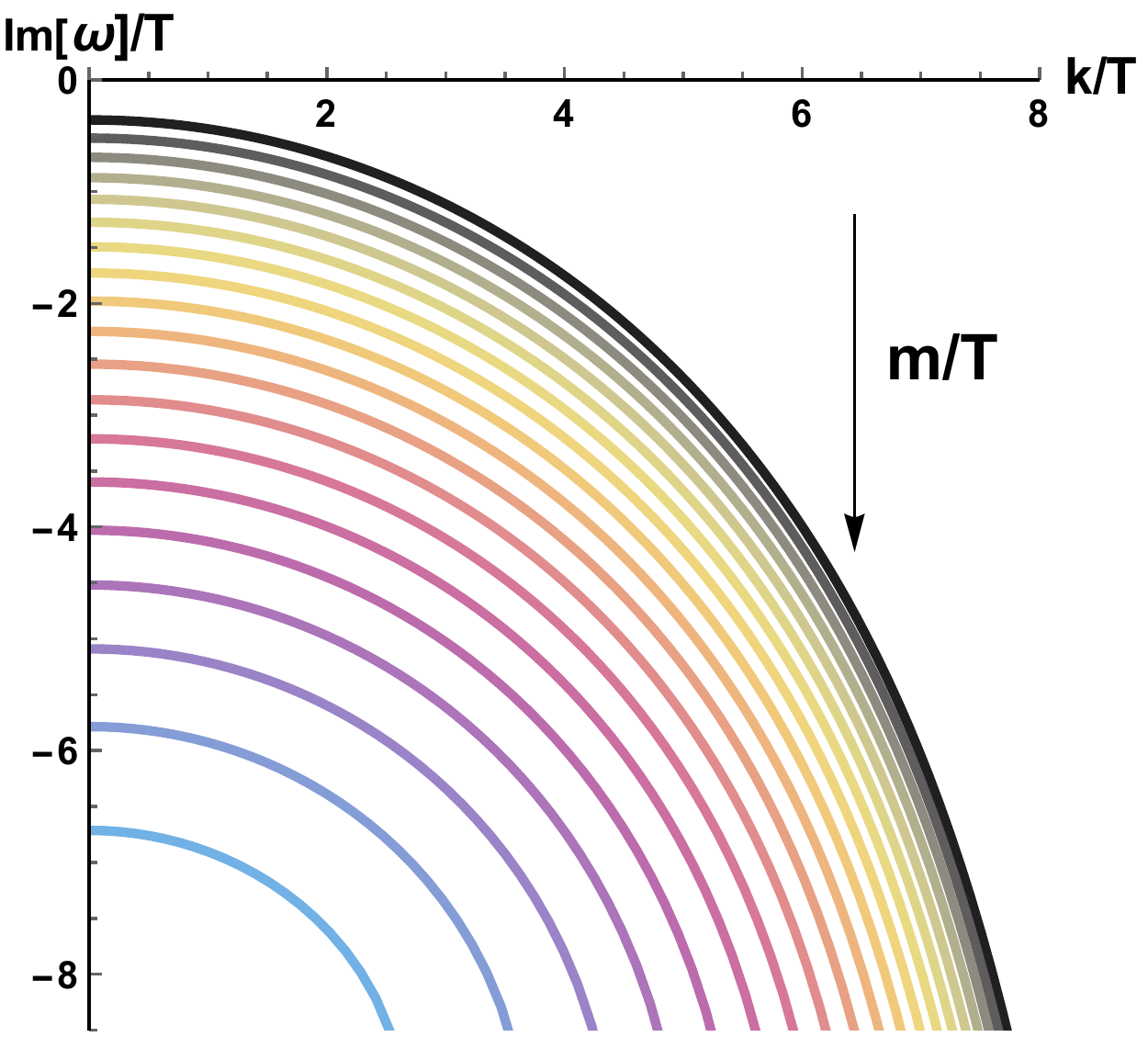}
    
    \vspace{0.5cm}

    \includegraphics[width=0.7\linewidth]{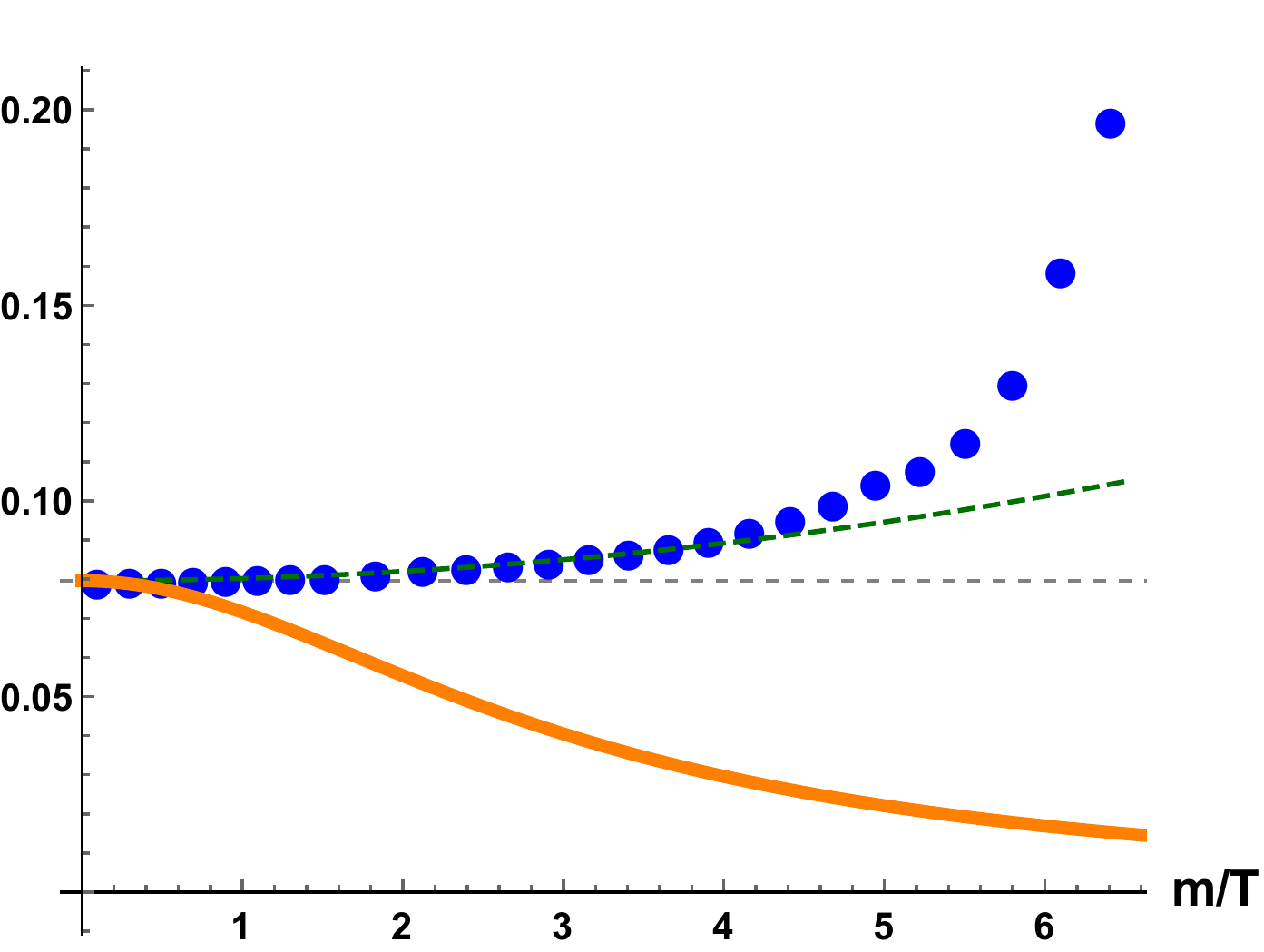}
    \caption{The shear mode in the simple ``\textit{linear axion model}'' corresponding to $V(X)=X$ is pseudo-diffusive. \textbf{Top: } The dispersion relation of the pseudo-diffusive mode $\omega=-i \Gamma -i D_\pi k^2$ for $m/T \in [1,6.5]$ (from black to light blue). \textbf{Bottom: } In orange the viscosity to entropy ratio $\eta/s$, in blue the dimensionless shear diffusion constant $D_\pi T$ obtained numerically and in  green the analytic formula~\eqref{form}. The horizontal dashed value is $1/4\pi$. Figure taken from~\cite{Baggiolili2005}.}
    \label{figd1-2}
\end{figure}

Computing the momentum-momentum Green's function holographically, one can check that the momentum diffusivity in the simplest linear axion model ($V(X)=X$), in the limit of slow momentum relaxation, reads~\cite{Ciobanu:2017fef}
\begin{equation}
D_\pi=\frac{1}{4\,\pi\,T}\left[1+\frac{1}{24}\left(9+\sqrt{3}\pi-9\log 3\right)\frac{m^2}{8\,\pi^2\,T^2}+\dots\right]\label{form}\,,
\end{equation}
where the $\dots$ indicate higher order corrections in the dimensionless parameter $m/T$. For arbitrary large values of $m/T$, one can obtain the diffusion constant by solving the QNMs of the black hole numerically. We show a summary of the results in Fig.~\ref{figd1-2}. From the bottom panel, it is evident that when we consider the momentum diffusivity, there is no violation of the bound at all even in presence of EXB of translations. 

This is the first indication, that in less symmetric (and more general) scenarios, the quantity to consider is $D_\pi T$ and not $\eta/s$. It is a mere coincidence that for a relativistic neutral fluid the two coincide. This argument is also confirmed by an independent analysis~\cite{Baggioli:2020lcf} which shows that the kinematic viscosity $\nu_m\equiv \eta/\varrho$ of the QGP is of the same order of all common liquids at their minimum, $\sim 10^{-7}$ m$^2/$s, and it is well approximated by the simple formula:
\begin{equation}
\nu_m\,=\,\frac{1}{4\,\pi}\,\frac{\hbar}{\sqrt{m_e\,m_p}}\,,
\end{equation}
with\,, $m_e,m_p$ the mass of the electron and the proton respectively.

The original discussion of~\cite{Hartnoll:2014lpa} is more general and was initially focused on the diffusion of energy and charge. In particular, given a generic diffusion constant $D_i$, where $i$ indicates the corresponding operator associated to the diffusive dynamics, Hartnoll conjectured that an equality of the type:
\begin{equation}
D_i\,\ge\,v^2\,\tau \,,
\label{boubou}
\end{equation}
has to be valid, where $v$ is a characteristic velocity scale and $\tau$ a minimal relaxation time. When the momentum is relaxed very rapidly (\emph{e.g.} in the bad metal case of~\cite{Hartnoll:2014lpa}), $m/T\gg 1$ (we also call it low temperature limit or \textit{incoherent limit}.), the dynamics associated with charge and heat transport become purely diffusive and they decouple:
\begin{equation}      
\mathcal{D}=\left(                
  \begin{array}{ccc}   
    D_C & 0\\  
    0 & D_T\\  
  \end{array}
\right)    \,.             
\end{equation}
Then, the matrix of the diffusivities gets completely diagonal.
In this case, the bound \eqref{boubou} should apply. Nevertheless the question is: which are the characteristic velocity and time scales? The timescale is naturally associated with the Planckian time $\tau_{pl}=\hbar/k_B T$. The discussion about the velocity is more subtle. This velocity cannot be the Fermi velocity, as it is in general not sharply defined in the strongly coupled systems without quasi-particles. 

\subsection{Butterfly velocity and chaos}
\begin{figure}[ht]
\centering
\includegraphics[width=0.9\linewidth]{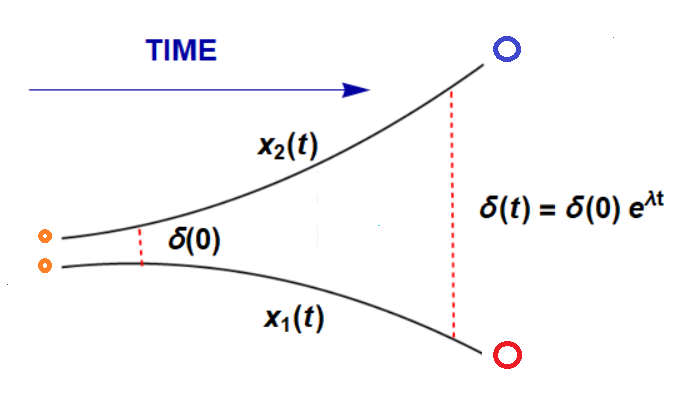}
\caption{The classical butterfly effect intended as the exponential sensitivity to the initial boundary conditions. Its quantum generalization can be formulated as a exponential growth of the OTOC~\eqref{oo}.}
\label{fig:butteff}
\end{figure}

In the attempt of making Hartnoll's proposal predictive, M. Blake proposed that the velocity scale appearing in the diffusivity bounds could be identified with the butterfly velocity $v_B$ \cite{Blake:2016wvh,Blake:2016sud}. $v_B$ measures the speed of propagation of information through a system and it can be generically extracted from the out-of-time-order correlator (OTOC)~\cite{1969JETP...28.1200L}:
\begin{align}
&\Big\langle \left[\hat{W}(t,x),\hat{V}(0,t)\right]^2\Big\rangle_\beta\sim e^{\lambda_L(t-t^*-|x|/v_B)}\,,\label{oo}\\ \nonumber
&\,\,\,\,\,\,\,\,\text{for}\,\,\,\,\,\,\,t_{\text{local}}\ll t \ll t^*\,,
\end{align}
where $\lambda_L$ is called the Lyapunov exponent, $t^*$ is the scrambling time and $t_{\text{local}}$ is the timescale that the system reaches local equilibrium. Here $\hat{W},\hat{V}$ are two generic hermitian operators. In analogy to the classical chaos (see Fig.~\ref{fig:butteff}), this exponential growth can be viewed as a quantum mechanical definition of it, originating from the non-trivial commutator of two operators set at different times.

\begin{figure}[ht]
\centering
\includegraphics[width=0.9\linewidth]{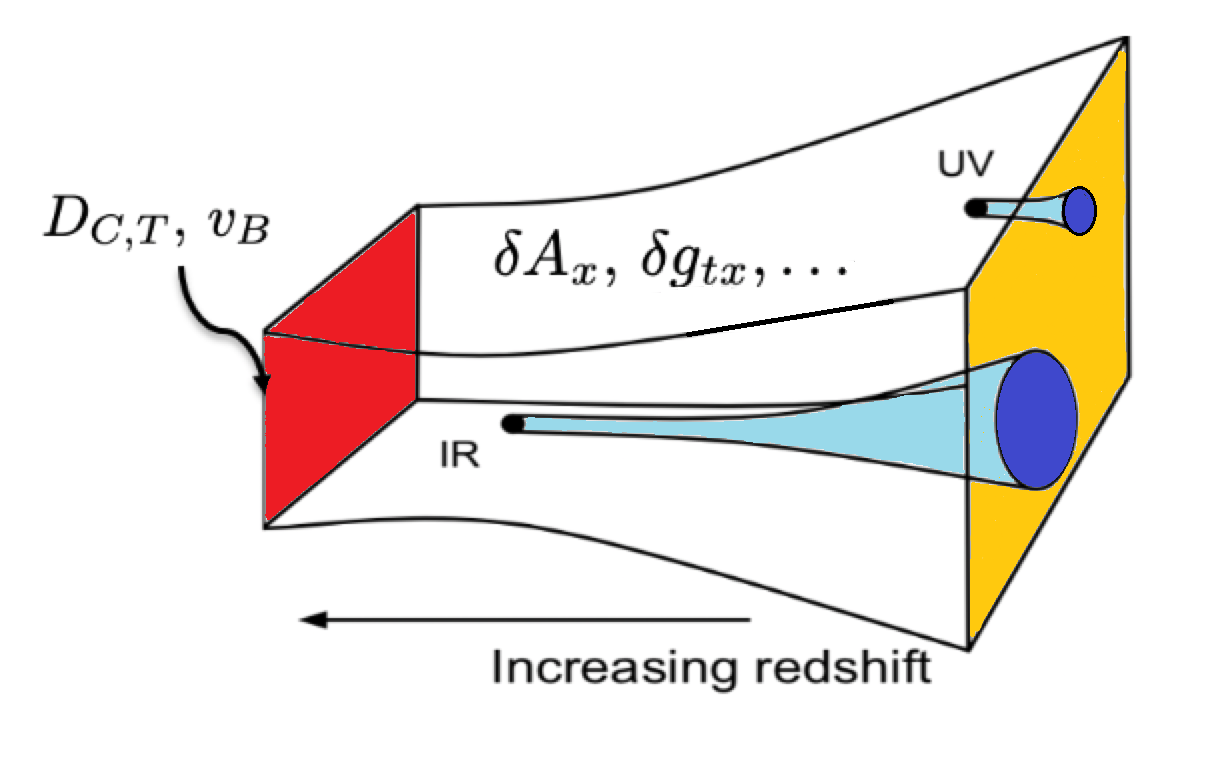}
\caption{The diffusivities and butterfly velocity can be related via the membrane paradigm.}
\label{figme}
\end{figure}

In holography, the butterfly
velocity can be easily calculated by considering a shock wave solution near the black hole horizon. In full generality, it only depends on the IR metric and is insensitive to the matter field configurations in the bulk. Given a metric of the form
\begin{equation}
ds^2=-f(r)dt^2+\frac{dr^2}{f(r)}+h(r)dx_{d-1}^2\,,
\end{equation}
the butterfly velocity is given by~\cite{Blake:2016wvh}
\begin{equation}
v_B^2=\frac{f'(r_h)}{d\, h'(r_h)}\,.
\end{equation}
Exploiting the \textit{membrane paradigm}~\cite{Damour:2008ji,Iqbal:2008by}, the transport coefficients can usually be expressed in terms of horizon data. Then, it is not hard to find the direct relations between the diffusivities and the butterfly velocity (see the illustration in Fig.~\ref{figme}):
\begin{equation}
D_{C,T}=C_{C,T}\,\frac{v_B^2}{2\pi T}\,, 
\label{reld}
\end{equation}
where $C_{C,T}$ are constants.

Note that these relations are only valid in the low temperature limit or strong momentum relaxation (incoherent) limit~\cite{Kim:2017dgz}, where the charge and thermal sectors decouple (see \cite{Kovtun:2014nsa} for a discussion about the general situation and possible more general bounds). Furthermore, \eqref{reld} should be viewed as a low energy IR statement, since both of $C_{C,T}$ are only determined by the scaling symmetry of the IR fixed point, irrelevant to any UV parameters, even though they are model-dependent quantities. For the Einstein-Maxwell-dilaton model, with a hyperscaling violating IR geometry~\cite{Blake:2016wvh,Blake:2017qgd}, one obtains
\begin{equation}
C_C=\frac{d-\theta}{\Delta_\chi}\,,\qquad C_T=\frac{z}{2z-2}\,. \label{bbb}
\end{equation}
Here, $\Delta_\chi$ is the scaling dimension of the charge susceptibility, $z$ is the dynamical critical exponent and $\theta$ the hyperscaling violation exponent. For anisotropic Q-Lattice models, $C_T$ has been computed in \cite{Ahn:2017kvc} and it has been shown to obey the lower bound \eqref{bbb}, proving that the latter it is not affected by anisotropy.

\subsection{Pole-skipping and the complex plane}

The chaotic properties of a dynamical system, such as the Lyapunov exponent ($\lambda_L$) and the butterfly velocity ($v_B$), are encoded in a specific {\it four} point function, out-of-time-order-correlator (OTOC) \eqref{oo} at {\it short} time scale. Interestingly, it has been observed that $\lambda_L$ and $v_B$ can be also detected by a  {\it two} point function, the retarded Green's function of the energy density operator at {\it long} time scale. This property has been dubbed \textbf{pole-skipping} phenomenon~\cite{Grozdanov:2017ajz, Blake:2017ris, Blake:2018leo} and $\lambda_L$ and $v_B$ have been proven to be related with the so-called "\textit{pole-skipping points}". Even though the pole-skipping phenomenon has been introduced in the context of quantum chaos, its mathematical concept is more general and physical applications may be wider. Thus, we start with a general definition of the pole-skipping points. 

Pole-skipping points are special points in the complexified momentum (complex frequency, complex wave number) space. At these points, two point retarded Green's functions of given operators are not uniquely defined. The non-uniqueness of the Green's function at some value of frequency/momentum is not very novel and there are prescriptions to define the Green's function at that point, making contact with transport properties via Kubo's formulas.  For example, see \eqref{Kubo123}, where the Green's functions are not well defined at $\omega=k=0$ and we specify the order of limit to define them uniquely. It turns out that the pole-skipping points occur at the integer/half integer values of $i \omega /(2\pi T)$ for bosonic/fermionic operators.\footnote{However, see \cite{Ahn:2020baf} for possibilities for non-integer $i\omega/(2\pi T)$.}

\begin{figure}[]
    \centering
    \includegraphics[width=0.7\linewidth]{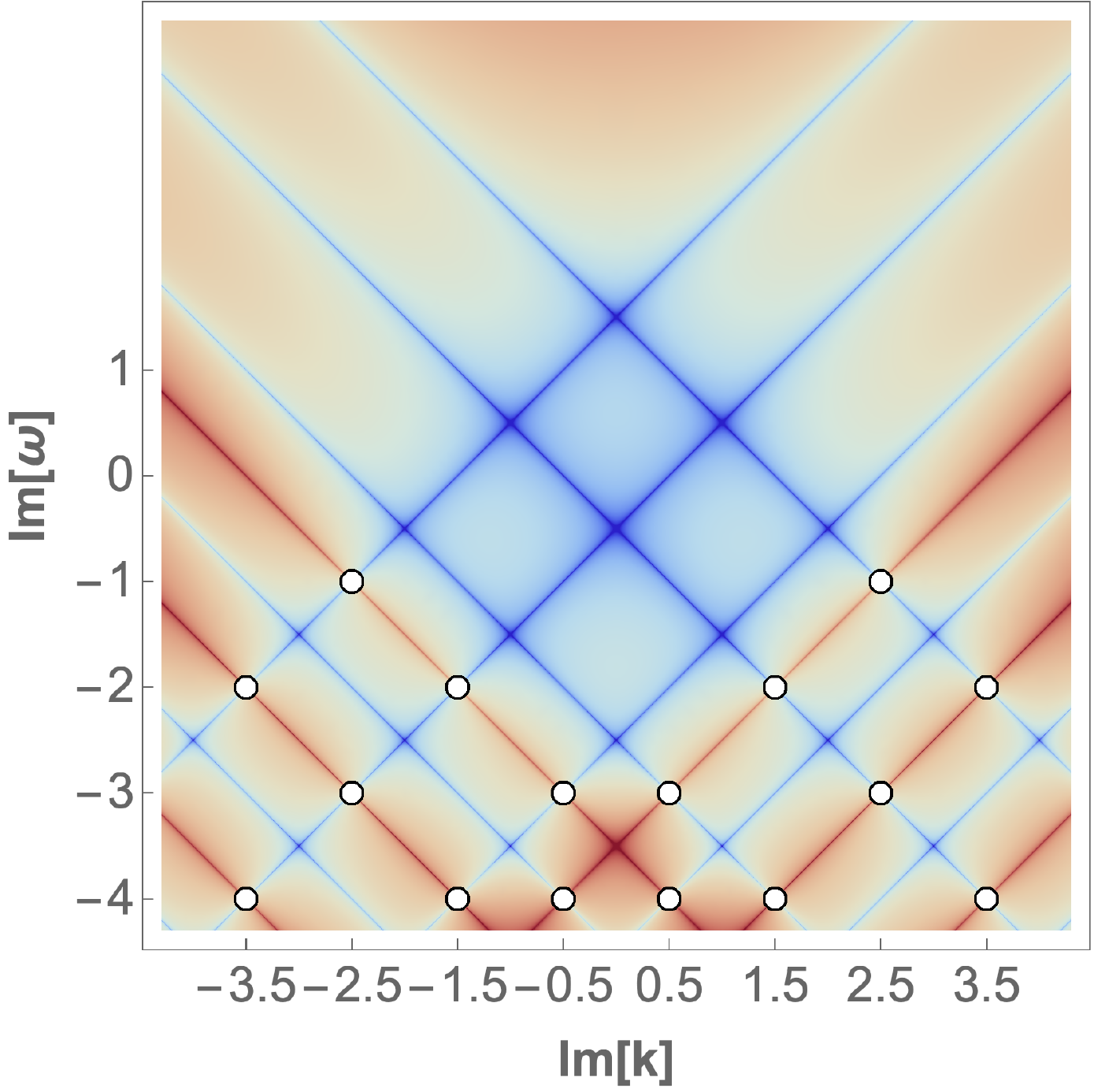}
    \caption{The pole-skipping points for a scalar field in the hyperbolic space. The red line indicates the curve on which $\mathcal{G}_{AB}(\omega,k)=\infty$ and the blue one the curve on which $\mathcal{G}_{AB}(\omega,k)=0$. The white circle locates the pole-skipping point $(\omega^*,k^*)$. Here, we set $2\pi T = 1$. Figure taken from~\cite{Ahn:2020bks}.}
    \label{fig:poleskip1}
\end{figure}

Another way to introduce the pole-skipping points is using its literal meaning: the points where the ``pole'' is ``skipped''. For this purpose, let us consider the expression for the retarded Green's function of the operators $A$ and $B$:
\begin{equation}
    \mathcal{G}_{AB}(\omega,k)\,\sim\, \frac{\mathcal{B}(\omega , k)}{\mathcal{A}(\omega ,k)}, \label{eq:poleskip1}
\end{equation}
where we suppress the operator dependence on the RHS not to clutter. The pole is defined by $\mathcal{A}(\omega ,k) = 0$, which gives the constraint between $\omega$ and $k$. If this constraint lives in the pure imaginary $\omega$ and pure imaginary $k$ space, it defines a curve in the space ($\mathrm{Im}[k], \mathrm{Im}[\omega]$), which corresponds to the red line in Fig.~\ref{fig:poleskip1} for example.\footnote{This example is chosen for the pedagogical purpose. In general, there is no \textit{a priori} reason that the condition $\mathcal{A}(\omega ,k) = 0$ lives in the space  ($\mathrm{Im}[k], \mathrm{Im}[\omega]$). For more complete and general discussion, we refer to~\cite{Ahn:2020bafxxx}. } However, there might be a second relevant curve coming from the condition $\mathcal{B}(\omega ,k) = 0$, which is this time the blue line in Fig. \ref{fig:poleskip1}. 
The intersection of these two curves, say $(\omega_*, k_*)$, is dubbed the pole-skipping point because there the would-be pole ($\mathcal{A}(\omega_* ,k_*) = 0$) is skipped ($\mathcal{B}(\omega_* ,k_*) = 0$), \emph{i.e.}
\begin{equation}
    \mathcal{G}_{AB}(\omega_*,k_*)\,\sim\, \frac{\mathcal{B}(\omega_* , k_*)}{\mathcal{A}(\omega_* ,k_*)} \,\sim\,  \frac{0}{0} \,, \label{eq:poleskip2}
\end{equation}
where the last expression $0/0$ indicates that the Green's functions is not uniquely defined and we need a prescription to define the Green's function there. The path dependence prescription was clarified and classified in~\cite{Ahn:2020baf}.
This is nothing but the viewpoint of the first paragraph of this subsection. 

Having this general introduction, let us come back to the original argument regarding chaos~\cite{Grozdanov:2017ajz, Blake:2017ris, Blake:2018leo}. The main point is that chaotic nature of the OTOC \eqref{oo} is related with the pole-skipping points of the retarded Green's function of energy density $\mathcal{G}^R_{T^{tt}T^{tt}}$. Thus, the first step is to compute the retarded Green's function.

Let us consider an Einstein action with the matter Lagrangian $\mathcal{L}_{M}$
\begin{equation}
\mathcal{S}=\int d^{d+1}x\sqrt{-g}\left[{R\over 2}-\Lambda + \mathcal{L}_{M}  \right]\,,
\label{actionps}
\end{equation}
which includes \eqref{action}. 
 According to holography, the retarded Green's function can be computed by the coupled equations for metric ($\delta g_{\mu \nu}(u)$) and matter (collectively, $\delta \varphi(u)$) perturbations that are regular at the horizon in ingoing Eddington-Finkelstein (EF) coordinates: $v = t + u_*$ with the tortoise coordinate $u_*$. Near the horizon the perturbations can be written as
\begin{equation}
\label{eq:irexp}
\begin{split}
\delta g_{\mu \nu}(u) &= \delta g_{\mu \nu}^{(0)} + \delta g_{\mu \nu}^{(1)} (u - u_h) + \cdots,  \\ 
\delta \varphi(u) &= \delta \varphi^{(0)} + \delta \varphi^{(1)} (u - u_h) + \cdots  \,,
\end{split}
\end{equation}
where we consider metric perturbation near the background \eqref{metric0} in $d+1$ dimension. 
After plugging in the expansion \eqref{eq:irexp} into Einstein's equations, one can organize the equation as an expansion about the horizon. The near horizon equation including $\delta g_{vv}^{(0)}$ is
\begin{align}
\label{eq:equationwithmatter}
& \left(- i \frac{d-1}{u_h}\omega  + k^2 \right) \delta g_{vv}^{(0)} - i (2 \pi T + i \omega) \left[  \omega \delta g_{x^ix^i}^{(0)} + 2 k \delta g_{vx}^{(0)} \right]  \nonumber  \\
&= \frac{2}{u_h^2}  \bigg[u_h^2T_{vu}(u_h)\delta g_{vv}^{(0)} + \delta T_{vv}(u_h) \bigg] \,,
\end{align}
where $T_{\mu \nu}(u_h)$ is the bulk stress-energy tensor of the background matter fields and $\delta T_{\mu\nu }(u_h)$ comes from the matter perturbations. Thus, the information of the various possible matter content is explicitly encoded in the last term of \eqref{eq:equationwithmatter}. However, interestingly, it turns out that this term vanishes identically for a large class of systems such as Einstein-Maxwell-Dilaton-Axion gravity theories including our models~\cite{Blake:2018leo}. 

What are the consequences of this simplification (vanishing the last term of \eqref{eq:equationwithmatter})? 
In general, \eqref{eq:equationwithmatter} provides a constraint relating the parameters $\delta g_{vv}^{(0)}, \delta g_{vx}^{(0)}$, and  $\delta g_{x^ix^i}^{(0)}$. However, if $(\omega, k) =(\omega_*, k_*) $, with
\begin{equation} \label{eq:psp1}
\omega_* = 2 \pi T i \,,  \quad k_* =  \pm \sqrt{i \frac{d-1}{u_h} \omega_* }= \pm i\sqrt{\frac{2\pi T (d-1)}{u_h}} \,,
\end{equation}
we lose such a constraint. It means we have one more degree of freedom than usual implying the Green's function can not be uniquely determined. This non-uniqueness of the Green's function at a special point  in complex momentum space is precisely the defining property of the pole-skipping point, so \eqref{eq:psp1} is a pole-skipping point. 

This observation is remarkable. This is an alternative holographic way of understanding the pole-skipping points. In general, computing explicitly the Green's function is not easy and usually requires numerics. Finding the points where the usual constraint {\it at the horizon} disappears is on the contrary much easier. It is not only a technical simplification but also has a conceptual importance: pole-skipping points are determined by the black hole horizon property, signaling possible universal properties independent of the UV details of the theory.

Here comes an example. Note that all information about the matter part of the action $\mathcal{L}_m$ is encoded in the location of the horizon $u_h$. Thus, $\omega_*$ is universal regardless of the matter action, while $k_*$ is not. 
Noting that the butterfly velocity from the shock-wave geometry~\cite{Blake:2016sud} is 
\begin{equation} \label{eq:butter1}
v_B = \sqrt{\frac{2\pi T u_h}{ d-1}} \,.
\end{equation}
we find that the pole-skipping point \eqref{eq:psp1} may be related with the chaotic properties as follows.\footnote{The pole-skipping phenomenon occurs also in a non-maximally chaotic system where the pole-skipping points capture only the stress tensor contributions to chaos~\cite{Choi:2020tdj}.}  
\begin{equation} \label{eq:psp2}
\omega_* = i \lambda_L \,, \qquad k_* = \pm \,i\frac{\lambda_L}{v_B} \,.
\end{equation}
From another perspective, if we knew \eqref{eq:psp2} somehow, we could have computed \eqref{eq:butter1} just by the pole-skipping points, without calculating nor the OTOC or the shock wave geometry.

There is another interesting observation related with transport. Let us consider a system with energy conservation but no momentum conservation, which corresponds to the case with a very big $\alpha$ in  \eqref{linear}. In this case, the retarded Green's function of the energy density will have a hydrodynamic diffusion pole, $\omega = - i D_T k^2 $ -- energy diffusion. Extrapolating this hydrodynamic relation to the pole-skipping point \eqref{eq:psp2}  in the non-hydrodynamic region, we have
\begin{equation} \label{eq:psp3}
D_T  = \frac{ i \omega_*}{k^2_*}  = \frac{v_B^2}{\lambda_L}\,.
\end{equation}
Interestingly, it turns out that this extrapolation works and indeed \eqref{eq:psp3} is the universal lower bound for the holographic models with AdS$_2 \times R^{d-1}$ IR geometry~\cite{Blake:2016jnn}. This bound is saturated at the infinite momentum relaxation limit ($\alpha \to \infty$ in \eqref{linear}). It can be understood from an effective field theory perspective~\cite{Crossley:2015evo}.
In short, the pole-skipping point has a potential to predict universal hydrodynamic properties. 

So far we have focused on the specific pole skipping point of the Green's function of energy density, which corresponds to the gauge invariant scalar mode of the metric perturbation. However, the pole skipping points are ubiquitous in all kinds of Green's functions. They were observed in other gauge invariant modes of the metric perturbation and other fields with different spins such as scalar, vector, and spinor fields~\cite{Grozdanov:2019uhi,Blake:2019otz,Natsuume:2019xcy,Ceplak:2019ymw,Ahn:2020bks,Ahn:2020baf,Ahn:2020bafxxx}. However, it is still not clear which physical observables are related with the pole-skipping points in the cases apart from energy density. This is an interesting and important open question. See \cite{Grozdanov:2020koi,Ahn:2020bafxxx} for this direction.

In general, there are infinite towers of pole-skipping points. The point  \eqref{eq:psp2} from the near horizon equation~\eqref{eq:equationwithmatter} is just one case which is relevant to the {\it nearest} horizon geometry. If we expand the equations to higher orders in $(u-u_h)$, we can obtain an infinite set of pole-skipping points whose imaginary frequencies coincide with the  Matsubara values $\omega=-i 2\pi T n$. For example, one can see part of towers in Fig.~\ref{fig:poleskip1}, where the pole-skipping points start from $\mathrm{Im}\omega=-1$ in units of $2\pi T=1$, which is the ``highest'' pole skipping frequency (on the vertical Im$\omega$ axis). The pole skipping points continue to appear at every integer frequency smaller than the highest one.
Thus, by coming down on the vertical Im$\omega$ axis, we are exploring the geometry away from the horizon. The highest pole skipping frequency is determined by the spin ($\ell$) of the operator~\cite{Haehl:2019eae, Ahn:2020bks}: 
\begin{equation}
\omega_*^{\mathrm{highest}} = i (\ell-1) \,.
\end{equation}
Note that the metric field ($\ell=2$) is the only case with the {\it positive} pole-skipping frequency, signaling an exponentially growing instability related to the chaotic behaviour. The other operators display only purely relaxational modes which therefore cannot be related in any clear way to a shock-wave solution and the onset of quantum chaos. Systematic methods to obtain these towers of pole skipping points have been developed in \cite{Blake:2019otz,Natsuume:2019xcy,Ahn:2020bks,Ahn:2020baf,Ahn:2020bafxxx}. 

\begin{figure}[ht]
    \centering
    \includegraphics[width=1.0\linewidth]{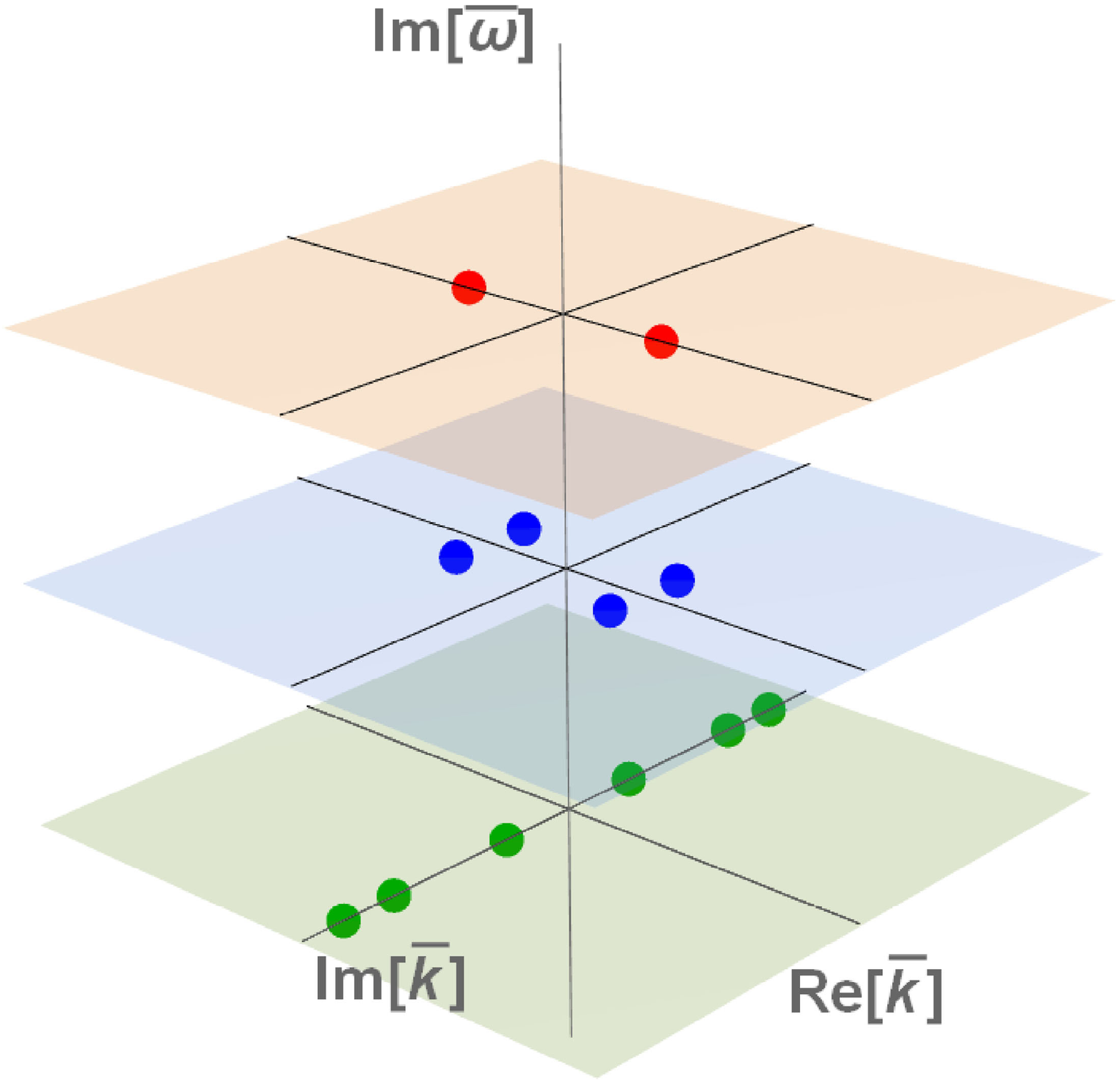}
    \caption{Pole-skipping wave number ($k_*$) can be real(red), complex(blue), and pure imaginary(green) for scalar field perturbation in $\text{AdS}_4$. The momentum relaxation parameter $\bar{\alpha}=2, 2, 1$ for $\text{Im}[\Bar{\omega}]=-1, -2, -3$ (from top to bottom) respectively. Here, the ``bar'' variables denote the quantities scaled by $2\pi T$. 
Figure taken from~\cite{Ahn:2020bafxxx}.}
    \label{fig:poleskip2}
\end{figure}

The pole-skipping frequency $\omega_*$ is universal in the sense it is independent of the matter action, but $k_*$ is not. The pole skipping wave number $k_*$ depends on the momentum relaxation parameter $\alpha$ in \eqref{linear}, chemical potential $\mu$, and other parameters in the matter action. In particular, $k_*$ can be real number or complex number (see Fig.~\ref{fig:poleskip2}). 

The pole skipping phenomena have been studied in the SYK models and the conformal field theory~\cite{Gu:2016oyy, Haehl:2018izb, Haehl:2019eae}. The relations between OTOC and pole skipping points (as well as the comparison between field theory and gravity analysis) have been investigated in~\cite{Ahn:2019rnq, Haehl:2019eae, Ahn:2020bks, Ahn:2020baf} where the hyperbolic space
was considered to have an analytic control. However, the final qualitative results are believed to hold in flat space too. The pole skipping points were studied also for rotating black holes and topologically massive gravity~\cite{Jahnke:2019gxr, Liu:2020yaf}.  For more general and detailed pole skipping analysis for our axion model, we refer to \cite{Grozdanov:2019uhi, Ahn:2020bafxxx}. \\

Let us conclude with a recent development. The idea of determining a property coming from a 4-points function using only a 2-points one it is suspicious and it can hardly be generic. Indeed, the relation discussed in this section is valid only for maximally chaotic systems (saturating the chaos-bound $\lambda_L\leq 2\pi T$ \cite{Maldacena:2015waa}), but it fails in general as recently shown in \cite{Choi:2020tdj}. In general, the pole skipping phenomenon determines only the stress tensor contribution to many-body chaos which encodes the full
chaotic dynamics at maximal chaos but can be decreased in non-maximally
chaotic theories or completely cancelled in integrable systems \cite{Perlmutter:2016pkf}.

\subsection{Bounds on thermal and crystal diffusion}
Violations of the bound \eqref{boubou} are reported in the case of the charge diffusivity $D_C.$ More precisely, $C_C$ can be made arbitrarily small when certain gauge-axion couplings are considered \cite{Baggioli:2016pia}. This is because the charge transport is controlled by the Maxwell equation while chaos, and more precisely the butterfly velocity, is controlled by the Einstein equations. A priori, it is therefore hard to envisage a universal connection between the two quantities. The case of the energy diffusivity is much stronger. Indeed, one can verify that
\begin{equation}
D_{T}=\frac{f'h^{d/2-1}}{(f'h^{d/2-1})'}\frac{h'}{h}\Big|_{r_h}\frac{v_B^2}{2\pi T}\,,
\end{equation}
with all quantities evaluated at the black hole horizon $r=r_h$. Since $C_T$ can always be resolved in terms of the horizon metric, there is no strong evidence so far to suggest that $C_T\sim \mathcal{O}(1)$ can be violated. We then take the position that there exists a universal lower bound for thermal diffusion:
\begin{equation}
D_{T}\ge C_T\frac{v_B^2}{2\pi T}\,.
\end{equation}
It is worth noting that this bound was checked recently in experiments\,\footnote{It is worth pointing out that the observation of a Planckian timescale in these experiments can be just the effect of electrons-phonons interactions as in any common metal~\cite{mousatov2020phonons}.}~\cite{mousatov2020planckian,zhang2017anomalous,xu2020thermal,zhang2019thermalization,behnia2019lower} and confirmed by several holographic and field theory computations. Only one subtle violation has been identified in a class of SYK chains~\cite{Lucas1608}, and its meaning is still under debate. Finally, it is interesting to notice that this bound can be formally derived using a mathematical property of the hydrodynamic expansion known as \textit{univalence}~\cite{Grozdanov:2020koi}.\\

All the diffusion bounds discussed in the previous section attain to the explicit breaking case. Now, let us move to the SSB case. For simplicity, we will only focus on the 4-dimensional axions model with $V(X)=X^N$ in the rest of this subsection. All the results can be directly generalized to the hyperscaling violating case by simply adding a dilaton scalar. As was pointed in the previous subsection~\ref{IVA}, in presence of SSBm the diffusive shear mode become sound-like (propagating) and an additional crystal diffusion mode, associated with the axions fluctuations in the longitudinal channel, appears. In this scenario, the only diffusive mode to which the Hartnoll bound~\eqref{boubou} could apply is the latter.

\begin{figure}[t]
\centering
\includegraphics[width=0.7\linewidth]{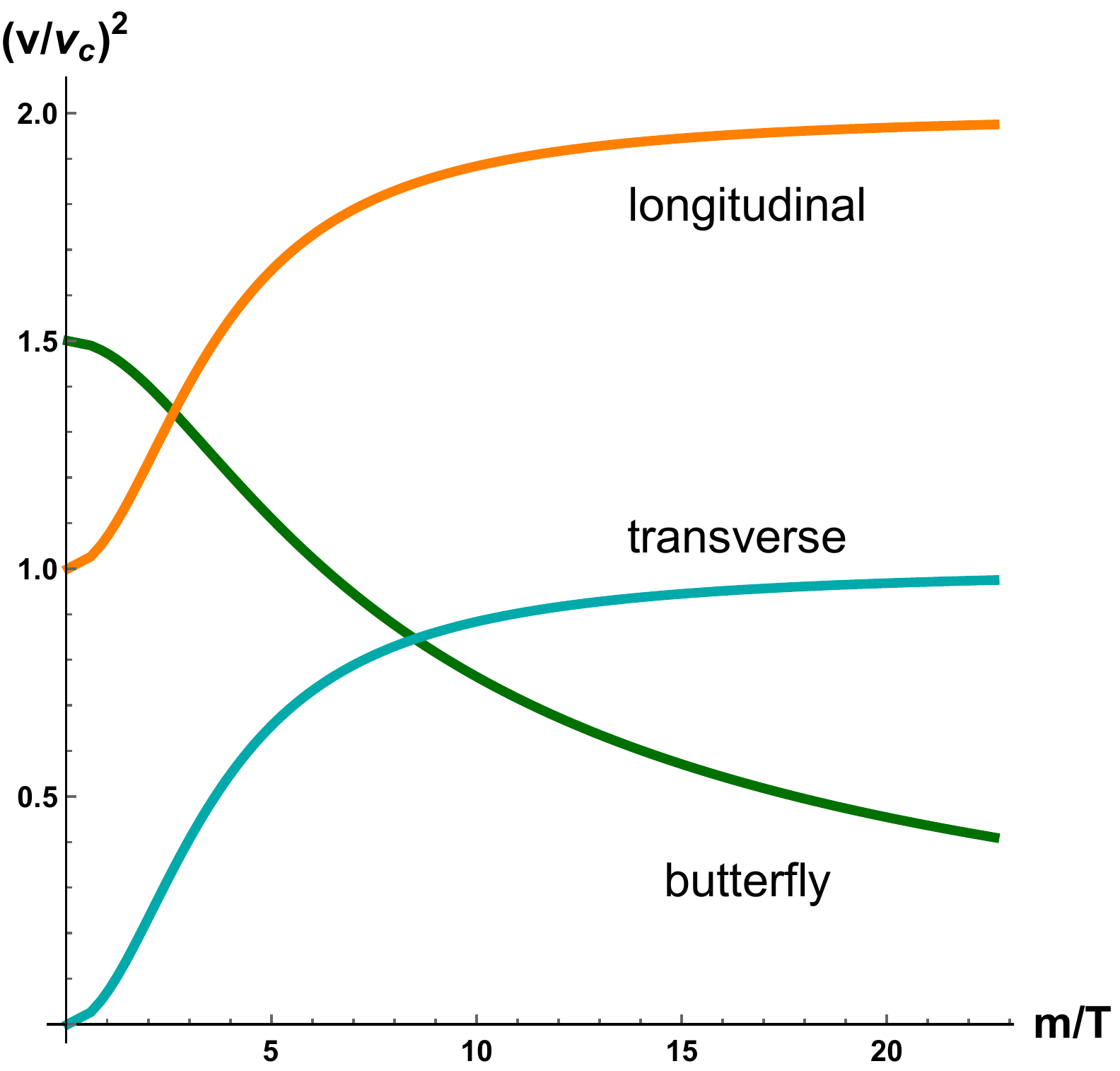}
    
\vspace{0.5cm}

\includegraphics[width=0.7\linewidth]{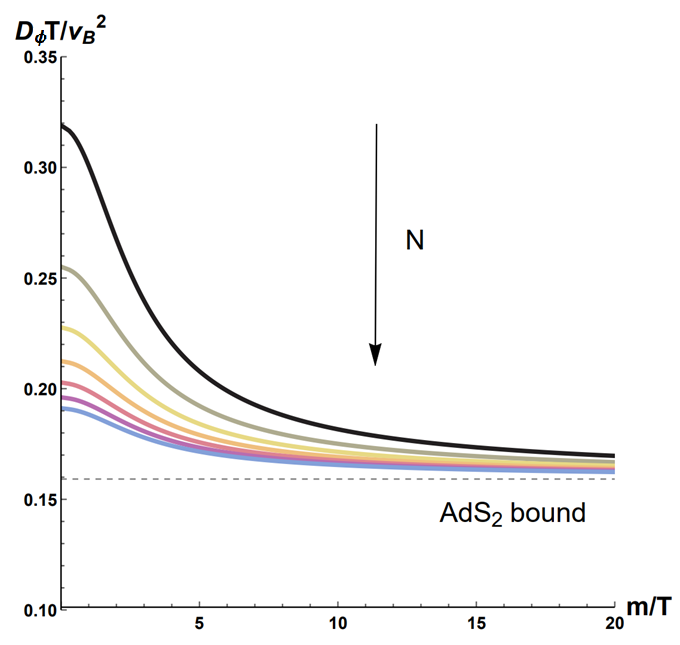}
\caption{\textbf{Top:} Various velocities in the holographic systems with SSB of translations. We consider $V(X)=X^N$ by  fixing $N = 3$. In orange the speed of longitudinal sound; in cyan the speed of transverse sound and in green the butterfly velocity. All the velocities are normalized by the conformal value $v_c^2 = 1/2$. \textbf{Bottom:} The dimensionless ratio $D_{\phi} T /v_B^2$ in function of m/T for various $N\in [3, 9]$ (from black to blue). The dashed line is the $AdS_2$ value $1/2\pi$. Figure taken from~\cite{Baggiolili2005}.}
\label{lowbound1}
\end{figure}

Let us start with the \textit{zero charge density} case. The crystal diffusive mode is decoupled from the charge sector. Its diffusion constant is given by~\cite{Armas:2019sbe,Ammon:2020xyv}:
\begin{equation}\label{hydroform}
D_\phi\,=\,\xi\,
 \frac{\left(K +G-\mathcal{P}\right)\,\chi_{\pi\pi}}{s'\,T^2\,v_L^2}\,.
\end{equation}
Note that this result perfectly matches the numeric data from the QNMs~\cite{Ammon:2020xyv}. Using this expression, one can easily check that the dimensionless quantity $D_{\phi} T$ decreases monotonously upon increasing $m/T$. On the other hand, there are three distinct velocities in the system, $v_L$, $v_T$ and $v_B$. The sound's ones both grow monotonically with $m/T$ approaching constant values at $m/T \rightarrow \infty$; on the contrary, the butterfly velocity decreases with $m/T$, exactly as the diffusion constant $D_\phi$. Using all this information, we can check that the dimensionless ratio $D_{\phi}T/v_B^2$ obeys a universal bound which is approached at infinite $m/T$, independently of the value of $N$ (see Fig.~\ref{lowbound1}):
\begin{equation}
D_\phi\,T\,\geq\,\frac{1}{2\,\pi}\,v_B^2\,.
\end{equation}

\begin{table}
\centering
\begin{tabular}{|d|b|}
\hline
\rowcolor{lightkhaki}
\textbf{Coefficient} & \textbf{Value} \\
\hline
\hline
 $\xi$ &$(3-m^2)^2/324$ \\
 \hline
  $K$ & $9/2$ \\
  \hline
   $G$ & $3/2$ \\
  \hline
   $\mathcal{P}$ & $3$ \\
  \hline
   $\chi_{\pi\pi}$ & $3$ \\
  \hline
   $T$ & $(3-m^2)/4\pi$ \\
  \hline
   $s'$ & $8\pi^2/9$ \\
  \hline
   $v_L$ & $1$ \\
  \hline
\end{tabular}
\label{tab2}
\caption{Various coefficients in the low temperature limit, $m/T\rightarrow \infty$, for $V(X)=X^3$ where the spatial translations are broken spontaneously. To achieve the data, we have assumed $u_h=1$.}
\end{table}

Furthermore, we find that, at least for $N=3$, the linearized field equations become simple enough to be solved analytically. In the low temperature limit, the near horizon geometry is $AdS_2\times R^2$. The butterfly velocity reads $v_B^2=\pi T/u_h$, going smoothly to be vanishing as $m/T\rightarrow \infty$. All the other coefficients appearing in Eq.~\eqref{hydroform} can be obtained analytically and they are summarized in Table~\ref{tab2}. For more details about the derivation, one refers to~\cite{Baggiolili2005}. Through a highly non-trivial cooperation of these coefficients, we finally achieve that
\begin{equation}
\frac{D_{\phi}T}{v_B^2}\rightarrow \frac{1}{2\pi}\equiv \text{AdS}_2\,\,
\text{bound}
\end{equation}
as $m/T\rightarrow \infty$, which is the same value as the thermal diffusion in the EXB case. It would be interesting to understand if this is just a coincidence or there is some deeper physics behind.

At \textit{finite charge density}, the crystal diffusion mode couples to the charge diffusion mode~\cite{Armas:2020bmo} and the model exhibits two diffusive modes in the longitudinal channel~\cite{Baggioli:2020edn}:
\begin{equation}
\omega_{1,2}=-i\,D_{1,2}\,k^2+\dots\,,
\end{equation}
where we denote with subscript ``1'' the mode related to crystal diffusion and with subscript ``2'' the one related to charge diffusion at zero density. Repeating the numeric calculations, one can check that $D_1$ is bounded below again but $D_2$ is not, a property reminiscent of the situation in presence of gauge-axion couplings or inhomogeneities. Here, we find a novel approach to violate the proposed bound on charge diffusion, \emph{i.e.} by introducing the phononic dynamics via the SSB of translational symmetry. 

To derive the lower bound above, it was assumed that the IR geometry has an $AdS_2$ sector. However, the conclusions should also be valid for the Lifshitz or hyperscaling violating case by using the IR scaling argument. Finally, we conclude that the crystal diffusion is similarly bounded below as the thermal one.
\begin{figure}
\centering
\includegraphics[width=0.65\linewidth]{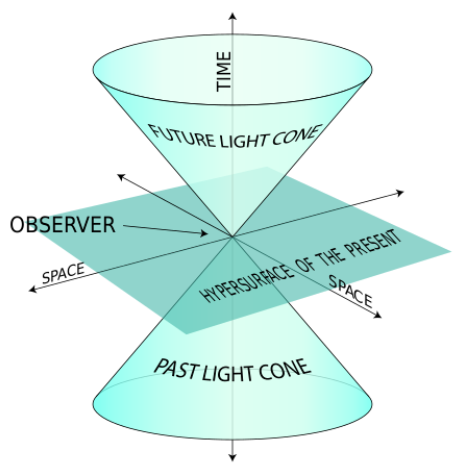}
\caption{In relativity, the causality allowed region is bounded by the lightcone with a slope $c$. Credits: Wikipedia \url{https://en.wikipedia.org/wiki/Light_cone}.}
\label{fig:light}
\end{figure}

\subsection{Diffusion bound from causality}
Now let us turn to investigate how another fundamental principle -- \textbf{causality} --  tha tmay limit diffusive processes by setting a universal upper bound on their diffusion constants. In a relativistic system, as shown in Fig.~\ref{fig:light}, any causal processes must happen in a region enclosed by a lightcone whose slope is set by the speed of light $c$. In non-relativistic systems, this is not necessarily true. However, there may also exist some ``emergent lightcone'' that limits the growth of operators. In these cases, the speed of operator growth defines an effective lightcone with speed $v_{lightcone}$, which in general does not equal to $c$. For instance, in Fermi liquids or graphene, the lightcone velocity can be identified as the Fermi velocity of the quasi-particle excitations, and $v_{lightcone}\equiv v_F\ll c $ (for graphene $\sim c/300$). 

\begin{figure}
\centering
\includegraphics[width=0.65\linewidth]{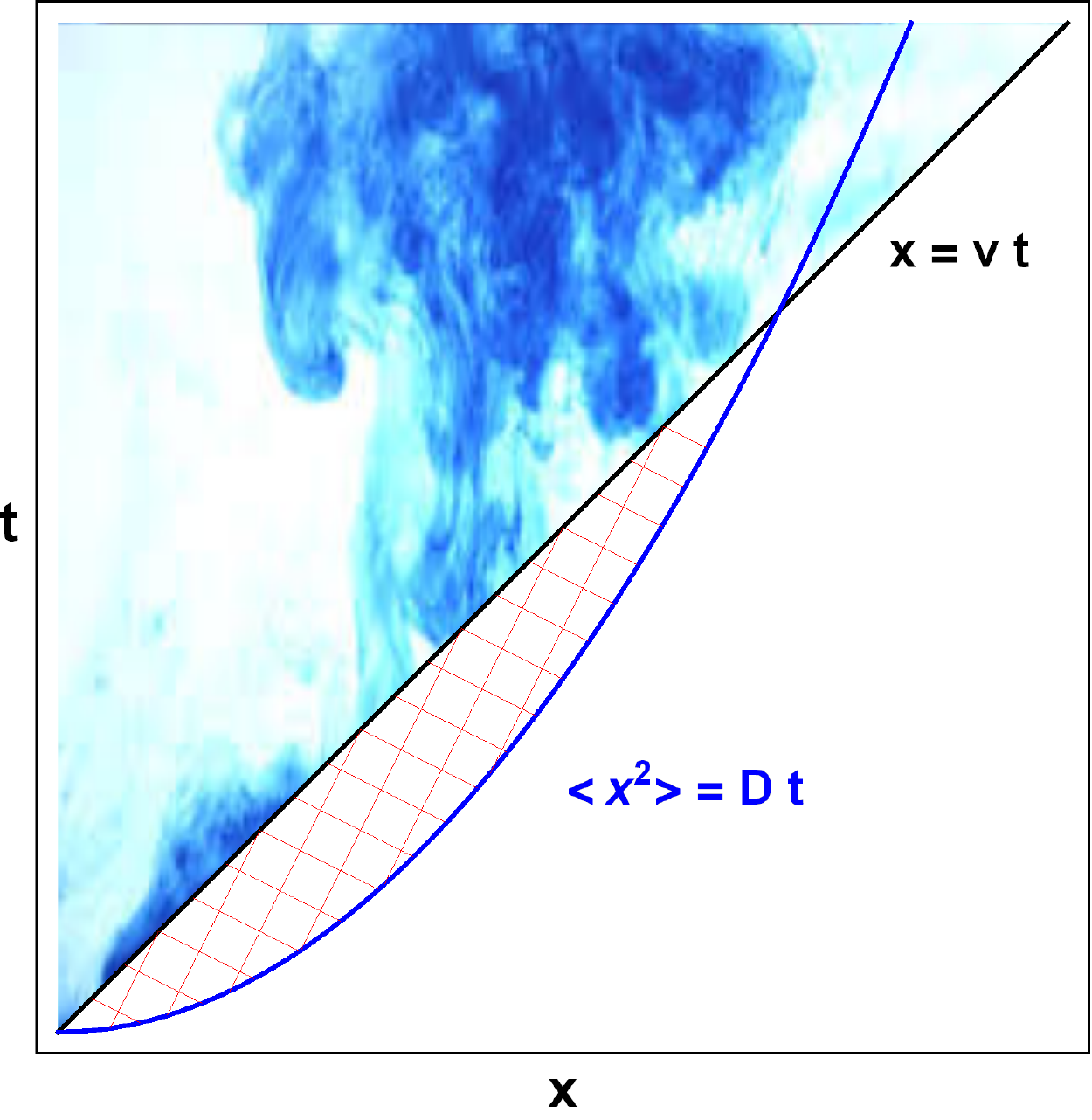}
\caption{A simple visual derivation of the upper bound on diffusion from causality. $v$ here is the lightcone speed. The region with grids is diffusion disallowed.}
\label{fig:lightcone}
\end{figure}

The transport of any local conserved quantities must happen inside this effective causal domain. Consider a diffusive process that obeys the Einstein-Stokes law:
\begin{equation}\label{ESeq}
\langle x^2 \rangle=D\,t\,.
\end{equation}
Consider a generic local fluctuation, its position should always be limited by $x\leq v_{lightcone} t$ for any later time. Combining this and~\eqref{ESeq}, we find that (see the illustration in Fig.~\ref{fig:lightcone}):
\begin{equation}
\sqrt{D\,t}\leq v_{lightcone}\,t\,.
\end{equation}
Here, we should remind the reader of the fact that diffusion sets in only after the local equilibration. It means that the equation~\eqref{ESeq} can be applied only after a timescale $\tau_{eq}$ at which the system reaches local equilibrium. Suppose that the diffusive process begins at $t=\tau_{eq}$. It is easy to see that
\begin{equation}\label{upper}
D\leq v_{lightcone}^2\,\tau_{eq}\,.
\end{equation}
This is an upper bound on any diffusion constant coming from causality and, in this respect, it has a more formal and well defined origin that the lower bound ~\eqref{boubou}. Since the two proposed bounds involve different characteristic velocities and timescales, some comments are in order:
\begin{itemize}
    \item The lightcone velocity is a microscopic velocity that in general depends on the UV parameters of the system. For instance, in a lattice fermionic system, we may do the identification like $v_{lightcone}\sim v_F\sim J\,a/\hbar$, where $J$ is the interaction scale and  $a$ is the lattice spacing. However, in some relativistic systems, the low energy excitations carrying conserved charges may also have a microscopic velocity that is not so UV-sensitive. For example, in relativistic hydrodynamic systems, the lightcone for collective modes is in general set by the sound speed $v_s$ which can be computed in terms of some macroscopic quantities, say, the pressure, energy density, etc. On the contrary, the butterfly velocity, $v_B$ is a completely IR quantity in any chaotic systems, characterizing the speed of scrambling the local perturbations initially set. It is a finite temperature effect, hence strongly depends on the temperature itself. For example, in a strongly coupled system with the hyperscaling violating IR fixed point, $v_B\sim T^{1-1/z}$ with $(1-1/z)>0$ at low temperatures~\cite{Blake:2016wvh,Roberts:2016wdl}. Therefore, it is in general much slower than the microscopic velocity in the IR region (see the top panel of Fig.~\ref{lowbound1}).
    \item The equilibrium timescale $\tau_{eq}$ characterizes how fast a \textit{generic} system gets local equilibration. The Planckian time $\tau_{pl}\sim \frac{\hbar}{k_B T}$ sets a universal minimum for $\tau_{eq}$ as $T\rightarrow 0$. In weakly coupled systems with long-lived quasi-particles, equilibration processes are slow.  Then, in general, $\tau_{eq}\gg \tau_{pl}$. However, for some strongly coupled systems, the equilibrium timescale may reach its minimum.
\end{itemize}

A manifest example where the upper bound Eq.~\eqref{upper} is obeyed is linearized relativistic hydrodynamics with a conserved stress tensor:
\begin{equation}
T^{\mu\nu}\,=\,\mathcal{E}\,v^\mu\,v^\nu\,+\,\mathfrak{p}\,\Delta^{\mu\nu}\,-\,\eta\,\sigma^{\mu\nu}\,\,.
\end{equation}
Using the hydrodynamics method, one can show that in the transverse channel there is a diffusive momentum mode with the dispersion relation:
\begin{equation}
\omega=-i \frac{\eta}{\mathcal{E}+\mathfrak{p}}\,k^2+\dots\,,
\end{equation}
which is the result of the momentum conservation. This, however, suffers the problem of superluminality, because its group velocity
\begin{equation}
v_g\equiv \Big|\frac{d\omega}{dk}\Big|\sim k\,,
\end{equation}
can be arbitrarily large if there is no cut-off on the momentum $k$. One well-known solution to this problem \footnote{It is actually well-known that a superluminal \underline{group} velocity does not imply any violation of causality \cite{doi:10.1119/1.15715}. This is a ``problem'' only for numerical simulation using linearized relativistic hydrodynamics as initial conditions. There is absolutely no fundamental nor physical problem here.} is to introduce a fictitious relaxation time $\tau_{\pi}$ which arises from the constitutive relation of the stress tensor to higher order. As a result, the diffusive equation gets modified as
\begin{equation}
\omega^2\,+\,i\,\omega\,\tau_{\pi}^{-1}\,\,=\,v_T^2\,k^2\,,\quad \quad v_T^2\,=\,\frac{\eta}{(\mathcal{E}+\mathfrak{p}   )\,\tau_{\pi}}\,,
\end{equation}
which is known as the \textbf{Israel-Stewart formalism}~\cite{ISRAEL1979341}. Solving this simple equation, we get
\begin{equation}
\omega=\frac{1}{2\tau_\pi}\left(-i\pm\sqrt{4v_T^2\,k^2\,\tau_\pi^2-1}\right)\,.
\label{bava}
\end{equation}
This time, the diffusive dynamics sets only when $k<k_g\equiv 1/(2v_T\tau_\pi)$. Here, $k_g$ is just the k-gap mentioned previously. Above the $k_g$ momentum, momentum propagates ballistically at the sound speed $v_T\leq c$. Identifying $v_T\sim v_{lightcone}$ and $\tau_\pi \sim \tau_{eq}$ and expanding~\eqref{bava}, one can observe that  $D_\pi=v_T^2\,\tau_\pi\sim v_{lightcone}^2\,\tau_{eq}$. Consequently, the causality requirement
\begin{equation}
    v_T^2\,\leq\,c^2\,,
\end{equation}
is equivalent to an upper bound on the diffusion constant
\begin{equation}
D_\pi \leq c^2\tau_\pi\,,
\end{equation}
which takes exactly the form of Eq.~\eqref{upper}, with $c$ being the relativistic lightcone velocity and $\tau_\pi$ the equilibration time.

\begin{figure}[h]
\centering
\includegraphics[width=0.75\linewidth]{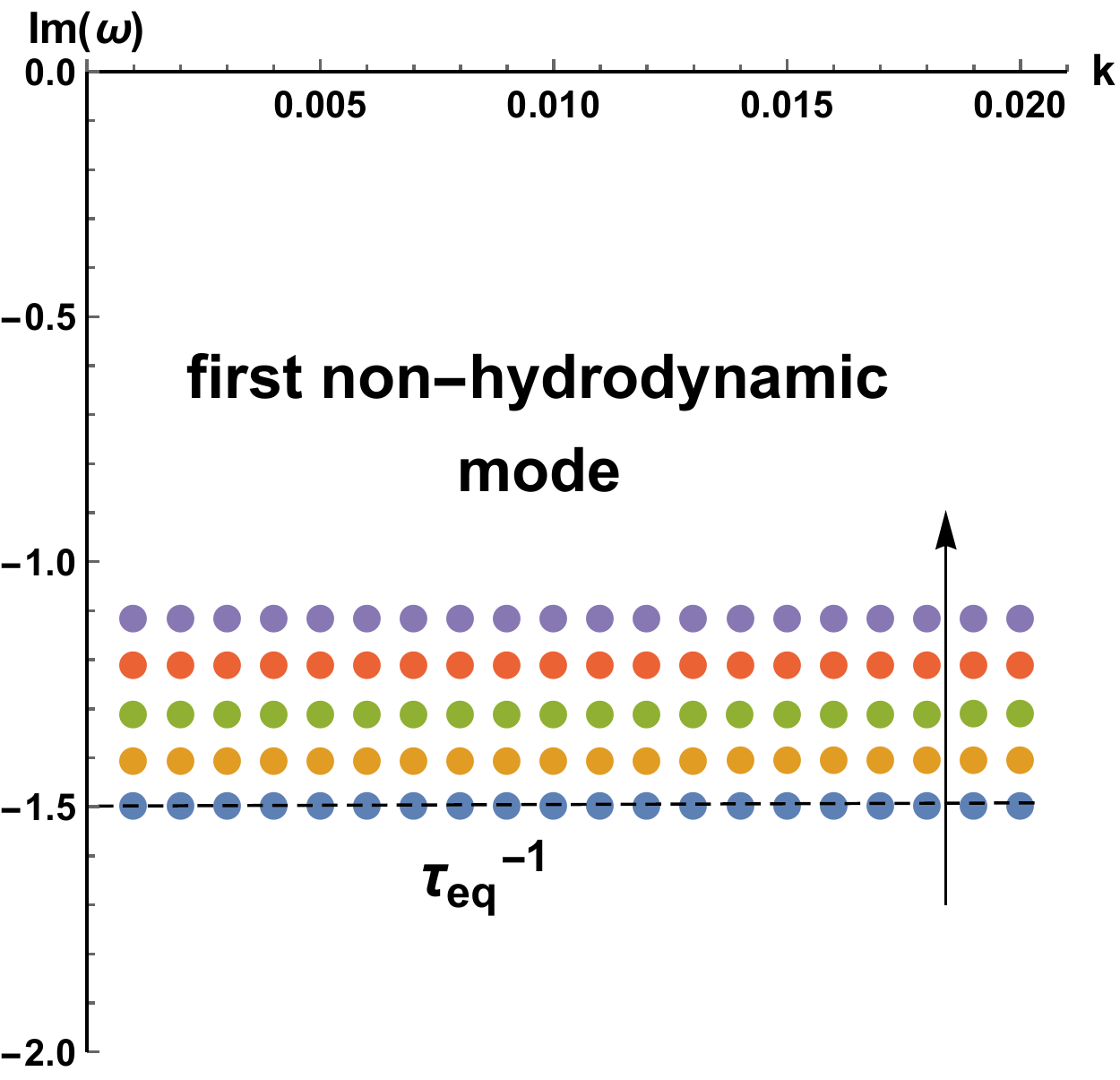}
\caption{The local equilibrium timescale is extracted from the imaginary part of the first non-hydrodynamic mode as in Eq.~\eqref{eqtime}. The arrow indicates its motion as the increase of $m/T$.}
\label{figup0}
\end{figure}
\begin{figure}[t]
\centering
\includegraphics[width=0.7\linewidth]{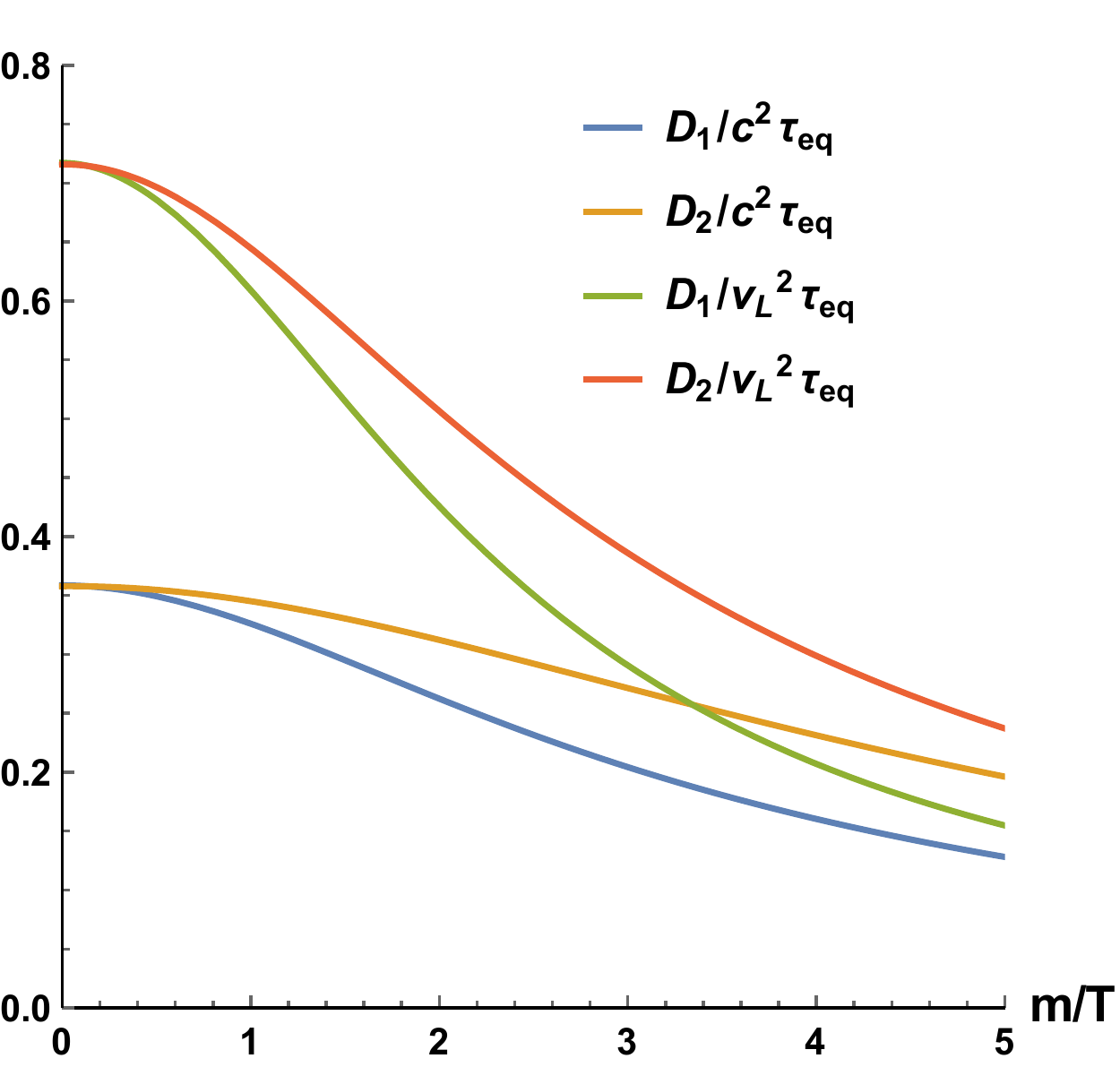}
\vspace{0.5cm}
\qquad\includegraphics[width=0.78\linewidth]{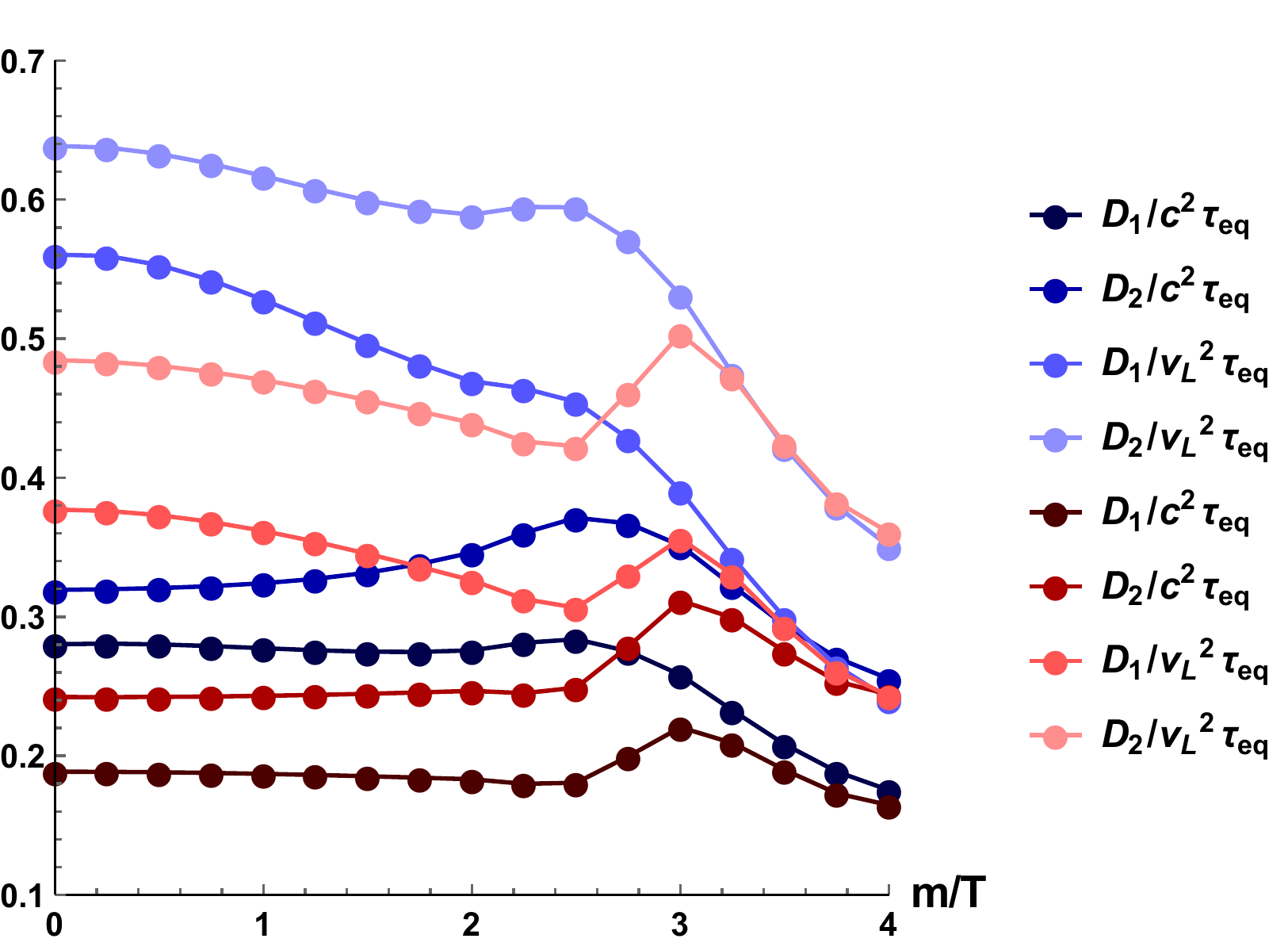}
\caption{The upper bound on diffusion.$c$ is the speed of light and $v_L$ is the speed of longitudinal sound. \textbf{Top:} The dimensionless ratio at zero density, where $D_1$ is the crystal diffusion and $D_2$ the charge diffusion. \textbf{Bottom:} The dimensionless ratio at finite density, where the blue
points are for $\mu/T = 3$ and the red for $\mu/T = 5$.}
\label{figup12}
\end{figure}

In holographic systems, the equilibration process is dominated by the first non-hydrodynamic modes.\,\footnote{The hydrodynamic modes are gapless, hence do not set any finite timescale for relaxations. Then, we should look at the gapped modes. Here, the lowest lying gapped QNM is what we term the first non-hydrodynamic mode.} Then, we may identify
\begin{equation}\label{eqtime}
\tau_{eq}\sim \frac{1}{\text{Im}\left[\omega_{\text{QNM}}\right]}\,, 
\end{equation}
where $\omega_{\text{QNM}}$ is the complex frequency of the first non-hydrodynamic QNM. Back to the holographic axion model, the longitudinal sound speed plays the role of the lightcone velocity, \emph{i.e.} $v_L\sim v_{lightcone}$. Then, one can directly check the upper bound by computing the black hole QNMs (see how to extract the local equilibrium timescale $\tau_{eq}$ in Fig.~\ref{figup0}). The results for zero density and charged case have been shown in Fig.~\ref{figup12}, respectively. It is obvious that the upper bound on diffusion from causality~\eqref{upper} is respected for all the cases.

\begin{figure}
\centering
\includegraphics[width=0.95\linewidth]{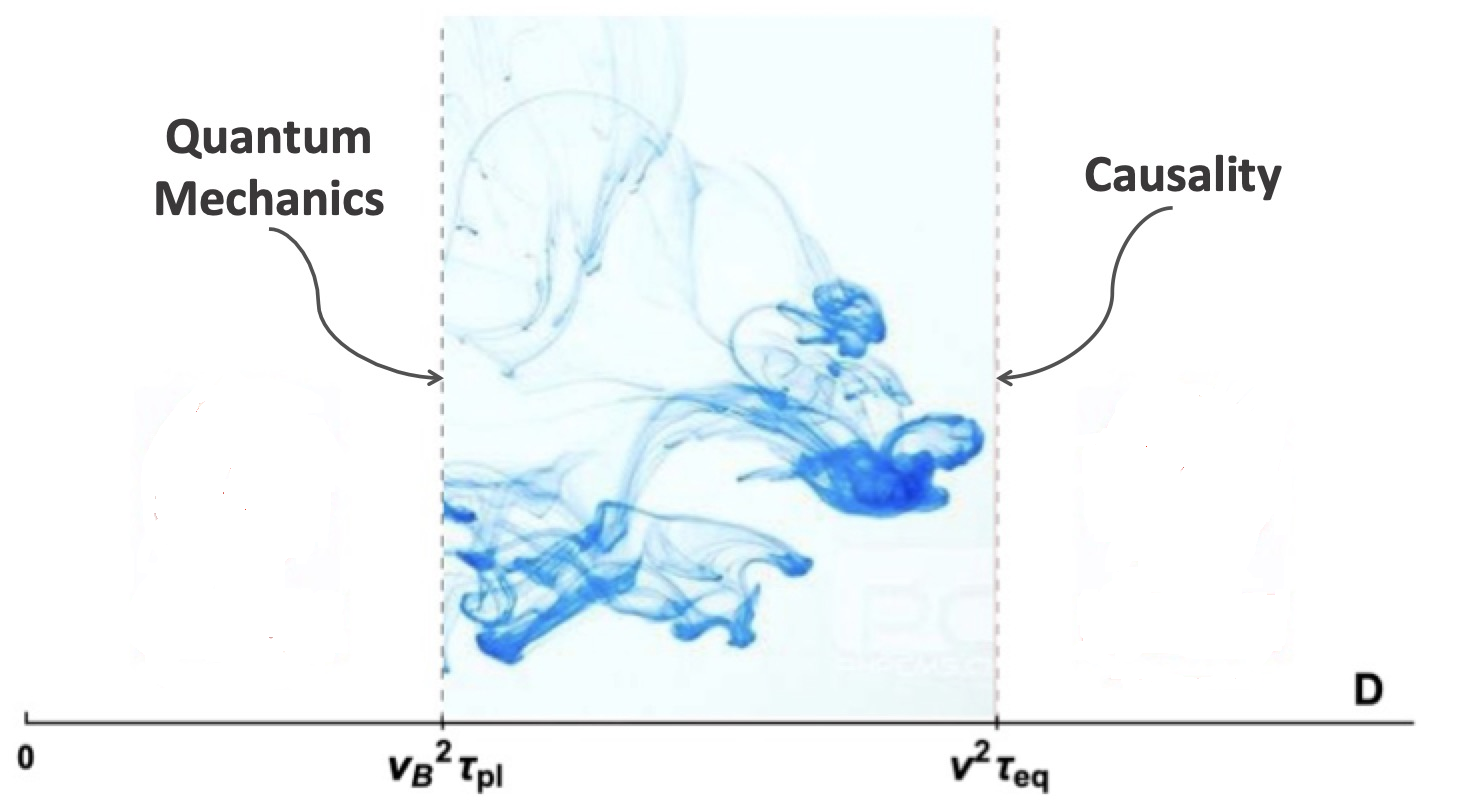}
\caption{The diffusion allowed region is bounded by quantum mechanics and causality.}
\label{figsand}
\end{figure}

Then, above analysis suggests that the diffusion constants may always be constrained in an intermediate range:
\begin{equation}
v_B^2\,\tau_{pl}\,\leq\,D\,\leq\,v_{ligthcone}^2\,\tau_{eq}\,.
\end{equation}
In Fig.~\ref{figsand} we show a "sandwich-like" illustration for this feature. Note that for many cases of holographic systems or quantum critical systems, the equilibrium time reaches the minimum, $\tau_{eq}\sim \tau_{pl}$ and the butterfly velocity determines the lightcone speed (assuming that there is no quasiparticle or well-defined microscopic velocity), for which we may have that $D\sim v_B^2\tau_{pl}$, \emph{i.e.} the diffusion allowed region shrinks to a single point. Recently,~\cite{Arean:2020eus} proposed to define the velocity scales in the diffusive bound as the ratio between the frequency and the momentum setting the radius of convergence of the linear hydrodynamic theory.

\subsection{A bound on stiffness}
Now, let us move to check if there is also some constraint on the propagating speed of the longitudinal sound modes. In some previous studies~\cite{Hohler:2009tv,Cherman:2009tw}, it has been proposed from the holographic duality that there exists an upper bound on speed of sound modes, which is set by its conformal value, $v_c\equiv 1/(d-1)$ for $d$ dimensional system. This proposal has been later related to the physics of neutron stars, very compact objects which display an extremely stiff equation of state~\cite{Bedaque:2014sqa}. This bound however has been later checked in various holographic models and its violation has been observed both at finite charge density case or in presence of multi-trace deformations~\cite{Hoyos:2016cob,Anabalon:2017eri}. 

Note that all these previous discussions were confined to fluid systems with no long-range order, where the longitudinal sounds are simply pressure waves, \emph{i.e.}
\begin{equation}
v_L^2=\frac{\partial \mathfrak{p}}{\partial \mathcal{E}}\,.
\label{cc}
\end{equation}
Note that the r.h.s. of Eq.~\eqref{cc} also defines the stiffness of the system
\begin{equation}
\kappa\equiv \frac{\partial \mathfrak{p}}{\partial \mathcal{E}}\,.
\end{equation}
This then means that the bound on the sounds is simultaneously a bound on the stiffness:
\begin{equation}
\kappa \leq \kappa_c, \quad \kappa_c\equiv \frac{1}{d-1}\,.
\end{equation}

In the holographic axion model, the SSB of translations results in a non-zero shear elasticity $G$ and the dual systems on boundary is a solid (see also Eq.~\eqref{eqforG}). The presence of a finite $G$ speeds up the longitudinal sounds, which becomes~\cite{Esposito:2017qpj}:
\begin{equation}
v_L^2=v_c^2+v_T^2\ge v_c^2\,.
\end{equation}
The first equivalence is a direct consequence of conformal invariance, \emph{i.e.} $\langle T^\mu_\mu\rangle=0$.
Again, this is a new way to break the proposed bound on the sound speed. Nevertheless, we will verify that the result above does not imply a violation of the stiffness bound. That is to say the bounds conjectured in~\cite{Hohler:2009tv,Cherman:2009tw} has to be viewed as a limit to the stiffness instead of a bound on the speed of sound.
\begin{figure}
\centering
\includegraphics[width=0.75\linewidth]{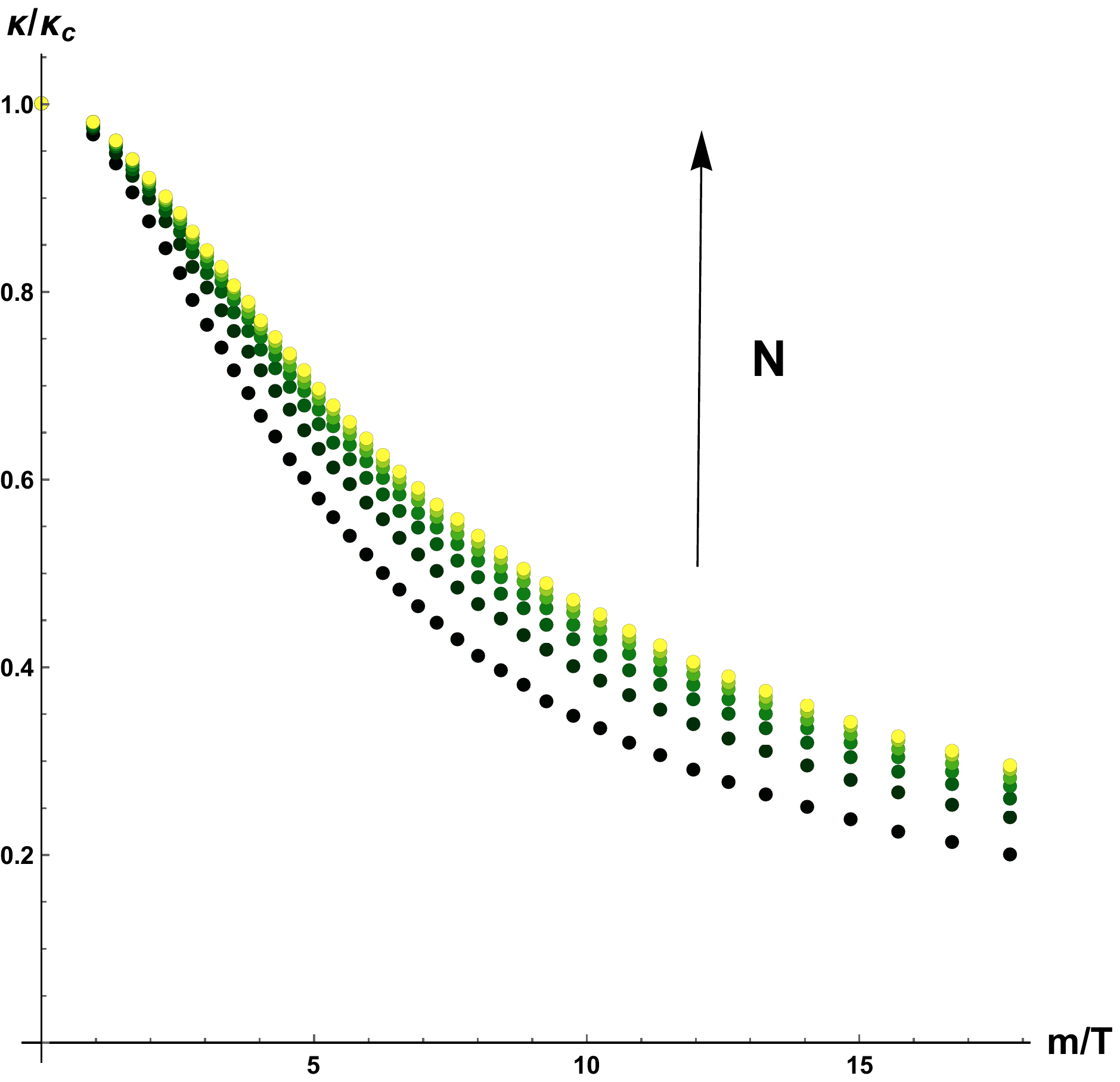}
\caption{The stiffness $\kappa$ for $N \in [3,9]$ (from black to yellow) in function of $m/T$. Figure taken from~\cite{Baggiolili2005}.}
\label{figkappa}
\end{figure}

As was revealed earlier, in presence of the finite strain pressure $\mathcal{P}$, the total pressure $\langle T^{xx}\rangle $ is no longer equal to the thermodynamic pressure $\mathfrak{p}$. For the conformal solid, the traceless stress tensor implies that
\begin{equation}
\langle T^{xx}\rangle= \mathfrak{p}+\mathcal{P}=\frac{1}{d-1}\mathcal{E},\,\,\,\mathcal{P}>0\,.
\end{equation}
Applying the definition of the stiffness, we obtain 
\begin{equation}\label{stiffness}
\kappa\equiv\frac{\partial \mathfrak{p}}{\partial \mathcal{E}}=\frac{1}{d-1}-\frac{\partial \mathcal{P}}{\partial \mathcal{E}}.
\end{equation}
For the holographic model $V(X)=X^N$ with $N>3/2$ and $d=3$, one can easily show that the second term in the last step is always positive by using 
\begin{align}
    &\mathcal{P}\,=\,\frac{m^2\, N\, u_h^{2N}}{(2 \,N-3) \,u_h^3}\,,\quad \mathcal{E}\,=\,\frac{1}{u_h^3}-\frac{m^2 \,u_h^{2N}}{(3-2 N) \,u_h^3}\,.
\end{align}
To make it clearer, we plot the final result of~\eqref{stiffness} in Fig.~\ref{figkappa}. It confirms our statement that
\begin{equation}
\kappa\leq \kappa_c\,,
\end{equation}
independently of $N$. For $N<3/2$, the phonons are destroyed by the external source that breaks spatial translations, and there is no low energy modes propagating at the speed set by the stiffness. Note also that when $N<3/2$, both of $G$, $\mathcal{P}$ become negative, implying the existence of a dynamical instability. Hence, we will not view this as a violation of the stiffness bound.

\section{Holographic pinned structures}\label{pinned}
\subsection{Pseudo-Goldstone modes}
So far, we have focused on two very distinct symmetry patterns: (A) the explicit breaking of translations, giving rise to the physics of momentum dissipation and the Drude model, and (B) the spontaneous symmetry breaking of translations, related to the physics of elasticity and the dynamics of phonons. In the previous sections, we have explained how to technically achieve these two limits using the holographic axion model.

Now, we want to make a step forward and combine the two in what is called the \textbf{pseudo-spontaneous regime}. As we will see, the holographic axion model is rich enough to encompass also this different situation. From the physical point of view, this regime is realized in QCD for the \textbf{Pions}~\cite{Burgess:1998ku}, where chiral symmetry is both broken spontaneously and explicitly, giving a small mass to the corresponding pseudo-Goldstones and in the so-called \textbf{pinned charge density waves}~\cite{RevModPhys.60.1129}, where impurities produce a pinning frequency for the corresponding phason modes.

Indicating with $\langle EXB \rangle$ the explicit breaking scale, and with $\langle SSB \rangle$ the spontaneous one, we want to work in the limit of:
\begin{equation}
\langle EXB \rangle/T\,\ll\,1\,\,,\quad \frac{\langle EXB \rangle}{\langle SSB \rangle}\,\ll\,1\,,
\end{equation}
such that the corresponding charge (momentum in this case) is ``approximately'' conserved (corresponding to a weak explicit breaking mechanism) and a pseudo-Goldstone mode can be still defined. Let us first address the question of how to realize this limit within holography and later describe in detail which are the phenomenological consequences. First, let us give an intuitive argument taken from~\cite{Alberte:2017cch}. What is the substantial difference between the explicit breaking and the spontaneous one? The explicit breaking appears at the level of the fundamental action of the theory and it relates to the presence of an operator whose source breaks the specific symmetry. The spontaneous breaking does not relate to the action of the system but rather to its ground state, the preferred solution in which the system wants to sit. In a sense, the EXB can be thought as an ultraviolet (UV) breaking, while the SSB as an infrared (IR) breaking. This idea is beautifully encoded in the holographic picture by considering that the UV dynamics is localized close to the AdS boundary, meanwhile the IR dynamics nearby the black hole horizon. One could think, for example, about the famous holographic model for superconductivity~\cite{Hartnoll:2008vx} where the mass of the gauge field (determining the breaking of the $U(1)$ symmetry) is localized close to the horizon and it is zero at the boundary (\emph{i.e.} no explicit breaking).

\begin{figure}[h]
\centering
\includegraphics[width=0.7\linewidth]{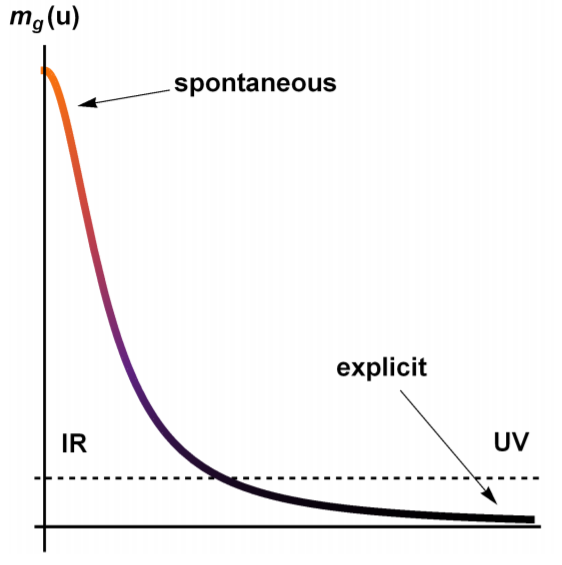}
\caption{The intuitive picture relating the spontaneous and explicit breaking of translations with the profile of the graviton mass in the holographic bulk. Picture taken from~\cite{Alberte:2017cch}.}
\label{exmass}
\end{figure}

The idea here is exactly analogous but this time, since we are dealing with translations, is the mass of the graviton that has to be considered\,\footnote{See~\cite{Amoretti:2016bxs} for a detailed analysis of the Ward's identities in the pseudo-spontaneous regime.}. In particular, one would naively connect the explicit breaking of translations with the presence of a finite UV mass, and the spontaneous breaking with the presence of a growing mass peaked in the IR. Therefore, in this picture, the pseudo-spontaneous regime should be accomplished by considering a bulk configuration where the graviton mass in the UV is small and it rapidly grows in the IR (see Fig.~\ref{exmass}).

In order to have such a situation, one could consider a potential of the form:
\begin{equation}
V(X)\,=\,X\,+\,\beta\,X^N\,,\quad \beta\gg\,1\,,
\end{equation}
since the graviton mass is proportional to $m_g^2(u)\,\sim\,V_X(X)$ and the argument $X\sim u^2$ vanishes at the boundary and grows towards the black hole horizon.
    
This intuition turns out to be correct and it can be formalized better using the language of the previous sections. Given a potential $V(X)=X^N$, we know now that the breaking of translations is explicit if $N<5/2$ and spontaneous if $N>5/2$. Therefore, to reach the pseudo-spontaneous regime, we need to utilize a potential of the type:
\begin{equation}
V(X)\,=\,X^N\,+\,\beta\,X^M\,,\qquad N<5/2,\,M>5/2\,.
\end{equation}
This is exactly what has been done in~\cite{Alberte:2017cch} which first\,\footnote{The same day, a work~\cite{Andrade:2017cnc} studying this regime in the context of holographic helical lattices was posted in arXiv.} studied the dynamics of pinning phonons in holography. Curiously\,\footnote{It is also funny to notice that historically this regime was achieved before the fully spontaneous one~\cite{Alberte:2017oqx}, which for technical reasons has been the most difficult to construct.}, this model was considered long time before, in~\cite{Baggioli:2014roa}, but the connection with the pseudo-spontaneous breaking of translations has not been given therein. A full understanding of this regime has appeared later in~\cite{Ammon:2019wci}. A slightly different way to realize this regime can also be found in~\cite{Li:2018vrz}.
    
\begin{figure}[ht]
\centering
\includegraphics[width=0.9\linewidth]{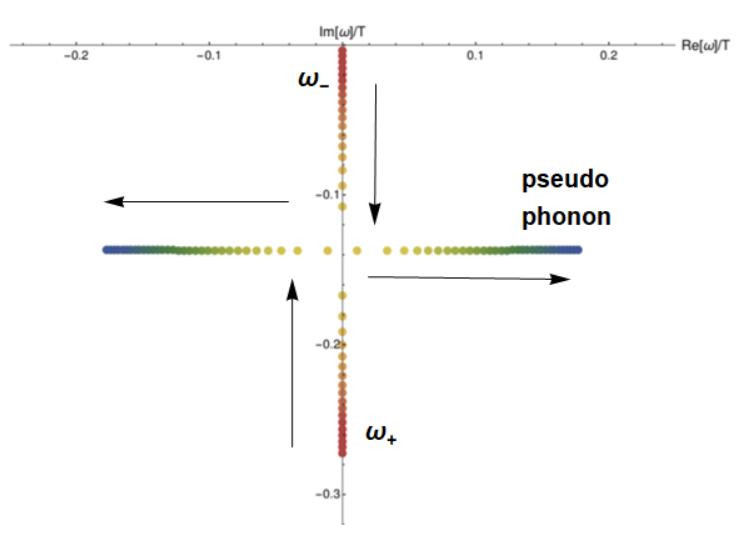}
\caption{The dynamics of the low-energy hydrodynamic modes in the pseudo-spontaneous regime. The dance of the modes is perfectly described by Eq.~\eqref{modes}. Figure adapted from~\cite{Ammon:2019wci}.}
\label{figcoll}
\end{figure}

Let us analyze in detail what one would expect in this regime. Following the hydrodynamic description of~\cite{Delacretaz:2017zxd}, in presence of both explicit and spontaneous breaking of translations, the low-energy hydrodynamic spectrum at zero momentum is described by the solutions of the equation:
\begin{equation}
\left(\Omega\,-\,i\,\omega\right)\,\left(\Gamma\,-\,i\,\omega\right)\,+\omega_0^2\,=\,0\,,
\label{modes}
\end{equation}
where $\Gamma$ is the \textbf{momentum relaxation rate} (how fast momentum is dissipated), $\Omega$ the \textbf{phase relaxation rate} (measuring the lifetime of the Goldstones) and $\omega_0$ the so-called \textbf{pinning frequency} -- the mass of the (not anymore) Goldstone modes. The two modes involved in the quadratic equation \eqref{modes} are the transverse momentum $\pi^\perp$ and the transverse component of the Goldstone field. Solving Eq.~\eqref{modes} we obtain a pair of excitations whose frequencies are given by
\begin{equation}
\omega_{\pm}\,=\,-\,\frac{i}{2}\,\left(\Gamma+\omega\right)\,\pm\,\frac{1}{2}\sqrt{4\,\omega_0^2\,-\,\left(\Gamma-\Omega\right)^2}\,.
\end{equation}
For $4\, \omega_0^2-(\Gamma-\Omega)^2<0$, the two modes lie along the imaginary axes; they are purely decaying modes. Exactly at $4\, \omega_0^2-(\Gamma-\Omega)^2=0$, the two modes collide on the imaginary axes and for $4\, \omega_0^2-(\Gamma-\Omega)^2>0$ they move onto the complex plane acquiring a finite and growing real part. This dynamics is perfectly obeyed by the holographic models, see, for example, Fig.~\ref{figcoll}.

\begin{figure}[ht]
\centering
\includegraphics[width=0.9\linewidth]{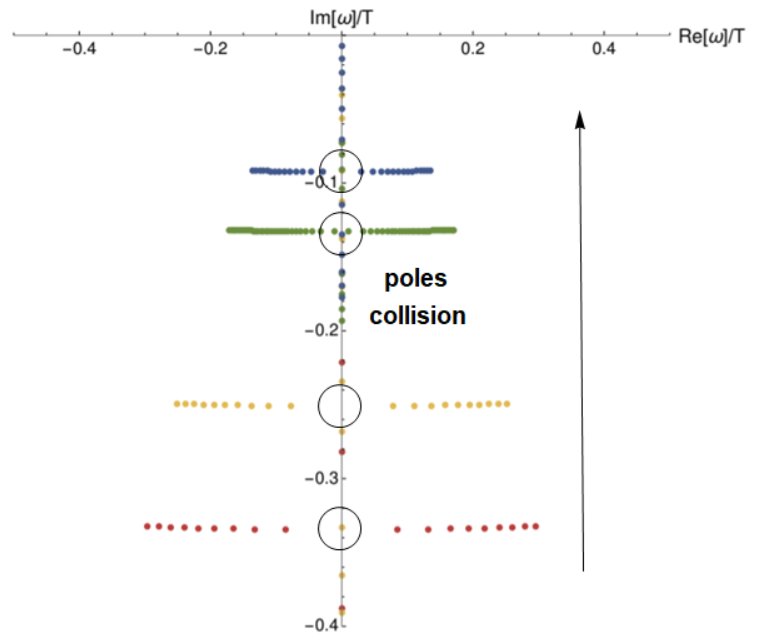}
\caption{The collision between the two modes by moving towards the pseudo-spontaneous regime. Figure adapted from~\cite{Ammon:2019wci}. Notice how this dynamics was already present in~\cite{Baggioli:2014roa}.}
\label{pp}
\end{figure}

Interestingly, this collision is not an exclusive feature of the pseudo-spontaneous regime. Indeed, this collision is the same giving rise to the incoherent-coherent transition in the models with pure explicit breaking~\cite{Davison:2014lua}. Rather, a peculiar characteristic of the pseudo-spontaneous regime is the fact that such collision happens at low frequency, within the so-called hydrodynamic regime. Increasing the amount of spontaneous breaking, both $\Gamma$ and $\Omega$ become smaller and the collision happens close to the origin $\omega=0$. This tendency is shown explicitly in Fig.~\ref{pp}. Notice that in the purely spontaneous regime, both modes would collapse at the origin and form the propagating shear sound.

\begin{figure}[ht]
\centering
\includegraphics[width=0.7\linewidth]{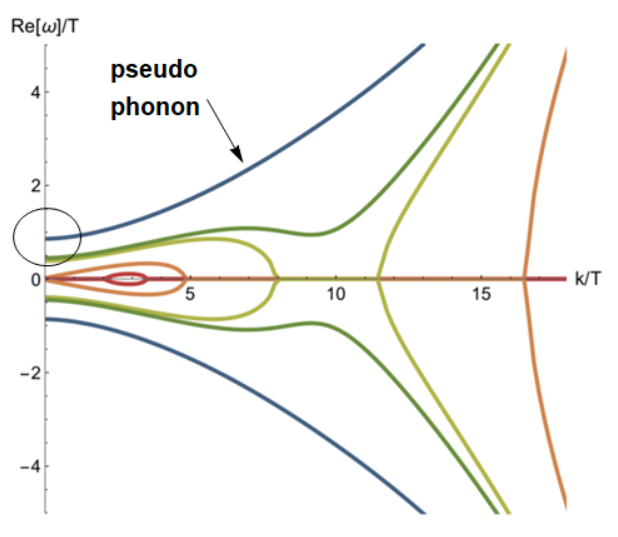}
\caption{The dispersion relation of the pseudo-phonon. Figure adapted from~\cite{Alberte:2017cch}.}
\label{cpp}
\end{figure}

Finally, one can perform also the computation at finite momentum and verify the dispersion relation of the various modes. As shown in Fig.~\ref{cpp}, the pseudo-phonon mode shows a characteristic dispersion:
\begin{equation}
\mathrm{Re}\left[\omega\right]=\,\sqrt{\omega_0^2\,+\,v^2\,k^2}\,,
\end{equation}
typical of massive modes (cf. Pions).

\begin{figure}
\centering
\includegraphics[width=0.47\linewidth]{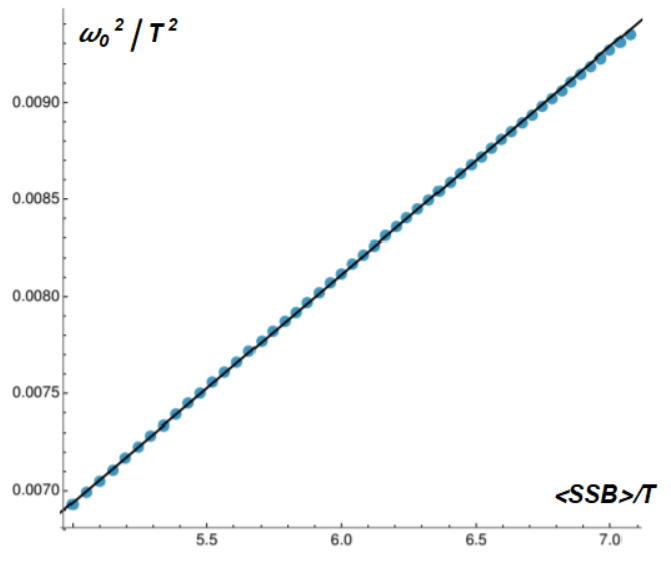}\qquad
\includegraphics[width=0.45\linewidth]{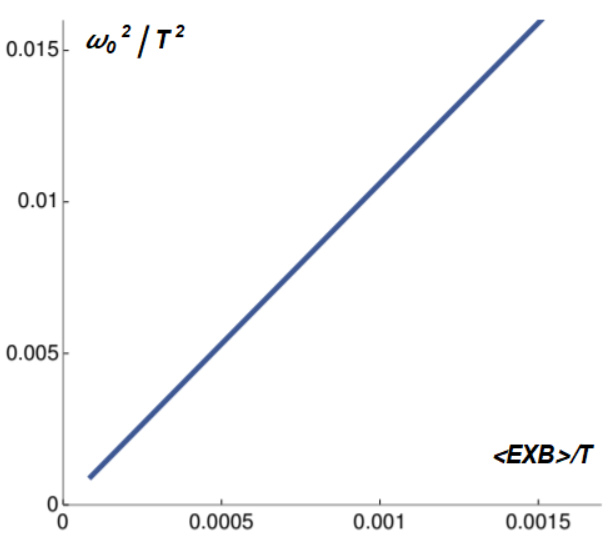}
\caption{The numerical confirmation of the GMOR relation \eqref{gmoreq}. Figure adapted from~\cite{Ammon:2019wci}.}
\label{gmor}
\end{figure}

Before moving to the next subsection, it is interesting to analyze how the parameters entering in Eq.~\eqref{modes} depend on the explicit and spontaneous breaking scales, $\langle EXB \rangle, \langle SSB \rangle$. Combining hydrodynamic and holographic arguments, we can derive a simple general formula:
\begin{align}
&\Gamma\,+\,\frac{\omega_0^2}{\Omega}\,=\,m^2\,\frac{V_X}{2\,\pi\,T}\,+\,\mathcal{O}(m^4)\,.
\label{f1}
\end{align}
Assuming the benchmark potential for the pseudo-spontaneous regime $V(X)=X+\beta X^N$ we can also obtain that
\begin{align}
& \Gamma\,=\,\frac{m^2}{2\,\pi\,T}\,\sim\,\langle EXB \rangle^2+\dots\,,\\ &\frac{\omega_0^2}{\Omega}\,=\,\frac{m^2\,\beta\,N}{2\,\pi\,T}\sim \langle SSB \rangle^2+\dots\,.
\label{io}
\end{align}
Moreover, using the numerical data (see Fig.~\ref{gmor}), we can prove robustly that
\begin{equation}
\omega_0^2\,=\,\langle EXB \rangle\,\langle SSB \rangle\,, \label{gmoreq}
\end{equation}
as imposed by the Gell-Mann-Oakes-Renner (GMOR) relation~\cite{PhysRev.175.2195} (cf. Pions).
Notice that combining~\eqref{io} together with the GMOR relation~\eqref{gmoreq} we find immediately that
\begin{equation}
\Omega\,\sim\,\frac{\langle EXB \rangle}{\langle SSB \rangle}\,\,!
\label{wow}
\end{equation}
This is surprising for various reasons and it will be the topic of the next subsection.

\subsection{Phase relaxation and universality}
The fact that the phase relaxation rate is proportional to the explicit breaking scale is \textit{per se} very surprising. In general, phase relaxation is induced by the presence of elastic defects such as dislocations and disclinations and it has nothing to do with the explicit breaking of momentum. It comes from the fact that the displacement vectors are not anymore single-valued and a non trivial Burgers vector appears.

First, let us notice how we have already encountered part of the scalings in Eq.~\eqref{wow} in the previous section, when we were discussing the collision of the modes. As also shown in Fig.~\ref{figcoll}, the collision moves towards the origin increasing the strength of the SSB and one of the reason is exactly that the second pole, approximately located at $\omega=-i\Omega$, moves upwards. This is in perfect agreement with the fact that $\Omega$ is inversely proportional to such strength and it becomes smaller by increasing the amount of SSB. Second, it is important to emphasize that the relation~\eqref{wow} has been confirmed numerically in a large class of holographic axion models~\cite{Ammon:2019wci,Baggioli:2019abx} (and in similar models~\cite{Donos:2019txg,Amoretti:2018tzw,Amoretti:2019cef,Andrade:2018gqk,Andrade:2020hpu}). This confirms the universal character of this scaling.

The story is even more fascinating, since the authors of~\cite{Amoretti:2018tzw} motivated and proposed an even more universal relation:
\begin{equation}
\Omega\,=\,\mathrm{M}^2\,\xi\,=\,\frac{\omega_0^2\,\chi_{\pi\pi}}{G}\,\xi \label{universe}\,,
\end{equation}
where $\mathrm{M}$ is the mass of the pseudo-Goldstone mode and $\xi$ the dissipative parameter determining the diffusion constant of the Goldstone mode in the un-relaxed theory. Using the fact that at leading order the Goldstone diffusion is given by $D_\phi=G\,\xi$ and that $v_T^2=G/\chi_{\pi\pi}$, we can re-write Eq~\eqref{universe} as:
\begin{equation}
\Omega\,=\,\frac{\omega_0^2\,D_\phi}{v_T^2}\,.
\end{equation}
Notice how Eq.~\eqref{universe} is compatible with the scalings discussed and derived in the previous paragraphs:
\begin{equation}
\underbrace{\frac{\langle EXB \rangle}{\langle SSB \rangle}}_{\Omega}\,\sim\,\underbrace{\langle EXB \rangle \langle SSB \rangle}_{\omega_0^2}\,\underbrace{\frac{1}{\langle SSB \rangle^2}}_{\xi/G}\,.
\end{equation}
\begin{figure}
\centering
\includegraphics[width=0.7\linewidth]{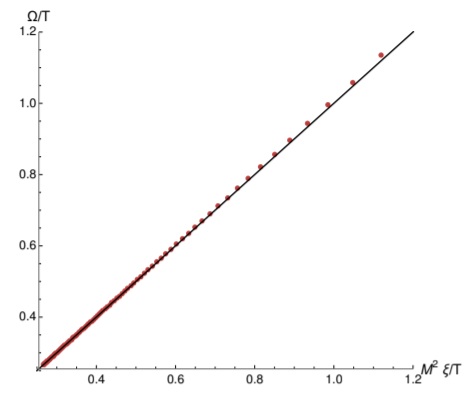}
\caption{Numerical verification of the universal relation \eqref{universe}. Figure taken from~\cite{Ammon:2019wci}.}
\label{fig:uni}
\end{figure}
In the context of axions models, the universal relation~\eqref{universe} has been confirmed numerically for various choices of the potential $V$ in~\cite{Ammon:2019wci,Baggioli:2019abx} (see Fig.~\ref{fig:uni}).

Importantly, the dissipative coefficient $\xi$ can be derived analytically in terms of horizon data and it generally reads:
\begin{equation}
\frac{\xi}{G}\,=\,\frac{4\,\pi\,s\,T^2}{2\,m^2\,\chi_{\pi\pi}^2\,V_X}\,.
\end{equation}
Combining this result, with the previous equations~\eqref{f1} and~\eqref{io}, one can immediately prove that for the benchmark model $V(X)=X+\beta X^N$:
\begin{equation}
\frac{\Omega\,G}{\omega_0^2\,\xi\,\chi_{\pi\pi}}\,=\,\frac{1\,+\,N\,\beta}{N\,\beta}\,\underbrace{\rightarrow}_{\beta \rightarrow \infty}\,1\,,
\end{equation}
where the pseudo-spontaneous limit $\beta \rightarrow \infty$ has been taken. In summary, the universal relation~\eqref{universe} can be proven explicitly, in agreement with the numerical data (see Fig.~\ref{fig:uni}).

Finally, the universal relation~\eqref{universe} has been formally derived in the context of dissipative EFT using Keldysh-Scwhinger techniques in~\cite{Baggioli:2020haa} (see also~\cite{Baggioli:2020nay}). It turns out that this universal relation arises directly from the intertwined symmetry breaking pattern between spacetime translations and internal shifts. From the holographic point of view, this intertwined dynamics has been also discussed in similar models in~\cite{Amoretti:2019kuf,Donos:2019hpp}.

\subsection{Optical conductivity and pinning}
Very importantly, in presence of a finite charge density, the physics just described couples to the dynamics of the electric current and induces important effects in the electric conductivity. In particular, the low frequency behaviour of the electric conductivity is given by
\begin{equation}
\sigma(\omega)\,=\,\sigma_0\,+\,\frac{\frac{\rho^2}{\chi_{\pi\pi}}\,(\Omega-i \omega)-\omega_0^2\,\gamma\,[2\,\rho+\gamma \chi_{\pi\pi}(\Gamma-i \omega)]}{(\Gamma-i \omega)\,(\Omega-i \omega)\,+\,\omega_0^2}\,,
\label{condu}
\end{equation}
where $\sigma_0$ is the incoherent conductivity~\cite{Davison:2015taa} (the one coming from the incoherent current and not sensitive to the dynamics of momentum), $\rho$ the charge density and $\gamma$ a dissipative parameter coming from the coupling between the electrical current and the Goldstone mode. Beyond the complicated expression for the optical conductivity, the most important feature arising from Eq.~\eqref{condu} is the spectral weight transfer from zero frequency to an intermediate IR frequency which is governed by $\omega_0$, the mass of the Goldstone mode. More specifically, the real part of the optical conductivity displays a peak at a certain real frequency $\omega^*=\sqrt{4 \omega_0^2-(\Gamma-\Omega)^2}$. In other words, the Drude peak moves towards higher frequency and, because of the sum-rule, the DC component of the conductivity decreases at the same time. 

\begin{figure}
\centering
\includegraphics[width=0.75\linewidth]{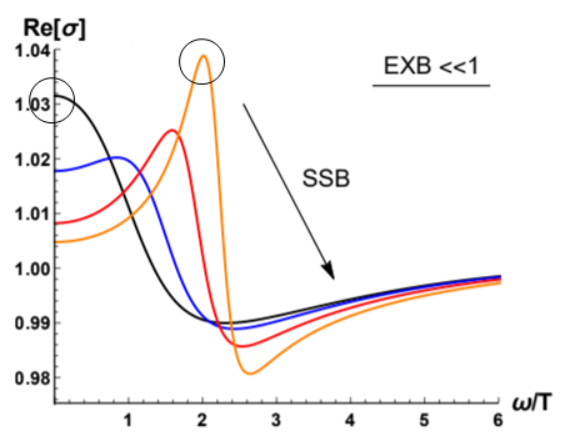}
    
\vspace{0.5cm}
    
\includegraphics[width=0.75\linewidth]{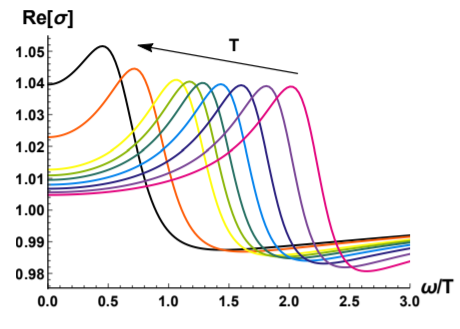}
\caption{\textbf{Top: } The shift of the Drude peak to finite intermediate frequencies as a consequence of the pseudo-spontaneous dynamics. \textbf{Bottom: } The temperature dependence of the conductivity peak. Figures taken from~\cite{Ammon:2019wci}.}
\label{opti}
\end{figure}
This dynamics is shown in the top panel of Fig.~\ref{opti}. The position of this peak plays an important role in the proposal that bad metals can be understood as strongly coupled material with fluctuating charge density waves relics. More precisely, in~\cite{Delacretaz:2016ivq}, it has been proved that in order for this theory to describe the optical properties of strange metals the peak must move to larger frequencies increasing the temperature. Not only that, but it has been proposed that the energy scale controlling such peak is the same as the one controlling the linear in $T$ resistivity of those materials, the Planckian time $\tau=\hbar/k_B T$. Accordingly to their analysis of the experimental data, the position of the peak in the optical conductivity should increase linearly with the temperature. 

Unfortunately, this feature is not recovered in the holographic axion model~\cite{Ammon:2019wci}. It was found that the optical conductivity follows a more standard and natural transition towards an insulating state. Increasing the temperature, the system tends to become more metallic, the DC conductivity wants to grow and the optical conductivity wants to return to its original Drude shape (see the bottom panel of Fig.~\ref{opti}). This drawback can be cured by a fine-tuned generalization, involving a dilatonic coupling, and displaying this behaviour in a very small range of temperatures (see the inset of Fig.\,3 in~\cite{Amoretti:2018tzw}).

\section{Phenomenology}\label{sec:Pheomen}
\subsection{Metal-Insulator transitions}\label{MITsection}
The \textbf{metal-insulator transition} (MIT) is one of the oldest as well as the fundamentally least understood problems in condensed matter. Although many theories have been proposed, the mechanisms toward the metal-insulator transition remain controversial and somewhat incomplete (see~\cite{  Imada:1998zz,2011arXiv1112.6166D} for reviews). It is obvious that a good metal and a good insulator are very different physical systems, characterized by quite different elementary excitations. In particular, in the intermediate regime of the transition, different types of excitations coexist and simple theoretical tools prove of little help. Since the discovery of high temperature superconductivity, the study of metal-insulator transition came to the strong correlation era, for which physical pictures based on weak-coupling approaches prove insufficient or even misleading. 

Holography provides a new approach to tackle states of quantum matter without quasiparticle excitations, for which the transport properties deviate strongly from conventional approach described, in particular, by Fermi liquid theory. From an EFT point of view, the holographic axion model concerns broken translation invariance and implements non-perturbative renormalization group flows. To realise a holographic metal-insulator transition, one should first overcome the obstruction found in~\cite{Grozdanov:2015qia} where it was proven that in some simple holographic theories with arbitrary spatial inhomogeneity (disorder) the electrical conductivity is bounded from below by a universal minimal conductance. Therefore, one can not obtain an insulating phase where the electric DC conductivity at zero temperature is very small or eventually zero. It was then found in~\cite{Baggioli:2016oqk} that it is possible to introduce additional couplings between the charge and translation breaking sectors allowed by the symmetries (see also~\cite{Gouteraux:2016wxj,Baggioli:2016pia} ). Then a clear disorder-driven metal-insulator transition was observed~\cite{Baggioli:2016oqk}.

The minimal holographic model of a metal-insulator transition is described by~\eqref{action} with $Z=0$
\begin{equation}\label{MIaction}
\mathbf{\mathcal{S}}=\int d^4 x \sqrt{-g}\left[\frac{R}{2}-\Lambda-\frac{Y(X)}{4e^2} F_{\mu\nu}F^{\mu\nu}-m^2 V(X)\right]\,.
\end{equation}
The consistency of a theory imposes some constraints on the couplings that appear in the Lagrangian. For the theory~\eqref{MIaction}, it was shown that $V(X)$ and $Y(X)$ should satisfy the following constraints~\cite{Baggioli:2016oqk}:
\begin{align}\label{oldconstraint}
V^{\prime}(X)>0, \quad Y(X)>0, \quad Y^{\prime}(X)<0\,,
\end{align}
in which $Y'(X)<0$ plays a key role in triggering a metal-insulator transition.\footnote{See \emph{e.g.}~\cite{Donos:2012js,Mefford:2014gia,Rangamani:2015hka,Baggioli:2016oju,Ling:2014saa,Cremonini:2017qwq,Andrade:2017cnc} for other holographic realizations of metal-insulator transitions driven via other mechanisms.}  More recently, a much stringent constraint was found for $Y$ in~\cite{An:2020tkn}. Without loss of generality, one can parametrize the couplings $Y$ and $V$ in the following expansion as $X\rightarrow 0$ (weak momentum dissipation): 
\begin{equation}\label{YVexpansion}
Y(X)=1-\gamma\,X+\mathcal{O}(X^2),\quad V(X)=\frac{1}{2 m^2}X+\mathcal{O}(X^2)\,,
\end{equation}
with $\gamma$ a constant. By requiring a positive definite longitudinal conductivity in the presence of charge density and magnetic field restricts the allowed parameter space of theory parameters. 
\begin{equation}\label{constk}
0\leqslant \gamma\leqslant1/6\quad\Rightarrow -1/6\leqslant Y'(0)\leqslant 0\,.
\end{equation}

\begin{figure}
\centering
\includegraphics[width=0.8\linewidth]{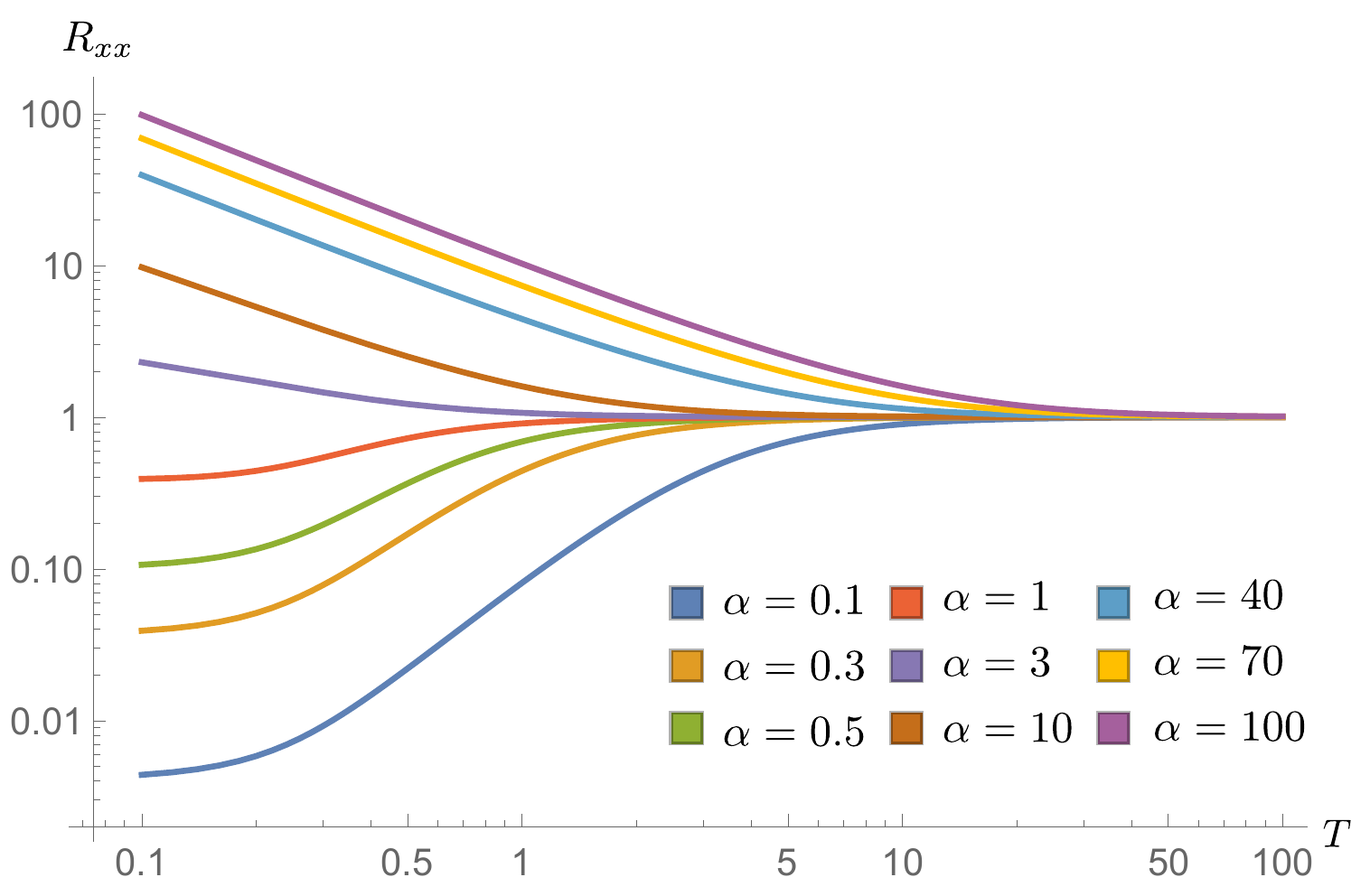}
    
\vspace{0.2cm}
    
\includegraphics[width=0.8\linewidth]{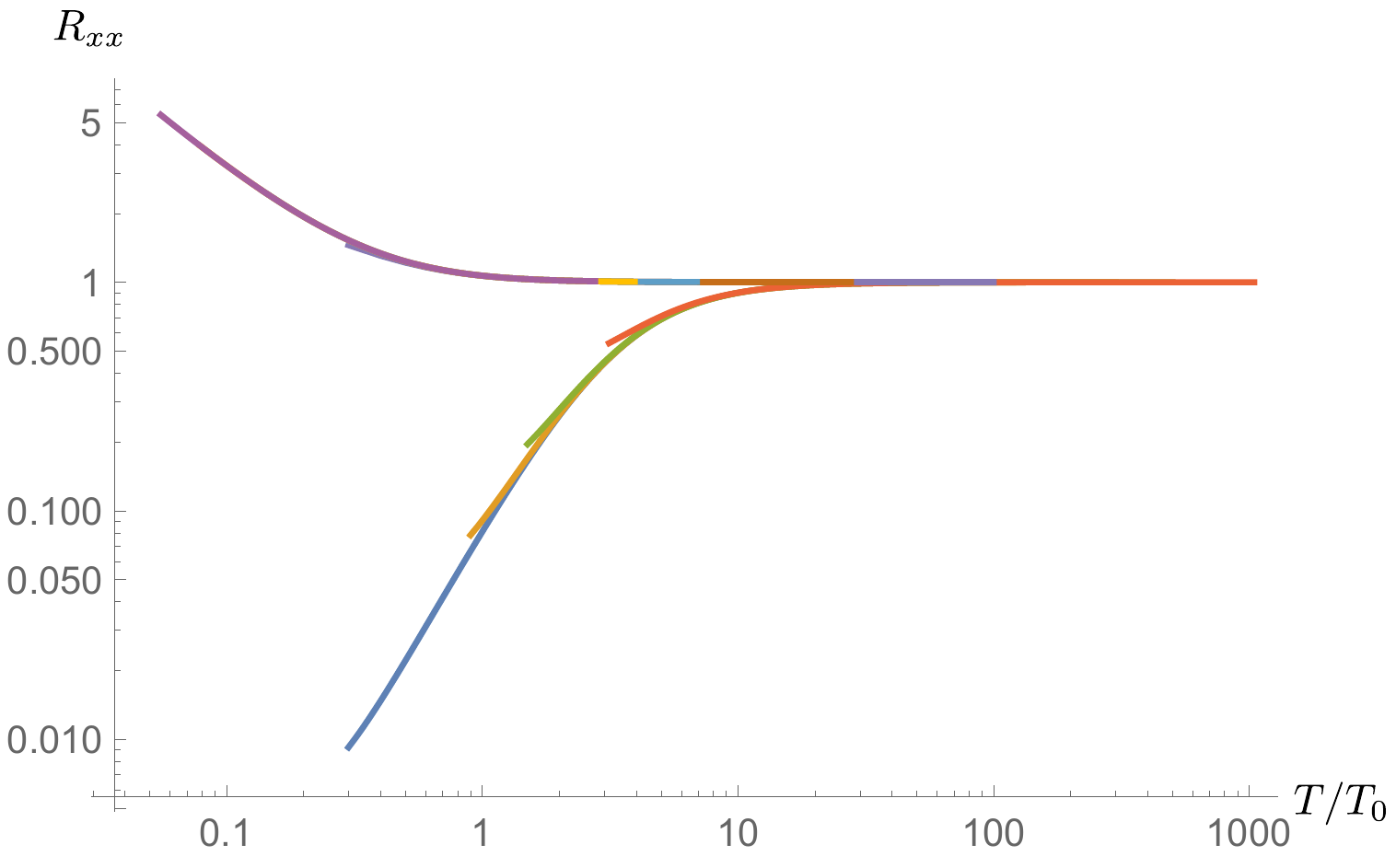}
\caption{Temperature dependence of the resistivity versus disorder strength at zero magnetic field for the model with $Y=1-\frac{X}{6}, V=\frac{X}{2m^2}$. Top: The metal-insulator transition driven by the disorder strength $\alpha$. Bottom: Scaling of resistivity with scaled temperature $T/T_0$. The collapse of data into two separated curves both in the metallic and insulating sides is manifest. Other parameters are chosen by $e=\rho=1$. Figure taken from~\cite{An:2020tkn}.}
\label{fig:scalingMI}
\end{figure}

The working and phenomenological definition of a metal versus an insulator behavior is given by
\begin{equation}\label{definition}
\text{metal:}\quad \frac{d\,R_{xx}}{d\, T}>0,\quad \text{insulator:}\quad \frac{d\,R_{xx}}{d\, T}<0\,,
\end{equation}
where $R_{xx}$ is the longitudinal DC resistivity. Some generic features without being concerned with details of the holographic theory was uncovered in~\cite{An:2020tkn}. In particular, the temperature dependence of resistivity is found to be well scaled with a single parameter $T_0$, which approaches zero at some critical charge density $\rho_c$, and increases as a power law $T_0\sim|\rho-\rho_c|^{1/2}$ both in metallic $(\rho>\rho_c)$ and insulating $(\rho<\rho_c)$ regions in the vicinity of the transition. Similar features also happen by changing the disorder strength $\alpha$ as well as magnetic field (see Fig.~\ref{fig:scalingMI}).
It was also found that the metallic and insulating curves are mirror symmetry in the high temperature regime:
\begin{equation}
R_{xx}(\rho-\rho_c,T)=1/R_{xx}(\rho_c-\rho, T)\,. 
\end{equation}
These results suggest that the mechanism responsible for the temperature dependence of conductivity on both insulating and metallic sides of the transition would be the same, or it would originate with some fundamental feature that is common to both.

The holographic results are reminiscent of the scaling behaviors for resistivity near the transition point reported in some two dimensional samples and materials, which shows the collapse of data into two separated curves and displays remarkable mirror symmetry over a broad interval of temperatures~\cite{1995PhRvB..51.7038K,1996PhRvL..77.4938K,1997PhRvB..5612764C,1998PhRvL..80.1292S}. This observation has been interpreted as evidence that the transition
region is dominated by strong coupling effects characterizing the insulating phase~\cite{1997PhRvL..79..455D}. 
Nevertheless, the dependence of $T_0$ near $\rho_c$ is a power law with the power that is different from holographic result $1/2$, suggesting that holographic theory~\eqref{MIaction} falls into a different universality class from those materials. A natural extension is to consider a holographic setup that is asymptotically Lifshitz with a dynamical exponent $z$ which parametrizes the relative scaling of space and time, making the model compatible with experimental data. 

\begin{figure}
\includegraphics[width=0.86\linewidth]{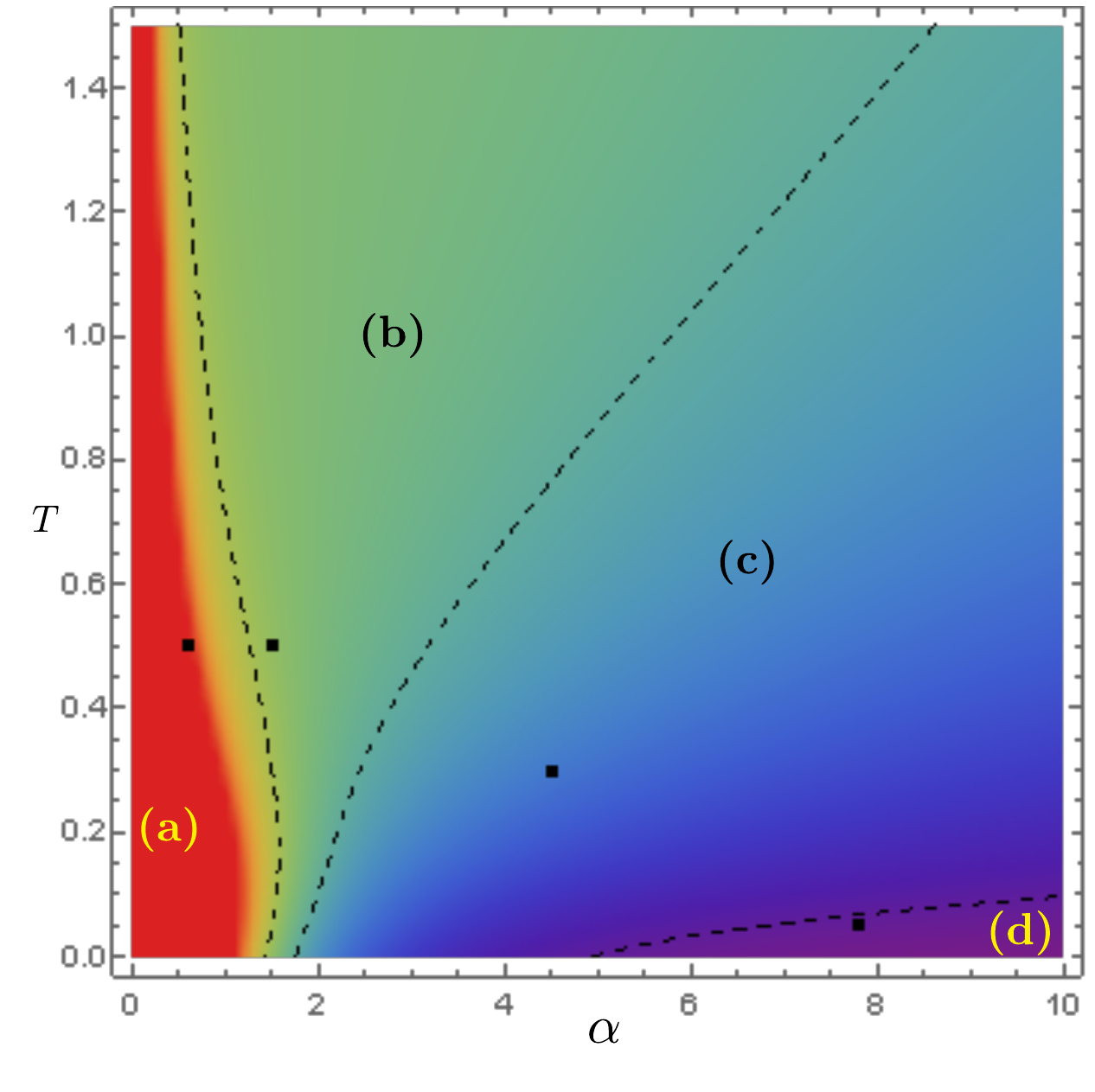}
\includegraphics[width=0.11\linewidth]{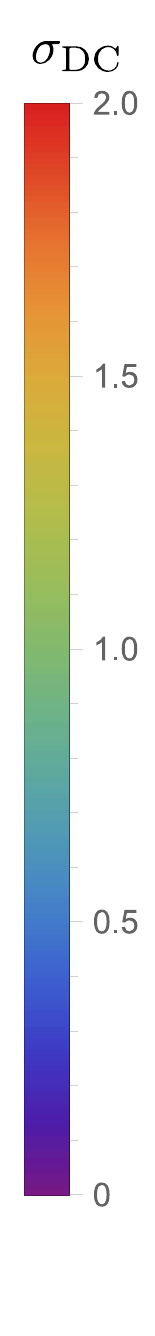}
\caption{Phases diagram for the model with $Y=1-\frac{X}{6}, V=\frac{X}{2m^2}$ in the absence of magnetic field. Four regions are denoted by (a) good metal, (b) incoherent metal, (c) bad insulator and (d) good insulator, respectively. Other parameters are chosen by $e=\rho=1$. The figure is updated from~\cite{An:2020tkn}.}
\label{fig:phaseMI}
\end{figure}

\begin{figure}
\includegraphics[width=0.8\linewidth]{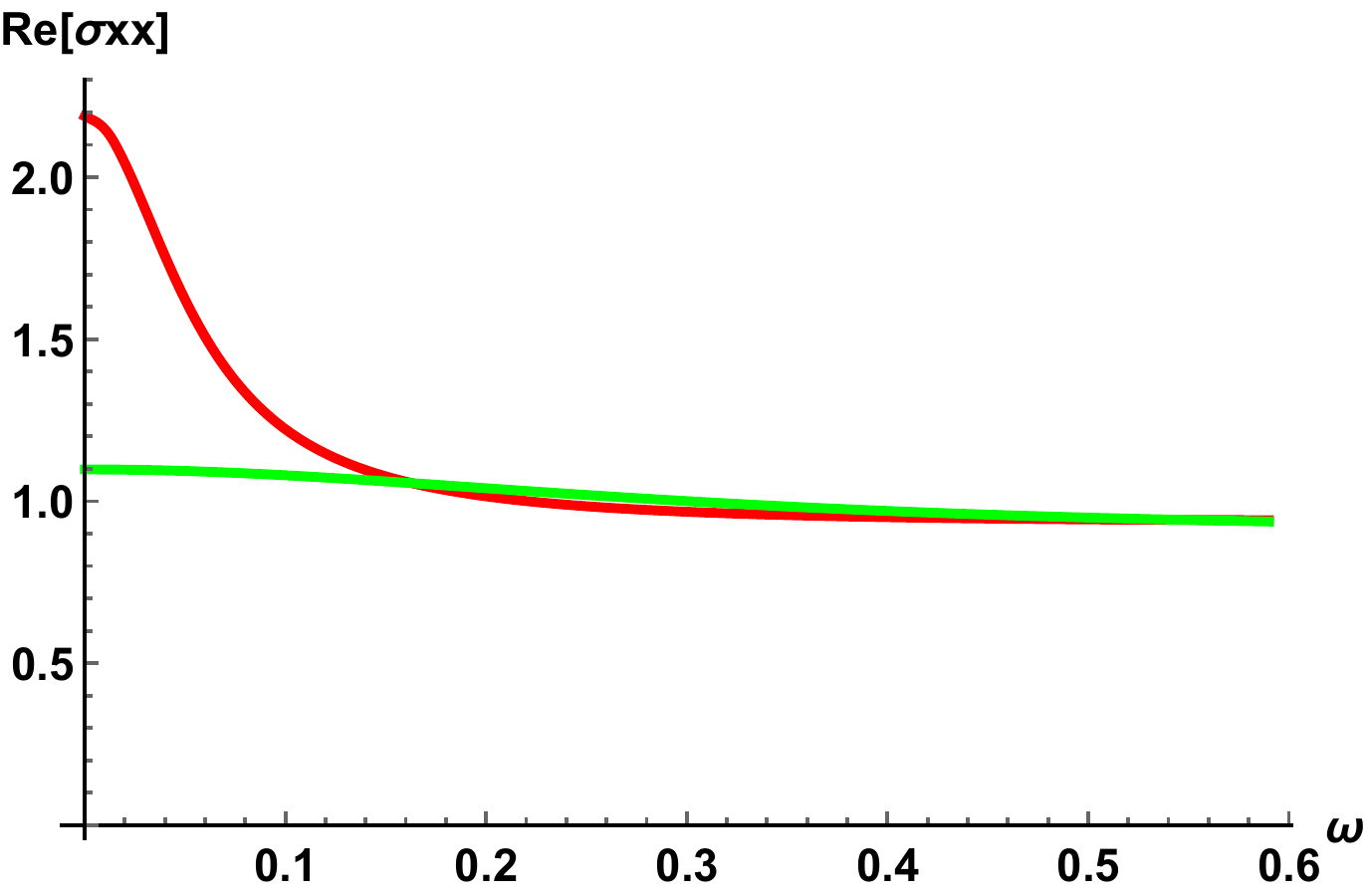}

\vspace{0.2cm}

\includegraphics[width=0.8\linewidth]{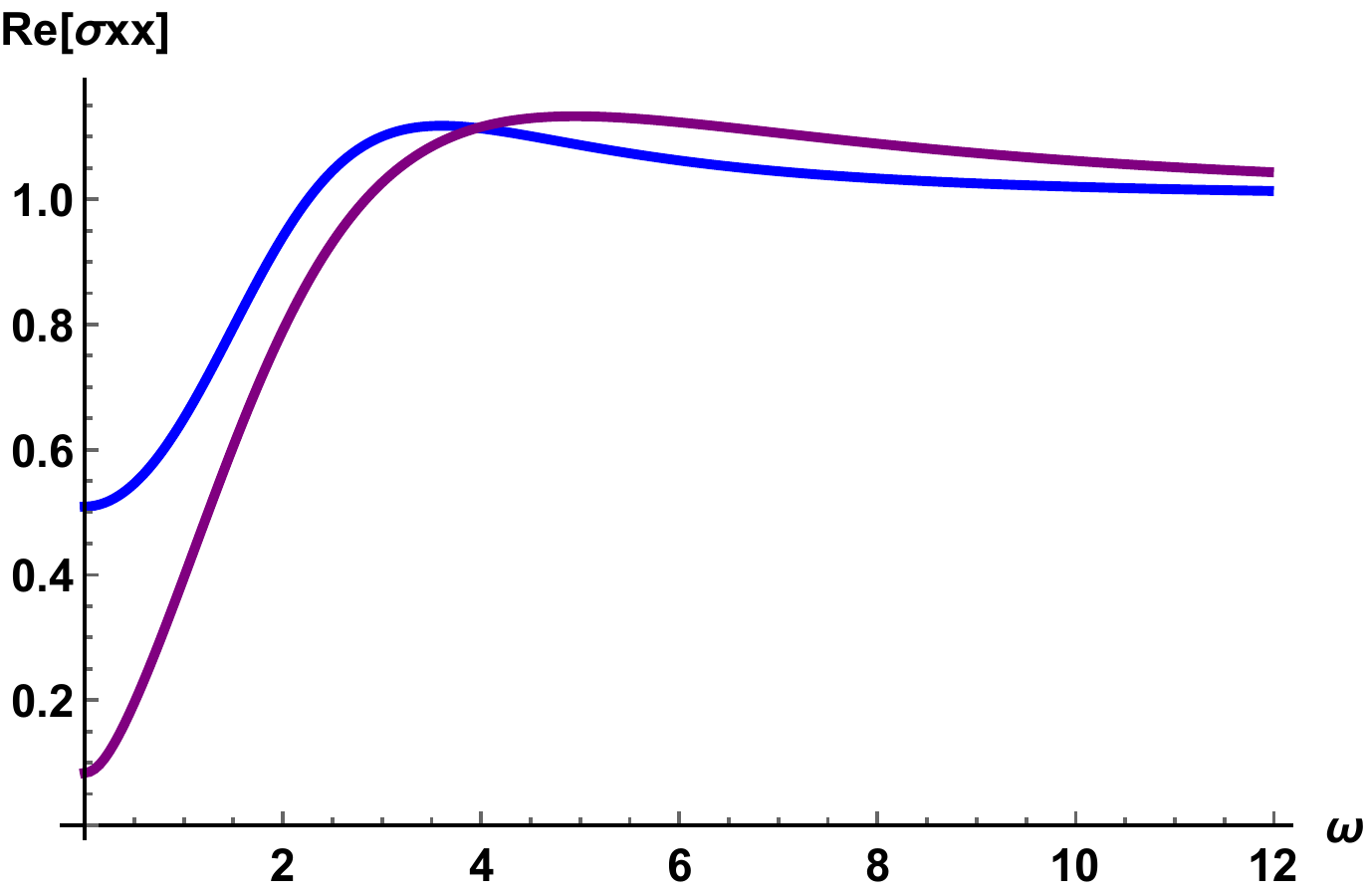}
\caption{Representative examples of the AC electric conductivity $\sigma_{xx}$ with unitary charge density $\rho=1$. There are four phases: (a) good metal (red) with $(\alpha=0.6, T=0.5)$, (b) incoherent metal (green) with $(\alpha=1.5, T=0.5)$, (c) bad insulator (blue) with $(\alpha=4.5, T=0.3)$ and (d) good insulator (purple) with $(\alpha=7.8, T=0.05)$. Figures taken from~\cite{An:2020tkn}.}
\label{fig:ACMI}
\end{figure}

The phase diagram gets richer and incorporates several phases of matter depending on the parameters. As shown in Fig.~\ref{fig:phaseMI}, there are as many as four different phases in the temperature-disorder phase diagram: good metal (a), bad or incoherent metal (b), bad insulator (c) and good insulator (d). Some representative examples of the AC electric conductivity for each phase are shown in Fig.~\ref{fig:ACMI}. A coherent metallic phase with a sharp Drude peak is obtained for small $\alpha$. As $\alpha$ increases, the Drude peak is suppressed and one arrives at an incoherent metallic phase where there is no clear and dominant localized long lived excitation, see green curves of Fig.~\ref{fig:ACMI}. When the disorder is strong enough, the spectral weight transfers to the mid-infrared, resulting in an insulating behavior. As a consequence, the DC conductivity keeps decreasing, and then triggers the transition from bad insulator to good insulator. Therefore, there is a clear disorder-driven transition from a coherent metal with a sharp Drude peak to a good insulator with a tiny or vanishing DC conductivity at zero temperature. 

The investigation of transport in this minimal holographic setup of a metal-insulator transition uncovered some interesting features, shedding light on this interesting transition and the physical mechanism that drives it.
There are still many interesting questions. There are as many as four different phases in the temperature-disorder phase diagram of Fig.~\ref{fig:phaseMI}, but all phases share the same symmetries of the underlying theory, and thus beyond a simple Ginzburg-Landau description. Some observables were examined in order to characterise different phases, but failed~\cite{An:2020tkn}. It is still an open question to find a good probe to the metal-insulator transitions. The scaling exponent of $T_0$ from the holographic setup is different from the experiments which yielded scaling exponents between 1.25 and 1.6~\cite{1995PhRvB..51.7038K,1996PhRvL..77.4938K,1997PhRvB..5612764C,1998PhRvL..80.1292S}. It is interesting to generalise the model to obtain an exponent that is compatible with experimental observation. It is also worth studying the thermal response and the mechanical response.

\subsection{The scalings of strange metals} \label{secssm}
\begin{figure}[h]
\includegraphics[width=0.8\linewidth]{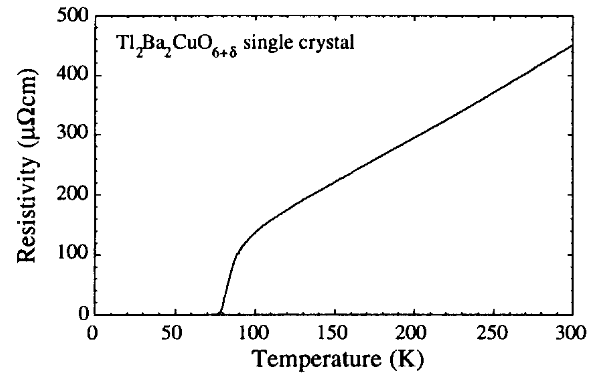}

\vspace{0.2cm}

\includegraphics[width=0.8\linewidth]{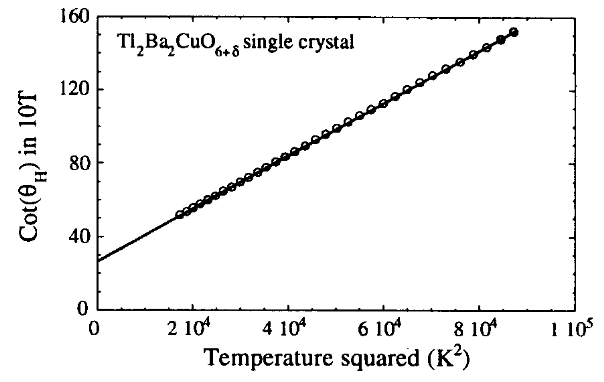}
\caption{Experimental observation of the cuprates scalings for the in-plane resistivity $R_{xx}\sim T$ and inverse Hall angle $\cot(\Theta_H)\sim T^2$. Figures taken from~\cite{1997PhyC..282.1185T}.}
\label{fig:SMetal}
\end{figure}

The anomalous metallic transport in the high-temperature superconducting cuprates is one of the most remarkable puzzles in condensed matter physics. The so-called \textbf{strange metal phase} is characterized by universal temperature scalings which are robust across widely different systems, and are believed to be controlled by an underlying strongly interacting quantum critical sector. Its anomalous features include a linear temperature dependence for the resistivity $R_{xx}\sim T$ and
the scaling of the Hall angle $\cot(\Theta_H)\sim T^2$ (see Fig.~\ref{fig:SMetal}). Realizing the anomalous temperature dependence of both $R_{xx}$ and $\cot(\Theta_H)$ at once within a holographic model has proven to be a challenge. Although many efforts were made, standard Einstein-Maxwell-dilaton (EMD) theories have thus far been unable to give the expected anomalous scalings of the strange metal (see \emph{e.g.}~\cite{Blake:2014yla,Amoretti:2015gna,Amoretti:2016cad}). As we are dealing with strongly correlated electron matter, it may be crucial to take into account the nontrivial dynamics between the charge degrees of freedom to reliably capture transport in these phases~\cite{Hartnoll:2009ns,Kim:2010zq,Karch:2014mba}. 

The first consistent holographic realization of the strange metal scalings of the resistivity and Hall angle was given in~\cite{Blauvelt:2017koq} by working with a string-theory-motivated gravitational model encoded by the Dirac-Born-Infeld (DBI) action.\footnote{Note that the strangle metal scalings have also been realized in the Einstein-Maxwell-dilaton-axion model with a hyperscaling IR geometry only for a special value of $B$ \cite{Zhou:2015dha}.} The motivation is to describe a strongly coupled quantum theory containing a sector of dilute charge carriers that interact amongst themselves as well as with a quantum critical bath. The charge degrees of freedom is treated as a probe when compared to the larger set of neutral quantum critical degrees of freedom. The gravitational theory takes the generic form
\begin{equation} \label{imp1234}
\mathcal{S}=\int d^4x\sqrt{-g}[\mathcal{L}_{\text{bath}}+\mathcal{L}_{\text{charge}}]\,.
\end{equation}
The bath sector $\mathcal{L}_{\text{bath}}$ is supported, for example, by a neutral scalar and axionic scalars. The charge sector $\mathcal{L}_{\text{charge}}$ describes the dynamics of a $U(1)$ gauge field, taking into account non-linear interactions between the charged degrees of freedom. In~\cite{Blauvelt:2017koq} the bath geometry is nonrelativistic and hyperscaling-violating supported by a neutral scalar field and two axions $\phi^I$. In the so-called probe limit for which the backreaction of the DBI interactions on the geometry can be safely neglected, the nonlinear dynamics of the gauge field sector allows a clean scaling regime for the cuprate strange metals:
\begin{equation} \label{Hall123}
R_{xx}\sim T,\quad \cot(\Theta_H)=\frac{R_{xx}}{R_{yx}}\sim T^2\,,
\end{equation}
with $R_{yx}$ the Hall resisitivity. 

Because of its richness, the DBI theory can support a wide spectrum of temperature scalings. Therefore, one can use similar construction to realize other scaling behaviors observed in strange metals. Recently, novel strange metal behavior was observed in the pnictides~\cite{2016NatPh..12..916H}, for which the magnetic field $B$ plays the same role as the temperature $T$ (see also the observation of linear-in-field resistivity in cuprates~\cite{2018Sci...361..479G}). The measurements imply that the in-plane resistivity behaves as~\cite{2016NatPh..12..916H}
\begin{equation}
R_{xx}=\sqrt{\gamma\, T^2+\eta\, B^2}\,,
\end{equation}
with $\gamma$ and $\eta$ two constants. This striking behavior
was realized from holography in~\cite{Kiritsis:2016cpm}. The holographic theory also predicts a Hall resistivity in the same temperature regime that is linear in the magnetic field and approximately temperature independent. This idea has been generalised to include a generic nonlinear gauge field sector in~\cite{Cremonini:2018kla}. A particularly simple nonlinear model whose structure is natural from the point of view of the DBI action was found to be able to realise the temperature scalings of the entropy $\sim T$, resistivity $\sim T$, Hall angle $\sim T^2$ and weak-field magneto-resistance $\sim T^{-4}$ observed in cuprates. 

The underlying mechanism for above holographic construction relies on having a quantum critical IR fixed point and on the nonlinear structure of the interactions between the charges.
It would be desirable to understand the underlying dynamics and to match the holographic description with the expected interactions of electrons in real materials, building intuition for the mechanisms underlying the unconventional behavior of strange metals. The first step towards connecting to phases and critical points of Hubbard model~\cite{Sachdev:2010uz} was addressed~\cite{Sachdev:2010um}. The realisation of strange metal scalings using holography beyond the probe approximation is still an open question. It is fair to say that a consensus about the strange metals phenomenology is far from being reached both from the holographic point of view \cite{khveshchenko2020die} and the condensed matter one \cite{singh2020leading}.

\subsection{Superconductivity}
The \textbf{holographic superconductors} model~\cite{Hartnoll:2008vx,Hartnoll:2008kx} is one of the first and most popular frameworks in Applied Holography (see~\cite{Cai:2015cya} for a recent review). Its simplicity comes nevertheless with a price. In particular, because of the translational invariance of the background, both the normal state and the broken SC phase exhibit an infinite conductivity:
\begin{equation}
\sigma(\omega)\,=\,\sigma_0\,+\,\frac{i}{\omega}\,\left(\frac{\rho_s}{\mu}\,+\,\frac{\rho_n^2}{\mu \rho_n+s T}\right)\,,
\label{c1}
\end{equation}
with $\rho_n$ and $\rho_s$ respectively the normal and superfluid densities. The first infinity comes from the superfluid flow while the second from translational invariance. This drawback makes it impossible to distinguish the two phases at the level of electric transport without a very careful analysis. To avoid this problem, the power of the holographic axion model was originally used to dissipate momentum and have a normal state with finite DC conductivity. The holographic superconductors model was endowed with the axion fields in~\cite{Andrade:2014xca,Kim:2015dna,Baggioli:2015zoa} and further explored in~\cite{Baggioli:2015dwa,Kim:2016hzi,Ling:2016lis,Gouteraux:2019kuy}.

\begin{figure}[ht]
\centering
\includegraphics[width=0.8\linewidth]{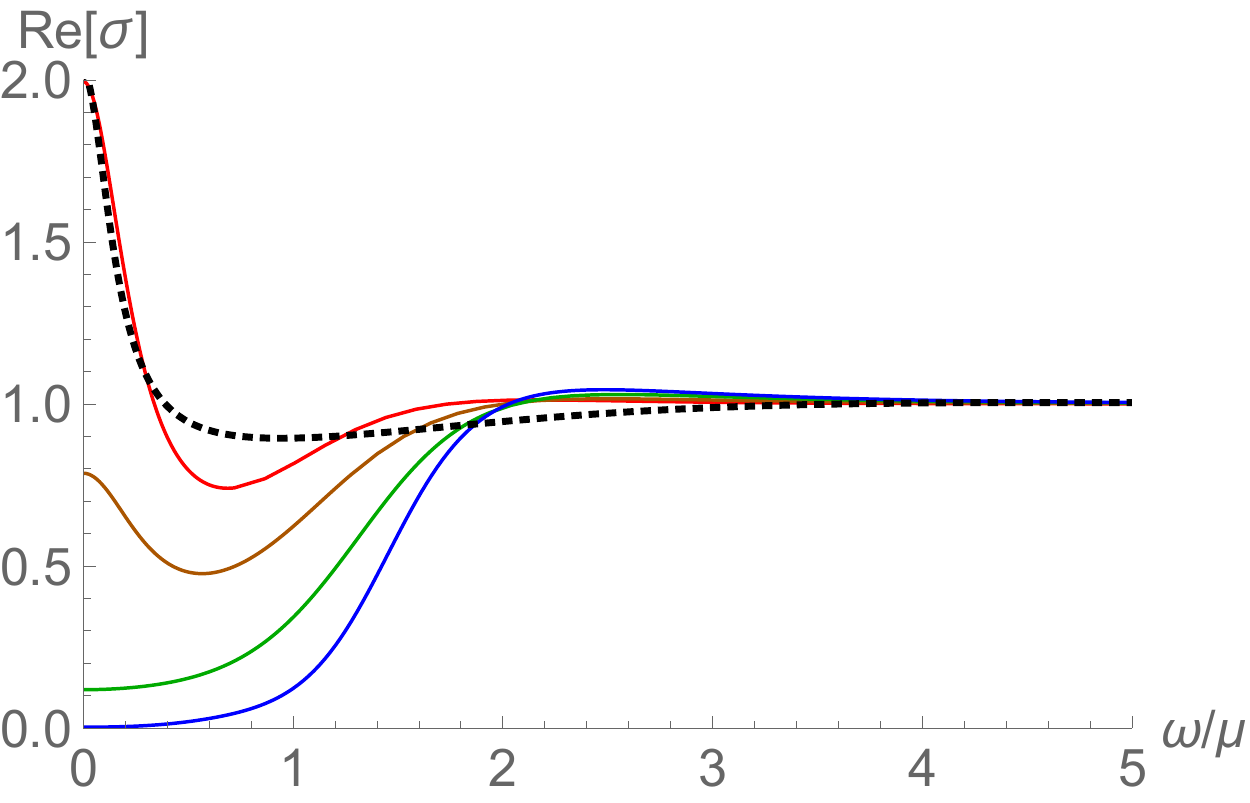}
    
\vspace{0.2cm}
    
\includegraphics[width=0.8\linewidth]{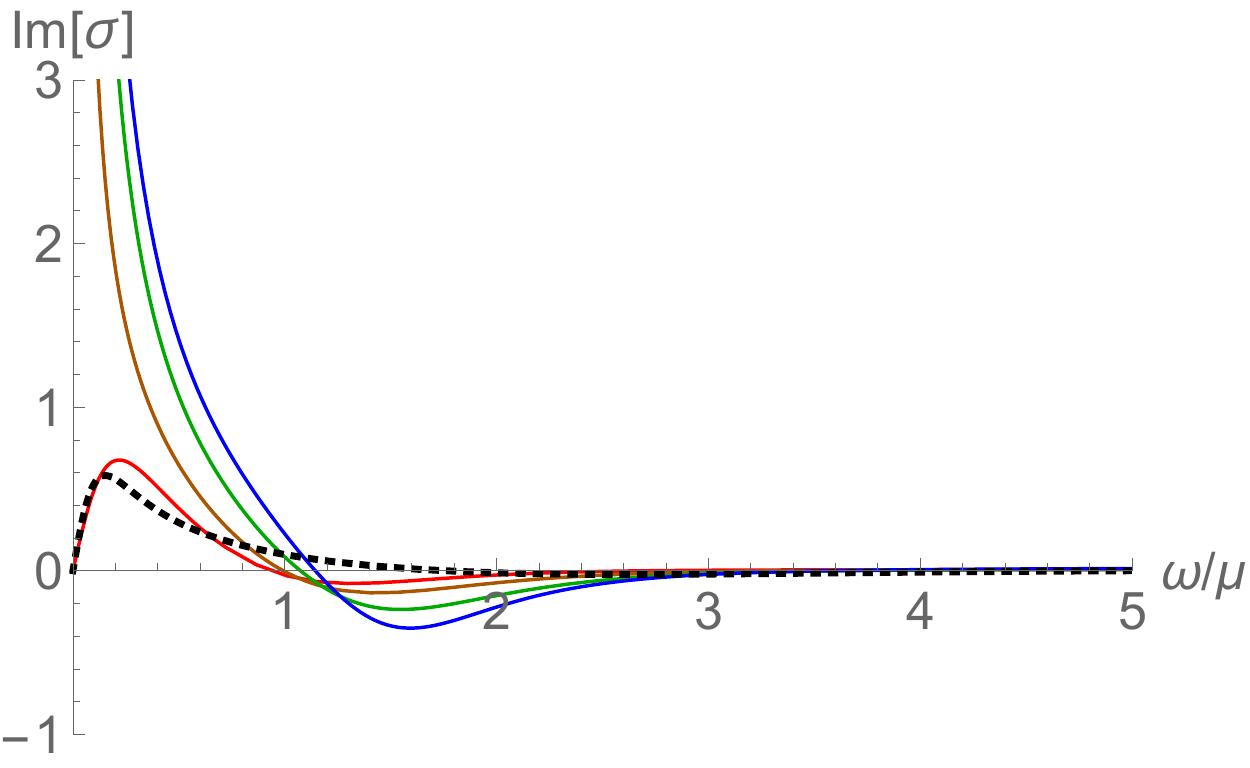}
\caption{ $\alpha/\mu = 1$, $T/T_c = 3.2, 1,0.89,0.66, 0.27$ (dotted, red, orange, green, blue). Figures adapted from~\cite{Kim:2015dna}  }
\label{ACcon123}
\end{figure}

Once momentum dissipation is introduced, via the axion scalars, the expression~\eqref{c1} gets modified as:
\begin{equation}
\sigma(\omega)\,=\,\sigma_0\,+\,\frac{i}{\omega}\,\frac{\rho_s}{\mu}\,+\,\frac{\rho_n^2}{\mu \rho_n+s T}\,\frac{i}{\omega\,+\,i\,\Gamma}\,,
\label{c2}
\end{equation}
where $\Gamma$ is the usual momentum relaxation rate. In the normal phase, $\rho_s=0$, the DC conductivity is finite and it becomes infinite only for $T<T_c$. 
The corresponding AC finite frequency conductivity is shown in Fig.~\ref{ACcon123}. 
Above the critical temperature, there is no $1/\omega$ pole in the imaginary part of conductivity thanks to a finite $\alpha$. Below the critical temperature, a simple pole (corresponding to the second-sound in superconductors) appears in the imaginary part of the conductivity, yielding - through Kramers-Kroenig relations - the appearance of a delta function at zero frequency in the real part. In other words, the DC conductivity goes to infinity, as we would expect for a superconductor.

\begin{figure}[ht]
\centering
\includegraphics[width=0.7\linewidth]{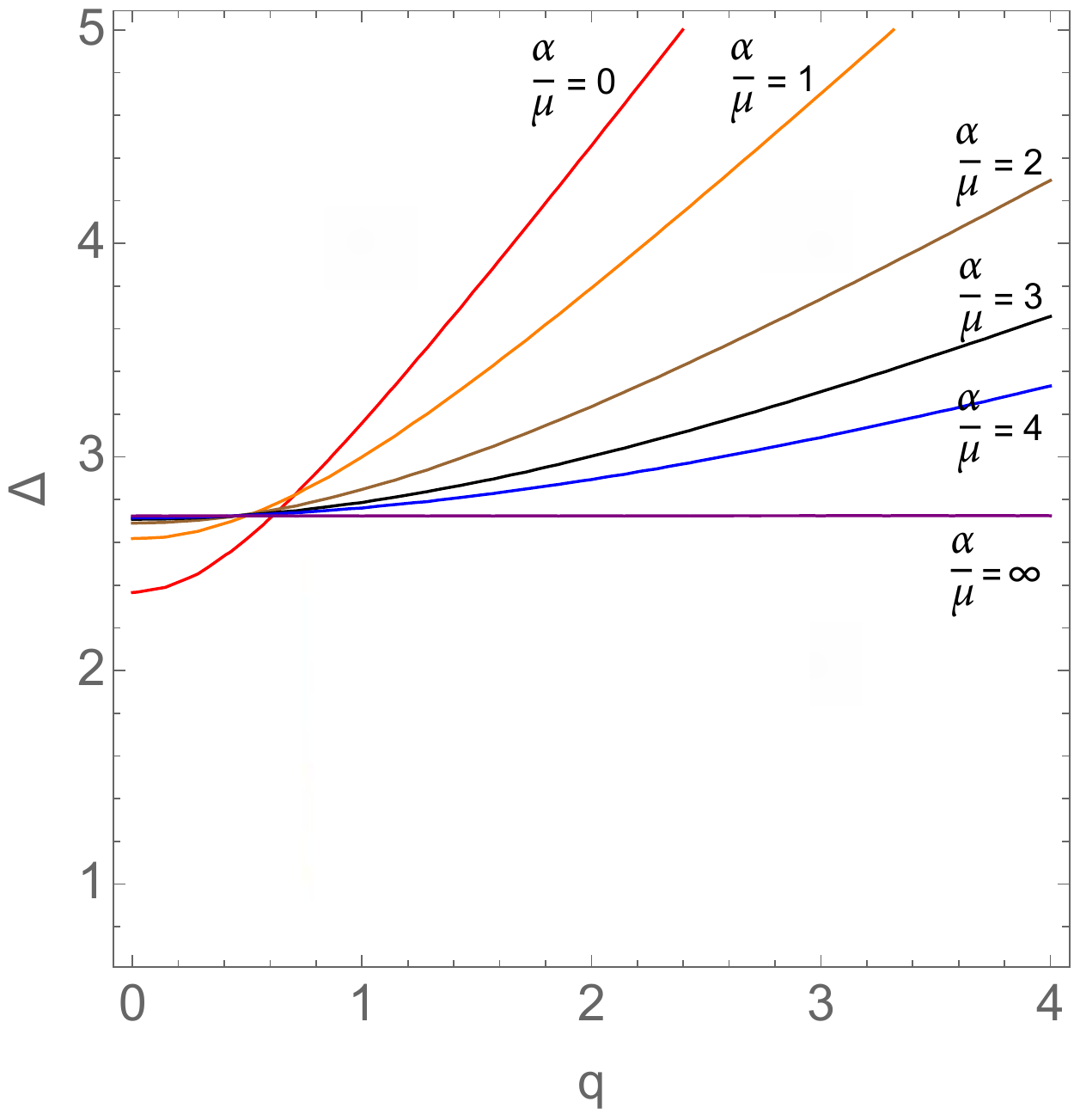}
    
\vspace{0.2cm}
    
\includegraphics[width=0.85\linewidth]{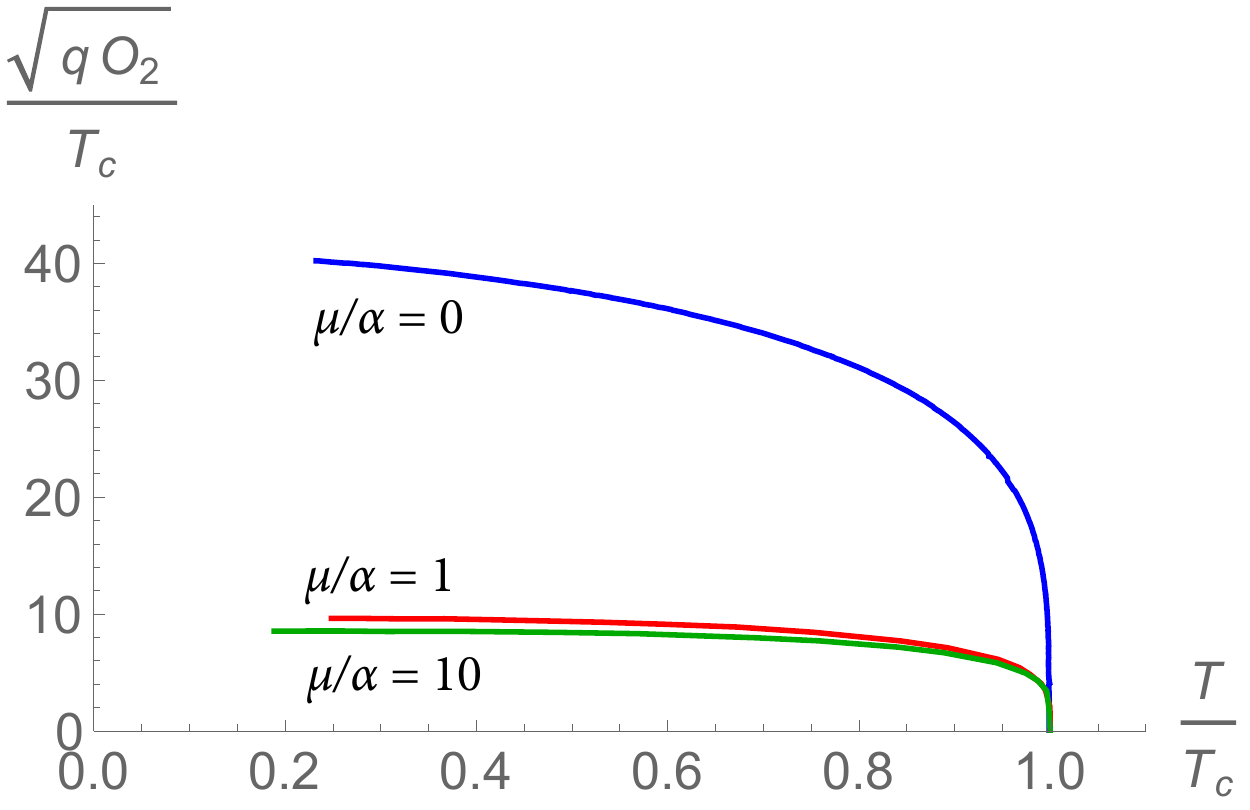}
\caption{The effects of momentum dissipation on the phase diagram of the holographic superconductors model and on the SC condensate. $\Delta$ is the conformal dimension of the complex bulk scalar and $q$ its charge. Figures adapted from~\cite{Kim:2015dna}.}
\label{SC1}
\end{figure}

Interestingly, the presence of momentum dissipation has two important effects (see Fig.~\ref{SC1}): (I) it shrinks the area in the phase diagram where the superconducting instability can appear, (II) it decreases the value of the superconducting condensate.

There is an interesting open problem regarding holographic superconductors. 
In high-temperature superconductors and some conventional superconductors, there is a universal property called \textbf{Homes' law}~\cite{Homes_2004}.  It relates three quantities: the superfluid density at zero temperature ($\rho_{s}$),  the phase transition temperature ($T_c$), and the DC conductivity in the normal phase close to $T_c$ ($\sigma_{\mathrm{DC}}$) with a material independent universal number ($C$). i.e.
\begin{equation} \label{H1mmm}
\rho_{s}(T = 0) = C \sigma_{\mathrm{DC}}(T_{c}) \, T_{c} \,.
\end{equation}
Here, the DC conductivity is involved, so momentum relaxation is necessary to study the Homes' law and the axion model is obviously the simplest to consider. In the context of the axion model, the universality of $C$ in \eqref{H1mmm} means $C$ is independent of $\alpha$, the strength of momentum relaxation, which can be interpreted as a parameter effectively specifying the microscopic material properties. However, it turns out the Homes' law does not work in the holographic superconductor-axion model~\cite{Kim:2016hzi}. The Homes' law have been studied also in other models, \emph{e.g.} helical lattices~\cite{Erdmenger:2015qqa} and Q-lattices~\cite{Kim:2016jjk}. In these cases, the Homes' law hold for a window of momentum relaxation parameters, but there is still not a good understanding of this mechanism from the holographic (geometric) viewpoint. To remedy this situation, it would be advantageous to have a holographic model with robust linear-$T$-resistivity up to high temperature~\cite{Jeong:2018tua,Ahn:2019lrh} because in this case the factor $T_c$ in in \eqref{H1mmm} would cancel. It seems that the strong momentum relaxation is also a necessary ingredient~\cite{Homeswip}.\\

Finally, let us emphasize that all the homogeneous models do not display any commensurability effects \cite{Andrade:2015iyf}. In fact, because of their homogeneous nature, and contrarily to standard periodic lattices, they do not select any preferred wave-vector. In order to introduce such effects in the holographic framework, one should consider more complicated and fully inhomogeneous models \cite{Andrade:2017leb} which go far beyond the scope of this review.\color{black}

\subsection{Conductivities at finite magnetic field}

As explained in subsection~\ref{secssm}, conductivities at finite magnetic field such as the Hall conductivity, the Nernst effect, and the Hall angle, play important roles in understanding strongly correlated electron systems such as cuprates. 
Indeed, transport in strongly correlated material has been one of the leading themes of the early AdS/CMT era~\cite{Hartnoll:2007ai,Hartnoll:2007ih,Hartnoll:2007ip}. 
Here, a constant magnetic field $B$ is introduced by  the following background gauge potential
\be\label{ansatz2}
A_\mu \mathrm{d} x^\mu = \frac{B}{2}(x  \,\mathrm{d} y -y  \,\mathrm{d} x) \,.
\ee
However, in these pioneering works, momentum relaxation was lacking or not treated in a full manner. To remedy it, the axion model was employed~\cite{Amoretti:2015gna,Blake:2015ina, Lucas:2015pxa, Kim:2015wba}, where the electric, thermoelectric, and thermal conductivity at finite magnetic field have been computed.  For a general class of  Einstein-Maxwell-Dilaton-Axion theories all DC conductivities were expressed in terms of the black hole horizon data.
In particular, for the dyonic black hole modified by axions, the background solution was analytically obtained and the AC electric, thermoelectric, and thermal conductivity were numerically computed.

For the dyonic black hole, the Hall angle \eqref{Hall123} is computed as~\cite{BlakeHall, Kim:2015wba}
\be \label{Hall2}
\cot({\Theta_H}) = \frac{ \alpha^2}{u_h\mu B} \frac{u_h^2B^2 + (\mu^2 + \alpha^2)}{u_h^2B^2 + (\mu^2 +2 \alpha^2)} \,,
\ee
where $\mu$ is chemical potential, $u_h$ is horizon location, and $\alpha$ is the momentum relaxation parameter. 
Because $u_h$ is a complicated function of $T, B,\mu, \alpha$ it is not easy to figure out the $T$ dependence of the Hall angle. By numerical analysis it was found that the $T$ dependence of the Hall angle ranges between $T^0$ and $T^1$. In the large $T$ regime, $u_H \sim 1/T$ so the Hall angle always scales as $T$. Therefore, this standard but simple model does not exhibit the characteristic Hall angle behavior $\sim T^2$. For an improved version displaying a $T^2$-Hall angle, see \eqref{imp1234} and discussions therein. 

\begin{figure}[ht]
    \centering
    \includegraphics[width=0.8\linewidth]{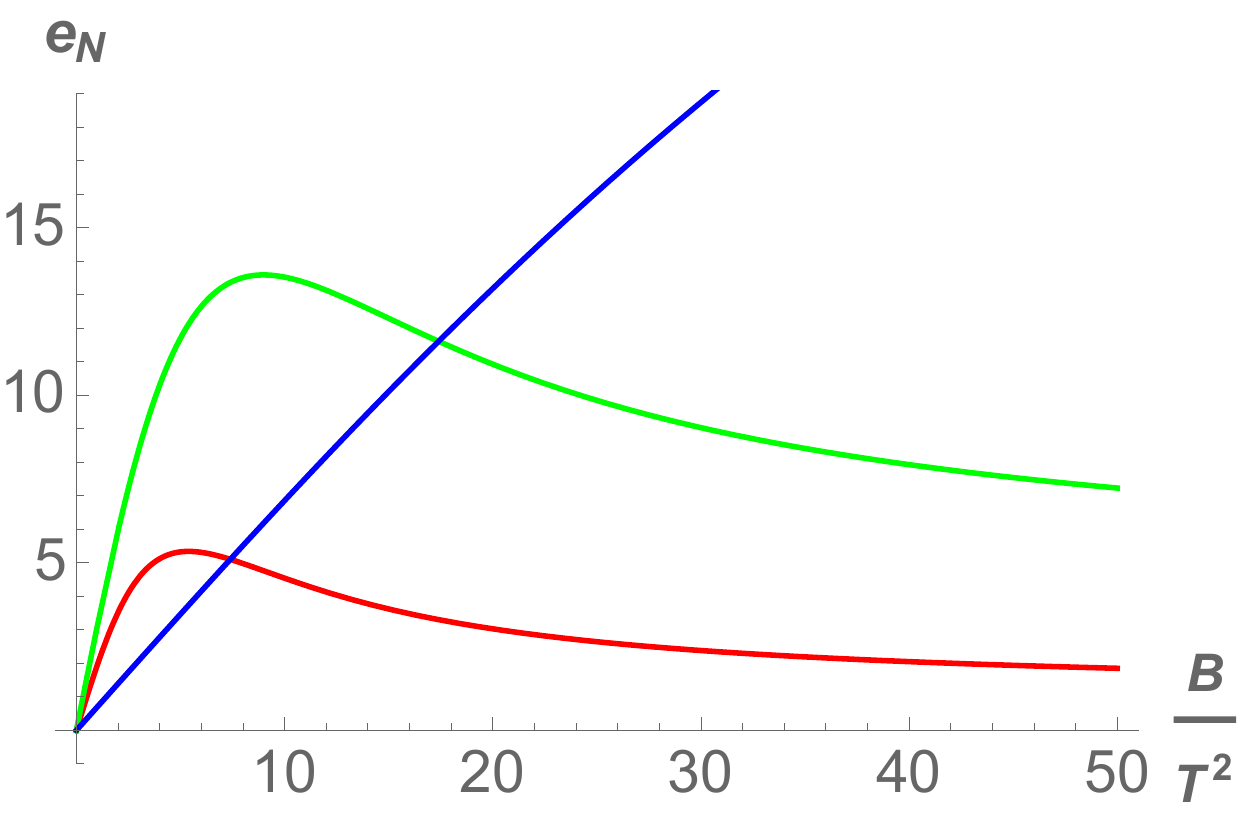}
    \caption{Nernst signal for the dyonic black hole. $\alpha/T=0.5, 1, 4$(red, green, blue). For a large $\alpha$ it is linear to magnetic field (conventional-metal-like) and for a small $\alpha$ it is bell-shaped (cuprate-like). 
Figure taken from~\cite{Kim:2015wba}.}
    \label{fig:Nernst}
\end{figure}

There is another important phenomenon to consider, where both the magnetic field and momentum relaxation are essential. It is the Nernst effect. In the presence of a magnetic field,  a transverse (say, $x$ direction) electric field can be generated by a longitudinal (say, $y$ direction) or transverse thermal gradient. The former is  the Nernst effect and  the latter is the Seebeck effect. 
The Nernst effect is characterized by the the Nernst signal $e_N$
\be
e_N = -{(\sigma^{-1}\cdot \mathscr{A})_y}^{x}\,,
\ee
where $\sigma$ is electric conductivity and $\mathscr{A}$ is the thermoelectric conductivity ($2 \times 2$) matrix. 
Note that  it is zero if there is no momentum relaxation ($\alpha=0$), because the electric conductivity becomes infinite. Thus, a finite $\alpha$ is essential for the holographic model of the Nernst effect.
For the dyonic black hole, the Nernst signal yields~\cite{Kim:2015wba}
\be\label{Nernst01}
e_N = \frac{4\pi  \alpha^2 B}{u_h^2\mu^2 B^2 +(\mu^2 +\alpha^2)^2} \,.
\ee
We want to see the $B$ dependence of the Nernst signal $e_N$ since it displays a different behaviour in cuprates with respect to conventional metals. For example, in the normal state of a cuprate it is bell-shaped as a function of $B$, while in conventional metals it is linear in $B$~\cite{Wang_2006}.
Because $u_h$ in \eqref{Nernst01} is a complicated function of $T, B,\mu, \alpha$ it is not easy to figure out the $B$ dependence of the Nernst signal.  Thus, we make a plot of $e_N$ as a function of $B$ at a fixed $\alpha$ in Fig.~\ref{fig:Nernst}.
Interestingly, by looking at the Nerst signal, this model shows the transition from a conventional metal (blue line) to a cuprate-like  state (green and red) as $\alpha$ decreases.

In general, the AC conductivities with non vanishing $B$ display a peak at the finite $\omega$.
This pick is related to a pole of the conductivity, in complex $\omega$ plane, dubbed the \textbf{cyclotron resonance} pole~\cite{Hartnoll:2007ih,Hartnoll:2007ip}.
\be \label{cycpole}
\omega_* \equiv \omega_c -i\gamma  \,,
\ee
where the ``cyclotron frequency'' $\omega_c$ is the relativistic hydrodynamic analog of the free particle case, $\omega_f = eB/mc$, even though here it should be understood as coming from a collective fluid motion. A damping $\gamma$ could be due to interactions between the positively charged current and the negatively charged current of the fluid, which are counter-circulating. Momentum relaxation $\alpha$ shifts both $\omega_c$ and $\gamma$. For small $B$ these shifts scale as $\sim\alpha^2 B$ and $\sim \alpha^2$~\cite{Kim:2015wba}. It is natural that $\alpha$ increases the damping effect. We refer to \cite{Kim:2015wba} for more detailed analysis of AC conductivities and the effect of $\alpha$ on them.

\subsection{Magnetophonons}
In presence of an external magnetic field $B$, together with the SSB of translations, the dynamics of the low-energy Goldstone modes become richer. In particular, the transverse and longitudinal phonon modes mix together and give rise to the so-called \textbf{magnetophonons} and \textbf{magnetoplasmons} ~\cite{Chen2005QuantumSO,PhysRevB.18.6245,PhysRevB.46.3920}, with dispersion relations:
\begin{equation}
\begin{split}
    &\mathrm{Re}\,[\omega_-]\,=\,\frac{v_\perp\,v_\parallel}{\omega_c}\,k^2\,+\,\dots \label{vv2}\,,\\
   & \mathrm{Re}\,[\omega_+]\,=\,\omega_c\,+\,\frac{(v_\parallel^2+v_\perp^2)}{2\,\omega_c}\,k^2\,+\,\dots\,,
\end{split}
\end{equation}
with $\omega_c$ being the cyclotron frequency. The fundamental reason is that the Poincar\'e algebra is now modified into:
\begin{equation}
     \left[P_i\,,\,P_j\right]\,=\,-\,i\,\epsilon_{ij}\,B\,\mathcal{Q}\,,
\end{equation}
where $\mathcal{Q}$ is the electric charge operator. This implies that translations do not commute anymore with each other and the effective low energy description for the Goldstone fluctuations $\pi^i$ associated with translations can contain a new term:
\begin{equation}
    \mathcal{L}\,=\,\epsilon^{ij}\,\pi_i\,\partial_t\pi_j\,+\,\dots\,.
\end{equation}
In accordance with the Watanabe-Brauner formalism~\cite{Watanabe:2014fva}, the system will display the presence of a \textbf{type-B Goldstone} mode -- the magnetophonon.

\begin{figure}[h]
\centering
\includegraphics[width=0.7\linewidth]{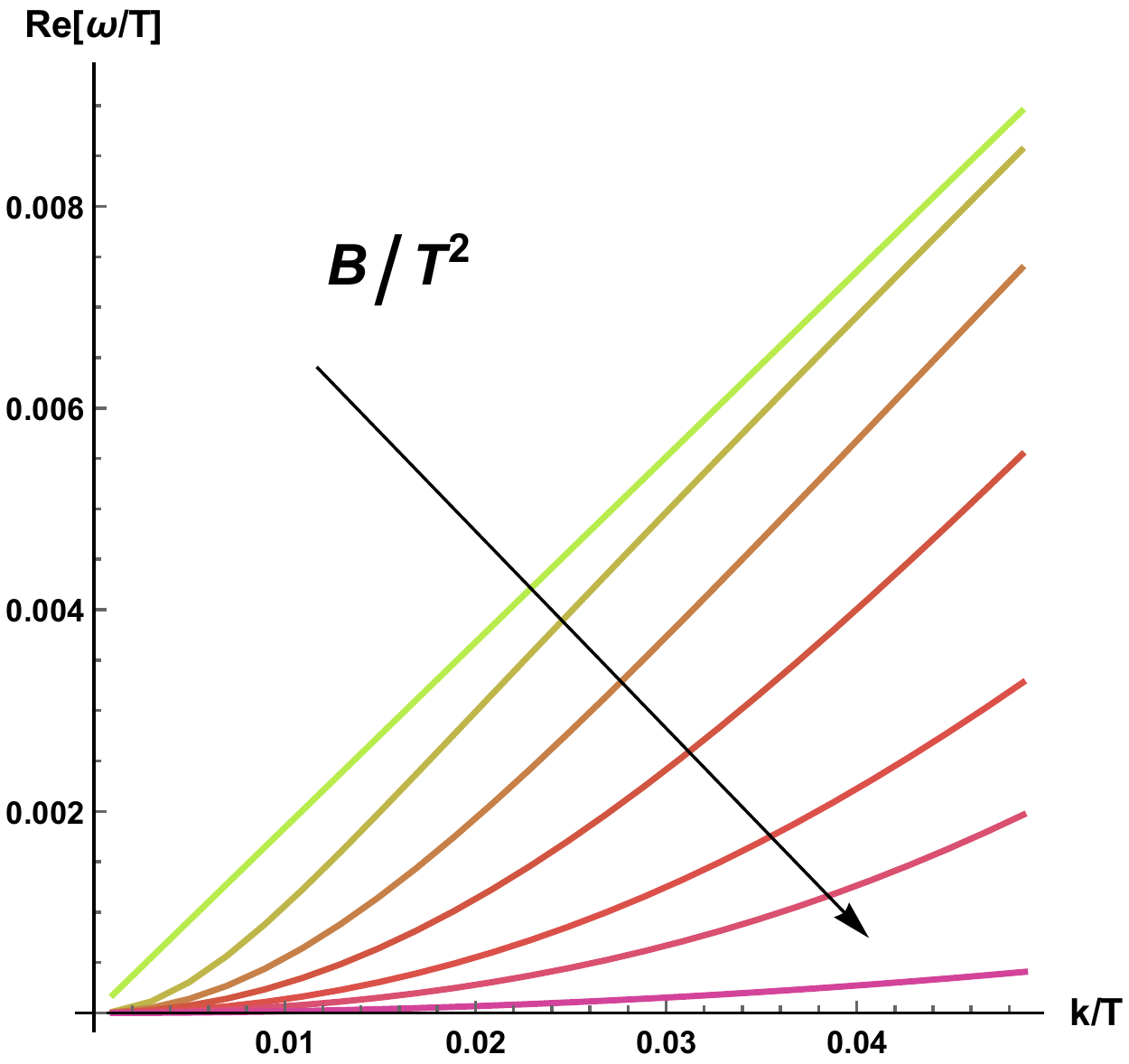}
    
\vspace{0.2cm}
    
\includegraphics[width=0.7\linewidth]{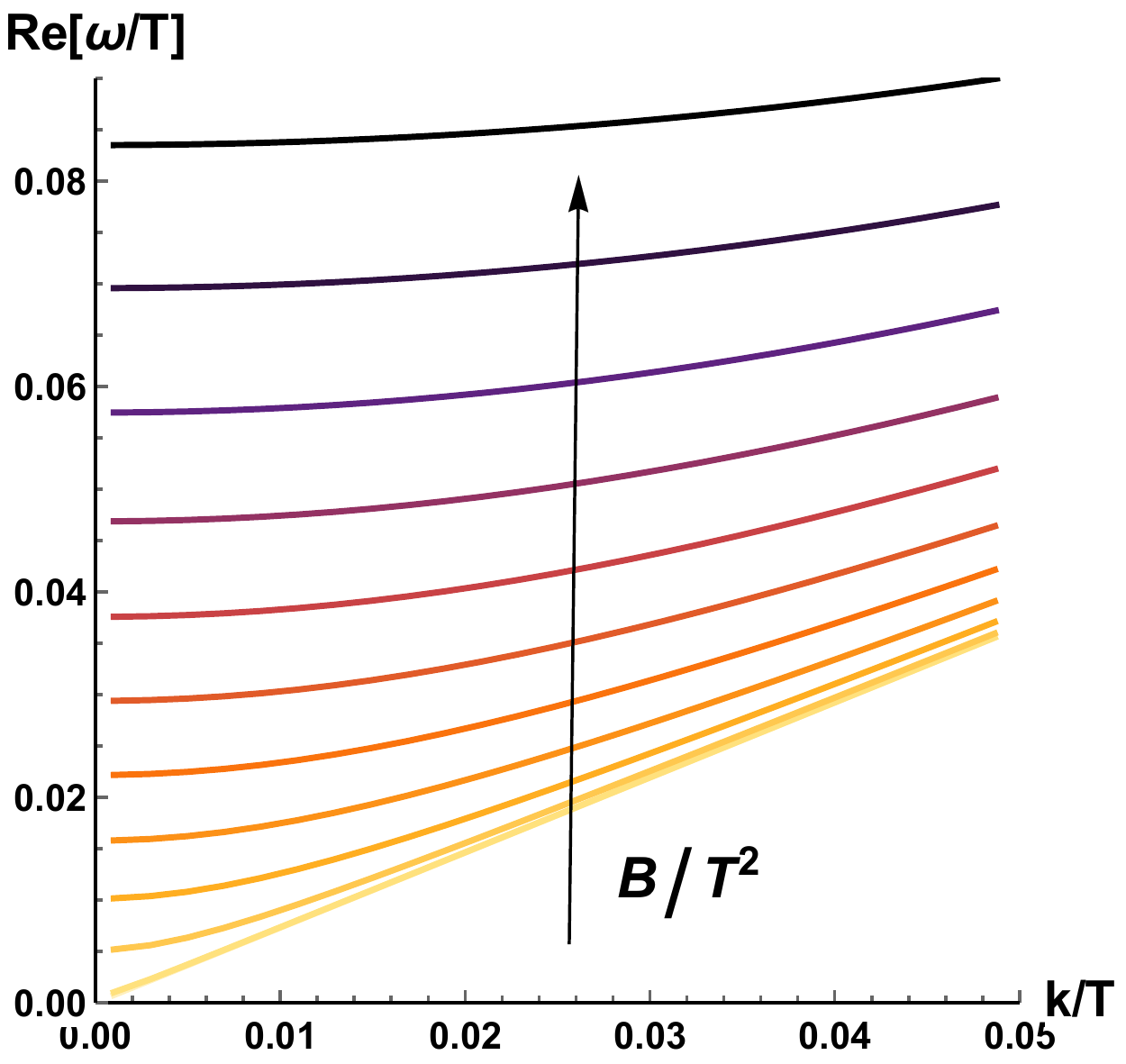}
\caption{The dispersion relations of the magnetophonon and magnetoplasmon in the holographic axion model with an external magnetic field. The arrow indicates the direction of growth of $B$. Figures adapted from~\cite{Baggioli:2020edn}.}
\label{mag}
\end{figure}

This mechanism was successfully verified within the holographic axion model in~\cite{Baggioli:2020edn}. See Fig.~\ref{mag} for the dispersion relations of the modes just mentioned. Interestingly, it was observed that the imaginary part of the magnetophonon is compatible with a quadratic diffusive behavior which is not envisaged from EFT methods~\cite{Hayata:2014yga}. Actually, field theory approaches suggest a $\sim k^4$ behavior of imaginary part for quadratic type-B Goldstone modes manifesting the quasiparticle nature of excitation. In contrast, a quasiparticle excitation in the holographic axion model~\cite{Baggioli:2020edn} is not guaranteed. This is a good example to show that holography is able to describe strongly coupled quantum matter without quasiparticle excitations. It was verified explicitly that the number of type-B phonons and the number of gapped partners sum up to the number of broken generators. For the holographic model the broken generators are the two momenta. One has two linear phonon modes at zero magnetic field. At finite magnetic field, there are one massless type-B magnetophonon and its gapped partner -- the magnetoplasmon.

The situation becomes even more interesting when a small amount of disorder -- EXB of translations -- is introduced in the system. In this case, the magnetophonon gets pinned producing a characteristic peak feature in the optical transport. Despite several theoretical frameworks~\cite{PhysRevB.59.2120,PhysRevB.62.7553,PhysRevB.65.035312,Delacretaz:2019wzh}, a concrete understanding of this phenomenon and in particular of the $B$ dependence of the pinning frequency is still lacking. The dependence of the magnetophonon peak $\omega_{pk}$ as a function of magnetic field is easy to measure accurately and can give useful information on the feature of disorder in the material. In more detail, the dependence is sensitive to whether the material is in a classical or quantum regime. In the classical regime the classical treatment of the pinning mechanism predicts $\omega_{pk}\sim 1/B$, while in the quantum regime, the results can be quite different and the peak can increase with the magnetic field.

\begin{figure}[h]
\centering
\includegraphics[width=0.7\linewidth]{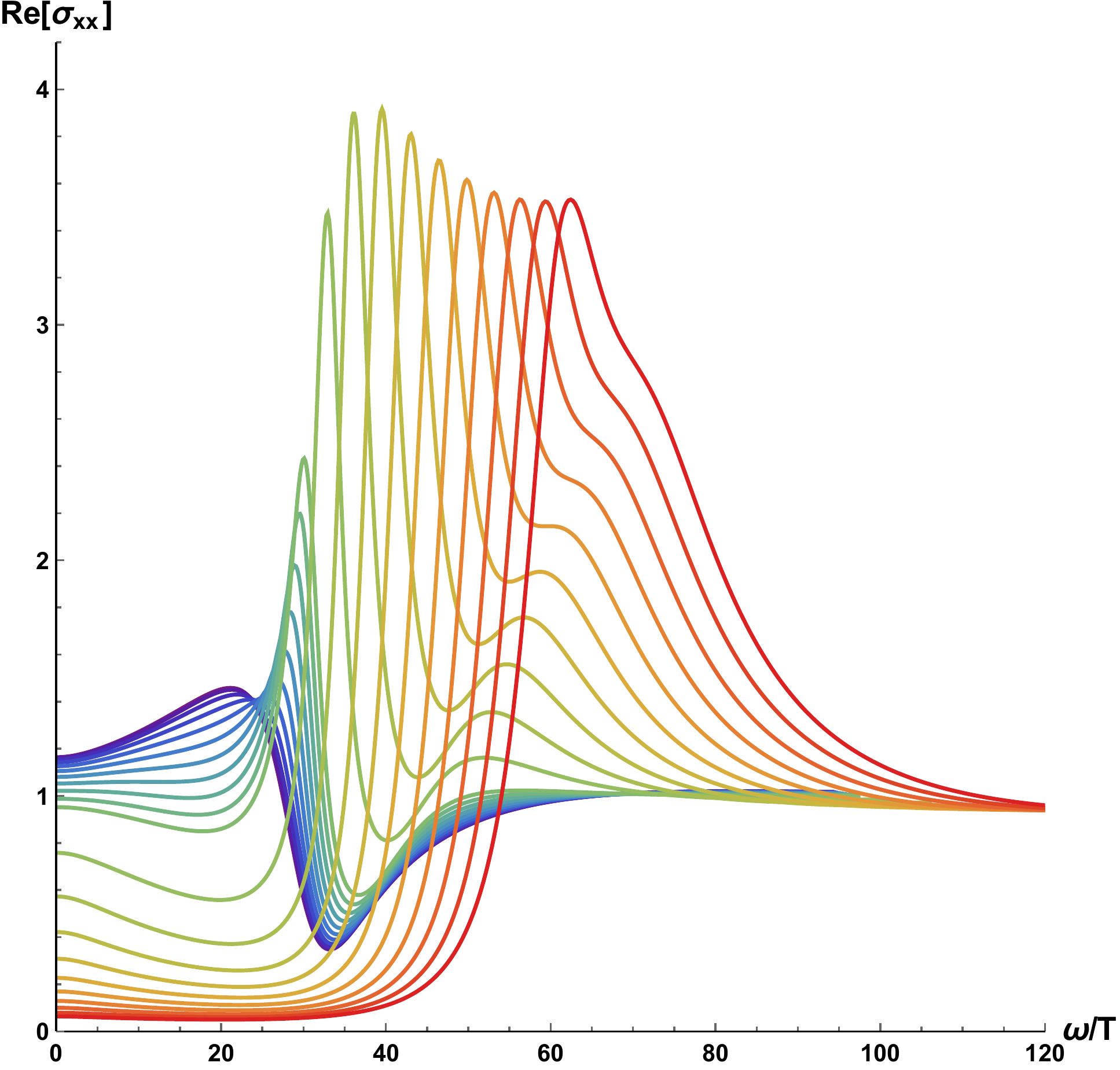}
    
\vspace{0.2cm}
    
\includegraphics[width=0.7\linewidth]{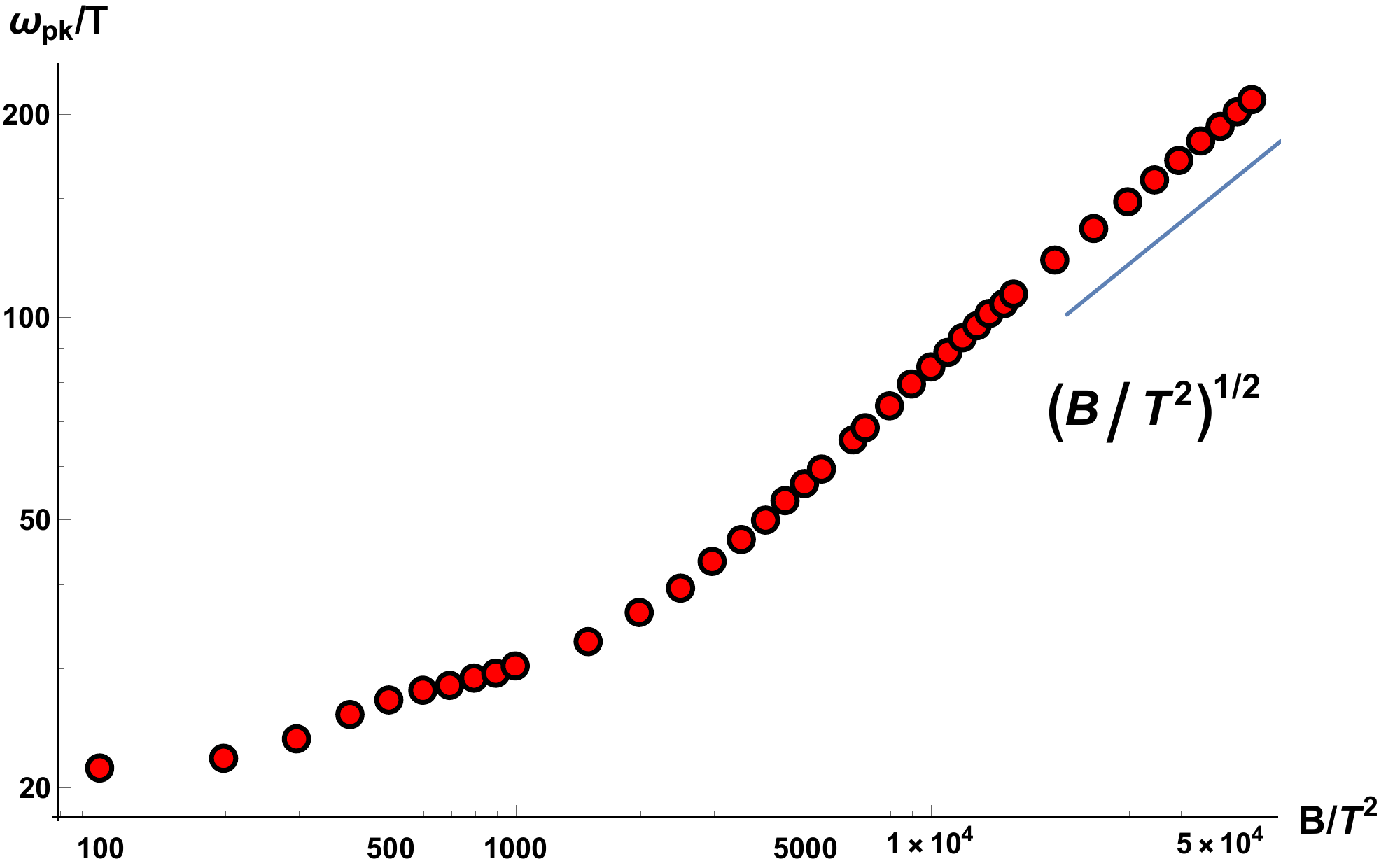}
\caption{The optical conductivity moving the magnetic field $B$ and the scaling of the peak. The position of the peak increases monotonically with the magnetic field. Figures adapted from~\cite{Baggioli:2020edn}.}
\label{mag2}
\end{figure}

In the holographic axion model, it was found~\cite{Baggioli:2020edn} that the pinning frequency increases with the magnetic field $B$ (in contrast to what discussed in~\cite{Delacretaz:2019wzh}) and at large magnetic field it scales like $\sim B^{1/2}$ (see Fig.~\ref{mag2}). Interestingly, this scaling is consistent with the experimental measurement in certain two-dimensional materials~\cite{Chen2005QuantumSO}. This does not imply that the holographic model describes any specific material, but rather that the scaling found from holography is consistent with realistic observation, while at odds with the discussion in~\cite{Delacretaz:2019wzh}. It would be interesting to understand what these results tell us about the nature of the ``disorder'' implemented by the holographic axion model.

\subsection{Non-linear elasticity and rheology}
\label{sec:rheology}
A basic aspect of matter is to understand and characterize the response of materials under mechanical deformations. So far, we have considered the elastic properties of the holographic axion model only in the linear regime, where the external strain is small, and the stress-strain relation can be linearized as:
\begin{equation}
\sigma^{ij}\,=\,C^{ijkl}\,\epsilon_{kl}\,.
\end{equation}
More in general, one could consider an arbitrarily large external strain, such that the stress-strain relation becomes highly non-linear\footnote{Here, $\sigma$ has not to be confused with the electric conductivity.}:
\begin{equation}
\sigma(\epsilon)\,.
\end{equation}
This scenario indicates the onset of \textbf{non-linear elasticity}, in which the higher order corrections:
\begin{equation}
\sigma \sim \epsilon\,+\,\underbrace{\epsilon^2+\epsilon^3+\epsilon^4+\dots}_{\text{higher-order}}\,,
\end{equation}
cannot be neglected anymore.

\begin{figure}[h]
\centering
\includegraphics[width=0.8\linewidth]{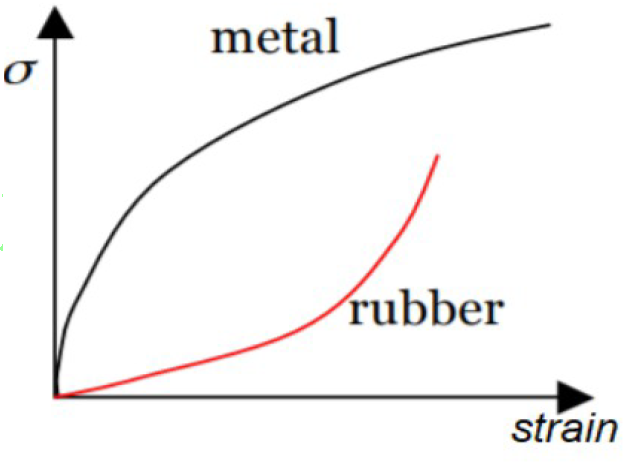}
\caption{The different nonlinear elastic behaviour between metals and rubbers.}
\label{diffig}
\end{figure}

The non-linear elastic features of the various solids can vary a lot and they can be very useful to characterize them. For example, metals and rubbers are very different in this respect (see Fig.~\ref{diffig}).
In particular, in one case (the metal) the stress-strain curve exponent becomes sub-linear at large strain. This phenomenon is called \textbf{strain-softening} and it indicates that the material becomes softer by increasing the deformation strain. In the second case (the rubber), the non-linear behaviour is faster than linear; the material becomes more rigid at finite deformations -- \textbf{strain hardening}.

From the field theory point of view, the non-linear elasticity theory has been recently implemented in~\cite{PhysRevD.100.065015}. In order to follow the same logic from the holographic perspective, a few ingredients must be changed. Since the deformation strain is now an $\mathcal{O}(1)$ external field, it must be endowed in the background configuration and, in particular, the scalars profile must be modified into:
\begin{equation}
\phi^I\,=\,O^I_j\,x^j\,,
\end{equation}
with
\begin{equation}\label{OIJ}
O^I_j\,=\,\alpha\begin{pmatrix} 
\sqrt{1+\varepsilon^2/4} & \varepsilon/2 \\
\varepsilon/2 & \sqrt{1+\varepsilon^2/4} 
\end{pmatrix}\,.
\end{equation}
For $\alpha=1$, we have $\mathrm{Det}\,O^I_j=1$, which means the deformation does not change the volume of the system, it is a \textbf{pure-shear} deformation, parametrized by the parameter $\varepsilon$. On the contrary, the $\alpha$ parameter accounts for the changes of volume: it is a \textbf{bulk-strain} deformation. Notice that the background configuration~\eqref{OIJ} breaks explicitly the isotropy of the system.
\begin{figure}[htp]
\begin{center}
\includegraphics[height=0.6\linewidth]{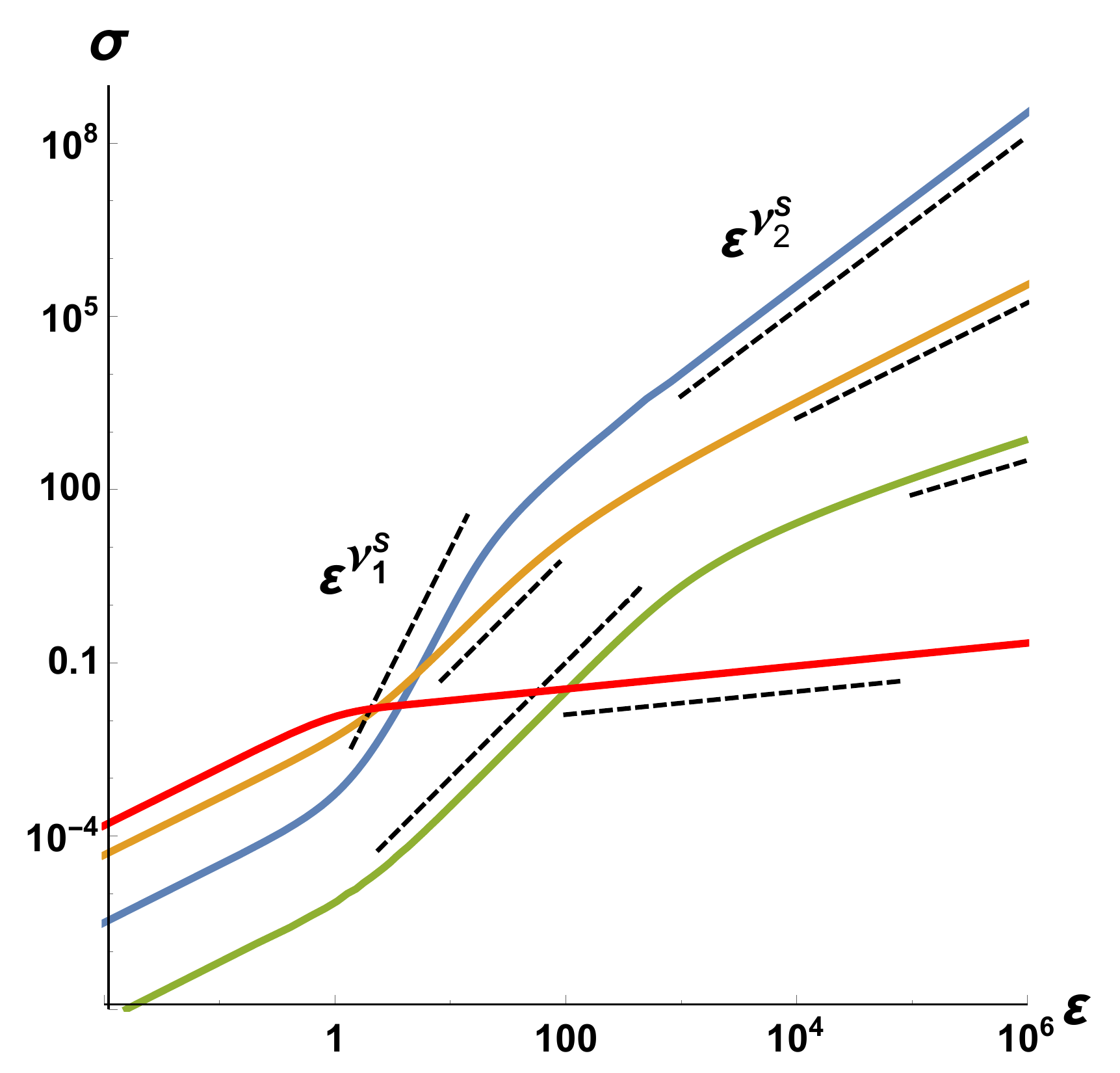}

\vspace{0.2cm}

\includegraphics[height=0.6\linewidth]{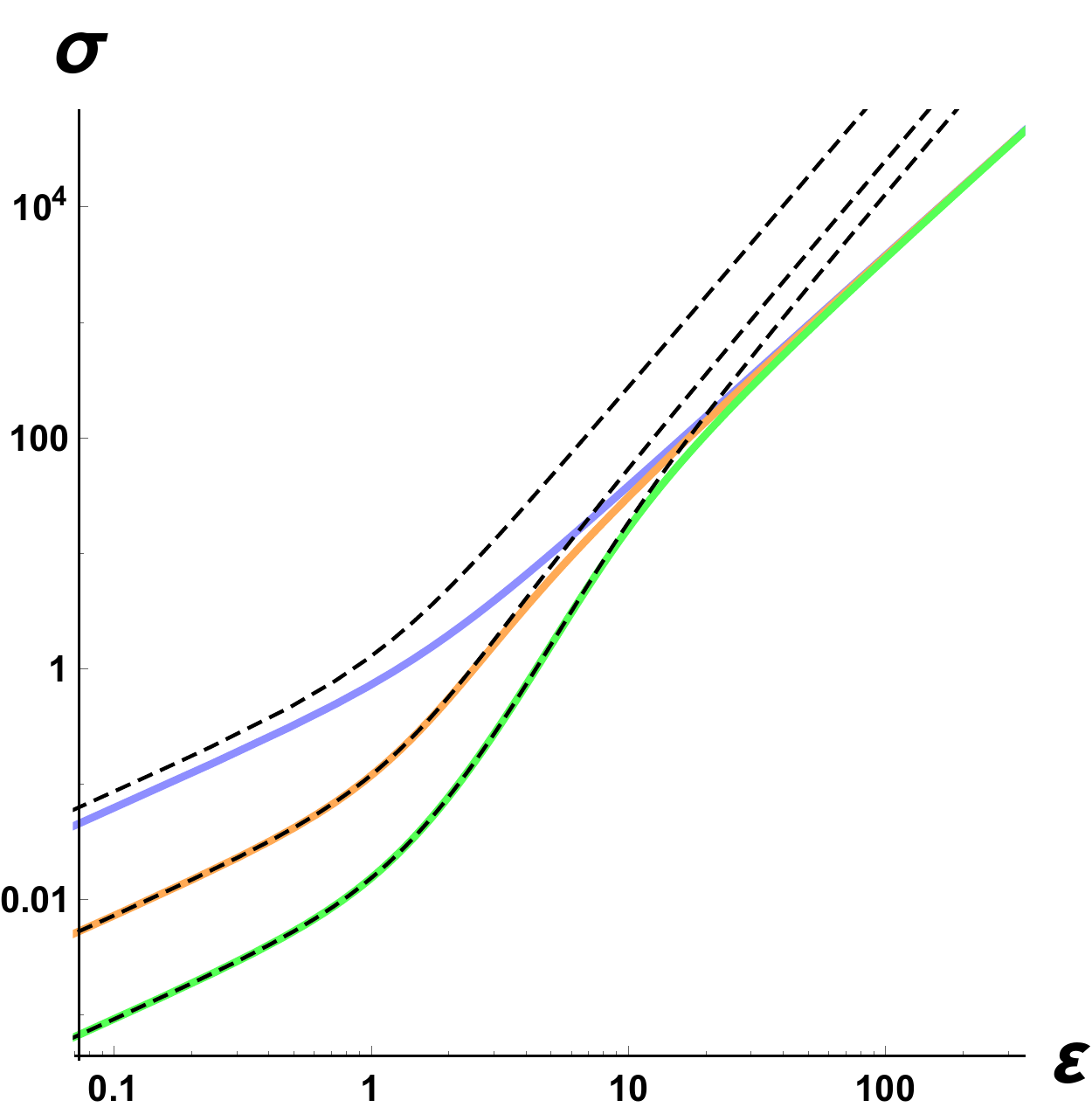}
 \caption{\textbf{Top panel: }Shear stress strain curve for various potentials and relative (dashed) large strain scaling.  \textbf{Bottom panel: }Shear stress strain curves for different temperatures and comparison with the analytic formula \eqref{integral} (dashed lines). As expected for $T/m \gg 1$ the formula gives a very good approximation. Figures taken from~\cite{Baggioli:2020qdg}.}
 \label{non1}
\end{center}
\end{figure}

In order to find a background solution with the scalar configuration of Eq.~\eqref{OIJ}, the metric ansatz must be modified into:
\begin{equation}
ds^2\,=\,\frac{1}{u^2}\left(-f(u)\,e^{-\chi(u)}\,dt^2+\frac{du^2}{f(u)}+\gamma_{ij}(u)\,dx^i dx^j\right)\,,
\label{geometry}
\end{equation}
where $\gamma_{ij}$ is a two dimensional spatial metric with unitary determinant. This time, the background has to be found numerically by solving a simple set of ordinary differential equations in the radial coordinate $u$. The stress-tensor $T^{ij}$ can be then obtained using standard methods~\cite{Balasubramanian:1999re} and it is a non-linear function of the background strains $\alpha$ and $\varepsilon$. In particular, for a generic potential $V=(X,Z)$, an analytic formula can be derived:
\begin{equation}
\sigma(\varepsilon)=\frac{1}{2}m^2\alpha^2\varepsilon  \sqrt{4+\varepsilon^2} \int_0^{u_h}\frac{V_X\left(\bar{X},\bar{Z}\right)}{\zeta^2}\,d\zeta\,,
\label{integral}
\end{equation}
which is valid at small graviton mass, $m/T \ll 1$. Here, we have defined $(\bar{X},\bar{Z})=(\alpha^2\frac{1}{2}(2+\varepsilon^2)\zeta^2,\zeta^4 \alpha^4)$ and $\sigma\equiv T^{x}_y$. The convergence of the integral in \eqref{integral} is equivalent to the positivity of the linear bulk modulus.
In order to make some more quantitative predictions, we will consider the benchmark potential:
\begin{equation}
V(X,Z)\,=\,X^\mathfrak{a}\,Z^{\frac{\mathfrak{b}-\mathfrak{a}}{2}}\,,
\label{bench}
\end{equation}
The non-linear stress-strain curves for different powers are shown in Fig.~\ref{non1} together with the comparison with the perturbative expression Eq.~\eqref{integral}. At intermediate strain, a power law scaling $\sigma \sim \varepsilon^{\nu_1^S}$ appears, with 
\begin{equation}
\nu_1^S = 2\mathfrak{a}\,.
\label{nuscal1}
\end{equation}
Additionally, at much larger values of the strain, a secondary scaling appears $\sigma\sim\varepsilon^{\nu_2^S}$ with a different exponent
\begin{equation}
\nu_2^S\,= \,3\,\frac{\mathfrak{a}}{\mathfrak{b}}\,.
\label{nuscal}
\end{equation}
The presence of two scaling regimes happens only at high enough temperature. At low temperature, the stress-strain curve directly interpolates from the linear regime to the $\nu_2^S$ scaling. A similar power law behaviour appears in the bulk stress-strain curve~\cite{Baggioli:2020qdg}, where there is a universal scaling
\begin{equation}
\sigma_L\,\propto\,\kappa^3\,,
\label{univ}
\end{equation}
with $\kappa\equiv \partial \cdot \phi$ and $\sigma_L=T_{xx}(\kappa)-T^{eq}_{xx}$.

Until now, we have discussed the realm of non-linear elasticity only in the context of static deformations. Nevertheless, the biggest interest in the field of \textbf{rheology} deals with time dependent deformations and the corresponding reaction of the system. In particular, a typical experiment -- oscillatory shear test -- consists in an external shear strain taking a simple sinusoidal form 
\begin{equation}
\gamma(t)\,=\,\gamma_0\,\sin(2 \pi\omega t)\,,
\end{equation}
where $\gamma_0$ is the strain amplitude and $\omega$ the characteristic frequency.

\begin{figure}[ht]
\centering
\includegraphics[width=0.9 \linewidth]{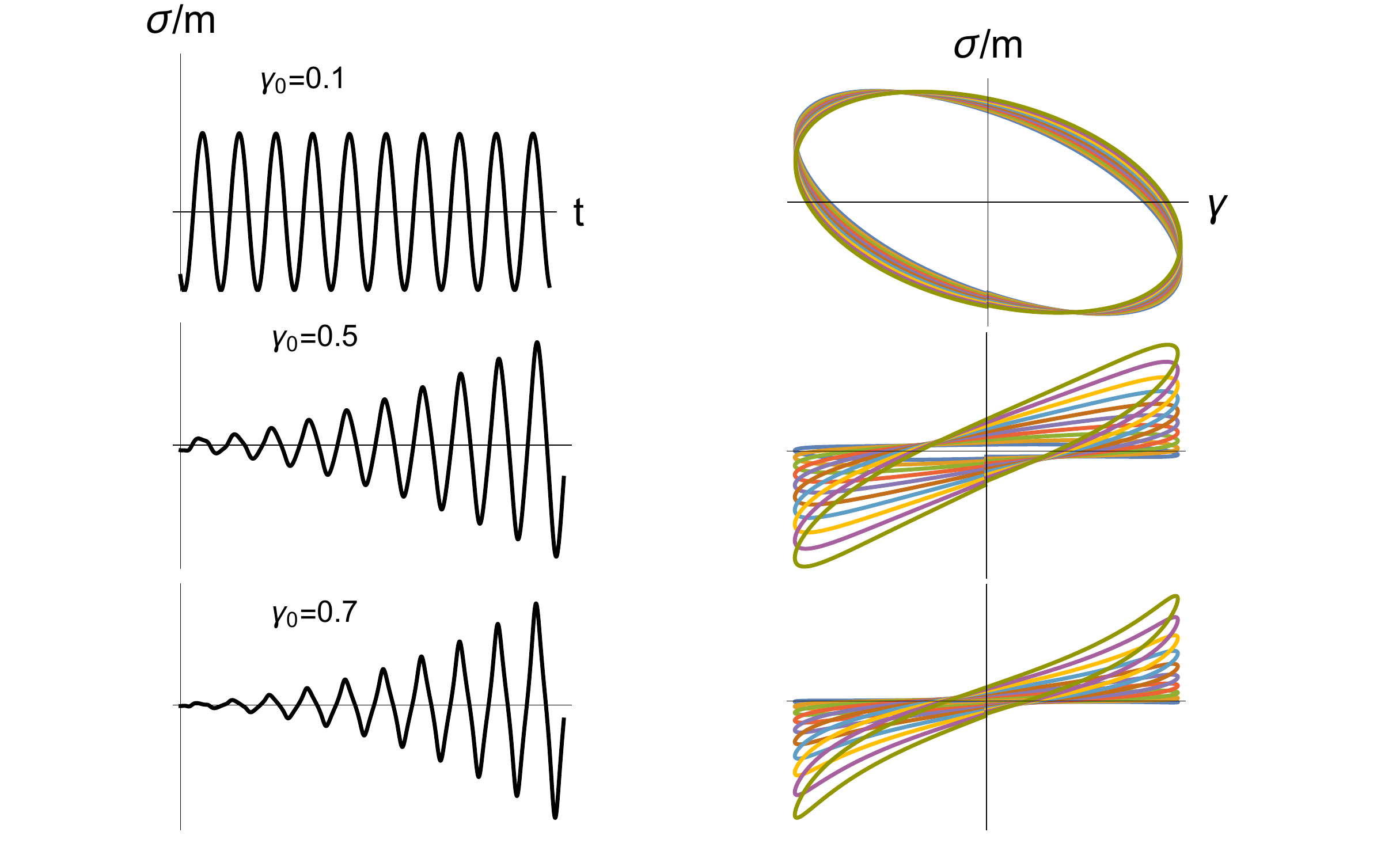}
    
\vspace{0.3cm}
    
\includegraphics[width=0.85 \linewidth]{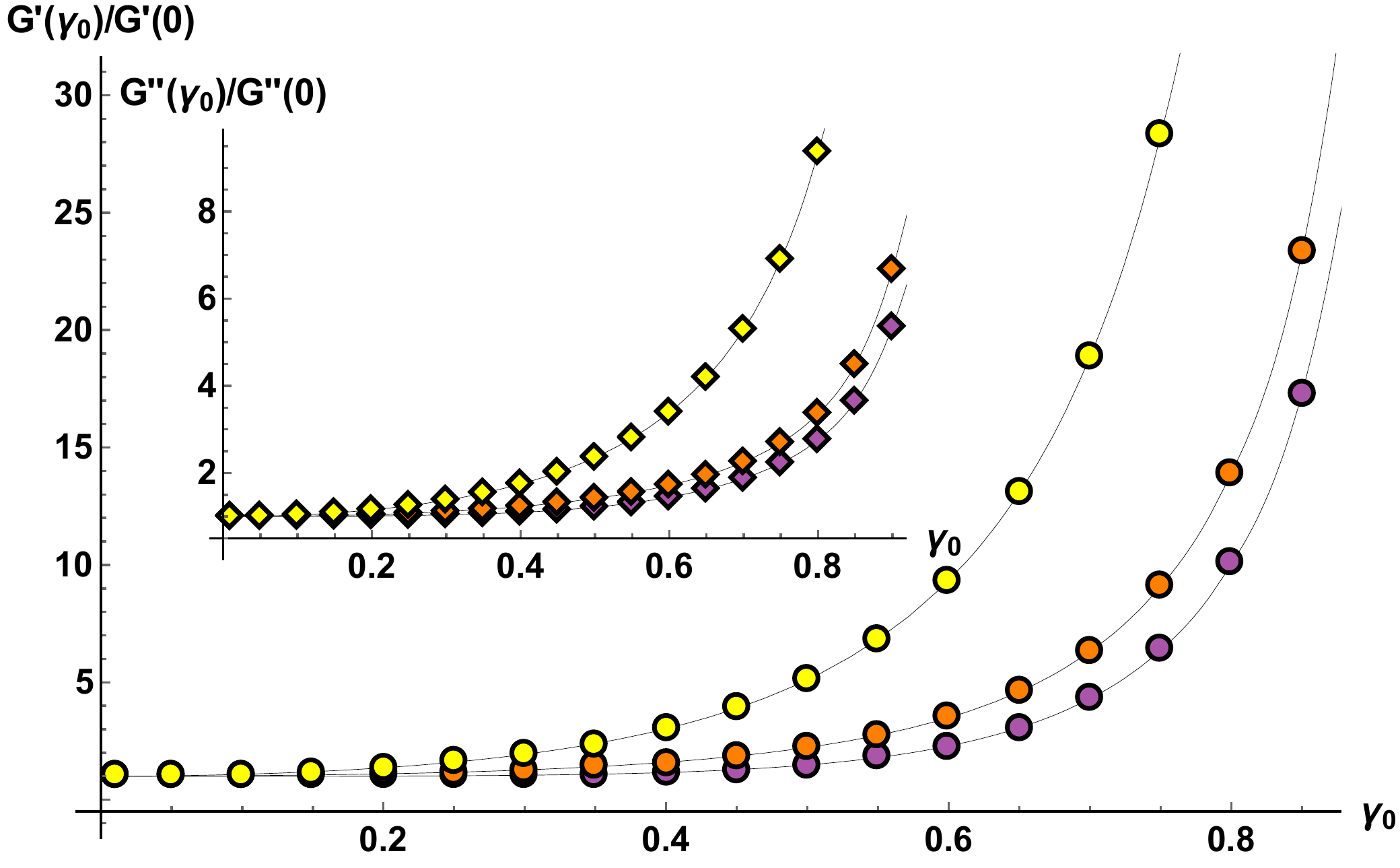}
\caption{The onset of nonlinear elasticity by increasing the strain amplitude. \textbf{Top panel:} the real time dynamics and the Lissajiou's figures. \textbf{Bottom panel: } the non-linear complex modulus. Figures updated from~\cite{Baggioli:2019mck}.}
\label{nonlinear}
\end{figure}

Particularly challenging is the regime where the amplitude of the external strain is large, $\gamma_0 = \mathcal{O}(1)$, which takes the name of large amplitude oscillatory shear regime (LAOS). In the LAOS regime, linear viscoelasticity is not applicable anymore and the response is fully nonlinear and very little is known ~\cite{pipkin1986lectures,HYUN20111697,4f5dcea3a806435da7b4b4cbadb06d63}. The LAOS regime has been recently studied in the holographic axion model in~\cite{Baggioli:2019mck} and the non-linear regime has been directly observed with several methods (see Fig.~\ref{nonlinear}).

\begin{figure}
\centering
\includegraphics[width=0.8\linewidth]{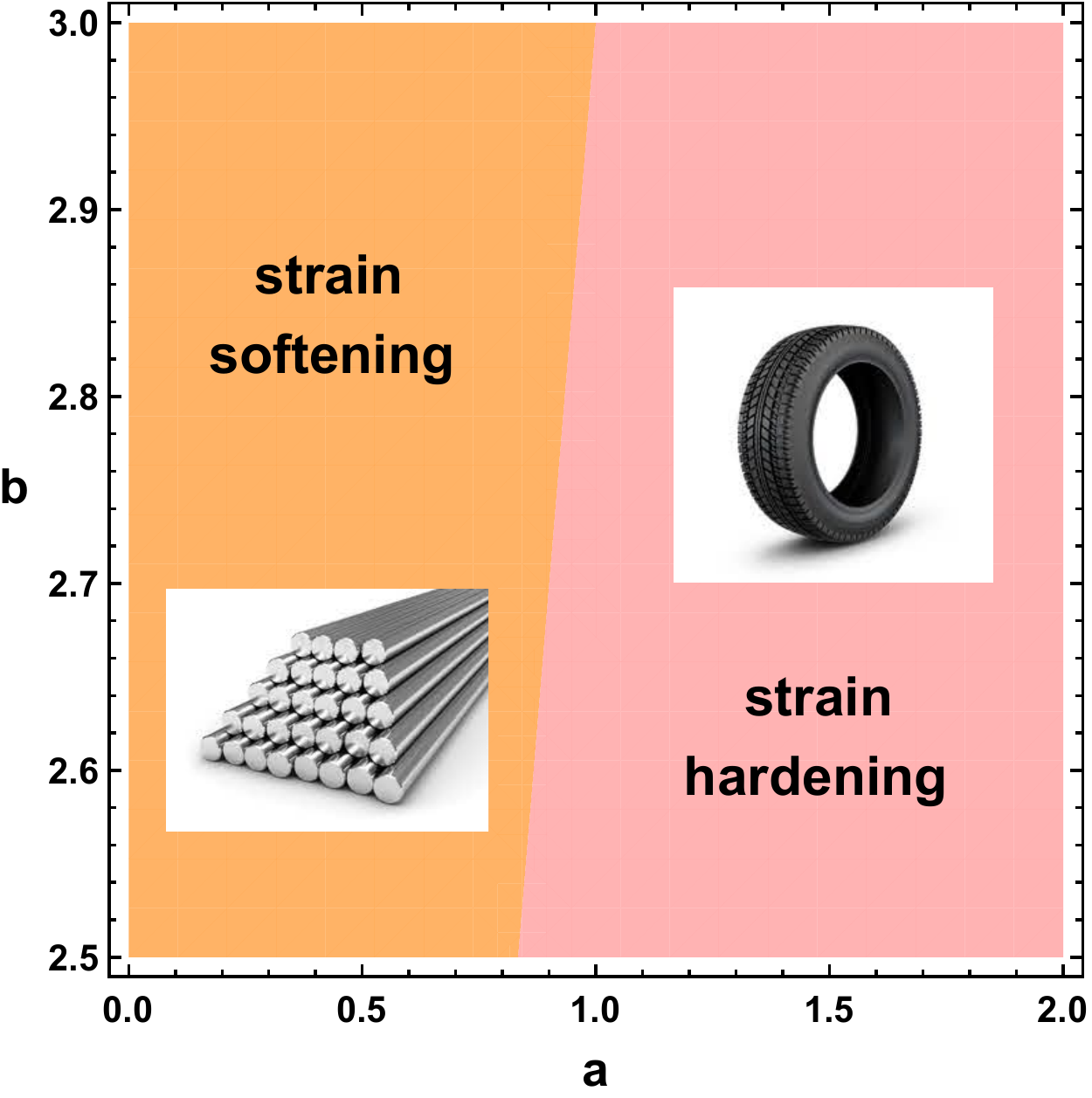}
\caption{The phase diagram of the holographic axion model with benchmark potential~\eqref{bench} according to the non-linear elastic properties.}
\label{soft}
\end{figure}

Moreover, it has been found that at least for the potentials considered, the holographic model displays a well-defined strain-hardening mechanism. More generally, depending on the potential chosen in Eq.~\eqref{bench} the holographic axion model can exhibit either strain-softening or strain hardening. See Fig.~\ref{soft} for a map of the two situations depending on the powers in Eq.~\eqref{bench}.

\subsection{Plasmons}
\begin{figure}
\centering
\includegraphics[width=0.9\linewidth]{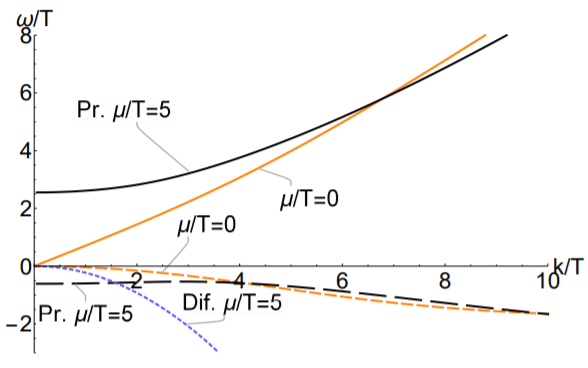}
\caption{The appearance of plasmons in the Reissner-Nordstrom background as a consequence of imposing the mixed boundary conditions~\eqref{mix}. Figure from~\cite{Aronsson:2017dgf}.}
\label{plas1}
\end{figure}

An interesting phenomenological direction which has been recently investigated is the dynamics of plasmon modes in strongly coupled materials and therefore holographic models. The seminal work can be found in~\cite{Aronsson:2017dgf} and it was motivated by the recent surprising experimental results of~\cite{mitrano2018anomalous,husain2019crossover} (see also~\cite{Krikun:2018agd,Andrade:2019bky} for discussions around this point). The main idea is to modify the boundary conditions of the gauge fields fluctuation to impose the Maxwell equations in the boundary dual field theory. This can be achieved by fixing:
\begin{equation}
    \omega^2\,\delta A_x\,+\,\lambda\,\delta A'_x\,=\,0\,, \label{mix}
\end{equation}
which is a mixed boundary condition and makes the gauge field dynamical at the boundary. The parameter $\lambda$ measures the strength of the emergent Coulomb force in the boundary theory. Doing so, a nice plasmon mode $\mathrm{Re}\left[\omega\right]=\sqrt{\omega_p^2\,+\,k^2}$ is obtained at finite charge (see Fig.~\ref{plas1}).

\begin{figure}
\centering
\includegraphics[width=0.8\linewidth]{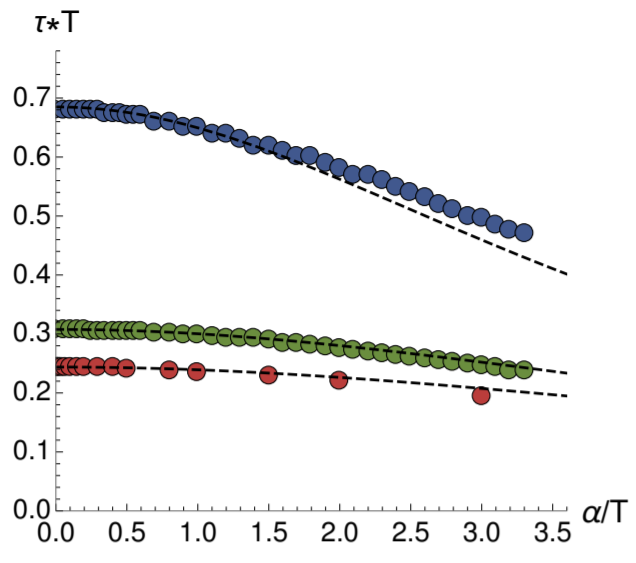}
\caption{The numerical confirmation of the inverse Matthiessen' rule for the plasmon lifetime. The dashed lines are Eq.\eqref{matt} and the colored dots the numerical data for various $\lambda$. Figure taken from~\cite{Baggioli:2019aqf}.}
\label{inv}
\end{figure}

The effects of the explicit and spontaneous breaking of translations, using the holographic axion model, have been studied in a series of follow-up works \cite{Baggioli:2019aqf,Baggioli:2019sio}. The first observation is that the lifetime of the plasmon mode obeys a inverse Matthiessen’ rule (see Fig.~\ref{inv}):
\begin{equation}
\tau^{-1}\,=\,\tau_{EM}^{-1}\,+\,\tau_M^{-1}\,,
\label{matt}
\end{equation}
where $\tau_{EM}$ is the contribution for the Coulomb interactions and $\tau_M$ is the contribution coming from momentum dissipation and equal to the inverse of the momentum relaxation rate $\Gamma$.
Additionally, in the transverse spectrum, momentum dissipation induces a peculiar modes repulsion dynamics which was discussed in~\cite{Baggioli:2019sio}.

\begin{figure}
\centering
\includegraphics[width=0.8\linewidth]{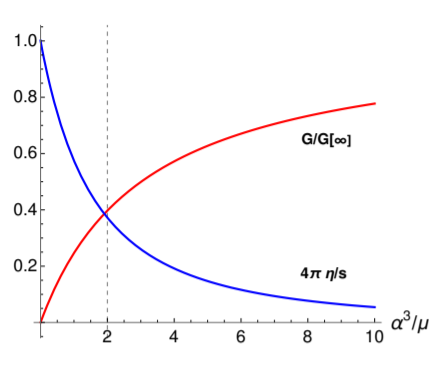}
    
\vspace{0.2cm}
    
\includegraphics[width=0.8\linewidth]{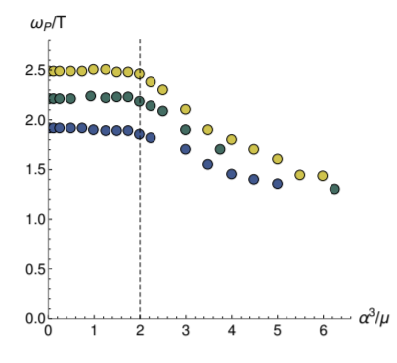}
\caption{The fluid to solid crossover and the depletion of the plasma frequency $\omega_p$. Figure taken from~\cite{Baggioli:2019aqf}.}
\label{plasmon2}
\end{figure}

Finally,~\cite{Baggioli:2019aqf} analyzed also the plasmons dynamics in presence of elasticity -- SSB of translations. An interesting transition appears at the point in which the shear viscosity becomes comparable with the shear modulus (see Fig.~\ref{plasmon2}). At that value, the fluid becomes to behave more like a solid and the plasma frequency -- the mass of the plasmon -- starts to rapidly decay as shown in Fig.~\ref{plasmon2}.

The physics of holographic plasmons is still highly unexplored and more work, specially in connection with a possible hydrodynamic description, is needed.

\section{Additional topics}\label{sec:add}
\subsection{SYKology}
The Sachdev-Ye-Kitaev (SYK) model is a many body quantum system which has become very popular  in the physics community because it is strongly coupled, exactly solvable, chaotic and nearly conformal invariant~\cite{Rosenhaus:2018dtp}. Moreover, it bears several interesting relations with AdS$_2$ gravity, black holes physics and strange metals~\cite{Trunin:2020vwy}.

The relation between the SYK model and axions-like holographic model was put forward in~\cite{Davison:2016ngz} with special emphasis on the thermodynamic, transport and quantum chaos properties. More specifically several connections between the two frameworks were successfully established and analyzed later in~\cite{PhysRevX.5.041025}.

Recently, several works~\cite{Choi:2020tdj,Arean:2020eus} computed exactly the energy-energy correlator in the SYK model and they compared it with that extracted from the linear axion model of~\cite{Andrade:2013gsa}. Again, in the strong coupling limit, the results obtained from the two frameworks were found to be very similar. These findings contribute to the question about which is the gravity dual of the disordered SYK model and if that has to do with our holographic axion model with momentum dissipation.

\subsection{Quantum information}

There are increasing evidences for the existence of a deep connection between quantum information in the boundary field theory and the spacetime geometry in bulk. This link was initially triggered by the definition of entanglement entropy and its holographic dual -- the Ryu-Takayanagi formula~\cite{Ryu_20061,Ryu_20062}. See \cite{Rangamani_2017} for a review. However, there are many other quantum informational quantities, which capture several aspects of quantum information different from the entanglement entropy. According to holography, these informational concepts "must" have their own dual geometric objects. For example, quantum information probes for mixed states have been proposed: the entanglement of purification, the logarithmic negativity, the odd entanglement entropy and the reflected entropy. Their holographic dual is related to the so-called entanglement wedge cross section.\footnote{For more description of the concepts we refer to \cite{Jeong:2019xdr} and references therein.} Quantum complexity is another important concept, because it is conjectured to explore the inside of the black hole horizon, while entanglement entropy can not~\cite{Susskind:2018pmk}. This line of research played a key role to achieve a possible resolution of the black hole information paradox~\cite{Almheiri:2020cfm}. 

It is natural to consider the axion model to study various quantum informational quantity, because momentum relaxation is ubiquitous and plays an important role in real quantum systems.  
The momentum relaxation effect on the holographic entanglement entropy~\cite{Mozaffara:2016iwm} and the complexity in the complexity-action conjecture~\cite{Babaei-Aghbolagh:2020vsz} have already been studied. Under thermal quench, holographic entanglement entropy~\cite{Li_2020},  subregion complexity in the complexity-volume conjecture~\cite{Zhou:2019xzc} have been investigated. Towards the entanglement measure for mixed states, holographic entanglement entropy, mutual information, and entanglement of purification have been considered in \cite{Huang:2019zph}.

\subsection{Fermionic response}
The fermionic spectral function is a very important observable, specially in strongly correlated materials, which can be directly probed experimentally with Angle Resolved Photoemission Spectroscopy (ARPES) or Scanning Tunneling Microscopy (STM). The fermionic spectral function has been considered in the realm of Holography in several pioneering works~\cite{Lee:2008xf,Liu:2009dm,Cubrovic:2009ye,Faulkner:2009wj} in relation to possible non-Fermi liquids signatures.

The holographic spectral function can be computed by solving the bulk Dirac equation. A class of fermion bulk action with the mass and the dipole coupling is given by
\begin{eqnarray}
\mathcal{S}_{\text{spinor}} &=&i \int d^4 x \sqrt{-g} \bar{\psi} \left( \Gamma^{M} D_{M} -m - \, \frac{i  p}{2}\Gamma^{MN} F_{MN}  \right)\psi  \,,\nonumber \\
\Gamma^{MN} &=& \frac{1}{2} [\Gamma^{M},\Gamma^{N}] \,,
\end{eqnarray}
with $\Gamma^{M}$ and $D_{M}$ the Gamma matrices and the covariant derivative in a curved spacetime, respectively. The dipole interaction drives the dynamical formation of a Mott-like gap in the absence of continuous symmetry breaking~\cite{Edalati:2010ww}. The study of holographic fermions in terms of fermion mass ($m$), dipole coupling ($p$) and the strength of momentum relaxation ($\alpha$) has been conducted in~\cite{Jeong:2019zab}. 

\begin{figure}
\centering
\includegraphics[width=1.0\linewidth]{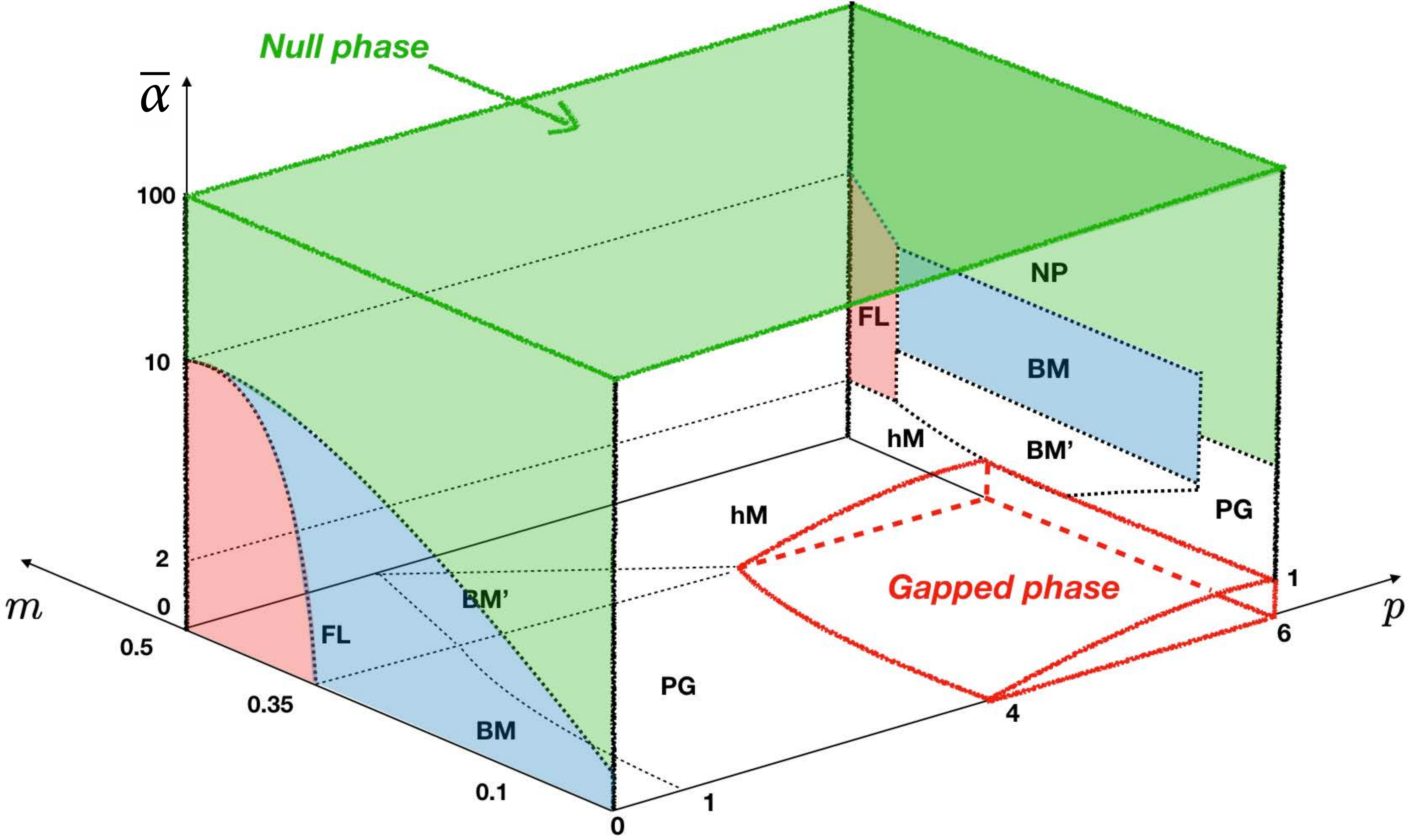}
\caption{Phase diagram in ($m, p, \bar{\alpha}$) space for the Einstein-Maxwell-linear axion model of~\cite{Jeong:2019zab}. Depending on the shape of spectral functions, one can classify different phases, such as fermi liquid like (FL), bad metal prime (BM'), bad metal (BM), pseudogap (PG), gapped (G) and so on. See~\cite{Jeong:2019zab} for more details. Figure taken from~\cite{Jeong:2019zab}. }
\label{fig:fermion}
\end{figure}

The holographic spectral function with momentum relaxation in two linear axion models was systematically investigated in~\cite{Jeong:2019zab}, where the momentum relaxation strength $\alpha$ is introduced via the bulk profile $\phi^I=\alpha \delta^I_i x^i$. By classifying the shape of spectral functions, the complete phase diagrams in ($m,p,\alpha$) space were constructed (see Fig.~\ref{fig:fermion}). Although the DC electric transport of two models are very different, the effects of momentum relaxation on the spectral function are similar. This may be due to the fact that holographic fermion does not back-react to geometry. 

Some common features were highlighted as follows~\cite{Jeong:2019zab}. First, it was found that for a given dipole coupling and momentum relaxation, the spectral functions tend to become sharper by increasing the mass of the bulk fermion. Second, a new peak at finite frequency can be generated as the dipole coupling increases. Third, in general the spectral function becomes more suppressed and broader as the strength of momentum relaxation is increased. Interestingly, the suppression of spectral weight and the gradual disappearance of Fermi surface along the symmetry breaking direction were also observed in the inhomogeneous holographic models by increasing the lattice strength~\cite{Cremonini:2018xgj,Cremonini:2019fzz,Balm:2019dxk,Iliasov:2019pav,Mukhopadhyay:2020tky}. The homogeneous holographic lattices can simulate the effects of translational symmetry breaking while retaining the homogeneity of the spacetime geometry. However, homogeneous lattices are unable to capture the physics of Umklapp, motivating the need to work with periodic lattices~\cite{Bagrov:2016cnr}.

\subsection{Modeling graphene}
Experimental measurements have uncovered evidence of the strongly coupled nature of the graphene. As a matter of fact, the Wiedemann-Franz law (the ratio of heat and electric conductivities, $L=\kappa/T\sigma$) is violated by up to a factor of 20 near the charge neutral point in extremely clean graphene~\cite{2016Sci...351.1058C}. It has been argued that graphene near charge neutrality forms a strongly coupled Dirac fluid without well-defined quasiparticle excitations. A fundamental reason for the appearance of the strong interaction in graphene is due to the smallness of the Fermi sea: electron-hole pair creation near the Dirac cone is insufficient to screen the Coulomb interaction, thus there should be a regime where electrons are strongly correlated. A hydrodynamics description with disorder and a single conserved $U(1)$ current was adopted to explain experimental observations~\cite{Hartnoll:2007ih,Lucas:2015sya}, but still left room for improvement.

\begin{figure}
    \centering
    \includegraphics[width=0.75\linewidth]{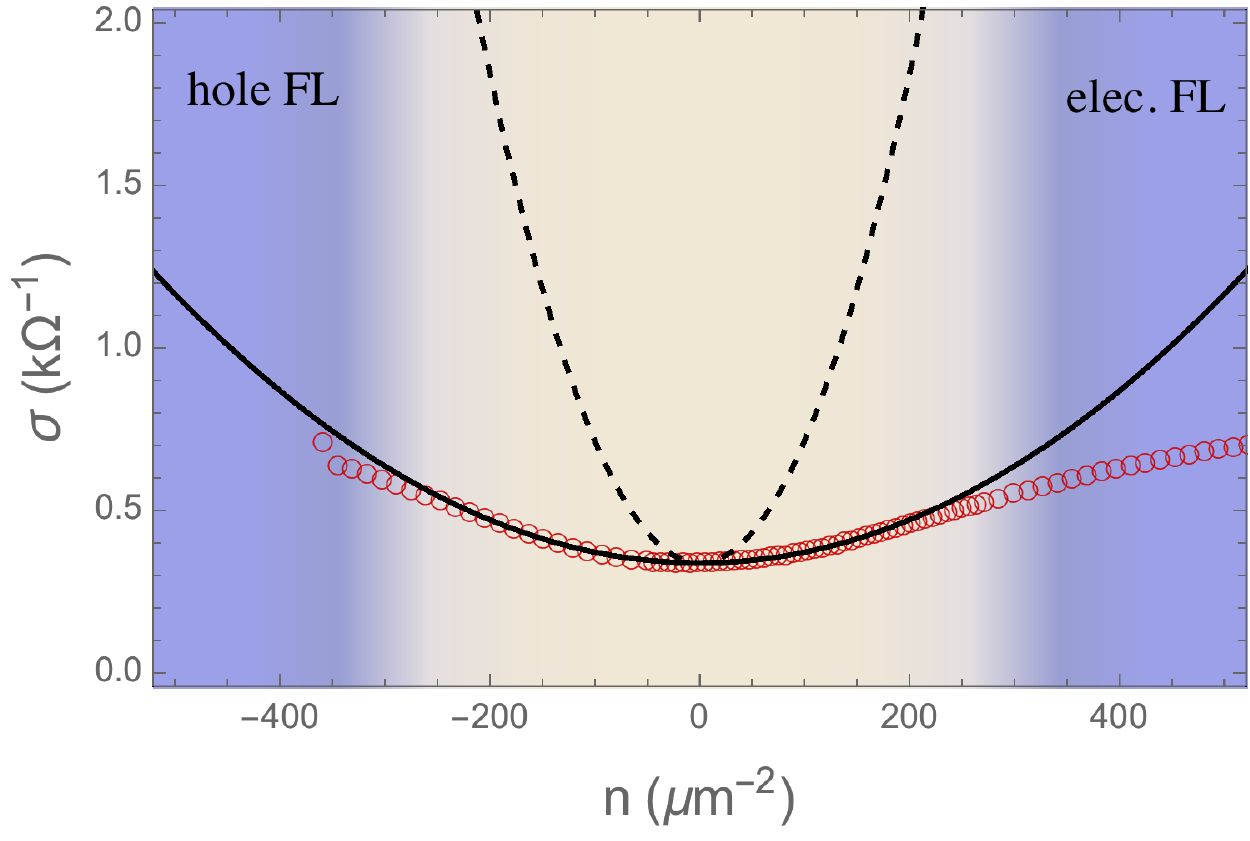}
    
    \vspace{0.2cm}
    
    \includegraphics[width=0.74\linewidth]{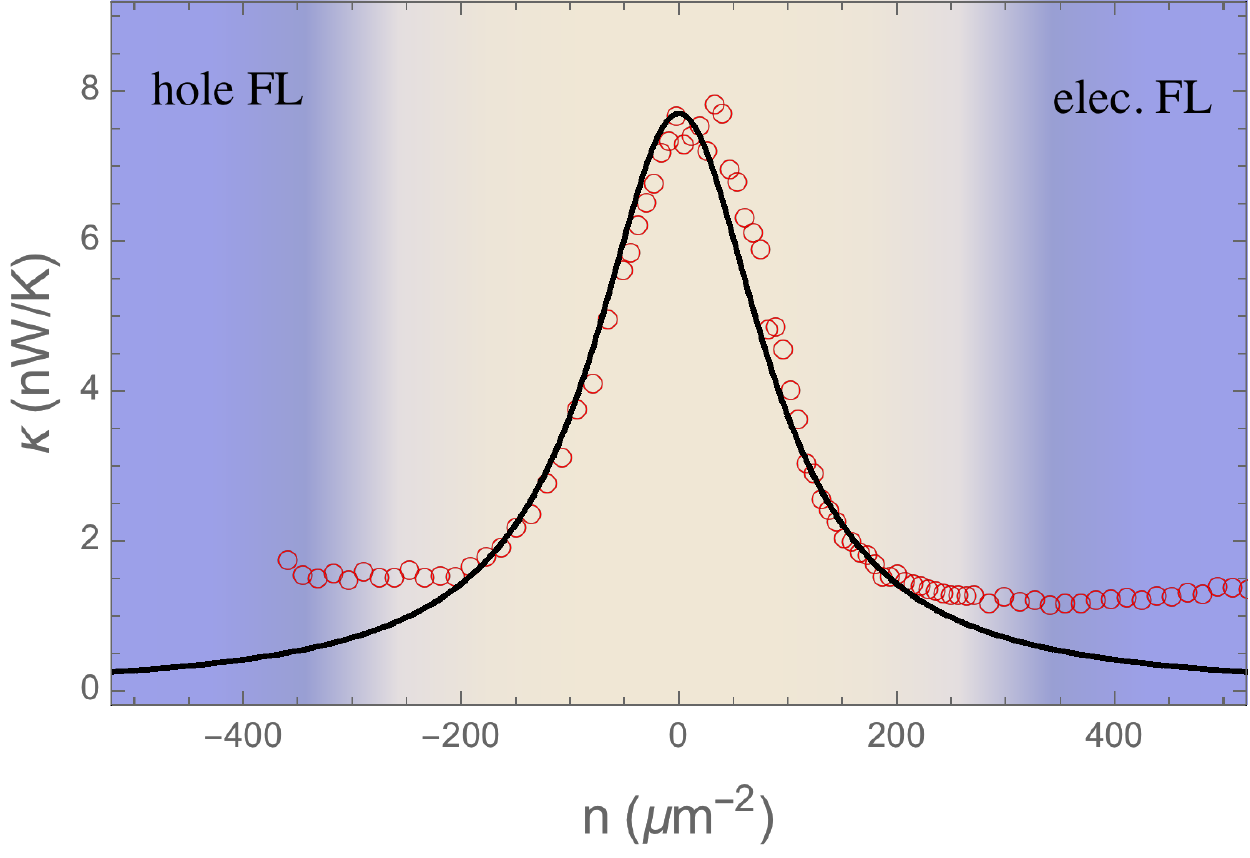}
    \caption{Comparison with experimental data. Density plot of electric conductivity $\sigma$ and of thermal conductivity $\kappa$. Red circles are for data used in~\cite{2016Sci...351.1058C} and black curves are for the holographic model with two currents. The regime marked in blue is for the Fermi Liquid (FL) that is far from the holographic theory. Figure taken from~\cite{Seo:2016vks}.}
    \label{fig:graphene}
\end{figure}

In contrast to the one current model, there are a few motivations including an extra current in the graphene~\cite{Seo:2016vks}. The first one is from the effect of imbalance between the electrons and holes due to the kinematic constraints of the Dirac cone, which also suggests the two conserved charges can be proportional. Other candidates include spin charge separation, valley currents, phonons and so on. The linear axion model with two distinct conserved $U(1)$ currents was proposed in~\cite{Seo:2016vks} to describe the experimental data. The electric, thermo-electric and thermal conductivities can be computed analytically. Then, under the assumption that the two conserved charges are proportional to each other, the holographic results for the density dependence of the electric and heat conductivities have a significantly improved match to the experimental data than the models with only one current (see Fig.~\ref{fig:graphene}). The holographic model also suggested an additive structure in the transport coefficients: the additivity of dissipative part of the inverse heat conductivity.
\begin{equation}
D[1/\kappa]=\sum_i D[1/\kappa_i],\quad \bar{D}[1/\sigma]=\sum_i \bar{D}[1/\sigma_i]\,,
\end{equation}
where $\kappa_i$ and $\sigma_i$ are the heat conductivity and electric conductivity for the $i$-th current. $D[f]$ denotes the dissipative part of $f$ and $\bar{D}[f]=f-D[f]$.

\begin{figure}
\centering
\includegraphics[width=0.71\linewidth]{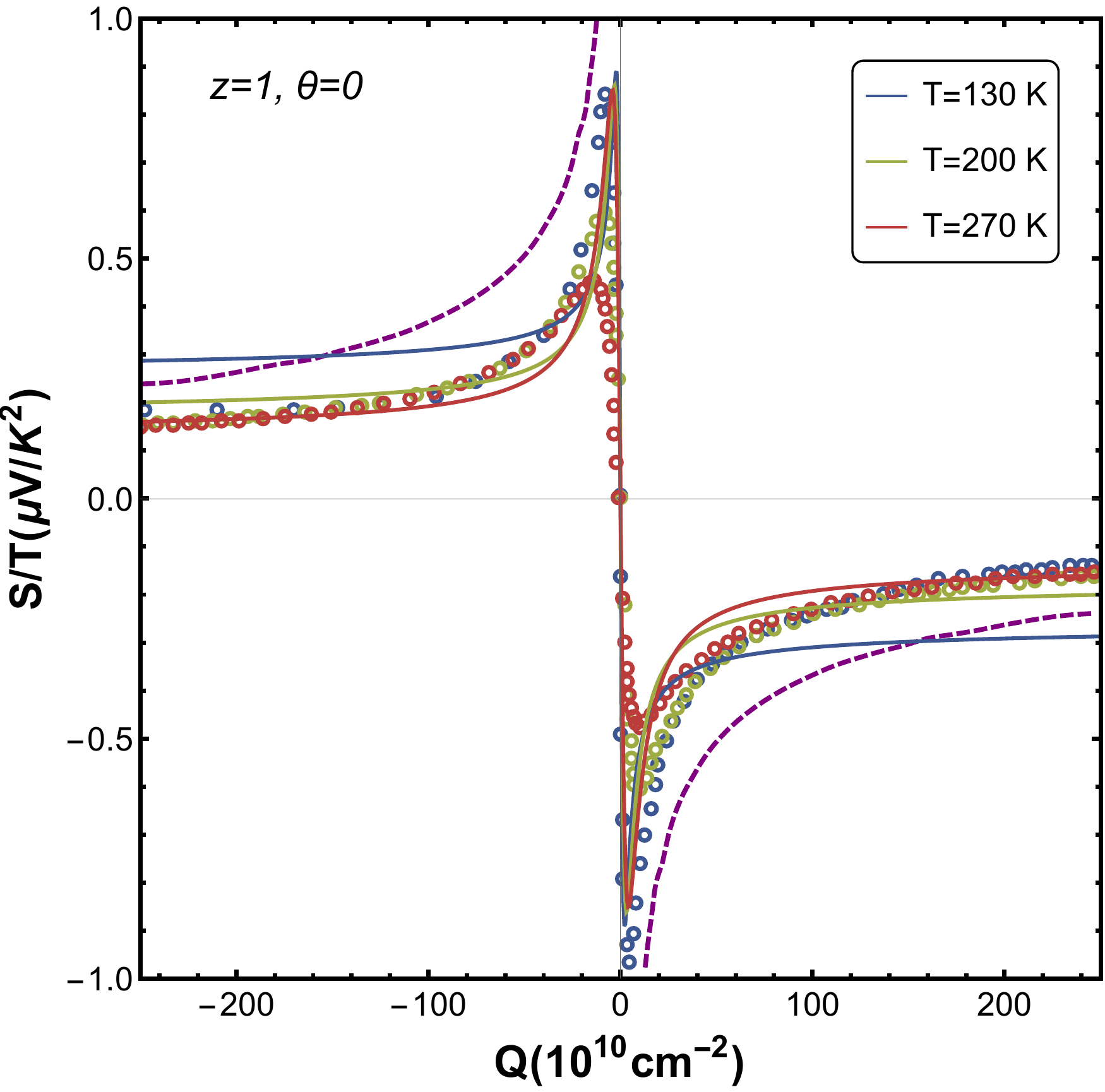}
    
\vspace{0.2cm}
     
\includegraphics[width=0.70\linewidth]{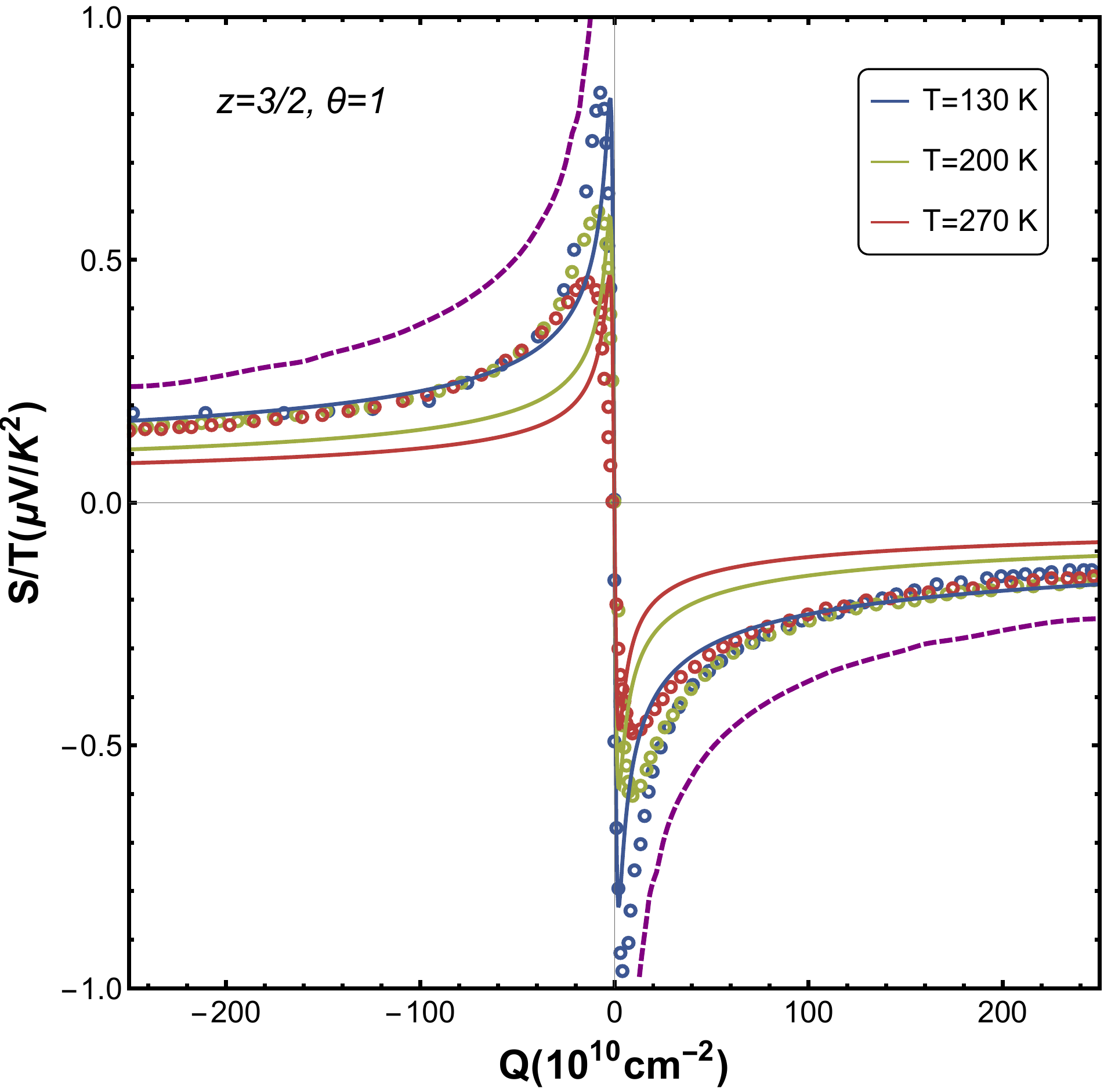}
\caption{Seebeck coefficient as a function of charge density $Q$. Circles are for experimental data used in~\cite{2016PhRvL.116m6802G} and dashed line for hydrodynamics result. Seebeck coefficient at low temperature fits well with experiment for the holographic two currents model with $(z=3/2, \theta=1)$. Figure updated from~\cite{Song:2020ojc}.}
\label{fig:graphenesk}
\end{figure}

Quantum criticality has been argued to be crucial for interpreting a wide variety of experiments. A large class of quantum critical points can be characterized by two scaling exponents, known as the dynamical critical exponent $z$ and the hyperscaling violation exponent $\theta$. The case for the holographic model of~\cite{Seo:2016vks} corresponds to $(z=1,\theta=0)$. A different set of dynamical exponents was considered in~\cite{Song:2020ojc} where it was found that the graphene data can be fit much more naturally by considering $(z=3/2, \theta=1)$. Furthermore, the Seebeck coefficient can also be fit using $(z=3/2, \theta=1)$ (see Fig.~\ref{fig:graphenesk}). In contrast, the previous model with $(z=1, \theta=0)$ fails to describe features of the experimental data at large density. The fact that this model does not fit with the experimental data for large temperature was argued to be due to the absence of phonon effect that is important for large temperature~\cite{Song:2020ojc}.

\subsection{Topological effects}

The {\it bulk} Chern-Simons terms play an important role in holography since it contributes to various new effects and new physics. For example, the Chern-Simons term $A\wedge F\wedge F$ can yield a charge density wave instability~\cite{Nakamura:2009tf,Donos:2013gda}, a metal-insulator transition~\cite{Donos:2012js}, and the presence of a non-trivial chiral magnetic conductivity \cite{Yee:2009vw, Kim:2010pu}. 

On the contrary, what if we want to study the {\it boundary} Chern-Simons term, $A \wedge F$? 
The boundary $A \wedge F$ term is important because it may be interpreted as the spin-orbit coupling. The spin-orbit coupling in $2+1$ dimension is relevant to interesting phenomena in topological insulators and Weyl semi-metals~\cite{Qi_2011}.\footnote{The spin-orbit interaction involves fermions but after integrating out fermions we may effectively deal with the Chern-Simons term $A\wedge F$ \cite{PhysRevD.31.2137,PhysRevD.32.1020}.}  The boundary $A \wedge F$ term can be  holographically lifted to $F\wedge F$ in bulk~\cite{Seo:2015pug}. 
To have a non-trivial dynamical effect we may couple $F\wedge F$ with some scalar operators.

Related with the axion model, one possible coupling is 
\begin{equation} \label{eq:XFF}
q_\chi X  F \wedge  F  \,,
\end{equation}
where $X$ is the kinetic term of the axion fields and $q_\chi$ is introduced to quantify the strength of this interaction.  
For the model \eqref{conven} with $Y=1, V=X$ together with the interaction \eqref{eq:XFF}, thermodynamic properties of the system and electric and thermal conductivities have been computed~\cite{Seo:2015pug,Seo:2017oyh,Song:2019rnf}. 

From the structure of \eqref{eq:XFF}, which is schematically $\sim q_\chi  \alpha^2  \rho  B$, we see that there can be a magnetization ($\sim q_\chi  \alpha^2  \rho$) even without explicit magnetic field. In this sense, the axion charge $\alpha$ may be interpreted as a magnetic impurity. This magnetic impurity induces a Hall current without an external magnetic field, so it may explain the presence of an anomalous Hall effect, which is ten times larger  than the one observed in non-magnetic materials.

Regarding the AC electric conductivity, the interaction  \eqref{eq:XFF} induces a new quasi-particle pole~\cite{Song:2019rnf}.  
This excitation is attributed to a new coupling between two gauge field fluctuations $a_x$ and $a_y$ by \eqref{eq:XFF}: $q_\chi \alpha^2 \partial_u a_y \partial_t a_x$. This quasi-particle pole may be considered as a kind of cyclotron pole \eqref{cycpole} induced this time by a magnetic impurity, not by an external magnetic field. Note that, for a cyclotron pole, 
$a_x$ and $a_y$ are connected indirectly by a metric fluctuation. 

As another interaction, we may consider~\cite{Kim:2019lxb,Kim:2020ozm}
\begin{equation} \label{eq:XFF1}
\varphi  F \wedge  F  \,,
\end{equation}
where $\varphi$ is a {\it real} scalar in the holographic superconductor model~\cite{Hartnoll:2008kx} with axion~\cite{Kim:2016hzi}. The basic idea is to use spontaneous $\mathbb{Z}_2$ symmetry breaking to induce spontaneous magnetization.  Because $\varphi$ is a real scalar (the symmetry is not $U(1)$ but $\mathbb{Z}_2$) the system is not superconducting and the conductivity is finite below the symmetry breaking transition~\cite{Hartnoll:2008kx}. The essential structure of \eqref{eq:XFF1} is $\sim \langle \mathcal{O} \rangle \rho B$, where $\langle \mathcal{O} \rangle$ is the (spontaneous) condensate of the operator dual to $\varphi$. Thus, even without $B$ the magnetization can be finite due to spontaneous $\mathbb{Z}_2$ symmetry breaking~\cite{Kim:2019lxb}. If $\alpha$ increases, the magnetization increases so $\alpha$ can be interpreted as magnetic impurity. This model exhibits magnetic hysteresis as well. Interestingly this model finds its applications in topological insulators,~\cite{bao123,bao1234}, where it is observed that the magnetoconductance starts showing hysteresis behavior similar to magnetization as magnetic doping increases. This is called hysteric magnetoconductance phase. The holographic model with the new coupling \eqref{eq:XFF1} qualitatively reproduces this phase as $\alpha$, which is identified with magnetic impurity strength, increases.

\subsection{Non-equilibrium physics and thermalization}
This review so far has mainly focused on the nature of holographic quantum matter at equilibrium and on the consequence of perturbing states very near equilibrium, for which the linear response theory applies. The non-equilibrium physics of strongly coupled quantum matter is an important but largely unexplored frontier. Non-equilibrium phenomena are general problems in ultra-relativistic heavy-ion collisions, cold atom systems, condensed matter physics and so on, for which there are almost no techniques from standard field theory to apply. By mapping the physics of quantum matter into a dual gravitational theory, one is able to study difficult non-equilibrium process by solving tractable non-stationary general relativity problems~\cite{Chesler:2013lia}.

A simple way to drive a system far from equilibrium is through a quantum quench by turning on a time dependent source $s(t)$. Two typical examples are as follows. Typically, $s(t)$ can either interpolate between  an initial and a final state $s(-\infty)=0$ and $s(\infty)=1$ or alternatively oscillate between them. The far from equilibrium dynamics will tend to drive quantum matter to a finite temperature states. This thermal effect is natural from bulk perspective as the energy will be absorbed by the black hole.

\begin{figure}
\centering
\includegraphics[width=0.95\linewidth]{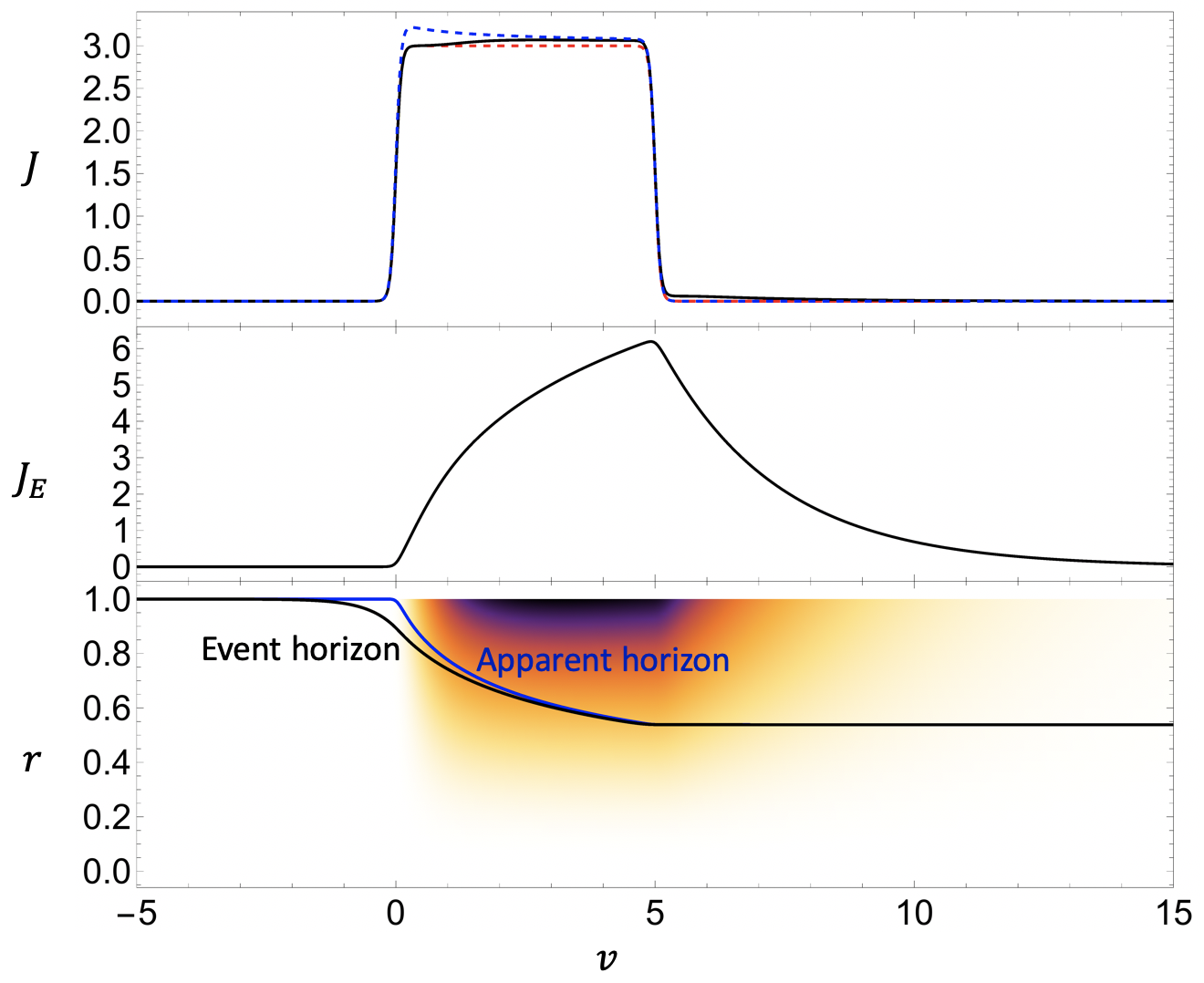}
\caption{Evolution for the case with top hat electric field. Top panel: The solid curve denotes the electric field $J$. The red dashed curve shows the quenched electric field $E(t)$ and the blue dashed curve gives the approximation to the electric conductivity. Middle panel: The energy current. Bottom panel: The evolution of event horizon and apparent horizon. The bulk distribution of the axion field with the linear $x$-dependence subtracted is illustrated in color. The charge density has been set to one. Figure updated from~\cite{Withers:2016lft}.}
\label{fig:noneqlm_bulk}
\end{figure}

Transport properties of strongly coupled systems from holographic duality have been the subject of much recent interest. The nonlinear response of a finite charge density system resulting from an electric field quench in a simple Einstein-Maxwell-axion model was investigated in~\cite{Withers:2016lft,Bagrov:2017tqn}. For the finite-time pulsed quench, the electric field is smoothly turned on, held for some time and then turned off again, see the red dashed curve in top panel of Fig.~\ref{fig:noneqlm_bulk}. As can be seen from Fig.~\ref{fig:noneqlm_bulk}, the system returns to equilibrium after turning the electric field off. It was found in~\cite{Withers:2016lft} that the system returns to equilibrium with the approach governed by the longest lived QNMs of the final black brane whose spectrum depends on the strength of momentum relaxation $k$ via $\phi^I=\alpha\delta^I_i x^i$. By dialing $\alpha$, one can see a qualitative change in the relaxation of currents, due to the pole collision and the presence of off-axis mode. In the small $\alpha$ coherent regime, the relevant QNMs are purely decaying, while in a large $\alpha$ incoherent regime, the heat current acquires enhanced contributions from a branch of QNMs that oscillate and decay.

\begin{figure}
    \centering
    \includegraphics[width=1.0\linewidth]{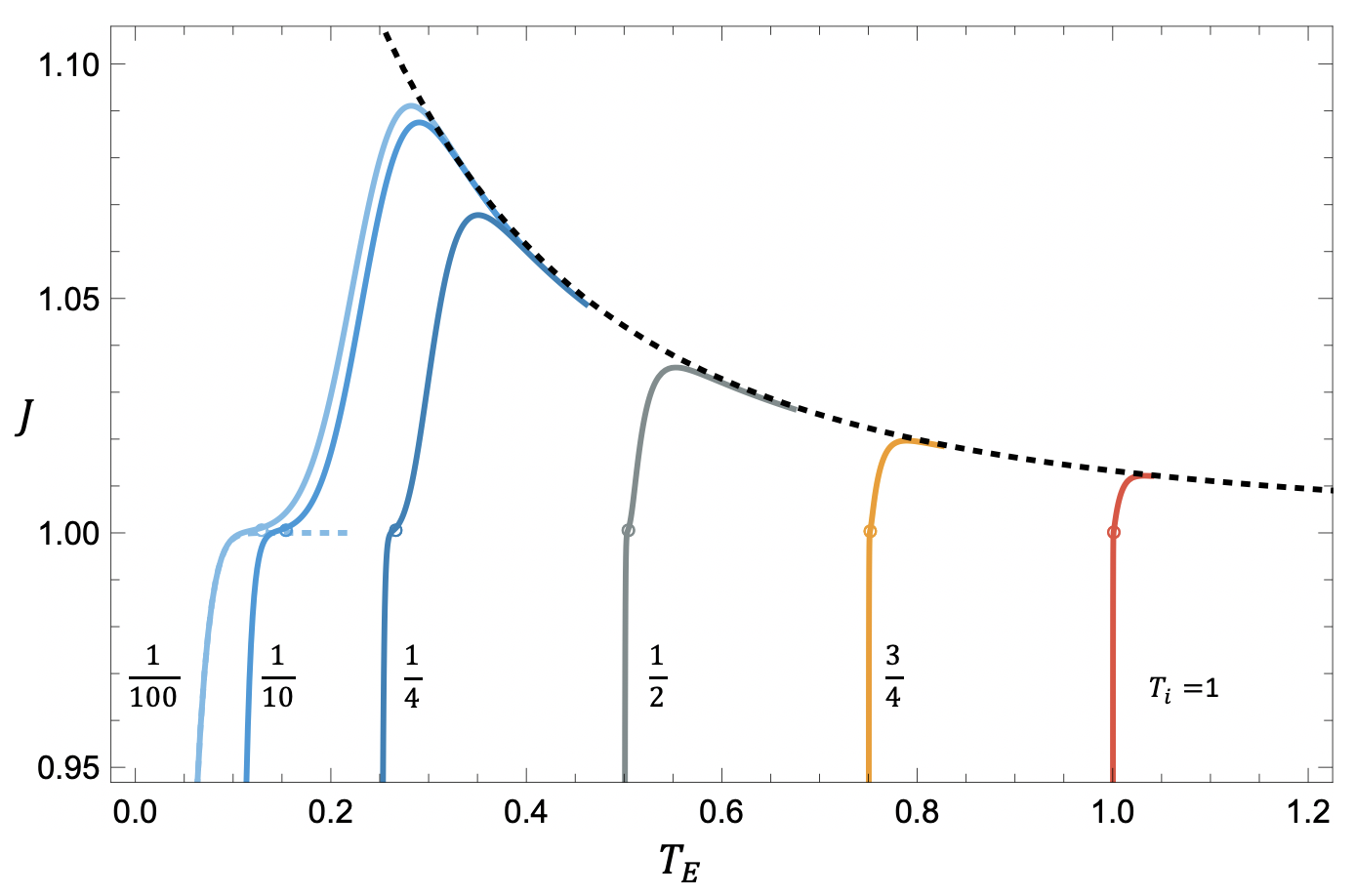}
    \caption{Nonlinear electric current response as a function of effective temperature $T_E$. Curves from left to right represent runs of different initial temperatures $T_i=i0^{-2}, 10^{-1}, 1/4, 1/2, 3/4, 1$, respectively. The black dashed line shows the DC linear response conductivity after the equilibrium temperature is promoted to $T_E$. The charge density has been set to one. Figure updated from~\cite{Withers:2016lft}.}
    \label{fig:noneqlm_J}
\end{figure}

The nonlinear thermoelectric response induced by holding the electric field constant was also discussed in~\cite{Withers:2016lft}. For small electric field, there is a steady state described by DC linear response, due to a balance between the driving electric field and the momentum sink. When the electric field is large, Joule heating will introduce significant time dependence on the bulk geometry. Nevertheless, in a regime where the rate of temperature increase is small, the nonlinear electric conductivity can be well approximated by a DC linear response calculation, once an appropriate effective temperature $T_E$ is taken into account (see Fig.~\ref{fig:noneqlm_J}). In contrast, the linear response result for the thermoelectric DC conductivity $\bar{\alpha}$ does not give good agreement over the same timescales, which means $\bar{\alpha}$ should have an explicit dependence on the electric field.

Another interesting case is driving a system with a very short and intense coherent electromagnetic pulse, after which the time evolution of the system is monitored by a linear response probe. The study of this pump-probe experiment from holography was presented in~\cite{Bagrov:2017tqn}. A holographic state at finite density with mildly broken translation invariance through linear axions was excited by oscillating electric field pulse. The thermalization process was numerically investigated by varying the pulse frequency $\omega_P$. It was found that in all circumstances the thermalization continues to be instantaneous for pump pulses devoid of a zero frequency component. For pump electric field with a significant DC component, the full time evolution is governed by a single thermalization time which is precisely half of the equilibrium momentum relaxation time at the final temperature. This feature can be understood from the fact that the metric component corresponding to momentum appears squared in the computation of the time-dependent conductivity. It was conjectured in~\cite{Bagrov:2017tqn} that large class of systems with a holographic dual will exhibit the phenomenon of instantaneous thermalization. These holographic findings can be tested by experiments which are in principle feasible in condensed matter laboratories.

The holographic axion model has been also used to study out-of-equilibrium dynamics in presence of anomaly ~\cite{Fernandez-Pendas:2019rkh,Morales-Tejera:2020xuv}. In this last context, it has been proven before~\cite{Copetti:2017ywz} that the chiral magnetic and chiral vortical conductivities are completely independent of the momentum relaxation rate introduced via the axions.

\section{Outlook}\label{sec:outlook}
\subsection{Open questions}
Before concluding, we find useful to collect the main open questions related to the holographic axion model discussed in this review. We will take a very direct attitude and list them one by one.
\begin{enumerate}
    \item \textbf{The viscosity of holographic solids}. Why does the viscosity decrease by increasing the amount of SSB of translations? Going from a fluid to a solid the viscosity rapidly increases in Nature, but it is not the case here. Is this connected to the fact that in most of the holographic systems (see~\cite{Cremonini:2012ny} for a counterexample) the viscosity grows with temperature? This is again not the case in liquids, but only in gases.
    \item \textbf{Holographic fluids}. We always sell the idea that AdS Schwarzchild is the dual of a relativistic fluids at strong coupling. That cannot be since fluids clearly have a finite electric conductivity. This suggests that a realistic holographic fluid must be encoded in the axions model with $V(Z)$~\cite{Alberte:2015isw,Baggioli:2019abx}, in agreement also with the effective theories expectations~\cite{Dubovsky:2011sj,deBoer:2015ija}. What is really the difference between these two setups? Is this really dual to a fluid or a gas?

    \item \textbf{The cost of homogeneity}. How do these homogeneous models compare with more realistic inhomogeneous setups? It seems that at low energy, \emph{i.e.} low frequency and low momentum, there is absolutely no difference. We did not learn anything about DC transport coefficients from these very complicated inhomogeneous models. We have recently proved that even the low energy spectrum is the same~\cite{Andrade:2020hpu}. Where can we find differences? It is clear that one has to go to more microscopic features, related to higher momenta. One case is the property of commensurability~\cite{Andrade:2015iyf}. Anything else? Does the numerical effort really pay back?
    \item \textbf{Phase relaxation vs. pseudo-spontaneous breaking}. We now know that the  pseudo-spontaneous breaking of translations produces an effective phase relaxation term which is fundamentally different from the one usually considered and coming from the presence of elastic defects such as dislocations. Why nevertheless do we not see any Drude peak in the frequency dependent viscosity?  Can we understand why the $\Omega$ pole is somehow hidden by the presence of explicit breaking?
    \item \textbf{Holographic dislocations}. How can we introduce in a simple way elastic defects in the holographic picture? And is their phenomenology what we do expect?
    \item \textbf{Holographic Glasses}. We have discussed in details fluids and solids. What about glasses? Have the axions model anything in common with glasses? Is there a Boson peak in the spirit of~\cite{Baggioli:2018qwu}? Similar ideas appeared already in~\cite{Anninos:2013mfa,Facoetti:2019rab}.
    \item \textbf{Phonons Hydrodynamics}. There has been recent effort in linking the idea of electron hydrodynamics~\cite{Lucas:2017idv} with holography~\cite{Erdmenger:2018svl}. What about phonons hydrodynamics~\cite{Cepellotti2015}? How to implement such limit?
    \item \textbf{Physical nature of holographic axions}. Despite a lot of work on the axions model, the physical nature of the dual system is still not well understood. To what extent can this simple homogeneous setup be trusted? And which phase of matter are we actually describing? These questions remain unanswered.
     \item \textbf{Negative energy and possible instability}. {A common feature of many simple linear axion models is that the energy density $\mathcal{E}$ becomes negative at large $\alpha$. It was argued that there might be some instability signalled by the appearance of negative energy density, but such instability was never found and all the linearized excitations are well behaved. How to understand this negative energy issue? Does it have any important consequence or limitation for some of the simplest holographic quantum matters?}
     \item \textbf{The residual entropy density}. {Another common feature of linear axion models is the residual entropy density at zero temperature, corresponding to the $AdS_2$ IR extreme geometry. What's the nature of this residual entropy? Is there any possible relation to localization or glasses? Recent developments in the SYK model taught us that such feature does not come from a degeneracy of the ground state but from the piling up of a very close excited states \cite{PhysRevX.5.041025}. Can we understand it from the gravitational picture? Is it again related to the highly unstable character of the $AdS_2$ geometry?}
      \item \textbf{Make it useful.} It is a truth acknowledged by several researchers in (and specially outside) the community, that AdS-CMT has not produced yet a strong smoking-gun result able to justify its usefulness for realistic condensed matter systems. Can we push the framework further, connect it to experiments, predict new phenomena and explain unresolved ones? This seems the only way for the tool to survive without becoming a niche product for a small group of enthusiasts.

\end{enumerate}

\subsection{Conclusions}\label{theend}
In conclusion, we hope to have convinced the Reader that, despite the apparent simplicity, the holographic axion model displays an incredibly rich set of features and applications which go far beyond the idea of dissipating momentum and make the DC electric conductivity finite.

In any case, whether you want to use them just to avoid annoying infinities or if you want to dig deeper in the physics of solids and fluids at strong coupling, this review is made for you. At the cost of resulting rather lengthy, we have made the effort of being as comprehensive as possible and discuss all the different faces of the model. We hope that any of you, in one way or another, will benefit from this read and will learn something new you were not aware of before. We also wish to have inspired new thoughts on the topic and the incentive for future developments in the field.

\section*{Acknowledgments}
We are grateful to the uncountable number of colleagues which participated with us in the process of understanding all the secrets of the holographic axion model. We thank Teng Ji, Giorgio Frangi, Hyun-Sik Jeong, Xi-Jing Wang and Yongjun Ahn for useful comments and helping proof-reading an early version of this manuscript.

M.B. acknowledges the support of the Shanghai Municipal Science and Technology Major Project (Grant No.2019SHZDZX01) and of the Spanish MINECO “Centro de Excelencia Severo Ochoa” Programme under grant SEV-2012-0249. K.K. was supported by Basic Science Research Program through the National Research Foundation of Korea (NRF) funded by the Ministry of Science, ICT \& Future Planning (NRF-2017R1A2B4004810) and the GIST Research Institute(GRI) grant funded by the GIST in 2020. L.L. is supported in part by the National Natural Science Foundation of China (NSFC) Grants No.12075298, No.11991052 and No.12047503. W.J.L. is supported in part by the NSFC under grant No.11905024 and DUT under grant No.DUT19LK20.
\\
\\
\appendix 
\section{Notations and conventions}\label{sec:notation}
In order to avoid confusion, in this appendix we describe in detail all the symbols and notations used in this review.

Greek letters $\mu,\nu,... $ run over spacetime indices, while Latin letters $i, j,...$ denote spatial ones. The axion flavor indices $I,J,...$ run over the number of broken translations. In this review, we also omit the summation symbol over the axion flavor indices, and use the Einstein convention for them too.

We always utilize a mostly plus metric $(-1,1,1,1)$ and we define the Fourier transform of a field $\Psi$ using the plain-wave $e^{-i \omega t+i k x}$. Finally, we indicate in Table I all the symbols used.
\begin{table}
\centering
\begin{tabular}{|a|b|}
\hline
\rowcolor{lightkhaki}
\textbf{Symbol} & \textbf{Meaning} \\
\hline
\hline
 $\eta$ & shear viscosity \\
 \hline
  $G$ & shear modulus \\
\hline
 $\zeta$ & bulk viscosity \\
 \hline
  $K$ & bulk modulus \\
\hline
$E$ & electric field \\
\hline
$B$ & magnetic field \\
\hline
$\tau$ & relaxation time \\
\hline
 $\Gamma$ & momentum dissipation rate \\
\hline
$\chi_{AB}$ & susceptibility\\
\hline
 $\chi_{\pi\pi}\,\text{or}\,\chi_{pp}$ & momentum susceptibility \\
\hline
$\sigma_{ij}\,\text{or}\,T_{ij}$ & stress \\
\hline
 $\varepsilon_{ij}$ & strain \\
\hline
 $\varrho$ & mass density \\
\hline
$v_T\,\text{or}\,v_\perp$ & shear sound speed \\
\hline
 $v_L\,\text{or}\,v_\parallel$ & longitudinal sound speed \\
\hline
 $u_i$ & displacements \\
\hline
 $\omega_0$ & pinning frequency \\
\hline
 $\Omega$ & phase relaxation rate \\
\hline
 $\mathcal{E}$ & energy density \\
\hline
 $\Gamma_L\,\text{or}\,\Gamma_\parallel$ & longitudinal sound attenuation \\
\hline
 $\Gamma_T\,\text{or}\,\Gamma_\perp$ & transverse sound attenuation \\
\hline
 $D_\phi$ & crystal diffusion constant \\
\hline
 $\xi$ & Goldstone diffusion parameter \\
\hline
 $\mathfrak{p}$ & pressure \\
\hline
 $\mathcal{P}$ & crystal pressure \\
\hline
 $m_g$ & graviton mass \\
\hline
 $v_B$ & butterfly velocity \\
\hline
$u$ or $r$ & radial coordinate \\
\hline
$u_h$ & horizon radius \\
\hline
$L$ & AdS radius \\
\hline
 $\Lambda$ & cosmological constant \\
\hline
 $T$ & temperature \\
\hline
 $s$ & entropy density \\
\hline
 $\rho$ & charge density \\
\hline
 $\mu$ & chemical potential \\
\hline
 $k_g$ & k-gap \\
\hline
 $z$ & Lifshitz exponent \\
\hline
 $\theta$ & hyperscaling parameter \\
\hline
 $\langle EXB \rangle $ & explicit breaking scale \\
\hline
 $\langle SSB \rangle $ & spontaneous breaking scale \\
 \hline
 $D_\pi $ & momentum diffusion constant \\
\hline
 $\omega $ & frequency \\
\hline
 $k $ & momentum in Fourier space or wave-number \\
\hline
 $H $ & Hamiltonian \\
\hline
 $\mathcal{G}_{AB} $ & Green's function \\
\hline
 $\omega_p $ & plasma frequency \\
\hline
\end{tabular}
\label{tab1}
\caption{Notations and symbols.}
\end{table}

\bibliographystyle{unsrt}
\bibliography{rev}

\end{document}